\documentclass[a4paper,11pt,fleqn]{article}

\usepackage{graphicx}
\usepackage{dcolumn}
\usepackage{bm}
\usepackage{amsmath}
\usepackage{amssymb}
\usepackage{amsfonts}
\usepackage{subfigure}
\usepackage{esvect}
\usepackage{hyperref}
\usepackage[dvipsnames]{xcolor}
\usepackage{fix-cm} 
\usepackage{cite}
\usepackage{soul}
\usepackage{mdframed}
\usepackage{lipsum}
\usepackage{multicol}
\usepackage{float}

\usepackage{empheq} 

\newcommand{\CL}{{\tt ${\mathcal C}$osmo${\mathcal L}$attice}~}
\newcommand{\CLns}{{\tt ${\mathcal C}$osmo${\mathcal L}$attice}}
\newcommand{\dx}{\ensuremath{\delta x}}
\newcommand{\dt}{\delta t}
\newcommand{\deta}{\delta\eta}
\newcommand{\bn}{{\bf n}}
\newcommand{\dd}{\text{d}}

\newcommand{\be}{\begin{equation}}
\newcommand{\ee}{\end{equation}}
\newcommand{\bea}{\begin{eqnarray}}
\newcommand{\eea}{\end{eqnarray}}

\newcommand{\mn}{{\mu\nu}}
\newcommand{\ab}{{\alpha\beta}}

\newcommand{\piSingl}{\tilde\pi_\varphi}
\newcommand{\piSinglpar}{\left(\piSingl\right)}
\newcommand{\piA}{\tilde\pi_A}
\newcommand{\piApar}{\left(\piA\right)}
\newcommand{\piDoubl}{\widetilde\pi_\Phi}
\newcommand{\piDoublpar}{\left(\piDoubl\right)}
\newcommand{\piB}{\tilde\pi_B}
\newcommand{\piBpar}{\left(\piB\right)}

\newcommand{\kersutwo}{\mathcal{K}_{B_i}}
\newcommand{\kersutwoComp}{\mathcal{K}_{B_i^a}}

\newcommand{\tableauII}{
\begin{array}{r|cc|l}
\, & {1\over4} & {1\over4}-{\sqrt{3}\over6} & \,\\
\, & \, & \, & \vspace*{-0.35cm}\\
\, & {1\over4}+{\sqrt{3}\over6} & {1\over4} & \,\\
\, & \, & \, & \vspace*{-0.35cm}\\
\hline
\vspace*{-0.35cm}\\
\, & {1\over2} & {1\over2} & \,
\end{array}
}

\newcommand{\tableauIII}{
\begin{array}{r|ccc|l}
\, & {5\over36} & {2\over9}-{\sqrt{15}\over15} & {5\over36} -{\sqrt{15}\over30} & \,\\
\, & \, & \, & \, & \vspace*{-0.35cm}\\
\, & {5\over36} + {\sqrt{15}\over24} & {2\over9} & {5\over36} - {\sqrt{15}\over24} & \,\\
\, & \, & \, & \, & \vspace*{-0.35cm}\\
\, & {5\over36} + {\sqrt{15}\over30} & {2\over9}+{\sqrt{15}\over15} & {5\over36} & \,\\
\, & \, & \, & \, & \vspace*{-0.35cm}\\
\hline
\vspace*{-0.35cm}\\
\, & {5\over18} & {4\over9} & {5\over18} & \,
\end{array}
}

\newcommand{\tableauIV}{
\begin{array}{r|cccc|l}
\, & \omega_1^- & \omega_1^+ - \omega_3^+ + \omega_4^-  & \omega_1^+ - \omega_3^+ - \omega_4^- & \omega_1^- - \omega_5^+ &\,\\
\, & \, & \, & \, & \, & \vspace*{-0.35cm}\\
\, & \omega_1^- - \omega_3^- + \omega_4^+ & \omega_1^+ & \omega_1^+ - \omega_5^- & \omega_1^- - \omega_3^- - \omega_4^+ &\,\\
\, & \, & \, & \, & \, & \vspace*{-0.35cm}\\
\, & \omega_1^- + \omega_3^- + \omega_4^+ & \omega_1^+ + \omega_5^- & \omega_1^+ & \omega_1^- + \omega_3^- - \omega_4^+ &\,\\
\, & \, & \, & \, & \, & \vspace*{-0.35cm}\\
\, & \omega_1^- + \omega_5^+  & \omega_1^+ + \omega_3^+ + \omega_4^- & \omega_1^+ + \omega_3^+ - \omega_4^- & \omega_1^-  &\,\\
\, & \, & \, & \, & \, & \vspace*{-0.35cm}\\
\hline
\vspace*{-0.35cm}\\
\, & 2\omega_1^- & 2\omega_1^+ & 2\omega_1^+ & 2\omega_1^- & \,
\end{array}
}

\newcommand{\tableauV}{
\begin{array}{r|ccccc|l}
\, & \omega_1^- & \omega_1^+ - \omega_3^+ + \omega_4^- & {32\over225} - \omega_5^- & \omega_1^+ - \omega_3^+ - \omega_4^- & \omega_1^- - \omega_6^+ &\,\\
\, & \, & \, & \, & \, & \, & \vspace*{-0.35cm}\\
\, & \omega_1^- - \omega_3^- + \omega_4^+ & \omega_1^+ & {32\over225} - \omega_5^+ & \omega_1^+ - \omega_6^- & \omega_1^- - \omega_3^- - \omega_4^+ &\,\\
\, & \, & \, & \, & \, & \, &\vspace*{-0.35cm}\\
\, & \omega_1^- + \omega_7^+ & \omega_1^+ + \omega_7^- & {32\over225} & \omega_1^+ - \omega_7^- & \omega_1^- - \omega_7^+ &\,\\
\, & \, & \, & \, & \, & \, &\vspace*{-0.35cm}\\
\, & \omega_1^- + \omega_3^+ + \omega_4^+  & \omega_1^+ + \omega_6^-  & {32\over225} + \omega_5^- & \omega_1^+ & \omega_1^- + \omega_3^+ - \omega_4^+  &\,\\
\, & \, & \, & \, & \, & \, &\vspace*{-0.35cm}\\
\, & \omega_1^- + \omega_6^+  & \omega_1^+ + \omega_3^+ + \omega_4^- & {32\over225} - \omega_5^+ & \omega_1^+ + \omega_3^+ - \omega_4^- & \omega_1^-  &\,\\
\, & \, & \, & \, & \, & \, &\vspace*{-0.35cm}\\
\hline
\vspace*{-0.35cm}\\
\, & 2\omega_1^- & 2\omega_1^+ & {64\over225} & 2\omega_1^+ & 2\omega_1^- & \,
\end{array}
}

\newcommand{\coeffTableauIV}{
\left[
    \begin{array}{c}
         \omega_1^\pm = {1\over8}\pm{\sqrt{30}\over144}\,,~~~~~~\omega_2^\pm = {1\over 2}\sqrt{15\,\pm\, 2\sqrt{30}\over35}\,,~~~~~~\omega_3^\pm = \omega_2^\pm\left({1\over6}\pm{\sqrt{30}\over24}\right)\,, \\
         \omega_4^\pm = \omega_2^\pm\left({1\over21} \pm {5\sqrt{30}\over168}\right)\,,~~~~~~\omega_5^\pm = \omega_2^\pm - 2\omega_3^\pm\,,
    \end{array}
    \right]
}

\newcommand{\coeffTableauV}{
\left[
    \begin{array}{c}
         \omega_1^\pm = {322 \,\pm\, 13\sqrt{70}\over3600}\,,~~~~\omega_2^\pm = {1\over 2}\sqrt{32\,\pm\, 2\sqrt{70}\over63}\,,~~~~\omega_3^\pm = \omega_2^\pm\left(452\,\pm\, 59\sqrt{70}\over3240\right)\,,~~~~ \omega_4^\pm = \omega_2^\pm\left(64\,\pm\, 11\sqrt{70}\over1080\right) \\
         \omega_5^\pm = 8\omega_2^\pm\left(23\,\mp\, \sqrt{70}\over405\right)\,,~~~~\omega_6^\pm = \omega_2^\pm - 2\omega_3^\pm - \omega_5^\pm\,,~~~~\omega_7^\pm = \omega_2^\pm\left(308\,\mp\, 23\sqrt{70}\over960\right)\,,
    \end{array}
    \right]
}

\newcommand{\addressIFIC}{\it Instituto de F\'isica Corpuscular (IFIC), Consejo Superior de Investigaciones \\Cient\'ificas (CSIC) and Universitat de Valencia (UV), Valencia, Spain.}
\newcommand{\addressEPFL}{\it Institute of Physics, Laboratory of Particle Physics and Cosmology (LPPC), \'Ecole \\ Polytechnique F\'ed\'erale de Lausanne (EPFL), CH-1015 Lausanne, Switzerland.}
\newcommand{\addressUNIBAS}{\it Department of Physics, University of Basel, \\ Klingelbergstr. 82, CH-4056 Basel, Switzerland.}

\newcommand{\addressSUNY}{Center for Nuclear Theory, Department of Physics and Astronomy, Stony Brook University \\ Stony Brook, New York 11794, USA.}

\begin{document}

\begin{figure}
\hspace{0.9cm}
\includegraphics[width = 0.9\textwidth]{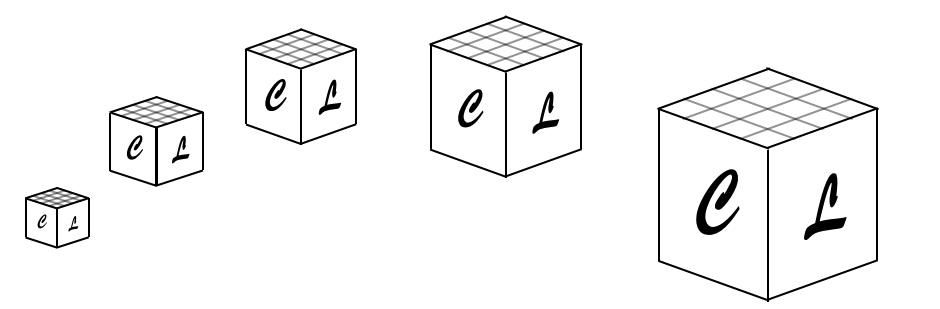}
\end{figure}

\begin{center}

\vspace*{0.25cm}

\vspace*{1.25cm}
{\fontsize{27.5}{0} \bf\textsf{The Art of Simulating the Early Universe}}
\\[0.15cm]
{\huge \it A dissertation on lattice techniques for the simulation of scalar and gauge field dynamics in an expanding Universe}
\vspace*{0.5cm}\\
{\bf \Large Part I: Integration Techniques and Canonical Cases}
\\[2.15cm]

{\Large  Daniel~G.~Figueroa\,\footnote[1]{daniel.figueroa@ific.uv.es}}\\
\addressIFIC
\\[0.5cm]
{\Large {\rm Adrien Florio}\,\footnote[2]{adrien.florio@stonybrook.edu}}\\
\addressSUNY
\\[0.5cm]
{\Large {\rm Francisco Torrenti}\,\,\footnote[3]{f.torrenti@unibas.ch}}\\ \addressUNIBAS
\\[0.5cm]
{\Large {\rm Wessel Valkenburg}\,\footnote[4]{wessel.valkenburg@epfl.ch}}\\
\addressEPFL

\vspace{.3cm}
\end{center}

\title{\bf The art of simulating the early Universe -- Part I}

\newpage

\author{Daniel G. Figueroa\,$^1$, Adrien Florio\,$^2$, Francisco Torrenti\,$^3$ and Wessel Valkenburg\,$^4$\vspace*{0.35cm}\\
$^1${\it
Instituto de F\'isica Corpuscular (IFIC), CSIC-Universitat de Valencia, Spain.}\\
$^{2,4}${\it Institute of Physics, Laboratory of Particle Physics and Cosmology (LPPC),} \vspace*{-0.1cm}\\{\it \'Ecole Polytechnique F\'ed\'erale de Lausanne (EPFL), CH-1015 Lausanne, Switzerland.}\\
$^2${\it Department of Physics and Astronomy, Stony Brook University, New York 11794, USA}\\
$^3${\it Department of Physics, University of Basel, Klingelbergstr. 82, CH-4056 Basel, Switzerland.}
}

\date{}
\maketitle

\begin{abstract}
We present a comprehensive discussion on lattice techniques for the simulation of scalar and gauge field dynamics in an expanding universe. After reviewing the continuum formulation of scalar and gauge field interactions in Minkowski and FLRW backgrounds, we introduce the basic tools for the discretization of field theories, including {\it lattice gauge invariant} techniques. Following, we discuss and classify numerical algorithms, ranging  from methods of $\mathcal{O}(\delta t^2)$ accuracy like {\it staggered leapfrog} and \hspace*{-1mm}{\it Verlet integration}, to {\it Runge-Kutta} methods up to $\mathcal{O}(\delta t^4)$ accuracy, and the {\it Yoshida} and {\it Gauss-Legendre} higher-order integrators, accurate up to $\mathcal{O}(\delta t^{10})$. We adapt these methods for their use in classical lattice simulations of the non-linear dynamics of scalar and gauge fields in an expanding grid in $3+1$ dimensions, including the case of
`self-consistent' expansion sourced by the volume average of the fields' energy and pressure densities. We present lattice formulations of canonical cases of: $i)$ {\it Interacting scalar fields}, $ii)$ {\it Abelian $U(1)$ gauge theories}, and $iii)$ {\it Non-Abelian $SU(2)$ gauge theories}. In all three cases we provide symplectic integrators, with accuracy ranging from $\mathcal{O}(\delta t^2)$ up to $\mathcal{O}(\delta t^{10})$. For each algorithm we provide the form of relevant observables, such as energy density components, field spectra and the Hubble constraint. Remarkably, all our algorithms for gauge theories respect the Gauss constraint to machine precision, including when `self-consistent' expansion is considered. As a numerical example we analyze the post-inflationary dynamics of an oscillating inflaton charged under $SU(2)\times U(1)$. The present manuscript is meant as part of the theoretical basis for {\tt ${\mathcal C}$osmo${\mathcal L}$attice}, a modern {\tt C++} MPI-based package for simulating the non-linear dynamics of scalar-gauge field theories in an expanding universe, publicly available at \href{http://www.cosmolattice.net}{\color{blue} http://www.cosmolattice.net}.
\end{abstract}

\tableofcontents

\section*{Conventions and notation}
\addcontentsline{toc}{section}{Conventions and notation}

Unless otherwise specified, throughout the document we use the following conventions. We use natural units $c=\hbar=1$ and choose metric signature $(-1,+1,+1,+1)$. We use interchangeably the Newton constant $G$, the full Planck mass $M_p \simeq 1.22\cdot 10^{19}$ GeV, and the reduced Planck mass $m_p \simeq 2.44\cdot 10^{18}$ GeV, all related through $M_p^2 = 8\pi m_p^2 = 1/G$. Concerning space-time coordinates, Latin indices $i, j, k, ... = 1,2,3$ are reserved for spatial dimensions, and Greek indices $\alpha, \beta, \mu, \nu,... = 0,1,2,3$ for space-time dimensions. We use the {\it Einstein convention} of summing over repeated indices only in the continuum, whereas in the lattice, unless otherwise stated, repeated indices do not represent summation. We consider a flat {\it Friedmann-Lema\^itre-Robertson-Walker} (FLRW) metric $ds^2 = -a^{2\alpha}(\eta)d\eta^2 + a^2(\eta) \, \delta_{ij} \, dx^i dx^j$ with $\alpha \in \mathcal{R}e$ a constant chosen conveniently in each scenario. For $\alpha = 0$, $\eta$ denotes the {\it coordinate time} $t$, whereas for $\alpha = 1$, $\eta$ denotes the {\it conformal time} $\tau = \int {dt'\over a(t')}$. For arbitrary $\alpha$, we will refer to the time variable as the {\it $\alpha$-time}. We reserve the notation $()^{\cdot}$ for derivatives with respect to cosmic time with $\alpha = 0$, and $()'$ for derivatives with respect to $\alpha$-time with arbitrary $\alpha$. Physical momenta are represented by ${\bf p}$, comoving momenta by ${\bf k}$, the $\alpha$-time Hubble rate is given by $\mathcal{H} = a'/a$, whereas the physical Hubble rate is denoted by $H = \mathcal{H}|_{\alpha = 0}$. Cosmological parameters are fixed to the CMB values given in \cite{Aghanim:2018eyx,Akrami:2018odb}. Our Fourier transform convention in the continuum is given by
\begin{eqnarray}\label{eq:FTcont}
\hspace*{2.5cm}f({\bf x}) = \frac{1}{(2 \pi)^3} \int d^3 {\bf k} \, f({\bf k}) \, e^{-i {\bf k} {\bf x}}\, ~~~ \Longleftrightarrow ~~~  f({\bf k}) = \int d^3 {\bf x} \, f ( {\bf x}) \, e^{+i {\bf k} {\bf x}}\,.
\end{eqnarray}


\section{Introduction}

\subsection{The {\it Numerical Early Universe}: a laboratory for non-linear high energy physics}

~~~~~ Significant evidence~\cite{Akrami:2018odb} supports the idea of {\it inflation}, a phase of accelerated expansion in the early universe, which provides both a solution to the shortcomings of the hot Big Bang framework~\cite{Guth:1980zm, Linde:1981mu, Albrecht:1982wi,Brout:1977ix,Starobinsky:1980te,Kazanas:1980tx,Sato:1980yn}, and an explanation for the origin of the primordial density perturbations~\cite{Mukhanov:1981xt, Guth:1982ec,Starobinsky:1982ee, Hawking:1982cz,Bardeen:1983qw}. Inflation is often assumed to be driven by a scalar field, the {\it inflaton}, with potential and initial conditions appropriately chosen to sustain a long enough period of accelerated expansion. To switch to the standard hot Big Bang cosmology, a {\it reheating} period must ensue after inflation, converting the energy available into light degrees of freedom ({\it dof}\,), which eventually thermalize and dominate the energy budget of the universe. This transition process is an integral part of the inflationary paradigm, although observationally is much less constrained than the period of inflation itself. For reviews on inflation and reheating, see~\cite{Lyth:1998xn,Riotto:2002yw,Bassett:2005xm,Linde:2007fr,Baumann:2009ds} and~\cite{Allahverdi:2010xz,Amin:2014eta,Lozanov:2019jxc,Allahverdi:2020bys}, respectively.

In many scenarios, following the end of inflation, the inflaton oscillates around the minimum of its potential, initially as a homogeneous condensate. Particle species coupled sufficiently strong to the inflaton are then created in energetic bursts. If the particles are Bosons, their production is driven by parametric resonance, what (depending on the coupling) can result in an exponential transfer of energy within few oscillations of the inflaton~\cite{Traschen:1990sw,Kofman:1994rk,Shtanov:1994ce,Kaiser:1995fb,Kofman:1997yn,Greene:1997fu,Kaiser:1997mp,Kaiser:1997hg}. If the particles are Fermions, there can also be a significant transfer of energy~\cite{Greene:1998nh,Greene:2000ew,Peloso:2000hy,Berges:2010zv}, albeit no resonance can be developed due to Pauli blocking. Particle production in this way, of either bosons or fermions, corresponds to a non-perturbative effect, which cannot be described with standard quantum field theory (QFT) perturbative techniques. Furthermore, particle species created by these effects are far away from thermal equilibrium, and in the case of bosonic species their production is exponential, so they eventually ‘backreact’ into the inflaton, breaking apart its initial homogeneous condition. The dynamics of the system becomes non-linear from that moment onward. All of these effects, from the initial particle production to the eventual development of non-linearities in the system, represent what is referred to as a {\it preheating} stage. In order to fully capture the non-perturbative, out-of-equilibrium, and eventual non-linearities of preheating, we need to study such phenomena in a lattice. This requires the use of classical field theory real-time simulations, an approach valid as long as the particle species involved in the problem have large occupation numbers $n_k \gg 1$, so that their quantum nature can be neglected~\cite{Khlebnikov:1996mc,Prokopec:1996rr}.

Parametric particle production can also be  developed in the early universe, in circumstances other than preheating. For instance, in the curvaton scenario~\cite{Enqvist:2001zp,Lyth:2001nq,Moroi:2001ct,Mazumdar:2010sa}, the initially homogeneous curvaton (a spectator field during inflation) may decay after inflation via parametric resonance, transferring abruptly all of its energy to other particle species~\cite{Enqvist:2008be,Enqvist:2012tc, Enqvist:2013qba, Enqvist:2013gwf}. If the Standard Model (SM) Higgs is weakly coupled to the inflationary sector, the Higgs can be excited either during inflation~\cite{Starobinsky:1994bd,Enqvist:2013kaa,DeSimone:2012qr}, or towards the end of it~\cite{Herranen:2015ima,Figueroa:2016dsc}, in the form of a condensate with a large amplitude. The Higgs then decays naturally into the rest of the SM species via parametric effects~\cite{Enqvist:2013kaa,Enqvist:2014tta,Figueroa:2014aya,Kusenko:2014lra,Figueroa:2015rqa,Enqvist:2015sua,Figueroa:2016dsc}, some time after inflation\footnote{Note that this differs from the Higgs-Inflation scenario~\cite{Bezrukov:2007ep,Bezrukov:2010jz}, where the Higgs also decays after inflation into SM fields via parametric effects~\cite{Bezrukov:2008ut,GarciaBellido:2008ab,Figueroa:2009jw,Figueroa:2014aya,Repond:2016sol,Ema:2016dny,Sfakianakis:2018lzf}, but as the Higgs plays the role of the inflaton, this scenario belongs to the category of preheating.}.  In supersymmetric (SUSY) extensions of the SM we encounter flat directions~\cite{Gherghetta:1995dv,Enqvist:2003gh}, configurations in field space where the renormalizable part of the scalar potential is exactly flat (as SUSY must be broken, the exact flatness is however typically uplifted by various effects~\cite{Dine:1995kz}). During inflation, due to quantum fluctuations, field configurations can be developed with a large expectation value along these directions~\cite{Affleck:1984fy,Enqvist:2003gh}. If such scalar condensates have a soft mass, its amplitude starts oscillating after inflation once the Hubble rate becomes smaller than its mass~\cite{Dine:1995kz,Gaillard:1995az}, possibly ensuing an explosive decay of the field condensate due to non-perturbative resonant effects~\cite{Olive:2006uw,Basboll:2007vt,Gumrukcuoglu:2008fk}.

In certain types of inflationary models where spontaneous symmetry breaking plays a central role, tachyonic effects can also lead to non-perturbative and out-of-equilibrium particle production, eventually driving the system into a non-linear regime. A paradigmatic example of this is Hybrid inflation~\cite{Linde:1993cn}, a family of models where the inflationary stage is sustained by the vacuum energy of a Higgs-like field. In these scenarios, the effective squared mass of the Higgs field is positive defined during inflation, but becomes negative when the inflaton eventually crosses around a critical point. The Higgs then sustains a tachyonic mass that leads into an exponential growth of the occupation number of its most infrared (IR) modes below its own tachyonic mass scale~\cite{Felder:2000hj,Felder:2001kt,Copeland:2002ku,GarciaBellido:2002aj}. This continues until the mass squared becomes positive again, due to the Higgs own self-interactions. In
Hilltop-inflation, inflation is sustained while the inflaton slowly rolls from close to a maximum of its potential (the `hilltop') towards its minimum, located at some non-vanishing scale. When the inflaton crosses a certain threshold amplitude, inflation ends, and the inflaton starts oscillating around its minimum. Its effective squared mass then alternates between positive and negative values, as the inflaton rolls back and forth between the minimum and the region of negative curvature of its potential around where inflation ended. Fluctuations of the inflaton then grow exponentially during successive oscillations~\cite{Antusch:2015nla,Antusch:2015vna,Antusch:2015ziz,Antusch:2019qrr}.

Preheating effects have also been studied in models with gravitationally non-minimal coupled fields~\cite{Bassett:1997az,Tsujikawa:1999jh,Tsujikawa:1999iv,Tsujikawa:1999me,Ema:2016dny,Crespo:2019src,Crespo:2019mmh}, and in particular, recently, in multi-field inflation scenarios~\cite{DeCross:2015uza,DeCross:2016fdz,DeCross:2016cbs,Nguyen:2019kbm,vandeVis:2020qcp}. In the latter, a single-field attractor behavior is developed during inflation, later persisting during preheating. Due to this, particle production after inflation becomes more efficient than in multi-field models with minimal couplings, where a {\it de-phasing} effect of the background fields' oscillations leads to a damping of the resonances~\cite{Battefeld:2008rd,Battefeld:2008bu,Battefeld:2009xw,Braden:2010wd,Amin:2014eta}.

Furthermore, as gauge fields are naturally present in the SM and in many of its extensions, their presence in inflationary scenarios has also been considered. Due to their bosonic nature, gauge fields can exhibit highly nonlinear dynamics during preheating. 
For instance, if the inflaton enjoys a shift-symmetry, a topological coupling to a gauge sector is allowed. In the case of $U(1)$ gauge fields, preheating effects have been studied in axion-inflation scenarios~\cite{Adshead:2015pva,Adshead:2016iae,Adshead:2018doq,Cuissa:2018oiw,Adshead:2019lbr,Adshead:2019igv}, showing that an interaction $\phi F\tilde F$ leads to an extremely efficient way to reheat the universe, as well as to very interesting (potentially observable) phenomenology. In~\cite{Figueroa:2017qmv,Cuissa:2018oiw,Figueroa:2019jsi} an improved lattice formulation of an interaction $\phi F\tilde F$ between an axion-like field and a $U(1)$ gauge sector was constructed, demonstrating that the topological nature of $F\tilde F$ as a total derivative $\partial_\mu {\mathcal{K}}^\mu$, can be actually realized exactly on a lattice (hence preserving exactly the shift symmetry at the lattice level). Interactions between a singlet inflaton and an Abelian gauge sector, via
$f(\phi)F^2$, or a non-Abelian $SU(2)$ gauge sector, via $f(\phi)\rm{Tr \,G^2}$, have also been explored in the context of preheating~\cite{Deskins:2013lfx,Adshead:2017xll}.

In Hybrid inflation models, the presence and excitation of gauge fields have also been addressed extensively, both for Abelian and non-Abelian scenarios, obtaining a very rich phenomenology, see e.g.~\cite{Rajantie:2000nj,Copeland:2001qw,Smit:2002yg,GarciaBellido:2003wd,Tranberg:2003gi,Skullerud:2003ki,vanderMeulen:2005sp,DiazGil:2007dy,DiazGil:2008tf,Dufaux:2010cf,Tranberg:2017lrx}. The case of preheating via parametric resonance, with a charged inflaton under a gauge symmetry, has however not been considered very often in the literature\footnote{Possibly, this is partially due to the fact that there is no particular need to `gauge' the inflationary sector, and partially because of the potential danger that gauge couplings may induce large radiative corrections in the inflaton potential, spoiling the conditions to sustain inflation.}. Nothing is wrong {\it per se} about considering an inflaton charged under a gauge group [and hence coupled to some gauge field(s)], as long as one constructs a viable working model, respecting all observational constraints. In such a case, when the inflaton starts oscillating following the end of inflation, the corresponding gauge bosons will be parametrically excited. This has been studied in detail in Ref.~\cite{Lozanov:2016pac}, for both Abelian $U(1)$ and non-Abelian $SU(2)$ gauge groups. Actually, in this manuscript we also consider a similar model for which we compute the preheating stage via parametric resonance effects into $U(1)$ and $U(1)\times SU(2)$ sectors. A natural realization of an inflationary set-up where the inflaton is charged under a gauge group is the Higgs-Inflation scenario~\cite{Bezrukov:2007ep,Bezrukov:2010jz}, where the SM Higgs is the inflaton. There the electroweak gauge bosons and charged fermions of the SM are coupled to the Higgs, and thus they experience parametric excitation effects during the oscillations of the Higgs after inflation~\cite{Bezrukov:2008ut,GarciaBellido:2008ab,Figueroa:2009jw,Figueroa:2014aya,Repond:2016sol,Ema:2016dny,Sfakianakis:2018lzf}. In the case where the SM Higgs is rather a spectator field during inflation, the post-inflationary decay of the Higgs into SM fields has also been considered in~\cite{Figueroa:2015rqa,Enqvist:2015sua,Kohri:2016wof,Figueroa:2017slm,Ema:2017loe}.

The non-linear dynamics characteristic of preheating scenarios, and in general of non-perturbative particle production phenomena, are interesting not only by themselves, but also because they may lead to cosmologically relevant and potentially observable
phenomena. Among these, we highlight:

\begin{itemize}

\item The generation of scalar metric perturbations~\cite{Bassett:1998wg,Bassett:1999mt,Bassett:1999ta,Finelli:2000ya,Chambers:2007se,Bond:2009xx,Imrith:2019njf,Musoke:2019ima,Giblin:2019nuv,Martin:2020fgl}, possibly leading to the formation of primordial black holes~\cite{Cotner:2019ykd,Martin:2019nuw,GarciaBellido:1996qt,Green:2000he,Hidalgo:2011fj,Torres-Lomas:2014bua,Suyama:2004mz,Suyama:2006sr,Cotner:2018vug}.

\item The production of stochastic gravitational wave backgrounds by parametric effects~\cite{Khlebnikov:1997di,Easther:2006gt,Easther:2006vd,GarciaBellido:2007af,Dufaux:2007pt,Dufaux:2008dn,Dufaux:2010cf,Zhou:2013tsa,Bethke:2013aba,Bethke:2013vca,Antusch:2016con,Antusch:2017flz,Antusch:2017vga,Liu:2018rrt,Figueroa:2017vfa,Fu:2017ero,Lozanov:2019ylm,Adshead:2019lbr,Adshead:2019igv,Armendariz-Picon:2019csc}. For a recent review see~\cite{Caprini:2018mtu}.

\item The creation of topological defects, like cosmic string networks~\cite{Hindmarsh:1994re,Felder:2000hj,Copeland:2009ga,Dufaux:2010cf,Lozanov:2019jff}, and their evolution during the scaling regime~\cite{Vincent:1997cx,Bevis:2006mj,Hindmarsh:2014rka,Daverio:2015nva,Lizarraga:2016onn,Hindmarsh:2018wkp,Hindmarsh:2019csc} and corresponding emission of GWs~\cite{Figueroa:2012kw,Figueroa:2020lvo}.

\item The creation of soliton-like structures such as oscillons~\cite{Copeland:2009ga,Amin:2011hj,Zhou:2013tsa,Antusch:2016con,Antusch:2017flz,Lozanov:2017hjm,Amin:2018xfe,Liu:2018rrt,Kitajima:2018zco,Lozanov:2019ylm,Antusch:2019qrr,Kasuya:2020szy} and other field configurations~\cite{Enqvist:2002si,Amin:2019ums,Niemeyer:2019gab,Musoke:2019ima}.

\item The realization of magnetogenesis~\cite{DiazGil:2005qp,DiazGil:2007qx,DiazGil:2007dy,DiazGil:2008tf,Fujita:2016qab,Adshead:2016iae,Vilchinskii:2017qul} and baryogenesis mechanisms~\cite{Kolb:1996jt,Kolb:1998he,GarciaBellido:1999sv,Allahverdi:2000zd,Cornwall:2001hq,Copeland:2001qw,Rajantie:2000nj,Smit:2002yg,GarciaBellido:2003wd,Tranberg:2003gi,Smit:2002yg,Tranberg:2009de,Kamada:2010yz,Lozanov:2014zfa,Adshead:2016iae}.

\item The determination of the post-inflationary equation of state, and its implications for the CMB inflationary observables~\cite{Podolsky:2005bw,Lozanov:2016hid,Figueroa:2016wxr,Lozanov:2017hjm,Krajewski:2018moi,Antusch:2020iyq}, or for the dark matter relic abundance~\cite{Garcia:2018wtq}.

\end{itemize}

In general, the details of nonlinear phenomena are difficult, when not impossible, to be grasped by analytic calculations. In order to fully understand the non-linearities developed in  a given model, the use of numerical techniques becomes necessary. The results arising from the non-linear dynamics of early universe high-energy phenomena, represent an important perspective in determining the best observational strategies to probe the unknown physics from this era. It is therefore crucial to develop numerical techniques, as efficient and robust as possible, to simulate these phenomena. Numerical algorithms developed for this purpose must satisfy a number of physical constraints (e.g.~energy conservation), and keep the numerical integration errors under control. It is actually useful to develop as many techniques as possible, to validate and double check results from simulations. Only in this way, we will achieve a certain robustness in the predictions of the potentially observational implications from non-linear high energy phenomena. Furthermore, the techniques developed for studying nonlinear dynamics of classical fields are common to many other non-linear problems in the early universe, like the dynamics of phase transitions~\cite{Hindmarsh:2001vp,Rajantie:2000fd,Hindmarsh:2001vp,Copeland:2002ku,GarciaBellido:2002aj,Figueroa:2017hun,Brandenburg:2017neh,Brandenburg:2017rnt,Figueroa:2019jsi} and their emission of gravitational waves~\cite{Hindmarsh:2013xza,Hindmarsh:2015qta,Hindmarsh:2017gnf,Cutting:2018tjt,Cutting:2019zws,Pol:2019yex,Cutting:2020nla}, cosmic defect formation~\cite{Hindmarsh:2000kd,Rajantie:2001ps,Rajantie:2002dw,Donaire:2004gp,Dufaux:2010cf,Hiramatsu:2012sc,Kawasaki:2014sqa,Fleury:2016xrz,Moore:2017ond,Lozanov:2019jff}, their later evolution~\cite{Vincent:1997cx,Bevis:2006mj,Hindmarsh:2014rka,Daverio:2015nva,Lizarraga:2016onn,Hindmarsh:2018wkp,Eggemeier:2019khm,Hindmarsh:2019csc,Gorghetto:2018myk} and gravitational wave emission~\cite{Dufaux:2010cf,Figueroa:2012kw,Hiramatsu:2013qaa,Figueroa:2020lvo}, axion-like field dynamics~\cite{Kolb:1993hw,Kitajima:2018zco,Amin:2019ums,Buschmann:2019icd,Fukunaga:2019unq,Patel:2019isj}, moduli dynamics~\cite{Giblin:2017wlo,Amin:2019qrx}, etc. These techniques can also be used in applications of interest not only to cosmology, but also to other high energy physics areas. For example, classical-statistical simulations have been used to compute quantities such as the sphaleron-rate \cite{Philipsen:1995sg,Ambjorn:1995xm,Arnold:1995bh,Arnold:1996dy,Arnold:1997yb,Moore:1997sn,Bodeker:1998hm,Moore:1998zk,Moore:1999fs,Bodeker:1999gx,Arnold:1999uy, Tang:1996qx,Ambjorn:1997jz,Moore:2000mx,DOnofrio:2012phz,DOnofrio:2015gop}, and to study the Abelian~\cite{Buividovich:2015jfa,Buividovich:2016ulp,Figueroa:2017hun,Figueroa:2019jsi,Mace:2019cqo,Mace:2020dkp} and non-Abelian~\cite{Akamatsu:2015kau} dynamics associated to the chiral anomaly. They have also been used to study spectral quantities~\cite{Boguslavski:2018beu,Schlichting:2019tbr}, and some properties of the quark-gluon plasma~\cite{Laine:2009dd,Laine:2013lia,Panero:2013pla,Boguslavski:2020tqz}.

\subsection{Purpose of this manuscript. Introducing \CL}

~~~~~As just reviewed in the previous section, the phenomenology of high-energy non-linear processes in the early universe is vast and very rich. In order to make reliable predictions of their potentially observable consequences, we need appropriate numerical tools. The {\it Numerical Early Universe}, i.e.~the study of high-energy non-linear field theory phenomena with numerical techniques, is an emerging field, increasingly gaining relevance, especially as a methodology to assess our capabilities to experimentally constrain (or even determine) the physics of this (yet) unknown epoch. It is because we recognize the importance of this, that we have created this dissertation, the content and purpose of which we explain next.

The present manuscript constitutes a theoretical basis for the code {\tt ${\mathcal C}$osmo${\mathcal L}$attice}, publicly available at \href{http://www.cosmolattice.net}{\color{blue} http://www.cosmolattice.net}. \CL is a modern multi-purpose MPI-based C++ package, developed as a user-friendly software for lattice simulations of non-linear dynamics of scalar and gauge field $dof$ in expanding backgrounds, including the possibility to `self-consistently' source the expansion of the universe by the fields themselves. Of course, exploring numerically the nonlinear dynamics of interacting fields during the early universe is not a new idea, as witnessed by the increasing number of lattice codes dedicated to this purpose that have appeared within the last years. With the exception of the recent \texttt{GFiRe} code~\cite{Lozanov:2019jff}, that includes an integrator for Abelian gauge theories, previous public packages were dedicated only to interacting scalar fields, either with finite difference techniques in real space, like \texttt{Latticeeasy}~\cite{Felder:2000hq}, \texttt{Clustereasy}~\cite{Felder:2007nz},  \texttt{Defrost}~\cite{Frolov:2008hy}, {\tt CUDAEasy}~\cite{Sainio:2009hm}, \texttt{HLattice}~\cite{Huang:2011gf},  \texttt{PyCOOL}~\cite{Sainio:2012mw} and \texttt{GABE}~\cite{Child:2013ria},
or pseudo-spectral codes like {\tt PSpectRe}~\cite{Easther:2010qz} and {\tt Stella}~\cite{Amin:2018xfe}. In most of the mentioned codes, metric perturbations (whenever present) are sourced passively, neglecting backreaction effects on the dynamics of the scalar fields. Notable exceptions to this are {\tt HLATTICE v2.0}, and especially the recent {\tt GABERel}~\cite{Giblin:2019nuv}, which allows for the full general relativistic evolution of non-linear scalar field dynamics. Given that all these codes are already available, one may wonder what is the point of releasing yet a new one. In order to answer this, let us explain the purpose of \CLns, which is actually twofold:

\begin{enumerate}

\item \CL is meant to be a `platform' for users to implement any system of equations suitable for discretization on a lattice. That is, \CL is not a code for doing one type of simulation with one specific integration technique, such as e.g.~the real-time evolution of interacting scalar fields sourcing self-consistently the expansion of the universe.  
The idea is rather something else: \CL is  a package that introduces its own \textit{symbolic language}, by defining field variables and operations over them. Therefore, once the user becomes familiar with the basic
`vocabulary' of the new language, they can write their own code: be it for the time evolution of the field variables in a given model of interest, or for some other operation, like a Monte-Carlo generator for thermal configurations. One of the main advantages of \CL is that it clearly separates the $physics$ (i.e.~fields living on a lattice and operations between them) from the $implementation \ details$, such as the handling of the parallelization or the Fourier transforms. For example, let us imagine a beginner user with little experience in programming, and with  no experience at all in parallelization techniques. With \CLns, they will be able to run a fully parallelized simulation of their favourite model (say using hundreds of processors in a cluster), while being completely oblivious to the technical details. They will just need to write a basic \textit{model file} in the language of \CLns, containing the details of the model being simulated. If, on the contrary, the user is rather an experienced one  and wants to look inside the core routines of \CL and modify, for example, the MPI-implementation, they can always do so, and perhaps even contribute to improving them. On top of this, \CL includes already a {\it library} of basic routines and field-theoretical operations. This constitutes a clear advantage of using \CL as a platform to implement a given scenario over writing your own code from scratch. In particular, \CL comes with symbolic scalar, complex and $SU(2)$ algebras, which allows to use vectorial and matrix notations without sacrificing performances. Furthermore, \CL is MPI-based and uses a discrete Fourier Transform parallelized in multiple spatial dimensions \cite{Pi13}, making it very powerful for probing physical problems with well-separated scales, running very high resolution simulations, or simply very long ones. \CL is publicly available in \href{http://www.cosmolattice.net}{\color{blue} http://www.cosmolattice.net}, and it comes with a detailed manual explaining its whole structure and the basic instructions to start running your own simulations.

\item \CL includes already a set of algorithms to evolve lattice scalar-gauge theories  in real-time, which can be selected with a single `switch' option. Part of this document can be actually considered as the theoretical basis for such algorithms. In fact, this manuscript is really meant to be a primer on lattice techniques for non-linear simulations, as we present a comprehensive discussion on such techniques, in particular for the simulation of scalar and gauge field dynamics in an expanding universe. In Section~\ref{sec:ContinuumFldDynamics} we review first the formulation of scalar and gauge field interactions in the continuum, both in a flat space-time and in {\it Friedmann-Lema\^itre-Robertson-Walker} (FLRW) backgrounds. In Section~\ref{sec:LatticeApproach} we introduce the basic tools for discretizing any bosonic field theory in an expanding background, including a discussion on {\it lattice gauge-invariant} techniques for both {\it Abelian} and {\it non-Abelian} gauge theories. Next, we introduce and classify a series of numerical algorithms, starting with the {\it staggered leapfrog} and \hspace*{-1mm}{\it Verlet integration} methods of $\mathcal{O}(\delta t^2)$ accuracy, passing through {\it Runge-Kutta} methods up to $\mathcal{O}(\delta t^4)$ accuracy, and finally covering higher-order integrators accurate up to $\mathcal{O}(\delta t^{10})$, such as the {\it Yoshida} and {\it Gauss-Legendre} methods. In the following Sections~\ref{sec:LatScalars}, \ref{sec:LatU1} and \ref{sec:LatSUN}, we adapt the previous algorithms to a specialized use for classical lattice simulations of scalar and gauge field dynamics in an expanding background. We include the possibility of
`self-consistent' expansion of the universe, sourcing the evolution of the scale factor by the volume average of the fields' energy and pressure densities, independently of whether the fields are scalars, Abelian gauge fields, or non-Abelian gauge fields. In Section~\ref{sec:LatScalars}, we present a variety of lattice formulations of {\it interacting scalar fields}, consisting in different integrators that can reproduce the continuum theory to an accuracy ranging from $\mathcal{O}(\delta t^2)$ to $\mathcal{O}(\delta t^{10})$. Analogously, in Sections~\ref{sec:LatU1} and~\ref{sec:LatSUN}, we present a set of algorithms for {\it Abelian $U(1)$ gauge theories} and {\it Non-Abelian $SU(2)$ gauge theories} respectively, again with an accuracy ranging between $\mathcal{O}(\delta t^2)$ and $\mathcal{O}(\delta t^{10})$. In the case of interacting scalar fields, we provide both symplectic and non-symplectic integrators, whereas for gauge fields only symplectic integrators are built. For every algorithm presented, we always provide the form of the most significant observables, such as the energy density components, relevant field spectra, and the Hubble constraint. The latter is verified by our symplectic algorithms with an accuracy that depends on the integrator order, reaching even machine precision in the case of the highest order schemes. Furthermore, it is worth noting that our integration algorithms for gauge theories always respect the Gauss constraint exactly, down to machine precision, independently of the order of the integrator. This remains true even in the case of self-consistent expansion, and independently of whether the gauge sector is Abelian or non-Abelian. We note that all the explicit-in-time algorithms presented in Sections~\ref{sec:LatScalars}\,-\,\ref{sec:LatSUN} are already implemented in \CLns, and have been made publicly available with the release of the code.

\end{enumerate}

It should also be noticed that this manuscript represents only {\it Part I} of our intended discussion on lattice techniques for the simulation of interacting fields in an expanding universe. In this document we focus on the presentation of general integration techniques (Section~\ref{sec:LatticeApproach}), and in their use to build explicit-in-time integration algorithms for {\it canonical} scalar-gauge theories, i.e.~for field theories with canonically normalized kinetic terms, standard scalar potentials (Section~\ref{sec:LatScalars}), and standard scalar-gauge Abelian (Section~\ref{sec:LatU1}) and non-Abelian (Section~\ref{sec:LatSUN}) interactions. We would like to highlight that we present higher-order integration algorithms for interacting scalar fields, similarly as in~\texttt{HLattice}~\cite{Huang:2011gf}. Whereas in~\texttt{HLattice} one family of algorithms with accuracy up to $\mathcal{O}(\delta t
^6)$ was built, we go a step beyond, building a variety of algorithm families, and introducing explicit implementations to all (even) orders, including also $\mathcal{O}(\delta t^{8})$ and $\mathcal{O}(\delta t^{10})$. Analogously, we present higher-order integration algorithms for Abelian $U(1)$ gauge theories, similar in spirit to the algorithm presented in~\texttt{GFiRe}~\cite{Lozanov:2019jff}, which was tested to $\mathcal{O}(\delta t^{4})$. We present a variety of flavours of these algorithms, and demonstrate explicitly their numerical implementation for all (even) accuracy orders for the first time, including now $\mathcal{O}(\delta t
^{6})$, $\mathcal{O}(\delta t^{8})$ and $\mathcal{O}(\delta t^{10})$. Furthermore, to the best of our knowledge, we also present for the first time an algorithm for non-Abelian $SU(N)$ gauge theories that is symplectic, explicit in time, and of arbitrary order, as well as preserving exactly the Gauss constraint while solving for the expansion of the universe self-consistently. As a numerical example to test our algorithms in scalar-gauge canonical theories, we analyze the post-inflationary preheating dynamics of an oscillating inflaton charged under $SU(2)\times U(1)$ in Section~\ref{sec:WorkingExample}. We postpone the discussion about methods for non-canonical scenarios for {\it Part II} of our dissertation on lattice techniques,
to be published elsewhere~\cite{PartII}, together with the public release of their implementation in \CLns. Non-canonical scenarios are theories e.g.~with non-minimal gravitational couplings, or more generally with kinetic terms with non-trivial field metrics, as considered e.g.~in~\cite{Child:2013ria,Nguyen:2019kbm,vandeVis:2020qcp}. Non-canonical scenarios  may also include interactions between field variables and their conjugate momenta, as naturally arising in exact derivative couplings between an axion-like field and gauge fields, as considered e.g.~in~\cite{Cuissa:2018oiw}. Non-canonical interactions can be numerically complicated to deal with, and usually require integration techniques which are either non-symplectic or simply more involved, typically with high memory requirements, and often not explicit in time. It is precisely because of these circumstances that we naturally separate the methods for canonical scalar-gauge theories presented here in {\it Part I} (in Sections~\ref{sec:LatScalars}\,-\,\ref{sec:LatSUN} of the present document), from the numerical integrators that we will present for non-canonical interactions in {\it Part II}~\cite{PartII}.
\vspace*{0.2cm}

To conclude this section, let us mention that precisely because \CL is a platform rather than a specialized code for certain type of scenarios or integration techniques, there is a number of extensions (beyond the routines currently discussed here in {\it Part I}, or planned to be presented in {\it Part II}), which we would like to add in \CL in the mid-term, as we go updating and improving the code in time. We hope to eventually consider (perhaps in collaboration with you?) the following aspects:\vspace*{0.1cm}

$\bullet$ Addition of fermions. Even though this is numerically very costly, one can simulate the out-of-thermal-equilibrium dynamics of classical bosonic fields coupled to quantum fermions. This has been done by~\cite{Saffin:2011kc} and successive works~\cite{Saffin:2011kn,Mou:2013kca,Mou:2015aia}, combining the lattice implementation based on the quantum mode equations proposed in~\cite{Aarts:1998td}, with the
`low cost' fermions introduced in~\cite{Borsanyi:2008eu}. \vspace*{0.1cm}

$\bullet$ Computation of metric perturbations. This could be done for scalar and vector perturbations following~\cite{Huang:2011gf}, whereas tensor perturbations representing gravitational waves (GW) can be obtained following~\cite{Figueroa:2011ye} (based on the idea originally proposed in~\cite{GarciaBellido:2007af}), as this allows for general GW sources built from either scalar and gauge fields (or even fermions if they were present).
\vspace*{0.1cm}

$\bullet$ Addition of {\it relativistic hydrodynamics}. This can be useful to describe scenarios where a classical scalar field, playing the role of an {\it order parameter} in a phase transition, is coupled to a relativistic fluid by means of a phenomenological friction term. This is the basis to describe numerically the dynamics of first order phase transitions~\cite{Hindmarsh:2001vp,Rajantie:2000fd,Hindmarsh:2001vp,Copeland:2002ku,GarciaBellido:2002aj,Figueroa:2017hun,Brandenburg:2017neh,Brandenburg:2017rnt,Figueroa:2019jsi} and their emission of gravitational waves~\cite{Hindmarsh:2013xza,Hindmarsh:2015qta,Hindmarsh:2017gnf,Cutting:2018tjt,Cutting:2019zws,Pol:2019yex,Cutting:2020nla}. \vspace*{0.1cm}

$\bullet$ Addition of new `initializer' routines. So far we have only considered the initialization of field fluctuations in Fourier space (on top of homogeneous field values), given a theoretical spectrum as an input. However, in order to simulate e.g.~the dynamics of a network of cosmic strings or other type of topological defects, different algorithms need to be used to create initially the defect network in configuration space, see e.g.~\cite{Vincent:1997cx,Bevis:2006mj,Figueroa:2012kw,Hindmarsh:2014rka,Daverio:2015nva,Lizarraga:2016onn,Lopez-Eiguren:2017dmc,Hindmarsh:2018wkp,Eggemeier:2019khm,Hindmarsh:2019csc,Gorghetto:2018myk}. \vspace*{0.1cm}

$\bullet$ Addition of `importance sampling' algorithms. Monte-Carlo algorithms and Langevin dynamics are used to generate fields according to some probability distributions. They can be used to set up thermal initial conditions to study e.g.~chiral charge dynamics in gauge theories at finite temperature~\cite{Figueroa:2017hun,Figueroa:2019jsi}. Alternatively, one could turn \CL into a  general platform to sample positive definite path integrals. While specific and highly optimized open-source codes exist to simulate lattice QCD \cite{OpenQCD,Clark:2009wm}, to the best of our knowledge, there is no truly versatile software to easily simulate other theories.

\section{Field dynamics in the continuum}
\label{sec:ContinuumFldDynamics}

~~~~~~In this section, we describe briefly the formulation of scalar and gauge field dynamics in the continuum. We review first, in Section~\ref{subsec:GaugeFldsMinkowski}, the case of interacting fields in a Minkowski background, starting with scalar fields only, and then introducing gauge symmetries and the corresponding gauge field degrees of freedom ({\it dof}). We then promote the background metric into a curved manifold, and specialize our study to the case of a spatially-flat, homogeneous, and isotropic space-time, described by the FLRW metric. We consider the dynamics of scalar and gauge fields living in a FLRW background in Section~\ref{subsec:DynamicsFLRW}, and the dynamics of the background itself, as sourced by the fields that live within it, in Section~\ref{subsec:DynamicsFLRW2}.

\subsection{Scalar and Gauge field interactions in flat space-time}
\label{subsec:GaugeFldsMinkowski}

Let us consider first a set of $N_{\rm s}$ relativistic interacting scalar fields with action in flat space-time
\begin{eqnarray} \label{eq:ScalarActionCont}
S_{\rm S} = - \int d^4 x \left\{{1\over2} \partial^{\mu} \phi_i \partial_{\mu}\phi_i + V(\lbrace \phi_j\rbrace)\right\}\,,
\end{eqnarray}
where $i, j = 1, ... N_{\rm s}$ label the fields, and the potential $V(\lbrace \phi_j\rbrace)$ characterizes the interactions between fields, including also possibly self-interactions. Because of the normalization constant $1/2$ in front of the kinetic term $\partial^{\mu} \phi_i \partial_{\mu}\phi_i$, we will refer to these fields as {\it canonically normalized scalar fields}. We note that space-time indices are raised with the Minkowski metric, e.g.~$\partial^\mu\phi \equiv \eta
^{\mu\nu}\partial_\nu\phi$. The equations of motion (EOM) of the system are obtained from minimizing Eq.~(\ref{eq:ScalarActionCont}). This leads to
\begin{eqnarray}
\label{eq:EOMflatScalarFlds}
-\Box_{\eta} \phi_i + \frac{\partial{V}}{\partial \phi_i} = 0\,,~~~~~~ {\rm with}~~\Box_{\eta} \equiv \eta^{\mu\nu}\partial_{\mu} \partial_{\nu} = \partial^{\alpha} \partial_{\alpha}\,.
\end{eqnarray}
In a more explicit form, the EOM can be written as follows
\begin{eqnarray}
\ddot\phi_i - {\vv \nabla}^{\,2}\hspace*{-1mm}\phi_i + \frac{\partial{V}}{\partial \phi_i} = 0\,~~~~ \Longleftrightarrow ~~~~ \left\lbrace
\begin{array}{rcl}
\dot\phi_i & \equiv & \pi_i\, , \vspace*{0.2cm}\\
 \dot\pi_i & = & {\vv \nabla}^{\,2}\hspace*{-1mm}\phi_i - \frac{\partial{V}}{\partial \phi_i}\, .
\end{array}
\right.\,
\end{eqnarray}


Let us now consider a general scalar-gauge theory in the continuum, including three types of (canonically normalized) scalar fields: a singlet $\phi$, a $U(1)$-charged field $\varphi$, and a $[SU(N)\times U(1)]$-charged field $\Phi$; as well as the corresponding Abelian $A_\mu$ and non-Abelian $B_\mu = B_\mu^a\,T_a$ gauge vector bosons. Here $\lbrace T_a \rbrace$ are the $N^2-1$ group generators of $SU(N)$, satisfying the properties of the $SU(N)$ Lie algebra
\bea \label{eq:TaProperties}
\begin{array}{llll}
[T_a, T_b] = i f_{abc} T_c \ , & {\rm Tr}(T_a) = 0 \ , & {\rm Tr}(T_a T_b) = \frac{1}{2} \delta_{ab} \ , & T^{\dagger}_a = T_a \ ,
\end{array}
\eea
with $f_{abc}$ the totally anti-symmetric {\it structure constants} of $SU(N)$. In the particular case of $SU(2)$, $T_a \equiv \sigma_a / 2$ ($a=1,2,3$), with $\sigma_a$ the {\it Pauli matrices}
\begin{equation}
    \sigma_1=\begin{pmatrix} 0& 1\\ 1 & 0 \end{pmatrix} \ \ , \ \ \sigma_2=\begin{pmatrix} 0& -i\\ i & 0 \end{pmatrix}  \ \ , \ \ \sigma_3=\begin{pmatrix} 1& 0\\ 0 & -1 \end{pmatrix}  \ .
\end{equation}
For later convenience we also write some of their properties,
\bea
\begin{array}{llll}
[\sigma_a, \sigma_b] = 2i \epsilon_{abc} \sigma_c \ , & {\rm Tr}(\sigma_a) = 0 \ , & {\rm Tr}(\sigma_a \sigma_b) = 2\delta_{ab} \ , & \sigma^{\dagger}_a = \sigma_a\,, 
\end{array}
\eea
with $\epsilon_{abc}$ the total anti-symmetric tensor. 

We can write a gauge invariant action as
\begin{eqnarray}
S \hspace*{-0.18cm}&=&\hspace*{-0.18cm} - \int d^4 x \left\{\frac{1}{2}\partial_{\mu} \phi \partial ^{\mu}\phi + (D_{\mu}^A \varphi)^{*}(D_A^{\mu} \varphi) +  (D_{\mu}\Phi )^{\dagger} (D^{\mu} \Phi) + \frac{1}{4} F_{\mu \nu} F^{\mu \nu} + \frac{1}{2}{\rm Tr}\{G_{\mu \nu}G^{\mu \nu}\} + V\right\}
\label{eq:Lagrangian}\\
&=&\hspace*{-0.18cm} \int d^4x \left\lbrace {{\dot \phi}^2\over2} - {|{\vec\nabla \phi}|^2\over2} + |D_0\varphi|^2 - |\vec D\varphi|^2 + |D_0\Phi|^2 - |\vec D\Phi|^2
+ {|\vec{\mathcal{E}}|^2\over2} - {|\vec{\mathcal{B}}|^2\over2} +
\sum_a \left({|{\vec{\mathcal{E}}_a}|^2\over2} -{|{\vec{\mathcal{B}}_a}|^2\over2}\right) - V
\right\rbrace,\nonumber
\label{eq:LagrangianII}
\end{eqnarray}
with a potential $V \equiv V(\phi,|\varphi|, |\Phi|)$ describing the interactions among the scalar fields,
\begin{eqnarray} \label{eq:ChargedScalars}
\begin{array}{ccccc}
\phi \in \mathcal{R}e & , &  \varphi \equiv {1\over\sqrt{2}}(\varphi_0 +i\varphi_1) & , & \Phi = \left(
\begin{array}{c}
\varphi^{(0)} \\ \varphi^{(1)} \\ \vdots \\ \varphi^{(N-1)}
\end{array}
\right) =
{1\over\sqrt{2}}
\left(
\begin{array}{c}
\varphi_0 +i\varphi_1 \vspace*{0.1cm}\\ \varphi_2 +i\varphi_3 \\ \vdots \\ \varphi_{2N -2} +i\varphi_{2N-1}
\end{array}
\right) \,,
\end{array}
\eea
and where we have introduced standard definitions of {\it covariant derivatives} (denoting $Q_{A}$ and $Q_B$ the Abelian and non-Abelian charges) and {\it field strength} tensors,
\bea
D_{\mu}^{\rm A} & \equiv &  \partial _{\mu} - i  g_AQ_AA_\mu \ ,
\label{eq:AbCovDerivCont}\\
D_{\mu} &\equiv & 
\mathcal{I}D^{\rm A}_\mu
- i g_B Q_B B_{\mu}^a \,T_a \ , \label{eq:CovDerivCont}\\
F_{\mu \nu} &\equiv & \partial_{\mu}  A_{\nu} - \partial_{\nu} A_{\mu} \ , \label{eq:FmnAbelian}\\
G_{\mu \nu} & \equiv & \partial_{\mu} B_{\nu} - \partial_{\nu} B_{\mu} - i[B_\mu,B_\nu] \ , \label{eq:GmnNonAb}
\eea
with $\mathcal{I}$ the $N\times N$ identity matrix. In the second line of (\ref{eq:Lagrangian}) we have used the properties of the generators, displayed in Eq.~(\ref{eq:TaProperties}), to obtain
\begin{eqnarray}
G_{\mu \nu} \equiv G_{\mu \nu}^a T_a  ~~~\Rightarrow ~~~
{1\over 2}{\rm Tr}(G_{\mu\nu}G^{\mu\nu}) \equiv  {1\over 2}G_{\mu\nu}^aG^{\mu\nu}_a\,;~~~  G_{\mu \nu}^a \equiv \partial_{\mu} B_{\nu}^a - \partial_{\nu} B_{\mu}^a + f^{a b c} B_{\mu}^b B_{\nu}^c \ ,
\end{eqnarray}
and introduced Abelian and non-Abelian electric and magnetic fields as
\be\label{eq:ElectricMagneticDefs}
\mathcal{E}_i \equiv F_{0i} , \,\,\,\,\,\,\,\,  \mathcal{B}_i = \frac{1}{2} \epsilon_{i j k} F^{j k} , \,\,\,\,\,\,\,\,   \mathcal{E}_i^a \equiv G_{0i}^a , \,\,\,\,\,\,\,\,  \mathcal{B}_i^a = \frac{1}{2} \epsilon_{i j k} G^{j k}_a \ . \ee

\noindent The equations of motion (EOM) of the system can be obtained from minimizing Eq.~(\ref{eq:Lagrangian}). They are 
\begin{eqnarray}
\label{eq:EOMflat}
&\begin{array}{rclr}
\partial^{\mu} \partial_{\mu} \phi  & = & \frac{\partial{V}}{\partial \phi} & [\text{Singlet}] \vspace*{0.1cm}\\

D_A^{\mu}D^A_{\mu} \varphi & = & \frac{1}{2}\frac{\partial{V}}{\partial |\varphi|}\frac{\varphi}{|\varphi |} & [U(1)\text{-charged}] \vspace*{0.1cm}\\

D^{\mu}D_{\mu} \Phi & = & \frac{1}{2}\frac{\partial{V}}{\partial |\Phi|} \frac{\Phi}{|\Phi|} & [U(1)\times SU(N)]
\vspace*{0.1cm}\\

\partial_{\nu} F^{\mu \nu} & = & J_A^\mu & [\text{Abelian vector}]
\vspace*{0.1cm}\\

(\mathcal{D}_{\nu})_{a b} G^{\mu \nu}_b & = & J_a^\mu & [\text{Yang-Mills vector}]
\end{array}\,,
\end{eqnarray}
where $(\mathcal{D}_{\nu}O)_a = (\mathcal{D}_{\nu})_{a b}O_b \equiv ( \delta_{a b}  \partial_{\nu} - f_{abc} B_{\nu}^c ) O_b$, and the currents are given by
\begin{eqnarray}
\label{eq:AbelianCurrent}
\hspace{1.8cm} J_A^\mu & \equiv & 2g_AQ_A^{(\varphi)} \mathcal{I}m [ \varphi^{*} ( D_A^{\mu} \varphi )] + 2g_AQ_A^{(\Phi)} \mathcal{I}m [ \Phi^\dag (D^{\mu} \Phi  )]\,,\\
\label{eq:NonAbelianCurrent}
\hspace{1.8cm} J_a^\mu & \equiv & 2g_BQ_B\mathcal{I}m [ \Phi^{\dag} T_a( D^{\mu} \Phi )]\,.
\end{eqnarray}

It is straightforward to show that both action (\ref{eq:Lagrangian}) and the EOM~(\ref{eq:EOMflat}) are invariant under the following set of gauge transformations,
\begin{eqnarray}
\begin{array}{rcl}
\phi(x) & \longrightarrow & \phi(x)\,,~~\text{[singlet]}\vspace*{0.13cm}\\
\varphi(x) & \longrightarrow & \omega(x)\varphi(x)\,,~~\omega(x) = e^{-i{g_A}Q_A^{(\varphi)}\alpha(x)}\,, \vspace*{0.13cm}\\
\Phi(x) & \longrightarrow & \omega(x)\Omega(x)\Phi(x)\,,~~ \Omega(x) \equiv e^{-ig_BQ_B\beta_a(x)T_a}\,,~~\omega(x) = e^{-i{g_A}Q_A^{(\Phi)}\alpha(x)}\,,
\vspace*{0.2cm}\\
A_\mu(x) & \longrightarrow & A_\mu(x) - \partial_\mu\alpha(x)\,,\vspace*{0.2cm}\\
B_\mu(x) & \longrightarrow & \Omega(x)B_\mu(x)\Omega^\dag(x) - {i\over g_B Q_B}
[\partial_\mu\Omega(x)]\Omega^\dag(x)\,,
\end{array} \label{eq:GaugeTrans}
\end{eqnarray}
with $\alpha(x)$ and $\beta_a(x)$ arbitrary real functions, $Q_A^{(\varphi)}$ and $Q_A^{(\Phi)}$ the Abelian charges of $\varphi$ and $\Phi$, and $Q_B$ the non-Abelian charge of $\Phi$. The transformation of the gauge fields imply that the field strengths transform as
\bea
\begin{array}{rcc}
F_\mn(x) & \longrightarrow & F_\mn(x)\,,~~\text{[invariant]}
\vspace*{0.2cm}\\
G_\mn(x) & \longrightarrow & \Omega(x)G_\mn(x)\Omega^\dag(x)\,.
\end{array}
\label{eq:GaugeTransII}
\eea


Using the definitions in Eq.~(\ref{eq:ElectricMagneticDefs}), we can also write the EOM in vectorial form, making more explicit the individual terms in each equation:
\begingroup
\allowdisplaybreaks
\begin{eqnarray}
\ddot\phi  
- {\vv \nabla}^{\,2}\phi  &=& - V_{,\phi} \ , \label{eq:singlet-eom_I} \vspace{0.1cm}\\
\ddot\varphi 
-{\vv D}_{\hspace{-0.5mm}A}^{\,2}\varphi &=& -\frac{V_{,|\varphi|} }{2}  \frac{\varphi}{|\varphi |}  \ , \label{eq:higgsU1-eom_I}\vspace{0.1cm}\\
\ddot\Phi 
- {\vv D}^{\,2}\Phi &=& -\frac{V_{,|\Phi|} }{2}\frac{\Phi}{|\Phi|} \ , \label{eq:higgsSU2-eom_I}
\vspace{0.1cm}\\
{\dot{\vv{\mathcal{E}}}} - {\vv \nabla} \times \vv{\mathcal{B}} &=& \vv{J_A} \equiv 2g_A Q_A^{(\varphi)} \mathcal{I}m [ \varphi^{*} \vv{D_A} \varphi] + 2g_A Q_A^{(\Phi)} \mathcal{I}m [ \Phi^\dag \vv{D} \Phi ] \ , \label{eq:U1vector-eom_I}
\vspace{0.1cm}\\
({\vv {\mathcal D}}_0{\vv{\mathcal{E}}})_a - (\vv{\mathcal D}\times {\vv{ \mathcal{B}}})_a &=& \vv{J_a} \equiv 2g_B Q_B^{(\Phi)} \mathcal{I}m [ \Phi^\dag T_a \vv{D} \Phi ] \ , \label{eq:SUNvector-eom_I}
\vspace{0.1cm}\\
-{\vv\nabla}{\vv{\mathcal{E}}} &=& J_0^A \equiv 2g_A Q_A^{(\varphi)} \mathcal{I}m [ \varphi^* (D_0^A \varphi)] + 2g_A Q_A^{(\Phi)} \mathcal{I}m [\Phi^\dag(D_0 \Phi)] \ , \,
\label{eq:U1vector-Gauss_I}
\vspace{0.1cm}\\
-({\vv {\mathcal D}}\vv{\mathcal{E}})_a &=& (J_0)_a \equiv 2g_B Q_B^{(\Phi)} \mathcal{I}m [ \Phi^\dag T_a (D_0 \Phi) ]\,.
\label{eq:SUNvector-Gauss_I}
\label{eq:EOMflatVect}
\end{eqnarray}
\endgroup
We note that Eqs.~\eqref{eq:U1vector-Gauss_I}-(\ref{eq:SUNvector-Gauss_I}) represent constraint equations, as they correspond to the equations associated with the temporal component of the gauge field, which is not dynamical. These constraints are equivalent to the standard \textit{Gauss Law} of electromagnetism $\vv\nabla\vv{ \mathcal{E}} = \rho$. In particular, they are the generators of gauge transformations \cite{Vlasov:1987vt}.

\subsection{Field dynamics in an expanding background}\label{subsec:DynamicsFLRW}

To describe the expansion of the universe we consider a flat {\it Friedmann-Lema\^itre-Robertson-Walker} (FLRW) metric, with line element
\be
\dd s^2 = g_{\mu\nu}\dd x^\mu\dd x^\nu = - a(\eta)^{2 \alpha} \dd \eta^2 + a(\eta)^2 \delta_{ij} \dd x^i \dd x^j \ , \label{eq:FLRWmetric}
\ee
where $a(\eta)$ is the scale factor, $\delta_{ij}$ is the Euclidean metric, and $\alpha$ is a constant parameter that will be chosen conveniently in a case by case basis. The choice $\alpha = 0 $ identifies $\eta$ with the {\it coordinate time} $t$, whereas $\alpha = 1$ makes $\eta$ the {\it conformal time} $\tau \equiv \int {dt'\over a(t')}$. For the time being, we will consider $\alpha$ as an unspecified constant, and we will refer to $\eta$ as the {\it $\alpha$-time} variable.\\

{\it Note -.} Recall that we reserve the symbol $\dot f \equiv {df/dt}$ for derivatives with respect to the coordinate time, whereas $f' \equiv {df/ d\eta}$ will indicate derivative with respect to any $\alpha$-time variable.\\

\noindent For later convenience we write the metric and inverse metric elements explicitly,
\bea
\label{eq:FLRWmetricElements}
  g_{00} = -a(\eta)^{2\alpha}~;~~~
  g_{ij} = a(\eta)^2 \delta_{ij}~;~~~
  g^{00} = -a(\eta)^{-2\alpha}~;~~~
  g^{ij} &= a(\eta)^{-2} \delta^{ij}\,.
\eea
\indent To obtain the EOM in curved space, we follow the \textit{minimal gravitational coupling} prescription, making the following replacements into the flat space-time equations,
\begin{eqnarray}
\eta_\mn &\rightarrow& g_\mn \ , \vspace*{0.1cm}\\
\partial_\gamma V^{\ab ..}_{\mn ..} \equiv V^{\ab ..}_{\mn ..\,,\gamma} &\rightarrow& \nabla_\gamma V^{\ab ..}_{\mn ..} \equiv V^{\ab ..}_{\mn ..\,;\gamma} =  V^{\ab ..}_{\mn ..\,,\gamma} + \Gamma^\alpha_{\gamma\sigma}V^{\sigma\beta ..}_{\mn ..} - \Gamma^\sigma_{\gamma\mu}V^{\ab ..}_{\sigma\nu ..} + ...\,,
\end{eqnarray}
where $V_{;\mu} = \nabla_\mu V$ represents a (gravitational) covariant derivative, with $\Gamma^\mu_{\alpha\beta}$ the Christoffel symbols, and $V^{\ab ..}_{\mn ..}$ an arbitrary tensor. Using the non-vanishing Christoffel symbols of the FLRW metric,
\bea
 \Gamma^{0}_{00} =   \alpha\frac{a'(\eta)}{a(\eta)} ~\, ,~~~
   \Gamma^{0}_{i j} = a^{-2\alpha + 2}\frac{a'(\eta)}{a(\eta)}\delta_{ij}\,~ , ~~~
  \Gamma^{i}_{i 0} = \frac{a'(\eta)}{a(\eta)}\,,
   \label{eq:FLRW_Cristoffel}
\eea
we can obtain, via the minimal coupling prescription, the EOM in an expanding universe. In practice, we can obtain directly the transformation of the derivative terms in the scalar and gauge field EOM, by making use of the following identities for the divergence of a vector and a rank-2 anti-symmetric tensor,
\bea \nabla_{\sigma} V^{\sigma}  &=& \frac{1}{\sqrt{g}} \frac{\partial (V^{\sigma} \sqrt{g})}{\partial x^{\sigma}} = \frac{1}{a^{3 + \alpha}} \frac{\partial(V^{\sigma} a^{3 + \alpha} (t) ) }{\partial x^{\sigma}} = \partial_{\sigma} ( g^{\sigma\lambda} V_{\lambda} ) +  (3 + \alpha) \frac{a'}{a} V^0 \ , \hspace{0.6cm} \\
\nabla_{\sigma} F^{\sigma \lambda} &=& \frac{1}{\sqrt{g}} \frac{\partial (F^{\sigma \lambda} \sqrt{g})}{\partial x^{\sigma}} = \frac{1}{a^{3 + \alpha}} \frac{\partial(F^{\sigma \lambda} a^{3 + \alpha} (t) ) }{\partial x^{\sigma} } = \partial_{\sigma}( g^{\sigma\mu}g^{\lambda \beta}F_{\mu \beta}) + (3 + \alpha) \frac{a'}{a} g^{0 \mu}g^{\lambda \beta}F_{\mu\beta} \,,
\eea
where $g = - det(g_{\mu\nu})$. This leads to
\bea
\partial_{\mu} \partial^{\mu} \phi & \longrightarrow & \nabla_{\mu} [ \partial^{\mu} \phi ] = - a^{-2\alpha} {\phi}'' + a^{-2} \partial_i \partial_i \phi - a^{-2 \alpha}  (3 - \alpha) \frac{{a'}}{a} {\phi'} \,,
\\*
\partial_{\mu} F^{\mu \nu} &\longrightarrow & \nabla_{\mu} F^{\mu \nu} = g^{\nu \nu} \left( - a^{-2 \alpha}\partial_0 F_{0 \nu} + a^{-2}\partial_i F_{i \nu} - (3 - \alpha) a^{-2 \alpha}\frac{{a'}}{a} F_{0 \nu} \right) - a^{-2 \alpha}F_{0 \nu}  \partial_0 g^{\nu \nu}\,.
\eea
Using these identities and the metric elements~(\ref{eq:FLRWmetricElements}), we obtain the EOM in an expanding background as
\begingroup
\allowdisplaybreaks
\begin{eqnarray}
\phi'' - a^{-2(1 - \alpha)} {\vv\nabla}^{\,2} \hspace{-1mm}\phi + (3 - \alpha)\frac{{a'}}{a} {\phi'} &=& - a^{2 \alpha} V_{,\phi} \ , \label{eq:singlet-eom} \\
\varphi'' - a^{-2(1 - \alpha)} {\vv D}_{\hspace{-0.5mm}A}^{\,2}\varphi + (3 - \alpha)\frac{{a'}}{a}  {\varphi'} &=& - \frac{a^{2 \alpha}V_{,|\varphi|} }{2} \frac{\varphi}{|\varphi |} \ , \label{eq:higgsU1-eom}\\
\Phi'' - a^{-2(1 - \alpha)} {\vv D}^{\,2}\Phi + (3 - \alpha)\frac{{a'}}{a}  {\Phi'} &=& - \frac{a^{2 \alpha} V_{,|\Phi|}}{2} \frac{\Phi}{|\Phi |} \ , \label{eq:higgsSU2-eom}
\\
\partial_0 F_{0i} - a^{-2(1 - \alpha )}\partial_j F_{ji} + (1 - \alpha) \frac{{a'}}{a} F_{0i} &=&
a^{2 \alpha}J^A_i \ , \label{eq:U1eom}
\\
(\mathcal{D}_0 )_{a b} (G_{0i})^b - a^{-2(1 - \alpha )} ( \mathcal{D}_j )_{a b} (G_{ji} )^b + (1 - \alpha) \frac{{a'}}{a} (G_{0i} )^b &=& a^{2 \alpha}(J_i)_a \ , \label{eq:SU2eom}
\\
\partial_i F_{0i} &=& a^2J^A_0 \ , \label{eq:GaussU1-eom}\\
(\mathcal{D}_i )_{a b} (G_{0i})^b &=& a^2(J_0)_a \,, \label{eq:GaussSU2-eom}
\end{eqnarray}
\endgroup
where the currents in the $rhs$ of the gauge field EOM are still given by Eqs.~(\ref{eq:AbelianCurrent})-(\ref{eq:NonAbelianCurrent}). We note that  Eqs.~(\ref{eq:GaussU1-eom}) and (\ref{eq:GaussSU2-eom}) are the generalization of the $U(1)$ and $SU(2)$ Gauss constraints in an expanding background. When discretizing the system of equations later on, we will use them as an indicator of the correctness of the discretization scheme, checking the degree of preservation of these constraints during the field evolution.

\subsection{Dynamics of the expanding background}\label{subsec:DynamicsFLRW2}

If the expansion of the universe is dictated by some external $dof$ different than the fields we are evolving, say e.g.~a fluid with a given equation of state, this corresponds to a {\it fixed background} case. If on the contrary, the matter fields (scalar and/or gauge) we are evolving are the ones sourcing themselves the expansion of the universe, we will refer to this as a {\it self-consistent expansion} case. In general, the evolution of the scale factor $a(\eta)$ is dictated by the stress-energy tensor of matter fields via the Friedmann equations. Denoting the background energy and pressure densities as $\bar\rho$ and $\bar p$, the stress-energy tensor of a background {\it perfect fluid} is given by
\be {\bar T}_{\mu \nu} \equiv (\bar\rho + \bar p )u_{\mu} u_{\nu} + \bar p g_{\mu \nu}  \ , \hspace{0.4cm} g_{\mu \nu} u^{\mu} u^{\nu} = -1 \hspace{0.4cm} \Longrightarrow \hspace{0.4cm}   \begin{cases}
      \bar\rho = a^{-2 \alpha}\,{\bar T}_{00} \ , \vspace{0.2cm}\\
      \bar p = {1\over 3a^2} \sum_j {\bar T}_{jj} \ ,
    \end{cases} \label{eq:stresstensor} \ee
where we have used $u_{\mu} = (a^{\alpha},0,0,0)$ and $u^{\mu} = - (a^{- \alpha},0,0,0)$. The evolution of the scale factor is then determined by the Friedmann equations, which, in $\alpha$-time, read as
\be
\mathcal{H}^2 \equiv \left({a'\over a}\right)^2 = a^{2 \alpha} \frac{\bar\rho}{3 m_p^2} \ , \hspace{0.6cm} {a''\over a} = \frac{a^{2 \alpha}}{6 m_p^2}[ (2 \alpha - 1) \bar\rho - 3 \bar p ]\,.\label{eq:Friedmann-full}
\ee

Let us consider first the case of a fixed background. If the expansion of the universe is created by an external fluid with equation of state $w \equiv p / \rho$, the two Friedmann equations can be combined into a single equation as
\be 2 {a''} + (1 + 3 \omega - 2 \alpha)  \frac{{a'}^2}{a} = 0 \ , \ee
with solution, for $w = const$,
\be
a(\eta) = a (\eta_{\rm i} ) \left(1 + \frac{1}{p}\mathcal{H}_i (\eta-\eta_{\rm i}) \right)^p \,,\hspace{0.3cm} \mathcal{H}(\eta) = {\mathcal{H}_{\rm i}\over \left(1 + \frac{1}{p}\mathcal{H}_i (\eta-\eta_{\rm i}) \right)} ={\mathcal{H}_{\rm i}\over \sqrt[p]{a(\eta)/a(\eta_{\rm i})}} \,,\hspace{0.3cm} p \equiv \frac{2}{3(1 + \omega) - 2 \alpha } \ , \ee
with $\eta_i$ some initial time. In order to solve the scalar/gauge field dynamics, we just need to plug-in the above expressions for $a(\eta)$ and $\mathcal{H}(\eta)$ into the EOM of the matter fields.

In the case of self-consistent expansion, the situation is however slightly more elaborated. We need first an expression for the energy momentum-tensor of the scalar/gauge matter fields, and then take a volume average of the corresponding local expressions of the energy and pressure densities, which source the Friedmann equations. From the Lagrangian in Eq.~(\ref{eq:Lagrangian}) we can actually derive a local expression for the stress-energy tensor of the scalar and gauge fields as
\bea T_{\mu \nu} &=& -\frac{2}{\sqrt{g}}\frac{\delta(\sqrt{g} \mathcal{L})}{\delta g^{\mu \nu}}
= g_{\mu \nu} \mathcal{L} - 2 \frac{\delta \mathcal{L}}{\delta g^{\mu \nu}}\\
&=& - g_{\mu \nu} \left(g^\ab\Big[(D_{\alpha} \Phi)^{\dagger} (D_{\beta} \Phi) + (D^A_{\alpha} \varphi )^* (D_{\beta}^A \varphi ) + \frac{1}{2} (\partial_{\alpha} \phi ) (\partial_{\beta} \phi )\Big] + \frac{1}{4}g^{\alpha\delta}g^{\beta\lambda} (G_{\alpha \beta}^a G_{\delta\lambda}^a  + F_{\alpha \beta} F_{\delta\lambda} ) + V \right) \nonumber \\
&& +~ \left[2 {\rm Re}\lbrace(D_{\mu} \Phi)^{\dagger}(D_{\nu} \Phi)\rbrace + 2 {\rm Re}\lbrace(D^A_{\mu} \varphi)^{*} (D_{\nu}^A \varphi )\rbrace + (\partial_{\mu} \phi) (\partial_{\nu} \phi ) \right]  + g^{\alpha\beta} \left(G_{\mu\alpha}^a G_{\nu\beta}^a + F_{\mu\alpha} F_{\nu\beta} \right) \ , \nonumber
\eea
where in the first equality we used\footnote{Had we wanted to obtain $T^\mn$ with the space-time indices above, then we should use instead   $\delta(\sqrt{g}) = + \frac{1}{2} g^{\mu \nu} \sqrt{g} \,\delta g_{\mu \nu}$\,.} $\delta(\sqrt{g}) = - \frac{1}{2} g_{\mu \nu} \sqrt{g} \,\delta g^{\mu \nu}$, and in the second we used  $\delta g^{\alpha \beta} = - g^{\alpha \mu} g^{\beta \nu} \delta g_{\mu \nu}$. Using $F^{\mu \nu} F_{\mu \nu} = - \frac{2}{a^{2 (1 + \alpha)}} \sum_i F_{0i}^2 + \frac{1}{a^4} \sum_{i,j} F_{ij}^2$ (similarly for $G_{\mu \nu}^a$), and $(D_{\mu} \Phi^{\dagger} ) (D^{\mu} \Phi) = - a^{-2 \alpha} (D_0 \Phi)^{\dagger} (D_0 \Phi) + a^{-2} (D_i \Phi)^{\dagger} (D_i \Phi)$ (similarly for the $U(1)$-charged and singlet scalar fields), we obtain for the energy and pressure densities,
\begin{eqnarray}
\rho &=& {K}_{\phi} + {K}_{\varphi} + {K}_{\Phi} + {G}_{\phi} + {G}_{\varphi} + {G}_{\Phi} + {K}_{U(1)} + {G}_{U(1)} + {K}_{SU(2)} + {G}_{SU(2)} + {V} \ ,  \label{eq:rhoLocal}\\
p &=& {K}_{\phi} + {K}_{\varphi} + {K}_{\Phi} -{1\over3}({G}_{\phi} + {G}_{\varphi} + {G}_{\Phi}) + {1\over3}({K}_{U(1)} + {G}_{U(1)}) + {1\over3}({K}_{SU(2)} + {G}_{SU(2)}) - {V} \ ,  \label{eq:pLocal}
\end{eqnarray}
with $V$ the interacting scalar potential, whereas the kinetic and gradient energy densities are
\bea
\begin{array}{lcl} \label{eq:energy-contrib}
{K}_{\phi} &=& \frac{1}{2 a^{2\alpha} } {\phi'}^2 \vspace{0.1cm}\\
{K}_{\varphi} &=& \frac{1}{a^{2\alpha} } (D_0^A \varphi)^*(D_0^A \varphi)
\vspace{0.1cm}\\
{K}_{\Phi} &=& \frac{1}{a^{2\alpha} } (D_0 \Phi )^\dag(D_0 \Phi)
\vspace{0.1cm}\\
\end{array}
\hspace{0.1cm};\hspace{0.4cm}
\begin{array}{lcl}
{G}_{\phi} &=& \frac{1}{2 a^2} \sum_i (\partial_i \phi)^2
\vspace{0.1cm}\\
{G}_{\varphi} &=& \frac{1}{a^2} \sum_i (D_i^A \varphi)^*(D_i^A \varphi)
\vspace{0.1cm}\\
{G}_{\Phi} &=& \frac{1}{a^2} \sum_i (D_i \Phi)^\dag(D_i \Phi)
\vspace{0.1cm}\\
\end{array}
\hspace{0.1cm};\hspace{0.4cm}
\begin{array}{lcl}
{K}_{U(1)} &=& \frac{1}{2 a^{2 + 2 \alpha}}  \sum_{i} F_{0i}^2
\vspace{0.1cm}\\
{K}_{SU(2)} &=& \frac{1}{2 a^{2 + 2 \alpha}}  \sum_{a,i} (G_{0i}^a)^2
\vspace{0.1cm}\\
{G}_{U(1)} &=& \frac{1}{2 a^4}  \sum_{i,j<i} F_{ij}^2
\vspace{0.1cm}\\
{G}_{SU(2)} &=& \frac{1}{2 a^4}  \sum_{a,i,j<i}  (G_{ij}^a)^2  \, . \vspace*{0.2cm}\\
\end{array}
\\
\text{(Kinetic-Scalar)} \hspace{2.5cm} \text{(Gradient-Scalar)} \hspace{2.75cm} \text{(Electric \& Magnetic)} \hspace{1cm} \nonumber
\eea
\indent Whenever dealing with scenarios with self-consistent expansion of the universe, we then need to take first a volume average of the local expressions in Eqs.~(\ref{eq:rhoLocal}), (\ref{eq:pLocal}), so that we obtain the background energy and pressure densities $\bar \rho$ and $\bar p$, within a given volume. Plugging back the background quantities into the Friedmann Eqs.~(\ref{eq:Friedmann-full}), will determine then the evolution of the universe within the chosen volume, namely
\begin{eqnarray}\label{eq:FriedmannHubble}
\left({a'\over a}\right)^2 &=&  \frac{a^{2 \alpha}}{3 m_p^2}\left\langle {K}_{\phi} + {K}_{\varphi} + {K}_{\Phi} + {G}_{\phi} + {G}_{\varphi} + {G}_{\Phi} + {K}_{U(1)} + {G}_{U(1)} + {K}_{SU(2)} + {G}_{SU(2)} + {V}\right\rangle \,,
\\
\label{eq:FriedmannDDa}
{a''\over a} &=& \frac{a^{2 \alpha}}{3 m_p^2}\left\langle (\alpha-2)({K}_{\phi} + {K}_{\varphi} + {K}_{\Phi}) + \alpha({G}_{\phi} + {G}_{\varphi} + {G}_{\Phi}) + (\alpha + 1)V \right.\\
&& \hspace{5.1cm}\left. +~ (\alpha-1)({K}_{U(1)} + {G}_{U(1)} + {K}_{SU(2)} + {G}_{SU(2)}) \right\rangle \,,\nonumber
\end{eqnarray}
where $\langle ... \rangle$ represents volume averaging. As long as the volume is sufficiently large compared to the scales excited in the matter fields, this approximation should lead to a well-defined notion of a 'homogeneous and isotropic' expanding background, within the given volume.

\section{Field dynamics in a computer: the lattice approach}
\label{sec:LatticeApproach}

\subsection{Lattice definition and discrete Fourier transform}
\label{subsec:LatticeDefinitionAndDFT}

~~~~ In order to simulate the dynamics of interacting fields, we will consider a cubic lattice with $N$ sites per dimension. As we are interested in three spatial dimensions, the lattice will have therefore $N^3$ points in total, labeled as
\begin{eqnarray}
{\bf n} = (n_1,n_2,n_3),~~~~ {\rm with}~~ n_i = 0,1,...,N-1 \,,~~~i = 1,2,3\,.
\end{eqnarray}
We will often refer to this set of points simply as the lattice, the {\it grid}, or even more colloquially, as the {\it box}. For convenience we define
\begin{eqnarray}\label{eq:unitLatticeVectors}
\hat 1 \equiv (1,0,0)\,, ~~~ \hat 2 \equiv (0,1,0)\,, ~~~\hat 3 \equiv (0,0,1)\,,
\end{eqnarray}
as unit vectors in the lattice, corresponding to positive displacements of length
\begin{eqnarray}
\dx \equiv {L\over N}\,,
\end{eqnarray}
in each of the independent directions in the continuum. The length $\dx$ is referred to as the {\it lattice spacing}.

A continuum function ${\tt f}(\bf x)$ in space is represented by a lattice function $f({\bf n})$, which has the same value as ${\tt f}(\bf x)$ at ${\bf x}={\bf n}\,\dx$. We note that whereas in a flat background, positions $\lbrace \bf x \rbrace$ and their corresponding lattice sites $\lbrace \bf n \rbrace$ represent physical spatial coordinates, in an expanding background they will rather represent {\it comoving} spatial coordinates. Unless specified otherwise, we will always consider {\it periodic boundary conditions} in the three spatial directions, so that $f({\bf n} + \hat \imath N) = f({\bf n})$, $i  = 1,2$ or $3$.

The periodic boundary conditions in coordinate space imply that momenta must be discretized, whereas the discretization of the spatial coordinates implies that any definition of a discrete Fourier transform must be periodic. For each lattice we can then consider always a reciprocal lattice representing $Fourier$ modes, with sites labeled as
\begin{eqnarray}
\tilde{\bf n} = (\tilde n_1, \tilde n_2, \tilde n_3), ~~~~{\rm with}~~
\tilde n_i = -\frac{N}{2}+1, -\frac{N}{2}+2, ... ,-1,0,1, ... , \frac{N}{2} - 1, \frac{N}{2}  \,,~~~ i  = 1,2,3\,.
\end{eqnarray}
We then define the discrete Fourier transform (DFT) as
\begin{eqnarray}\label{eq:FTdiscrete}
f({\bf n}) \equiv {1\over N^3}\sum_{\tilde {\bf n}} e^{-i{2\pi\over N} {\bf \tilde n n}} f({\bf \tilde n}) ~~~~ \Leftrightarrow ~~~~  f({\bf \tilde n}) \equiv \sum_{\bf n} e^{+i{2\pi\over N} {\bf n \tilde n} }f({\bf n})\,,
\end{eqnarray}
and we note the following identity
\begin{eqnarray}\label{eq:FTdiscreteDelta}
\sum_{\bf n} e^{i{2\pi\over N} {\bf n} \tilde {\bf n} } = N^3\delta_{{\bf 0}, \tilde {\bf n}}\,.
\end{eqnarray}
As expected, it follows that Fourier-transformed functions are  periodic in the reciprocal lattice, with periodic boundary conditions as $ f({\bf\tilde{n}} + {\hat \imath} N) =  f({\bf\tilde{n}})$, with ${ \hat \imath}$ analogous unit vectors as in Eq.~(\ref{eq:unitLatticeVectors}), but defined in the reciprocal lattice.

Let us emphasize that from the above discussion, it follows that we can only represent momenta down to a minimum infrared (IR) cut-off
\begin{eqnarray}
k_{\rm IR} = \frac{2\pi}{L} = \frac{2\pi}{N\dx}\,,
\end{eqnarray}
such that $\tilde{\bf n}$ labels the continuum momentum values ${\bf k} = (\tilde n_1, \tilde n_2, \tilde n_3)\, k_{\rm IR}$. Furthermore, there is also a maximum ultraviolet (UV) momentum
that we can capture in each spatial dimension,
\begin{eqnarray}
k_{i,\rm UV} = {N\over2}k_{\rm IR} = {\pi\over \dx}\,.
\end{eqnarray}
The maximum momentum we can capture in a three-dimensional reciprocal lattice is therefore
\begin{eqnarray}
k_{\rm max} = \sqrt{k_{1,\rm UV}^2+k_{2,\rm UV}^2+k_{3,\rm UV}^2} = {\sqrt{3}\over2}Nk_{\rm IR} = \sqrt{3}{\pi\over \dx}\,\,.
\end{eqnarray}

In many situations, it will be useful to define the $power~spectrum$ of ${\tt f}({\bf x})$, characterizing its ensemble average $\langle {\tt f}^2 \rangle$ in the continuum, as
\be
		\langle {\tt f}^2 \rangle = \int d\log k~\Delta_{\tt f}(k)~~, ~~~\Delta_{\tt f}(k) \equiv {k^3\over 2\pi^2}\mathcal{P}_{\tt f}(k)~~,~~~ \langle {\tt f}_{\bf k} {\tt f}_{{\bf k}^{\prime}} \rangle = (2\pi)^3 \mathcal{P}_{\tt f}(k) \delta (\mathbf{k}-\mathbf{k^{\prime}})~. \label{eq:continuumPS}
\ee
In a lattice, the ensemble average is substituted by a volume average,
\be
		\langle f^2 \rangle_V = \frac{\dx^3}{V}\sum_{\bf n} f^2({\bf n}) = \frac{1}{N^3}\sum_{\bf n} f^2({\bf n})~\,,
\ee
so that using the discrete Fourier transform defined above, we obtain
\be
		\langle f^2 \rangle_V = \frac{1}{2\pi^2}\sum_{|\tilde{\bf n}|}\Delta\log k(\tilde{\bf n}) ~k^3(\tilde{\bf n})\left(\frac{\delta x}{N}\right)^3 \big\langle \big|f(\tilde{\bf n})\big|^2\big\rangle_{R(\tilde{\bf n})}~\,,
		\label{eq:discretePSaux}
\ee
with $\langle ( ... ) \rangle_{R(\tilde{\bf n})} \equiv \frac{1}{4\pi|\tilde{\bf n}|^2}\sum_{\tilde{\bf n}^{\prime}\in R(\tilde{\bf n})}( ... )$ an angular average over the spherical shell of radius $\tilde{\bf n}^{\prime}\in \big[|\tilde{\bf n}|,|\tilde{\bf n}|+ \Delta\tilde{n}\big)$, with $\Delta\tilde{n}$ a given radial binning.  We also  defined $\Delta \log k(\tilde{\bf n}) \equiv \frac{k_{IR}}{k(\tilde{\bf n})}$. Identifying $\langle f^2 \rangle_V$ in Eq.~(\ref{eq:discretePSaux}) with $\langle {\tt f}^2 \rangle$ in Eq.~(\ref{eq:continuumPS}), we obtain the following expression for a discrete power spectrum
	\be
		 \Delta_f(k) \equiv \frac{k^3(\tilde {\bf n})}{2\pi^2}\left(\frac{\delta x}{N}\right)^3 \big\langle \big|f(\tilde{\bf n})\big|^2\big\rangle_{R(\tilde{\bf n})}\,,~~~~~ \mathcal{P}_f(k) \equiv \left(\frac{\delta x}{N}\right)^3 \big\langle \big|f(\tilde{\bf n})\big|^2\big\rangle_{R(\tilde{\bf n})}. \label{eq:discretePS}
	\ee

Finally, let us notice that we will be dealing in general with spatially dependent functions representing field amplitudes at a given time. As time goes by in the simulation, the amplitude of the functions will change. We can therefore think of the above functions depending not only on their coordinates $\bf n$ (or reciprocal coordinates $\tilde{\bf n}$), but also depending on a discrete variable $n_0 = 0, 1, 2, ...$ counting the number of time iterations in a simulation. In general, $n_0$ labels a time $\eta = \eta_* + n_0 \delta \eta$, where $\delta \eta$ is the temporal step chosen in the evolution, and $\eta_*$ denotes an initial time. We will think therefore of the above functions as 4-dimensional functions, and we will often write them as $f(n) = f(n_0,{\bf n})$, or $ f(\tilde n) =  f(n_0,\tilde {\bf n})$. We will use the notation $\hat 0$ to represent the advance of one time step, so e.g.~$f(n+\hat 0) = f(n_0+1,{\bf n})$.

\subsection{Lattice representation of differential operators}

\subsubsection{Derivative operators and lattice momenta}

~~~~ Either in an action or in the corresponding EOM, there are always continuum derivatives, so we need to replace these with lattice expressions that have the correct continuum limit, i.e.~that reproduce the continuum operations to some order in the lattice spacing/time step. A simple and symmetric definition of a lattice derivative is e.g.~the \textit{centered} or {\it neutral} derivative
\begin{equation}
\label{eq:neutrald}
[\nabla^{(0)}_\mu f] = \frac{f({n}+\hat\mu) - f({n}-\hat\mu)}{2\dx ^\mu} ~~\longrightarrow ~~ \partial_i{\tt f}({x})\big|_{{x}\,\equiv\, {\bf n}\dx+n_0\deta} + \mathcal{O}(\dx_\mu^2)\,,
\end{equation}
where in the case of spatial derivatives $\dx^\mu$ refers to the lattice spacing $\dx$, whereas for temporal derivatives it refers to the time step $\delta\eta$ (typically bounded to be smaller than $\dx$). The expression to the right-hand side of the arrow indicates where and to what order in the lattice spacing/time step the continuum limit is recovered. The neutral derivative in Eq.~(\ref{eq:neutrald}) has the drawback that it is insensitive to spatial variations at the smallest scale we can probe, $\sim \dx$, or temporal variations within a time of the order of the simulation time step $\sim \deta$. Because of this, a definition involving the nearest spatial/temporal neighbors is typically preferred. A standard way to do this, is to define the {\it forward} and {\it backward} derivatives
\begin{eqnarray}
\label{eq:forwardbackwardd}
[\nabla^\pm_\mu f] = \frac{\pm f({n}\pm \hat\mu) \mp f({n})}{\dx^\mu} ~~\longrightarrow ~~ \left\lbrace\begin{array}{l}
\partial_i{\tt f}({x})\big|_{{x}\,\equiv\, {\bf n}\dx+n_0\deta} + \mathcal{O}(\dx_\mu)\,.  \vspace*{0.2cm}\\
\partial_i{\tt f}({x})\big|_{{x}\,\equiv\, ({n} \pm \hat\mu/2)\dx^\mu} + \mathcal{O}(\dx_\mu^2)\,.
\end{array}\right.\,,
\end{eqnarray}
which recover the continuum limit to linear or to quadratic order in the lattice spacing/time step, depending on whether we interpret that the discrete operator lives in ${n}$ or in between the two lattice sites involved ${n} \pm \hat\mu/2$. This shows that in order to recover a continuum differential operation in the lattice, not only it is important to use a suitable discrete operator, but also to determine where it 'lives'. Depending on this choice, the operator might be symmetric or not with respect to the given location, hence recovering the continuum limit up to an even or an odd order in the lattice spacing/time step, respectively.

To improve accuracy, one could also consider lattice derivatives which involve more points, typically leading to definitions that have a symmetry either around a lattice site or around half-way between lattice sites, see e.g.~\cite{Frolov:2008hy}. Depending on the choice of lattice operator $\nabla_{i}$ for the spatial derivatives, the discrete Fourier transform will lead to different {\it lattice momenta}. In general, for any given derivative operator, the value of the derivative $[\nabla_i f]$ will be a linear combination of the field values at different lattice sites, $\left[\nabla_i f\right]({\bf l}) = \sum_{\mathbf m} D_i({\bf l},{\mathbf m})f({\mathbf m})$, with $D_i({\bf l},{\mathbf m})$ a real-valued function of two variables on the lattice. Since we want the derivative to be invariant under translations, $D_i({\bf l},{\mathbf m})$ can only be a function of the difference ${\bf l}-{\mathbf m}$, i.e.~$D_i({\bf l},{\mathbf m})=D_i({\bf l}-{\mathbf m})$, and we can write
\begin{eqnarray}
\left[\nabla_i f\right]({\bf l}) = \sum_{\mathbf m} D_i({\bf l},{\mathbf m})f({\mathbf m}) = \sum_{\mathbf m} D_i({\bf l}-{\mathbf m})f({\mathbf m})
 = \sum_{{\mathbf m}'} D_i({\mathbf m}')f({\bf l}-{\mathbf m}') \,.
\end{eqnarray}
For example, for the neutral derivative (\ref{eq:neutrald}),
\begin{eqnarray}
D^0_i({\mathbf m}')=\frac{\delta_{{\mathbf m}',-\hat\imath}-\delta_{{\mathbf m}',\hat\imath}}{2\dx}\,,
\end{eqnarray}
whereas for the nearest-neighbor derivative (\ref{eq:forwardbackwardd}),
\begin{eqnarray}
&& D^\pm_i({\mathbf m}') = \frac{\pm\delta_{{\mathbf m}',\mp\hat\imath/2}\mp\delta_{{\mathbf m}',\pm\hat\imath/2}}{\dx}\,,~~~{\rm if} ~{\bf l} = {\bf n} + {\hat{\imath}\over2}\,,\\
&& D^\pm_i({\mathbf m}')= {\pm\delta_{{\mathbf m}',\mp\hat\imath}\mp\delta_{{\mathbf m}',{\bf 0}} \over \dx}\,,~~~{\rm if} ~{\bf l} = {\bf n}\,.
\end{eqnarray}
The Fourier transform of the derivative $[\nabla_if]$  is
\begin{eqnarray}
{\nabla_i f}(\tilde{\bf n}) &=& \sum_{\bf n} e^{\frac{2\pi i}{N}\tilde{\bf n}\cdot{\bf n}}\sum_{\mathbf m} D_i({\bf n}-{\mathbf m})f({\mathbf m}) \\ &=& \sum_{{\bf n}'} e^{\frac{2\pi i}{N}\tilde{\bf n}\cdot{\bf n}'}D_i({\bf n}')\sum_{\mathbf m} e^{\frac{2\pi i}{N}\tilde{\bf n}\cdot{\mathbf m}}f({\mathbf m}) \\
&\equiv& -i{k}_{{\rm Lat},i}(\tilde{\bf n}) f(\tilde{\bf n})\,,
\end{eqnarray}
leading to define the {\it lattice momentum} ${\bf k}_{\rm Lat}(\tilde{\bf n})$ as
\begin{equation}
{\bf k}_{\rm Lat}(\tilde{\bf n})=i\sum_{{\bf n}'} e^{\frac{2\pi i}{N}\tilde{\bf n}\cdot{\bf n}'}\vec{D}({\bf n}').
\end{equation}
Conversely, any function ${\bf k}_{\rm Lat}(\tilde{\bf n})$ with the correct leading behaviour ${\bf k}_{\rm Lat}(\tilde{\bf n}) \approx \tilde{\bf n}\,k_{\rm IR}$ in the IR limit $|\tilde{\bf n}| \ll N$, defines a lattice derivative through the inverse Fourier transform.

In practice, for the neutral derivative (\ref{eq:neutrald}) we obtain
\begin{eqnarray}
\label{eq:neutralMomentum}
k_{{\rm Lat},i}^0=\frac{\sin(2\pi \tilde{n}_i/N)}{\dx}\,,
\end{eqnarray}
whereas for the forward/backward derivatives (\ref{eq:forwardbackwardd}),
\begin{eqnarray}
\label{eq:ForwardBackwardMomentum}
&& k_{{\rm Lat},i}^+ = k_{{\rm Lat},i}^- = 2\frac{\sin(\pi \tilde{n}_i/N)}{\dx}\,,~~~{\rm if} ~{\bf l} = {\bf n} \pm {\hat{\imath}\over2}\,,\\
\label{eq:ForwardBackwardMomentumII}
&& k_{{\rm Lat},i}^\pm = \frac{\sin(2\pi \tilde{n}_i/N)}{\dx} \pm i
\frac{1-\cos(2\pi \tilde{n}_i/N)}{\dx}\,,~~~{\rm if} ~{\bf l} = {\bf n}\,.
\end{eqnarray}
We note that for anti-symmetric lattice derivatives with $D_{i}(-{\bf m}')=-D_i({\bf m}')$, the lattice momentum ${\bf k}_{\rm Lat}$ must be real.

\subsubsection{Lattice gauge invariant techniques}
\label{subsubsec:latgaugeinvtech}

~~~~~Discretizing a gauge theory requires a special care in order to preserve gauge invariance at the lattice level. It is not enough to recover gauge invariance in the continuum, sending the lattice spacing/time step to zero, as gauge invariance is meant to remove spurious transverse degrees of freedom.  If we were to discretize a gauge theory substituting all ordinary derivatives in the continuum EOM by finite differences like those in Eqs.~(\ref{eq:neutrald}), (\ref{eq:forwardbackwardd}), the gauge symmetry would not be preserved and the spurious degrees of freedom would propagate in the lattice.

In order to understand this, let us consider the simplest possible case of a gauge theory, say an Abelian-Higgs model in flat space-time, with Lagrangian $-\mathcal{L} =$ $(\partial_\mu + ieA_\mu)\varphi^*(\partial^\mu - ieA^\mu)\varphi$ + $\frac{1}{4}F_{\mu\nu}F^{\mu\nu}$ + $V(\varphi^*\varphi)$, where $e=g_AQ_A$ in the notation of Sec.~\ref{sec:ContinuumFldDynamics}. This system is invariant under continuum gauge transformations $\varphi(x)\,\rightarrow\,e^{-ie\alpha(x)}\varphi(x)$, $A_\mu(x)\,\rightarrow\,A_\mu(x) - \partial_\mu\alpha(x)$ simply because the transformation of $\partial_\mu(e^{-ie\alpha(x)}\varphi(x))$ leads to a term $-ie\partial_\mu\alpha(x)e^{-ie\alpha(x)}\varphi(x)$, whereas the transformation of the gauge field in $-ieA_\mu e^{-ie\alpha(x)}$ leads to a term identical but of opposite sign, $+ie\partial_\mu\alpha(x)e^{-ie\alpha(x)}\varphi(x)$, which cancels out with the previous one. However, if we were to discretize the system by simply promoting continuum derivatives into finite differences, say replacing $\partial_\mu f(x)$ by $\Delta^+_\mu f(x)$, then $\Delta^+_\mu(e^{-ie\alpha(x)}\varphi(x)) \neq -i(e\Delta^+_\mu\alpha(x))e^{-ie\alpha(x)}\varphi(x) + e^{-ie\alpha(x)}\Delta^+_\mu\varphi(x)$, and hence the transformation of the field derivative would not produce a term to compensate the contribution from the gauge field transformation $A_\mu \,\rightarrow\,A_\mu(x) - \Delta^+_\mu\alpha(x)$. The
reason is simple, the \textit{Leibniz} rule $(fg)' = f'g + fg'$ does not hold for finite difference operators. The situation is no different in non-Abelian gauge theories.

\textit{How can we restore gauge invariance in the lattice?} To mimic a continuum gauge theory in the lattice, we must adopt a special discretization procedure that preserves some sort of discretized version of gauge transformations. Lattice gauge invariance is actually necessary in order to preserve constraints that follow from the EOM, in particular the Gauss laws. In order to introduce a general formalism valid for gauge theories (either Abelian or non-Abelian), let us consider the more general case of a $SU(N)$ gauge invariant theory. We introduce then a {\it parallel transporter}, connecting two points in space-time
\begin{eqnarray}
U(x,y) &=& Pexp\left\lbrace-ie\int_{x}^{y}dx^{\mu}A_\mu  \right\rbrace\,, 
\end{eqnarray}
where $Pexp\lbrace...\rbrace$ means \textit{path-ordered} along the trajectory. The crucial observation is that under a gauge transformation of the gauge fields, recall Eq.~(\ref{eq:GaugeTrans}),
the parallel transporter transforms as
\begin{eqnarray}\label{eq:LinkTrans}
U(x,y) \rightarrow \Omega(x)U(x,y)\Omega^\dag(y)\,,
\end{eqnarray}
which in the Abelian case reduces simply to $U(x,y) \rightarrow U(x,y)e^{-ie(\alpha(x)-\alpha(y))}$. Therefore, according to Eq.~(\ref{eq:LinkTrans}), a parallel transporter transforms exactly as the field strength transforms for $x = y$, see Eq.~(\ref{eq:GaugeTransII}). Thus, considering the  {\it minimal connector} between two space-time sites separated only by one lattice spacing/time step, $x(n) \equiv \bn \dx + n_0\delta t, x(n+\hat\mu) \equiv \bn \dx + n_0\delta t + \dx_\mu$, we define the \textit{link} variables as
\begin{eqnarray}
U_{0,n} \equiv Pexp\left\lbrace-ie\int_{x(n)}^{x(n+\hat0)}dt'A_0 \right\rbrace \approx  e^{-ie\delta t A_0}\,, ~~~~ U_{i,n} \equiv Pexp\left\lbrace-ie\int_{x(n)}^{x(n+\hat\imath)} dxA_i \right\rbrace \approx e^{-ie\dx A_i}\,,
\end{eqnarray}
where the gauge field, and hence the link, is considered to live in the point $n + {\hat\mu\over2}$. We also define $U_{-\mu,n} = U_{\mu,n-\mu}^\dag \equiv U_\mu^\dag(n-{1\over 2}\hat\mu)$. Before we continue, it will useful at this point to establish some conventions to simplify the notation of upcoming expressions.\vspace*{0.5cm}

{\it Convention -.} From now on, unless stated otherwise, a scalar field living in a generic lattice site $n = (n_o,\bn) = (n_o,n_1,n_2,n_3)$, i.e.~$\varphi_n = \varphi(n)$, will be simply denoted as $\varphi$. If the point is displaced in the $\mu-$direction by one unit lattice spacing/time step, $n + \hat\mu$, we will then use the notation $n+\mu$ or simply $+\mu$ to indicate this, so that the field amplitude in the new point is expressed as $\varphi_{+\mu} \equiv \varphi(n+\hat\mu)$. In the case of gauge fields, whenever represented explicitly in the lattice, we will automatically understand that they live in the middle of lattice points, i.e.~$A_{\mu} \equiv A_{\mu}(n+{1\over2}\hat\mu)$. It follows then that e.g.~$A_{\mu,+\nu} \equiv A_{\mu}\big(n + {1\over2}\hat\mu +  \hat\nu\big)$. In the case of links, we will use the notation $U_\mu \equiv U_{\mu,n} \equiv U_\mu(n+{1\over2}\hat\mu)$, and hence $U_{\mu,\pm\nu} = U_{\mu,n\pm\nu} \equiv U_\mu(n + {1\over2}\hat\mu \pm \hat\nu)$. Even though the lattice spacing $\dx$ and the time step $\delta t$ do not need to be equal, we will loosely speak of corrections of order $\mathcal{O}(\dx)$, independently of whether we are referring to the lattice spacing or the time step (the latter is actually always forced to be smaller than the former). In lattice expressions we will never consider summation over repeated indices. \vspace*{0.5cm}

\noindent In the continuum limit, we recover the gauge fields simply from
\begin{eqnarray}\label{Ap:gaugeField}
\frac{1}{e}{(\mathcal{I}- U_{\mu,n})\over i\delta x^\mu} ~~\longrightarrow ~~ A_\mu\big(n+{1\over2}\hat\mu\big) + \mathcal{O}(\dx)\,.
\end{eqnarray}
It turns out that we can actually build the action or EOM for any gauge theory, preserving a discretized version of the gauge symmetry, using only link variables and no gauge fields. That is known as the {\it compact formulation} of lattice gauge theories, which can be applied to both Abelian and non-Abelian gauge theories. Actually, in the case of non-Abelian theories, compact formulations are the only way to discretize them while respecting gauge invariance in the lattice. In Abelian gauge theories, however, it is still possible to make use of an explicit representation of the  gauge fields, in the so called {\it non-compact formulation}. In order to describe these formulations, we introduce below standard definitions of $links$, $plaquettes$ and {\it lattice covariant derivatives} (setting back $e=g_A Q_A$), specialized to both Abelian and non-Abelian gauge groups. These are simple objects in the lattice which suffice for writing down lattice actions or EOM invariant under (a discretized version) of gauge symmetries, see e.g.~\cite{Gattringer:2010zz}. We provide {\it toolkits} for Abelian and non-Abelian gauge theories, containing each the basic definitions of links, plaquettes and covariant derivatives for $U(1)$ and $SU(N)$ theories respectively, together with useful approximations, expressions and properties, that can be used to understand straightforwardly how a lattice action or a set of EOM approximate their continuum counterpart expressions (in the case of Abelian theories we include both compact and non-compact lattice formulations):\\

\begin{mdframed}
\begingroup
\allowdisplaybreaks
\begin{center}
\vspace*{0.5cm}
----- U(1) toolkit -----
\end{center}
\begin{eqnarray}
&&{\rm Links:}~ V_{\mu} \equiv e^{-i g_AQ_A \dx_{\mu} A_{\mu}} = \cos(g_AQ_A\dx_{\mu} A_{\mu}) - i \sin (g_AQ_A\dx_{\mu} A_{\mu}) ;~~~ V_{- \mu} \equiv V_{\mu,-\mu}^* ;~~~ V_{\mu}^* V_{\mu} = 1\,;\vspace*{0.4cm}\nonumber\\
&&{\rm Plaquettes}:~ V_{\mu \nu} \equiv V_{\mu} V_{\mu,+\mu} V_{\mu, +\nu}^* V_\nu^* \simeq e^{-i g_AQ_A\dx_{\mu} \dx_{\nu} [ F_{\mu \nu} + \mathcal{O}(\dx)] };~~~ V_{\mu\nu}^* = V_{\nu\mu}\,;
\vspace*{0.4cm}\nonumber\\
&&{\rm Covariant~Derivs.}:  (D_{\mu}^\pm\varphi)({\bf l}) = \pm{1\over \dx^\mu}(V_{\pm\mu}\varphi_{\pm\mu} - \varphi)\,,~~{\bf l} = {\bf n} \pm {1\over2}{\hat\mu}\,
\vspace*{0.6cm}\nonumber\\
&&{\rm Expansions}:
\left\lbrace
\begin{array}{rcl}
(D_{\mu}^\pm\varphi)({\bf l})  & \longrightarrow & (D_{\mu}\varphi)({\bf l}) + \mathcal{O}(\dx^2)\,~~{\bf l} = {\bf n} \pm {1\over2}{\hat\mu}\vspace*{0.2cm}\\
\mathcal{R}e\lbrace V_{\mu \nu} \rbrace  & \longrightarrow & 1 - \frac{1}{2} \dx_{\mu}^2 \dx_{\nu}^2g_A^2Q_A^2 F_{\mu \nu}^2 + \mathcal{O}(\delta x^5)\,,~~{\bf l} = {\bf n} + {1\over2}{\hat\mu} + {1\over2}{\hat\nu}\vspace*{0.2cm}\\ \mathcal{I}m\lbrace V_{\mu \nu} \rbrace & \longrightarrow & - \dx_{\mu} \dx_{\nu} g_AQ_AF_{\mu \nu} + \mathcal{O}(\delta x^3)\,,~~{\bf l} = {\bf n} + {1\over2}{\hat\mu} + {1\over2}{\hat\nu}
\end{array}\right.
\vspace*{0.6cm}\\
&&{\rm Expressions}:
\left\lbrace
\begin{array}{l}
\left.
\begin{array}{l}
\sum_n {1\over 4}F_{\mu \nu}^2 \cong -{1\over 2}\sum_n{\mathcal{R}e\lbrace V_{\mu \nu} \rbrace \over \dx_{\mu}^2 \dx_{\nu}^2g_A^2Q_A^2} = -{1\over4}\sum_n {(V_{\mu \nu}+V_{\mu \nu}^*)\over \dx_{\mu}^2 \dx_{\nu}^2g_A^2Q_A^2} + \mathcal{O}(\dx^2)\vspace*{0.2cm}\\
\sum_n {1\over4}F_{\mu \nu}^2 \simeq \sum_n {1\over4}{\mathcal{I}m^2\lbrace V_{\mu \nu} \rbrace \over \dx_{\mu}^2 \dx_{\nu}^2g_A^2Q_A^2} = -\sum_n {1\over4}{(V_{\mu \nu}-V_{\mu \nu}^*)^2\over \dx_{\mu}^2 \dx_{\nu}^2g_A^2Q_A^2} + \mathcal{O}(\dx^2)
\end{array}\right]~~({\tt Compact})\vspace*{0.5cm}\\
\left.
\begin{array}{l}
\sum_n {1\over4}F_{\mu \nu}^2 \simeq {1\over4}\sum_n (\Delta^+_\mu A_\nu - \Delta^+_\nu A_\mu)^2 + \mathcal{O}(\dx^2)
\end{array}\right]~~({\tt Non-Compact})
\end{array}\right.
\vspace*{0.6cm}\nonumber\\
&&{\rm Gauge~Trans.}:
\left.
\left\lbrace
\begin{array}{cll}
\phi  & \longrightarrow & e^{+ig_AQ_A\alpha}\phi\vspace*{0.2cm}\\
A_\mu & \longrightarrow & A_\mu - \Delta_\mu^+\alpha\vspace*{0.2cm}\\
V_{\pm \mu}  & \longrightarrow & V_{\pm \mu}e^{ig_AQ_A(\alpha_{\pm\mu}-\alpha)}
\end{array}\right.
\right] ~~\Longrightarrow ~~ \left\lbrace
\begin{array}{cll}
D_\mu^\pm\phi & \longrightarrow & e^{ig_AQ_A\alpha}(D_\mu^\pm\phi)\vspace*{0.2cm}\\
V_{\mu\nu}  & \longrightarrow & V_{\mu\nu} ~{\rm (gauge~inv.)}
\end{array}\right.
\nonumber
\label{eq:U1toolkit}
\end{eqnarray}
\endgroup
\\ \end{mdframed}
\bigskip
\bigskip
\bigskip
\bigskip
\bigskip

\begin{mdframed}
\begin{center}
\vspace*{0.5cm}
----- SU(N) toolkit -----
\end{center}
\begin{eqnarray}
&&{\rm Links}:~ U_{\mu} \equiv e^{-i g_B Q_B \dx B_\mu} = e^{-i g_B Q_B \dx B_{\mu}^a T_a} ;~~~ U_{- \mu} \equiv U_{\mu,-\mu}^{\dagger} ;~~~ U_{\mu}^{\dagger} U_{\mu} = \mathcal{I} \vspace*{0.4cm}\nonumber\\
&&{\rm Plaquettes}:~ U_{\mu \nu} \equiv U_{\mu} U_{\nu,+ \mu} U_{\mu, +\nu}^{\dagger} U_{\nu}^{\dagger} \simeq e^{-ig_B Q_B \dx_{\mu} \dx_{\nu} [ G_{\mu \nu}^a T_a + \mathcal{O} (\dx_{\mu} ) ] }\,; ~~~ U_{\mu \nu}^\dag = U_{\nu\mu} \vspace*{0.4cm}\nonumber\\
&&{\rm Covariant~Derivs.}:  (D_{\mu}^\pm\Phi)({\bf l}) = \pm{1\over \dx^\mu}(U_{\pm\mu}\Phi_{\pm\mu} - \Phi) ~\longrightarrow~ (D_{\mu}\Phi)({\bf l}) + \mathcal{O}(\dx^2),~~{\bf l} = {\bf n} \pm {1\over2}{\hat\mu}
\nonumber\vspace*{0.4cm}\\
&&{\rm Expansions}:
\left\lbrace
\begin{array}{ccl}
(D_{\mu}^\pm\Phi)({\bf l}) & \longrightarrow & (D_{\mu}\Phi)({\bf l}) + \mathcal{O}(\dx^2)\,,~~{\bf l} = {\bf n} \pm {1\over2}{\hat\mu}\vspace*{0.2cm}\\
(U_{\mu \nu} - U_{\mu \nu}^\dag ) & \longrightarrow & -2ig_B Q_B\delta x_{\mu} \delta x_{\nu}G_{\mu \nu} + \mathcal{O} (\dx_{\mu}^3)\,,~~{\bf l} = {\bf n} + {1\over2}{\hat\mu} + {1\over2}{\hat\nu} \vspace*{0.2cm}\\ {\rm Tr} [ U_{\mu \nu} ] & \longrightarrow & 2 - \frac{\dx_{\mu}^2 \dx_{\nu}^2g_B^2 Q_B^2}{4}\sum_a (G_{\mu \nu}^a)^2 + \mathcal{O} (\dx_{\mu}^5)\,,~~{\bf l} = {\bf n} + {1\over2}{\hat\mu} + {1\over2}{\hat\nu}
\end{array}\label{eq:SU2toolkit}
\right.
\vspace*{0.6cm}\\
&&{\rm Expressions}:
\left\lbrace
\begin{array}{l}
{1\over2}{\rm Tr}[G_{\mu\nu}G^{\mu\nu}] = {1\over4}\sum_a (G_{\mu\nu}^a)^2 \cong -{{\rm Tr} [ U_{\mu \nu} ]\over \dx_{\mu}^2 \dx_{\nu}^2 g_B^2 Q_B^2} + \mathcal{O}(\dx^2)\,, \vspace*{0.2cm}\\
G_{\mu \nu} = G_{\mu \nu}^aT_a \simeq \frac{i}{2\dx_{\mu} \dx_{\nu}g_B Q_B} (U_{\mu \nu} - U_{\mu \nu}^\dag) + \mathcal{O}(\dx^2) \,,\vspace*{0.2cm}\\
G_{\mu \nu}^a \simeq \frac{1}{\dx_{\mu} \dx_{\nu}g_B Q_B} {\rm Tr} [ (i T_a)  (U_{\mu \nu} - U_{\mu \nu}^\dag ) ] + \mathcal{O}(\dx^2)
\end{array}\right.
\vspace*{0.4cm}\nonumber\\
&&{\rm Gauge~Trans.}:
\left.
\left\lbrace
\begin{array}{cll}
\Phi  & \longrightarrow & \Omega\,\Phi\,,~~~ \Omega \equiv e^{+ig_B Q_B\alpha_aT_a}\vspace*{0.2cm}\\
U_{\pm \mu}  & \longrightarrow & \Omega \,U_{\pm \mu}\,\Omega^\dag_{\pm \mu}
\end{array}\right.
\right] ~~\Longrightarrow ~~ \left\lbrace
\begin{array}{cll}
D_\mu^\pm\Phi & \longrightarrow & \Omega\,(D_\mu^\pm\Phi)\vspace*{0.2cm}\\
U_{\mu\nu}  & \longrightarrow & \Omega \,U_{\mu\nu}\,\Omega^\dag
\vspace*{0.2cm}\\
{\rm Tr}\lbrace U_{\mu\nu} \rbrace  & \longrightarrow & {\rm Tr}\lbrace U_{\mu\nu} \rbrace
\end{array}\right.
\label{eq:SUNtoolkit}
\nonumber
\end{eqnarray}
\\ \end{mdframed}
\medskip

\noindent In the of case of $SU(2)$, any element can be written as
\begin{eqnarray}
U_{\mu} = c_{\mu 0} \mathcal{I} + \sum_{a=1}^3 i c_{\mu a} \sigma_a = \sum_{\nu=0}^3 c_{\mu\nu} \bar{\sigma}_a  \,,  \hspace*{0.5cm} \bar{\sigma}_a \equiv (1, i\vec{\sigma})\,, \hspace*{0.5cm} \sum_{\nu =0}^3 c_{\mu \nu}^2 = 1 \ , \label{eq:plaq-descomp}
\end{eqnarray}
or, in matrix form (in the gauge $U_0 = \mathcal{I}$)
\begin{eqnarray}
U_i \equiv U_i (c_{i0},c_{i1},c_{i2},c_{i3}) =
\left(\begin{array}{cc}
c_{i0} + i c_{i3}  &  c_{i2} + i c_{i1}\\
- c_{i2} + i c_{i1}  &  c_{i0} - i c_{i3}
\end{array}\right)\,, \hspace{0.3cm} U_i^{\dagger} = U_i (c_{i0}, -c_{i1}, -c_{i2}, -c_{i3} ) \,.\label{eq:plaq-descomp-mat} 
\end{eqnarray}
Useful expressions for the electric and magnetic fields are
\begin{eqnarray}
\mathcal{E}_i^a &=& G_{0i}^a \approx \frac{1}{\delta t \dx g_B Q_B} {\rm Tr} [ (i T_a) (U_{0i} - U_{i0} ) ]  \,, 
\\
\mathcal{B}_i^a &=& \frac{1}{2} \epsilon_{ijk} G_{jk}^a \approx \frac{\epsilon_{ijk}}{2 \dx^2 g_B Q_B} {\rm Tr} [(i T_a) (U_{jk} - U_{kj} ) ] \ .
\end{eqnarray}

\subsection{Evolution algorithms}

~~~~~ Solving the field dynamics in an expanding background in a lattice amounts to writing some appropriate discrete version of the continuum EOM [Eqs.~(\ref{eq:singlet-eom})-(\ref{eq:higgsSU2-eom}) for scalar fields, Eqs.~(\ref{eq:U1eom})-(\ref{eq:GaussSU2-eom}) for gauge fields, and Eqs.~(\ref{eq:FriedmannHubble})-(\ref{eq:FriedmannDDa}) for the scale factor], and then iterate the discrete EOM for a finite number of time steps. In general we will have to follow, on each spatial lattice site, the evolution of a number of $dof$ represented by real field amplitudes, say one per singlet, two per complex field, four per doublet, etc, as well as the Lorentz components of each gauge field considered. Let us denote these $dof$ collectively as the {\it field amplitudes} $\lbrace f_j \rbrace$, with $j$ some index labeling all the real field amplitudes involved in a given scenario, and $\lbrace \pi_j \rbrace$ their {\it conjugate momenta}. As the scale factor is only a homogeneous $dof$ (sourced by the volume averaged energy and pressure densities built from the matter fields), we will not include it in the previous numbered list of $dof$'s, and we will rather treat it as a separate variable $a(\eta)$, with conjugate momenta $\pi_a \equiv a'(\eta)$. For example, in a theory with two singlet scalar fields, say $\phi$ and $\chi$, and self-consistent expansion, we can consider $\lbrace f_1,f_2 \rbrace \equiv \lbrace \phi(x), \chi(x) \rbrace$ and $\lbrace \pi_1,\pi_2 \rbrace \equiv \lbrace \phi'(x), \chi'(x)\rbrace$, and then separately $a(\eta)$ and $\pi_a(\eta) = a'(\eta)$. Looking at the EOM in the continuum Eqs.~(\ref{eq:singlet-eom})-(\ref{eq:SU2eom}) and the scale factor Eqs.~(\ref{eq:FriedmannHubble})-(\ref{eq:FriedmannDDa}), we note the following  structure in the system of equations (independently of the nature of the fields involved),
\begin{eqnarray}\label{eq:SchemeContVirgin1}
\pi_a(\eta) &=& a'(\eta)\,,\\
\label{eq:SchemeContVirgin2}
\pi_a'(\eta) &=& \mathcal{K}_a[a(\eta), E_V(\eta), E_K(\eta), E_G(\eta)]\,,\\
\label{eq:SchemeContVirgin3}
\pi_i({\bf x},\eta) &=& \mathcal{D}_i[f_i'({\bf x},\eta),a(\eta),\pi_a(\eta);\lbrace f_{j}({\bf x},\eta) \rbrace, \lbrace f'_{j\neq i}({\bf x},\eta) \rbrace]\,,\\
\label{eq:SchemeContVirgin4}
\pi_i'({\bf x},\eta) &=& \mathcal{K}_i[f_i({\bf x},\eta),\pi_i({\bf x},\eta),a(\eta),\pi_a(\eta);\lbrace f_{j\neq i}({\bf x},\eta) \rbrace, \lbrace \pi_{j\neq i}({\bf x},\eta) \rbrace]\,,
\end{eqnarray}
where $\mathcal{D}_i[...]$ is a functional -- the {\it drift} -- that defines the conjugate momentum of the $i$th $dof$, $\mathcal{K}_i[...]$ is another functional -- the {\it kernel} or {\it kick} --, that determines the interactions of the $i$th $dof$ with the rest of $dof's$ (possibly including itself), and finally $\mathcal{K}_a[...]$ is given by the square root of the $rhs$ of Eq.~(\ref{eq:FriedmannDDa}), based on the volume averages $\langle ... \rangle$ of the different $dof$ contributions to the potential, kinetic and gradient energy densities, $E_V \equiv \langle V \rangle$, $E_{K} \equiv \langle  \sum_j K_{j}\rangle $ and $E_{G} \equiv \langle \sum_j  G_{j}\rangle$.

For canonical kinetic terms, the drift $\mathcal{D}_i$ actually does not depend on any field amplitude $f_{j}({\bf x},\eta)$, and whereas it does depend on $f_i'$, it will not depend on any other derivative $f'_{j\neq i}$. Note that we have separated within the argument of each kernel $\mathcal{K}_i$, the amplitude and momentum of the $i$th $dof$ itself, from the amplitudes and momenta of the rest of $dof$'s. The latter act in fact merely as 'instantaneous' parameters for an infinitesimal evolution of the i$th$ $dof$. Besides, without loss of generality, we can actually consider that $\mathcal{K}_i$ does not depend on the conjugate momenta $\lbrace \pi_{i\neq j}\rbrace$ of other $dof$'s\footnote{This is however not true for general non-canonically normalized kinetic theories, but we are not concerned with those here.}. Thus, in theories with canonically normalized kinetic terms, we will only care about the dependence of the kernel $\mathcal{K}_i$ on $f_i$ and $\pi_i$. Furthermore, we will encounter often that the time derivative $\pi_i'$ of a given $dof$ can (and often will) depend on its amplitude $f_i$, but not on $\pi_i$ itself. This is actually not a physical requisite, but rather a mathematical requisite that we will seek. In fact, the EOM in the continuum as written so far, rather lead to kernels $\mathcal{K}_{\rm i}$ that depend on $\pi_i$, see the friction terms in Eqs.~(\ref{eq:singlet-eom})-(\ref{eq:SU2eom}). However, from the point of view of the stability of the numerical algorithms used to solve the discrete EOM, it will be convenient to define appropriate conjugate momenta (or alternatively to massage appropriately the EOM), so that one arrives into effective kernels $\mathcal{K}_i$ that do not depend on $\pi_i$. We will see later how to do this in a case by case basis. For the time being, we will consider that this condition has been achieved, implicitly assuming that pertinent conjugate momenta definitions or EOM manipulations have been made to grant it. Taking into account all the above considerations, the typical system of equations we will want to solve (for a theory with canonical kinetic terms), looks as follows
\begin{eqnarray}\label{eq:SchemeContI}
\pi_a(\eta) &=& a'(\eta)\,,\\
\label{eq:SchemeContII}
\pi_a'(\eta) &=& \mathcal{K}_a[a(\eta), E_V(\eta), E_K(\eta), E_G(\eta)]\,,\\
\label{eq:SchemeContIII}
\pi_i({\bf x},\eta) &=& \mathcal{D}_i[f_i'({\bf x},\eta),a(\eta);...]\,,\\
\label{eq:SchemeContIV}
\pi_i'({\bf x},\eta) &=& \mathcal{K}_i[f_i({\bf x},\eta),a(\eta),\pi_a(\eta); ...]\,.
\end{eqnarray}
Let us note that, although any possible dependence of the drift $\mathcal{D}_i$ on $\pi_a$ would not pose a problem to the algorithms presented below, in practice we do not know of any theory that produces such dependence, so we have removed it as an explicit argument from $\mathcal{D}_i$. A discrete version of the EOM will then have a scheme similar to
\begin{eqnarray}\label{eq:SchemeDiscreteI}
\pi_a(\eta) &=& \Delta_0 a(\eta)\,,\\
\label{eq:SchemeDiscreteII}
\Delta_0 \pi_a &=& \mathcal{K}_a[a(\eta), E_V(\eta), E_K(\eta), E_G(\eta)]\,,\\
\label{eq:SchemeDiscreteIII}
\pi_i({\bf x},\eta) &=& \mathcal{D}_i[\Delta_0 f_i({\bf x},\eta),a(\eta);...]\,,\\
\label{eq:SchemeDiscreteIV}
\Delta_0\pi_i({\bf x},\eta) &=& \mathcal{K}_i[f_i({\bf x},\eta),a(\eta),\pi_a(\eta); ...]\,,
\end{eqnarray}
with $\Delta_0$ some discrete operator mimicking continuum time derivatives. As we will see in a moment, introducing time operators as simple as
\begin{eqnarray}
(\Delta_{r\hat0}^{\pm} f) = \frac{\pm f(n\pm r\delta t) \mp f(n)}{r\delta t} ~~~\longrightarrow ~~~
\left\lbrace
\begin{array}{ll}
(\Delta_{\hat 0}^{+} f) = \frac{f(n + \delta t) - f(n)}{\delta t} & ,\,\text{Standard Forward Deriv.}\vspace*{0.1cm}\\
(\Delta_{\hat 0}^{-} f) = \frac{f(n)-f(n - \delta t)}{\delta t} & ,\,\text{Standard Backward Deriv.}\vspace*{0.1cm}\\
(\Delta_{\hat0/2}^{+} f) = \frac{f(n + \delta t/2) - f(n)}{(\delta t/2)} & ,\,+{1\over2}~\text{Forward Deriv.}\vspace*{0.1cm}\\
(\Delta_{\hat0/2}^{-} f) = \frac{f(n) - f(n - \delta t/2)}{(\delta t/2)} & ,\,-{1\over2}~\text{Backward Deriv.}
\end{array}\right.
\end{eqnarray}
will actually enable us to address all basic algorithms to iterate coupled finite difference equations like (\ref{eq:SchemeDiscreteI})-(\ref{eq:SchemeDiscreteIV}), mimicking continuum coupled differential equations like (\ref{eq:SchemeContI})-(\ref{eq:SchemeContIV}). For simplicity, the $r = 1$ case corresponding to standard forward/backward derivative operators, will be often simply written as $(\Delta_0^{\pm} f)$, instead of $(\Delta_{\hat 0}^{\pm} f)$.

\subsubsection{(Staggered) Leapfrog}
\label{sec:staggered_LF}

One of the simplest methods for solving second order differential equations is the {\it leapfrog} algorithm. Let us illustrate it by solving a simple one-dimensional problem, consisting in one $dof$ $x(t)$ that depends only on a time variable $t$, with EOM
\begin{eqnarray}
\ddot x(t) = \mathcal{K}[x(t)]\,,
\end{eqnarray}
where $\mathcal{K}[...]$ is the kernel of $x(t)$. Taylor expanding the position at the next step we obtain
\begin{eqnarray}
x(t+\delta t) = x(t) + \dot x(t)\delta t + {1\over2}\mathcal{K}[x(t)]\delta t^2 + ... 
\equiv x(t) + \dot x(t+{\delta t/2})\delta t + ...\,, 
\end{eqnarray}
where in the second equality we have substituted the Taylor expansion of the velocity at half time step
\begin{eqnarray}
\dot x(t+\delta t/2) = \dot x(t) + {\delta t\over 2}\mathcal{K}[x(t)] + ... 
\equiv \dot x(t-\delta t/2) + \mathcal{K}[x(t)]\delta t + ...\,,
\end{eqnarray}
and where we have used that $\dot x(t) = \dot x(t-\delta t /2) + \mathcal{K}[x(t-\delta t/2)]\delta t + \mathcal{O}(\delta t^2)$ and $\mathcal{K}[x(t-\delta t/2)]\delta t + \mathcal{O}(\delta t^2) = \mathcal{K}[x(t)]\delta t + \mathcal{O}(\delta t^2)$. Applying recursively the above relations between velocity and position, we obtain
\begin{eqnarray}
x(t) &=& x(t - \delta t) + \dot x(t-\delta t/2)\delta t\,,\\
\dot x(t + \delta t/2) &=& \dot x(t - \delta t/2) + \mathcal{K}[x(t)]\delta t\,,\\
x(t + \delta t) &=& x(t) + \dot x(t+\delta t/2)\delta t\,,\\
\dot x(t + 3\delta t/2) &=& \dot x(t + \delta t/2) + \mathcal{K}[x(t+\delta t)]\delta t\,,\\
&....&\nonumber
\end{eqnarray}
The leapfrog method has an accuracy of order $\mathcal{O}(\delta t^2)$,
because each step advances $x$ or $\dot x$ in terms of its derivative at the middle of the step. This is better than a simpler {\it Euler} method, which has $\mathcal{O}(\delta t)$ accuracy. This can be demonstrated by simply noting the accuracy of the derivative expressions $(x(t+\delta t)- x(t))/\delta t \simeq \dot x(t+\delta t/2) + \mathcal{O}(\delta t^2)$ and $(\dot x(t+\delta t/2)- \dot x(t-\delta t/2))/\delta t \simeq \ddot x(t) + \mathcal{O}(\delta t^2)$. Let us label the initial time as $t_0$, and start with initial conditions $x_0 \equiv x(t_0)$ and $\dot x_0 \equiv \dot x(t_0)$. We can obtain first $\dot x(t_0 + \delta t/2) = \dot x_0 + 1/2 \mathcal{K}[x_0]\delta t$, and from then on, iterate as follows: $(x(t_0),\dot x(t_0+\delta t/2))$  $\longrightarrow$ $(x(t_1),\dot x(t_1+\delta t/2))$ $\longrightarrow ~~...~~ \longrightarrow$ $(x(t_n),\dot x(t_n+\delta t/2))$, with $t_n \equiv t_0 + n\delta t$, and $n$ the number of iterations.

In terms of the previously introduced time derivative operators, we can simply write the algorithm as
\begin{eqnarray}\label{eq:LeapFrogEq1}
\Delta_0^+x_n &=& \pi_{n+1/2}\,,\\
\label{eq:LeapFrogEq2}
\Delta_0^+\pi_{n+1/2} &=& \mathcal{K}[x_{n+1}]\,,
\end{eqnarray}
so that the first line defines the conjugate momentum $\pi_{n+1/2}$, and we understand that $x_n$ lives at 'integer' times $t_n \equiv t_0 + n\delta t$, whereas $\pi$ lives at 'semi-integer' times $t_{n+1/2} \equiv t_n + \delta t/2 = t_0 + (n+1/2)\delta t$. Correspondingly, this implies that $\Delta_0^+x_n \equiv (x_{n+1}-x_n)\delta t$ lives at $t_{n+1/2}$, and $\Delta_0^+\pi_{n+1/2} \equiv (\pi_{n+3/2}-x_{n+1/2})\delta t$ lives at $t_{n+1}$. Due to the fact that variable amplitudes live at integer times and conjugate momenta live at semi-integer times, sometimes this method is referred to as the 'staggered' leapfrog algorithm. We will content ourselves with simply referring to it as the leapfrog algorithm.

The leapfrog method, encapsulated in Eqs.~(\ref{eq:LeapFrogEq1})-(\ref{eq:LeapFrogEq2}) can be extended readily to multiple $dof$, simply labeling them with some index as $x^i_n$ and $\pi_{n+1/2}^i$, with $i = 1, 2, 3, ..., N_f$ counting the number of $dof$. Namely
\begin{eqnarray}\label{eq:LeapFrogMultiEq1}
\Delta_0^+x_n^i &=& \pi_{n+1/2}^i\,,\\
\label{eq:LeapFrogMultiEq2}
\Delta_0^+\pi_{n+1/2}^i &=& \mathcal{K}_i\left[x_{n+1}^i,\lbrace x_{n+1}^{j\neq i} \rbrace\right]\,,
\end{eqnarray}
where the kernels $\mathcal{K}_i$ represent the interaction of the $i$th $dof$ $x^i$ with the rest of $dof$s $\lbrace x^{j\neq i} \rbrace$, and possibly with itself. This method is however only applicable to ${\it conservative}$ forces\footnote{In reality nothing prevents you from applying it to {\it non-conservative} forces with $\mathcal{K} = \mathcal{K}[x(t),\dot x(t)]$, but then its stability properties and its $\mathcal{O}(\delta t^2)$ accuracy will be lost.}, i.e.~to systems with kernels that only depend on amplitude variables $\mathcal{K} \equiv \mathcal{K}[\lbrace x^i(t) \rbrace]$. This method can be therefore applied readily to our field theory EOM~(\ref{eq:SchemeDiscreteI})-(\ref{eq:SchemeDiscreteIV}) in a flat space-time background. If the expansion of the universe is switched off, i.e.~$a = 1$ and $\dot a = \ddot a = 0$, we can ignore the first two Eqs.~(\ref{eq:SchemeDiscreteI})-(\ref{eq:SchemeDiscreteII}) and take care of evolving only (\ref{eq:SchemeDiscreteIII})-(\ref{eq:SchemeDiscreteIV}), which represent the evolution of the matter field $dof$ in a flat background. Switching back to our $\alpha$-time variable, say introducing $\eta(n_0) \equiv \eta_0 + n_0\delta\eta$ with $n_0$ counting the number of time steps, we can solve Eqs.~(\ref{eq:SchemeDiscreteIII})-(\ref{eq:SchemeDiscreteIV}) with a leapfrog scheme simply as
\begin{eqnarray}
\label{eq:LeapFrogFlatEq1}
\Delta_0^+ f_i({\bf x},n_0) &=& \pi_i({\bf x},n_0+1/2) \ , \\
\label{eq:LeapFrogFlatEq2}
\Delta_0^+\pi_i({\bf x},n_0+1/2) &=& \mathcal{K}_i\left[\lbrace f_{j}({\bf x},n_0+1) \rbrace\right]\,,
\end{eqnarray}
with $i = 1, 2, ..., N_f$ counting our field theory $dof$, e.g.~scalar field real components and gauge field Lorentz components.

We note now that in any set of discrete EOM mimicking continuum EOM, the spatial coordinates ${\bf x}$ are discretized, being represented by a finite set of lattice sites $\bn = (n_1, n_2, n_3)$ with $n_i = 0, 1, 2, ..., N - 1$, and $N$ the number of lattice sites per spatial dimension (recall Section~\ref{subsec:LatticeDefinitionAndDFT} for definitions). This implies that spatial derivatives appearing in the discrete EOM, e.g.~the Laplacian operator for scalar fields $\nabla^2 f$, will be substituted by lattice derivative operators like that in Eq.~(\ref{eq:forwardbackwardd}). Due to this, kernels in the discretized EOM are not functions of the point $\bn$ only, but also of its nearest neighbours, e.g.~$\bn \pm \hat j$, with $j = 1,2,3$. The correct form of the discretized field EOM in a flat background will then look like
\begin{eqnarray}
\label{eq:LeapFrogFlatDiscreteEq1}
\Delta_0^+ f_i({\bf n},n_0) &=& \pi_i({\bf n},n_0+1/2) \,, \\
\label{eq:LeapFrogFlatDiscreteEq2}
\Delta_0^+\pi_i({\bf n},n_0+1/2) &=& \mathcal{K}_i\Big[\lbrace f_j({\bf m},n_0+1)\rbrace \Big]\,,
\end{eqnarray}
with ${\bf m}$ representing $\bn$ and its nearest neighbours (to be determined in each case depending on the choice of lattice spatial derivatives).

Note that the leapfrog algorithm cannot be applied directly to scenarios where the expansion of the universe is considered (either background or self-consistent expansion), without a careful choice of the $dof$ to evolve. Indeed, the EOM of matter fields in FLRW have kernels $\mathcal{K}[...]$ containing conjugate momenta $\pi_i$, due to the presence of the friction terms $\propto (a'/a)f'_i$ in the field EOM, c.f.~Eqs.~(\ref{eq:singlet-eom})-(\ref{eq:SU2eom}). Furthermore, the Friedmann equation $\pi_a' = \mathcal{K}_a[a,E_V, E_K, E_G]$, c.f.~Eq.~(\ref{eq:FriedmannDDa}), also contains the kinetic terms $E_K(\eta) \equiv \langle \sum_i K_i \rangle$, built out of the conjugate momenta of the fields. As in a leapfrog algorithm, conjugate momenta $\pi_i$ and $\pi_a$ live naturally at semi-integer times $\eta_{n+1/2}$, the leapfrog algorithm for kernels which contain conjugate momenta will not work, as they should rather live at integer times $\eta_n$ for the algorithm to be stable and order $\mathcal{O}(\delta t^2)$.  As we will show in Sections \ref{subsec:StaggeredLFaction} and  \ref{subsec:StaggeredLF_LE}, it is possible to overtake this problem by means of variable re-definitions and/or manipulations of the EOM, so that we can have a consistent iterative scheme with appropriate kernels, even in the presence of an expanding background. So for now, let us assume that we have managed to obtain $dof$ such that the kernels do not depend on the momenta. We can then write the following leapfrog algorithm in an expanding universe

\vspace{0.4cm}

\begingroup
\allowdisplaybreaks
\hspace*{1.0cm}{\it Leapfrog in an expanding background}
\begin{eqnarray}
\Delta_{0}^+\pi_i({\bf n},n_0-1/2)) &=& \mathcal{K}_i\Big[\lbrace f_j({\bf m},n_0)\rbrace,a(n_0)\Big]\,,
\label{eq:LFexpDiscreteEq1bis}\\
\Delta_{0}^+\pi_a({\bf n},n_0-1/2) &=& \mathcal{K}_a\Big[a(n_0), E_V(n_0), \overline{E_K}(n_0), E_G(n_0) \Big]\,,
\label{eq:LFexpDiscreteEq1}\\
\Delta_{0}^+a(n_0) &=& \pi_a(n_0+1/2)\,,
\label{eq:LFexpDiscreteEq2}\\
\Delta_0^+ f_i({\bf n},n_0) &=& \mathcal{D}_i[\pi_i({\bf n},n_0+1/2),a(n_0+1/2)]\,.
\label{eq:LFexpDiscreteEq2bis}
\end{eqnarray}
\endgroup
where in the first line $\overline{E_K}(n_0) \equiv {1\over2}(E_K(n_0-1/2)+E_K(n_0+1/2))$, and in the last line, $a(n_0+1/2) \equiv (a(n_0)+a(n_0+1))/2$. The concrete realizations of this algorithm specialized to the case of interacting scalar, Abelian and non-Abelian gauge fields, will be discussed in Sections~\ref{sec:LatScalars}, \ref{sec:LatU1} and \ref{sec:LatSUN}, respectively.

\subsubsection{Verlet integration (synchronized leapfrog)}
\label{sec:Verletint}

Let us study again the one-dimensional problem of a single $dof$ that depends only on time, $x(t)$, with EOM
\begin{eqnarray}
\ddot x(t) = \mathcal{K}[x(t)]\,,
\end{eqnarray}
say with initial condition $t_0 = 0$, $x(0) = x_0$ and $\dot x(0) = \pi_0$. Recall that in order to initiate the leapfrog algorithm just introduced in the previous section, we initially needed $x_0$ and $\pi_{1/2} \equiv \dot x(\delta t/2)$, so we proposed to obtain the initial half-time step displaced velocity as $\pi_{1/2} \simeq \pi_0 + {1\over2}\delta t\mathcal{K}[x_0] + \mathcal{O}(\delta t^2)$ (or equivalently, from applying $\Delta_{\hat 0/2}^+\pi_{0} = {1\over2}\mathcal{K}[x_0]$). Next, following the leapfrog prescription, we would apply $\Delta_0^+x_0 = \pi_{1/2}$ leading to $x_1$ at order $\mathcal{O}(\delta t^2)$, and then $\Delta_0^+\pi_{1/2} = \mathcal{K}[x_1]$ leading to $\pi_{3/2}$ at order $\mathcal{O}(\delta t^2)$, and so on and so forth with successive iterations. However, after obtaining $x_1$, we might as well apply $\Delta_{\hat 0/2}^+\pi_{1/2} = {1\over2}\mathcal{K}[x_1]$, leading to $\pi_1$, also at order $\mathcal{O}(\delta t^2)$. Essentially, by applying the velocity part of the leapfrog algorithm at two equal and successive half time steps (with one position update in between), we can simply jump from $(x_0,\pi_0)$ to $(x_1,\pi_1)$, and from there to $(x_2,\pi_2)$, and so on and so forth. In other words, we can actually obtain the position and velocity always at integer times, up at order $\mathcal{O}(\delta t^2)$, with a 'kick - drift - kick' scheme as
\begin{eqnarray}\label{eq:VVkickDrifKick}
\left.\begin{array}{rcl}
\pi_{n+1/2} &=& \pi_{n} + {1\over 2}\delta t\mathcal{K}[x_n]\,,\vspace*{0.15cm}\\
x_{n+1} &=& x_n + \pi_{n+1/2}\delta t\,,\vspace*{0.175cm}\\
\pi_{n+1} &=& \pi_{n+1/2} + {1\over2}\delta t\mathcal{K}[x_{n+1}]\,,
\end{array}\right\rbrace
~~~~~ \Longleftrightarrow ~~~~~
\left\lbrace\begin{array}{rcl}
\Delta^+_{\hat 0/2}\pi_{n} &=& \mathcal{K}[x_n]\,,\vspace*{0.15cm}\\
\Delta^+_{0}x_{n} &=& \pi_{n+1/2}\,,\vspace*{0.175cm}\\
\Delta^+_{\hat 0/2}\pi_{n+1/2} &=& \mathcal{K}[x_{n+1}]\,.
\end{array}\right.
\end{eqnarray}
In reality, this method is nothing else than the leapfrog algorithm, but adding an 'extra' computation of the conjugate momenta at integer times in each iteration,
\begin{eqnarray}\label{eq:VVdrifKickExtra}
\left.
\begin{array}{rcl}
x_{n+1} &=& x_n + \pi_{n+1/2}\delta t\,,\vspace*{0.15cm}\\
\pi_{n+3/2} &=& \pi_{n+1/2} + \delta t\mathcal{K}[x_{n+1}]\,,\vspace*{0.175cm}\\
\Big(\pi_{n+1} &=& \pi_{n+1/2} + {1\over2}\delta t\mathcal{K}[x_{n+1}]\Big)
\end{array}\right\rbrace
~~~~~ \Longleftrightarrow ~~~~~
\left\lbrace
\begin{array}{rcl}
\Delta^+_{0}x_{n} &=& \pi_{n+1/2}\,,\vspace*{0.15cm}\\
\Delta^+_{0}\pi_{n+1/2} &=& \mathcal{K}[x_{n+1}]\,,\vspace*{0.175cm}\\
\Big(\Delta^+_{\hat 0/2}\pi_{n+1/2} &=& \mathcal{K}[x_{n+1}]\Big)\,.
\end{array}\right.
\end{eqnarray}
Alternatively, since in this method we only care about the amplitudes and conjugate momenta at the same moment, say at integer times, the scheme can be put into a 'drift-kick' scheme, simply by
\begin{eqnarray}\label{eq:VVdrifKick}
\left.
\begin{array}{rcl}
x_{n+1} &=&  x_n + \pi_n\delta t + {1\over2}\mathcal{K}[x_n]\delta t^2\,,\vspace*{0.15cm}\\
\pi_{n+1} &=& \pi_{n} + {1\over2}(\mathcal{K}[x_{n}]+\mathcal{K}[x_{n+1}])\delta t\,,
\end{array}\right\rbrace
~~~~~ \Longleftrightarrow ~~~~~
\left\lbrace
\begin{array}{rcl}
\Delta^+_{0}x_{n} &=& \pi_{n} + {\delta t\over2}\mathcal{K}[x_n]\,,\vspace*{0.15cm}\\
\Delta^+_{0}\pi_{n} &=& {1\over2}(\mathcal{K}[x_{n}]+\mathcal{K}[x_{n+1}])\,.
\end{array}\right.
\end{eqnarray}
The method, represented by either scheme Eq.~(\ref{eq:VVkickDrifKick}), Eq.~(\ref{eq:VVdrifKickExtra}) or Eq.~(\ref{eq:VVdrifKick}), is known as the {\it velocity-Verlet} algorithm. This method is nothing else than a synchronized version of the leapfrog method, where the third extra computation at each time step, takes care of the synchronization between amplitudes and conjugate momenta. We note that the 2-step scheme in Eq.~(\ref{eq:VVdrifKick}) has actually no advantage versus the 3-step scheme in Eqs.~(\ref{eq:VVkickDrifKick}), (\ref{eq:VVdrifKickExtra}), as in reality the number of operations is the same: the 2-step scheme simply contains the 'third' step in the drift (i.e.~in the right hand side of the amplitude updates). The 2-step scheme is only a convenient way of writing the algorithm in a more compact way.

If instead we apply the coordinate part of the leapfrog algorithm at two equal and successive half time steps (with one velocity update in between), then the method turns into the {\it position-Verlet} algorithm, which in a 'drift-kick-drift' scheme, has the form
\begin{eqnarray}\label{eq:PVkickDrifKick}
\left.\begin{array}{rcl}
x_{n+1/2} &=& x_n + {1\over2}\pi_{n}\delta t\,,\vspace*{0.15cm}\\
\pi_{n+1} &=& \pi_{n} + \delta t\cdot\mathcal{K}[x_{n+1/2}]\,,\vspace*{0.175cm}\\
x_{n+1} &=& x_{n+1/2} + {1\over2}\pi_{n+1}\delta t\,,
\end{array}\right\rbrace
~~~~~ \Longleftrightarrow ~~~~~
\left\lbrace\begin{array}{rcl}
\Delta^+_{\hat 0/2}x_{n} &=& \pi_n\,,\vspace*{0.15cm}\\
\Delta^+_{0}\pi_{n} &=& \mathcal{K}[x_{n+1/2}]\,,\vspace*{0.175cm}\\
\Delta^+_{\hat 0/2}x_{n+1/2} &=& \pi_{n+1}\,.
\end{array}\right.
\end{eqnarray}
As before, this is nothing more than a synchronized version of the leapfrog algorithm, with an extra third computation at each time step to synchronize the variables at the same time. Position-Verlet is also an algorithm of order $\mathcal{O}(\delta t^2)$. The position-Verlet algorithm can also be put in a 2-step scheme like
\begin{eqnarray}\label{eq:PVdrifKick}
\left.
\begin{array}{rcl}
\pi_{n+1} &=&  \pi_n + \dt\cdot\mathcal{K}[x_n+{\dt\over2}\pi_n] 
\,,\vspace*{0.15cm}\\
x_{n+1} &=& x_{n} + {\delta t\over2}(\pi_{n}+\pi_{n+1})\,,
\end{array}\right\rbrace
~~~~~ \Longleftrightarrow ~~~~~
\left\lbrace
\begin{array}{rcl}
\Delta^+_{0}\pi_{n} &=& \mathcal{K}[x_n+{\dt\over2}\pi_n] 
\,,\vspace*{0.15cm}\\
\Delta^+_{0}x_{n} &=& {1\over2}(\pi_{n}+\pi_{n+1})\,,
\end{array}\right.
\end{eqnarray}
which again is just a more compact manner to write the algorithm, but the number of operations is still three at each time step, with the 'third' step now contained inside the argument of the kick.

The application of either Verlet algorithm to field theories in a flat space-time background is straightforward. Introducing again $\eta(n_0) \equiv \eta_{\rm i} + n_0\delta\eta$ as the discrete $\alpha$-time variable ($\eta_{\rm i}$ some initial time), and $i = 1, 2, 3, ..., N_f$ labeling the field theory $dof$ (namely scalar field real components and gauge field Lorentz components), the velocity-Verlet algorithm reads
\begin{eqnarray}\label{eq:VVflatDiscreteEq1}
\Delta_{\hat 0/2}^+\pi_i({\bf n},n_0) &=& \mathcal{K}_i\Big[\lbrace f_j({\bf m},n_0)\rbrace \Big]\,,\\
\label{eq:VVflatDiscreteEq2}
\Delta_0^+ f_i({\bf n},n_0) &=& \pi_i({\bf n},n_0+1/2)\,,\\
\label{eq:VVflatDiscreteEq3}
\Delta_{\hat 0/2}^+\pi_i({\bf n},n_0+1/2) &=& \mathcal{K}_i\Big[\lbrace f_j({\bf m},n_0+1)\rbrace \Big]\,,
\end{eqnarray}
whereas the position-Verlet algorithm is
\begin{eqnarray}
\label{eq:PVflatDiscreteEq1}
\Delta_{\hat 0/2}^+ f_i({\bf n},n_0) &=& \pi_i({\bf n},n_0)\,,\\
\Delta_{0}^+\pi_i({\bf n},n_0) &=& \mathcal{K}_i\Big[\lbrace f_j({\bf m},n_0+1/2)\rbrace \Big]\,,\\
\Delta_{\hat 0/2}^+ f_i({\bf n},n_0+1/2) &=& \pi_i({\bf n},n_0+1)\,.
\end{eqnarray}
Here, like in the staggered leapfrog algorithm, ${\bf m}$ on the $rhs$ represents $\bn$ and its nearest neighbours, which are determined by the choice of lattice spatial derivatives.

As in the case of leapfrog algorithms, in order to apply Verlet algorithms to the case of an expanding universe, a careful definition of variables needs to be made. The details of that will be discussed in Sections~\ref{sec:LatScalars}-\ref{sec:LatSUN}, where concrete realizations of the Verlet algorithms will be provided, specialized to the case of interacting scalar, Abelian and non-Abelian gauge fields. Assuming now that we have $dof$ variables such that the kernels are independent of the momenta, the velocity- and position-Verlet algorithms in an expanding background, in respective 'kick-drift-kick' and 'drift-kick-drift' schemes, read as:\\

\begingroup
\hspace*{1.0cm}{\it Velocity-Verlet in an expanding background}
\begin{eqnarray}
\Delta_{\hat 0/2}^+\pi_a({\bf n},n_0) &=& \mathcal{K}_a\Big[a(n_0), E_V(n_0), {E_K}(n_0), E_G(n_0) \Big]\,,
\label{eq:VVexpDiscreteEq1}\\
\Delta_{\hat 0/2}^+\pi_i({\bf n},n_0) &=& \mathcal{K}_i\Big[\lbrace f_j({\bf m},n_0)\rbrace,a(n_0)\Big]\,,\\
\Delta_{\hat 0}^+a(n_0) &=& \pi_a(n_0+1/2)\,,
\label{eq:VVexpDiscreteEq2}\\
a(n_0+1/2) &=& \frac{1}{2}(a(n_0)+a(n_0+1))\,,
\label{eq:VVexpDiscreteEq2bisII}\\
\Delta_0^+ f_i({\bf n},n_0) &=& \mathcal{D}_i[\pi_i({\bf n},n_0+1/2),a(n_0+1/2)]\,,
\label{eq:VVexpDiscreteEq2bis}\\
\Delta_{\hat 0/2}^+\pi_i({\bf n},n_0+1/2) &=& \mathcal{K}_i\Big[\lbrace f_j({\bf m},n_0+1)\rbrace,a(n_0+1) \Big]\,,
\label{eq:VVexpDiscreteEq3}\\
\Delta_{\hat 0/2}^+\pi_a({\bf n},n_0+1/2) &=& \mathcal{K}_a\Big[a(n_0+1), E_V(n_0+1), {E_K}(n_0+1), E_G(n_0+1) \Big]\,.
\end{eqnarray}\\
\endgroup

\begingroup
\allowdisplaybreaks
\hspace*{1.0cm}{\it Position-Verlet in an expanding background}
\begin{eqnarray}
\label{eq:PVexpDiscreteEq1}
\Delta_{\hat 0/2}^+ f_i({\bf n},n_0) &=& \mathcal{D}_i[\pi_i({\bf n},n_0),a(n_0)]\,,\\
\Delta_{\hat 0/2}^+ a(n_0) &=& \pi_a(n_0)\,,\\
\Delta_{\hat 0}^+\pi_i({\bf n},n_0) &=& \mathcal{K}_i\Big[\lbrace f_j({\bf m},n_0+1/2)\rbrace,a(n_0+1/2) \Big]\,,\\
\overline{E_K}(n_0+1/2) &\equiv&\frac{1}{2}\left( E_K(n_0+1) + E_K(n_0)\right)\\
\label{eq:PVflatDiscreteEq2bis}
\Delta_{0}^+\pi_a({\bf n},n_0) &=& \mathcal{K}_a\Big[a(n_0+1/2), E_V(n_0+1/2), \overline{E_K}(n_0+1/2), E_G(n_0+1/2)  \Big]\,,\\
\Delta_{\hat 0/2}^+a(n_0+1/2) &=& \pi_a(n_0+1)\,,\\
\Delta_{\hat 0/2}^+ f_i({\bf n},n_0+1/2) &=& \mathcal{D}_i[\pi_i({\bf n},n_0+1),a(n_0+1)]\,,
\end{eqnarray}
\endgroup

It is important to note that in both position- and velocity-Verlet algorithms for an expanding background, the kernels $\mathcal{K}_i[...]$ of the matter $dof$ must not depend on $\pi_a$, as the latter already depend on the conjugate momenta through the volume averaged kinetic energy $E_K[\lbrace \pi_j \rbrace]$. An advantage of the Verlet algorithm(s) is that they can readily be turned into more accurate schemes, as will be explained in Section~\ref{subsec:Yoshida}.

\subsubsection{Explicit Runge-Kutta methods}\label{subsec:RungeKutta}

Let us consider now a one-dimensional problem with a single $dof$, where the kernel of the EOM is also allowed to depend on the velocity,
\begin{eqnarray}
\ddot x(t) = \mathcal{K}[x(t),\dot x(t)]\, .
\end{eqnarray}
We take initial conditions $x(0) = x_0$ and $\dot x(0) = \pi_0$ at the initial time $t_0 = 0$. First-order Runge-Kutta algorithms are the {\it Euler method},
\begin{eqnarray}
x_{n+1} &=& x_n + \pi_n\dt\,,\\
\pi_{n+1} &=& \pi_n + \dt\mathcal{K}[x_n,\pi_n]\,,
\end{eqnarray}
and the {\it Euler-Cromer method},
\begin{eqnarray}
x_{n+1} &=& x_n + \pi_n\dt\,,\\
\pi_{n+1} &=& \pi_n + \dt\mathcal{K}[x_{n+1},\pi_n]\,.
\end{eqnarray}
Both methods have an accuracy of $O(\delta t)$. They are also less stable than Leapfrog/Verlet methods when integrated over many steps, as they are not {\it symplectic} algorithms, and hence they do not conserve well energy 
(see Section~\ref{subsec:IntegratorProperties} for further clarifications on this).

\noindent More accurate algorithms are the Runge-Kutta second-order or {\it modified Euler} algorithm [{\it RK2}],
\begin{eqnarray}
\left.
\begin{array}{rcl}
{\it k}_1 &\equiv& \mathcal{K}[x_n,\pi_n]\,,\vspace*{0.2cm}\\
{\it k}_2 &\equiv& \mathcal{K}[x_n + \pi_n\dt,\pi_n + {\it k}_1\dt]\,,
\end{array}
\right\rbrace
~~~ \Longrightarrow ~~~
\left\lbrace
\begin{array}{rcl}
x_{n+1} &=& x_n + \pi_n\dt + {1\over2}{\it k}_1\dt^2\,,\vspace*{0.2cm}\\
\pi_{n+1} &=& \pi_n + {\dt\over2}({\it k}_1+{\it k}_2)\,,
\end{array}\right.
\end{eqnarray}
accurate to $\mathcal{O}(\dt^2)$, and the Runge-Kutta fourth-order method [{\it RK4}]
\begin{eqnarray}
\left.
\begin{array}{rcl}
{\it k}_1 &=& \mathcal{K}[x_n,\pi_n]\,,\vspace*{0.2cm}\\
{\it k}_2 &=& \mathcal{K}[x_n + {1\over2}\pi_n\dt,\pi_n + {1\over2}{\it k}_1\dt]\,,\vspace*{0.2cm}\\
{\it k}_3 &=& \mathcal{K}[x_n + {1\over2}\pi_n\dt + {1\over4}{\it k}_1\dt^2,\pi_n + {1\over2}{\it k}_2\dt]\,,\vspace*{0.2cm}\\
{\it k}_4 &=& \mathcal{K}[x_n + \pi_n\dt + {1\over2}{\it k}_2\dt^2,\pi_n + {\it k}_3\dt]\,,
\end{array}
\right\rbrace
 \Longrightarrow
\left\lbrace
\begin{array}{rcl}
x_{n+1} &=& x_n + \pi_n\dt + {1\over6}\Big({\it k}_1 + {\it k}_2+ {\it k}_3\Big)\dt^2\,,\vspace*{0.2cm}\\
\pi_{n+1} &=& \pi_n + {1\over6}\Big({\it k}_1 + 2{\it k}_2 + 2{\it k}_3 + {\it k}_4\Big)\dt\,,
\end{array}\right.
\end{eqnarray}
accurate to $\mathcal{O}(\dt^4)$.

Adapting RG2 to field theory EOM, we obtain
\begin{eqnarray}
\left.
\begin{array}{r}
\left\lbrace
\begin{array}{llll}
{a}^{(1)} = a\,, & f_{i}^{(1)} = f_i\,, & \pi_{i}^{(1)} = \pi_i\,, & \pi_{a}^{(1)} = \pi_a\,,\vspace*{0.2cm}\\
a^{(2)} = a^{(1)} + {\deta}\pi_{a}^{(1)}\,, & f_j^{(2)} = f_j^{(1)} + {\deta}\pi_{j}^{(1)}\,, & \pi_{i}^{(2)} = \pi_i^{(1)} + {\deta}{\it k}_{1,i}\,, & \pi_{a}^{(2)} = \pi_a^{(1)} + {\deta}{\it k}_{1,a}\,,
\end{array}
\right.\vspace*{0.2cm}\\
\left\lbrace
\begin{array}{ll}
{\it k}_{1,i} = \mathcal{K}_i[a^{(1)},\pi_a^{(1)},\lbrace f_j^{(1)} \rbrace,\lbrace \pi_j^{(1)}\rbrace]\,,  & {\it k}_{1,a} = \mathcal{K}_a[a^{(1)}, E_K^{(1)},E_G^{(1)},E_V^{(1)}]\,,\vspace*{0.2cm}\\
{\it k}_{2,i} = \mathcal{K}_i[{a}^{(2)}, \pi_a^{(2)}, \lbrace f_j^{(2)}\rbrace, \lbrace \pi_{j}^{(2)}\rbrace]\,, & {\it k}_{2,a} = \mathcal{K}_a[{a}^{(2)}, E_K^{(2)},E_G^{(2)},E_V^{(2)}]\,,
\end{array}\right.
\end{array}
\right\rbrace\Longrightarrow \nonumber
\end{eqnarray}
\begin{eqnarray}
\Longrightarrow
\left\lbrace
\begin{array}{rcl}
\Delta_0^+f_i(\bn,n_0) &=& \pi_i(\bn,n_0) + {1\over2}{\it k}_{1,i}\deta\,,\vspace*{0.2cm}\\
\Delta_0^+a(n_0) &=& \pi_a(n_0) + {1\over2}{\it k}_{1,a}\deta\,,\vspace*{0.2cm}\\
\Delta_0^+\pi_i(\bn,n_0) &=& {1\over2}\Big({\it k}_{1,i} + {\it k}_{2,i}\Big)\,,\vspace*{0.2cm}\\
\Delta_0^+\pi_a(n_0) &=& {1\over2}\Big({\it k}_{1,a} + {\it k}_{2,a}\Big)\,,
\end{array}\right.
\end{eqnarray}
Similarly, RG4 leads to
\begin{eqnarray}\label{eq:RG4algorithm}
\left.
\begin{array}{r}
\left\lbrace
\begin{array}{llll}
{a}^{(1)} = a\,, & f_{i}^{(1)} = f_i\,, & \pi_{i}^{(1)} = \pi_i\,, & \pi_{a}^{(1)} = \pi_a\,,\vspace*{0.2cm}\\
a^{(2)} = a^{(1)} + {\deta\over2}\pi_{a}^{(1)}\,, & f_j^{(2)} = f_j^{(1)} + {\deta\over2}\pi_{j}^{(1)}\,, & \pi_{i}^{(2)} = \pi_i^{(1)} + {\deta\over2}{\it k}_{1,i}\,, & \pi_{a}^{(2)} = \pi_a^{(1)} + {\deta\over2}{\it k}_{1,a}\,,
\vspace*{0.2cm}\\
a^{(3)} = a^{(1)} + {\deta\over2}\pi_{a}^{(2)}\,, & f_j^{(3)} = f_j^{(1)} + {\deta\over2}\pi_{j}^{(2)}\,, & \pi_{i}^{(3)} = \pi_i^{(1)} + {\deta\over2}{\it k}_{2,i}\,, & \pi_{a}^{(3)} = \pi_a^{(1)} + {\deta\over2}{\it k}_{2,a}\,,
\vspace*{0.2cm}\\
a^{(4)} = a^{(1)} + {\deta}\pi_{a}^{(3)}\,, & f_j^{(4)} = f_j^{(1)} + {\deta}\pi_{j}^{(3)}\,, & \pi_{i}^{(4)} = \pi_i^{(1)} + {\deta}{\it k}_{3,i}\,, & \pi_{a}^{(4)} = \pi_a^{(1)} + {\deta}{\it k}_{3,a}\,,
\end{array}
\right.\vspace*{0.2cm}\\
\left\lbrace
\begin{array}{ll}
{\it k}_{1,i} = \mathcal{K}_i[a^{(1)},\pi_a^{(1)},\lbrace f_j^{(1)} \rbrace,\lbrace \pi_j^{(1)}\rbrace]\,,  & {\it k}_{1,a} = \mathcal{K}_a[a^{(1)}, E_K^{(1)},E_G^{(1)},E_V^{(1)}]\,,\vspace*{0.2cm}\\
{\it k}_{2,i} = \mathcal{K}_i[{a}^{(2)}, \pi_a^{(2)}, \lbrace f_j^{(2)}\rbrace, \lbrace \pi_{j}^{(2)}\rbrace]\,, & {\it k}_{2,a} = \mathcal{K}_a[a^{(2)}, E_K^{(2)},E_G^{(2)},E_V^{(2)}]\,,
\vspace*{0.2cm}\\
{\it k}_{3,i} = \mathcal{K}_i[{a}^{(3)}, \pi_a^{(3)}, \lbrace f_j^{(3)}\rbrace, \lbrace \pi_{j}^{(3)}\rbrace]\,, & {\it k}_{3,a} = \mathcal{K}_a[a^{(3)}, E_K^{(3)},E_G^{(3)},E_V^{(3)}]\,,
\vspace*{0.2cm}\\
{\it k}_{4,i} = \mathcal{K}_i[{a}^{(4)}, \pi_a^{(4)}, \lbrace f_j^{(4)}\rbrace, \lbrace \pi_{j}^{(4)}\rbrace]\,, & {\it k}_{4,a} = \mathcal{K}_a[a^{(4)}, E_K^{(4)},E_G^{(4)},E_V^{(4)}]\,,
\end{array}\right.
\end{array}
\right\rbrace\Longrightarrow \nonumber
\end{eqnarray}
\begin{eqnarray}
\Longrightarrow
\left\lbrace
\begin{array}{rcl}
\Delta_0^+f_i(\bn,n_0) &=& \pi_i(\bn,n_0) + {1\over6}\Big({\it k}_{1,i} + {\it k}_{2,i}+ {\it k}_{3,i}\Big)\deta\,,\vspace*{0.2cm}\\
\Delta_0^+a(n_0) &=& \pi_a(n_0) + {1\over6}\Big({\it k}_{1,a} + {\it k}_{2,a}+ {\it k}_{3,a}\Big)\deta\,,\vspace*{0.2cm}\\
\Delta_0^+\pi_i(\bn,n_0) &=& {1\over6}\Big({\it k}_{1,i} + 2{\it k}_{2,i} + 2{\it k}_{3,i} + {\it k}_{4,i}\Big)\,,\vspace*{0.2cm}\\
\Delta_0^+\pi_a(n_0) &=& {1\over6}\Big({\it k}_{1,a} + 2{\it k}_{2,a} + 2{\it k}_{3,a} + {\it k}_{4,a}\Big)\,.
\end{array}\right.
\end{eqnarray}

\subsubsection{Crank-Nicolson}

Let us now consider the specific problem of a one-dimensional single $dof$ with a kernel of the form
\begin{eqnarray}
\ddot x(t) = F(t)\dot x(t) + \mathcal{K}_x[x(t)]\, ,
\end{eqnarray}
where $F(t)$ is some external function depending on time. We will refer to $F(t)\dot x(t)$ as the friction term. We take initial conditions $x(0) = x_0$ and $\dot x(0) = \pi_0$ at the initial time $t_0 = 0$. The Crank-Nicolson method is based on representing the friction term by a semi-sum
\begin{eqnarray}
x_{n+1} &=& x_n + \pi_{n+1/2}\dt\,,\\
\pi_{n+1/2} &=& \pi_{n-1/2} + {\dt\over2}\left(F_{n+1/2}\pi_{n+1/2}+F_{n-1/2}\pi_{n-1/2}\right)+\dt\mathcal{K}_x[x_n]\,,
\end{eqnarray}
so that
\begin{eqnarray}
x_{n+1} &=& x_n + \pi_{n+1/2}\dt\,,\\
\pi_{n+1/2} &=& \left({1+{\delta t\over2}F_{n-1/2}\over1-{\delta t\over2}F_{n+1/2}}\right)\pi_{n-1/2} + {\dt\over(1-{\dt\over2}F_{n+1/2})}\mathcal{K}_x[x_n]\,.
\end{eqnarray}

The application of this algorithms to the case of field theories in an expanding universe is actually straight forward, as the friction term is played by the Hubble friction term naturally present in the kernels $\mathcal{K}_i[...]$ of the EOM of matter fields in FLRW, which contain terms $\propto (a'/a)f'_i$, c.f.~Eqs.~(\ref{eq:singlet-eom})-(\ref{eq:SU2eom}). We can write\\

\begingroup
\allowdisplaybreaks
\hspace*{1.0cm}{\it Crank-Nicolson in an expanding background}
\begin{eqnarray}
\pi_a(n_0+1/2) &=& \pi_a(n_0-1/2) + \delta t\mathcal{K}_a\Big[a(n_0), E_V(n_0), \overline{E_K}(n_0), E_G(n_0) \Big]\,,
\label{eq:CNexpDiscreteEq1}\\
\pi_i({\bf n},n_0+1/2) &=& \left({1-{3\delta t\over2a(n_0)}\pi_a(n_0-1/2)\over1+{3\delta t\over2a(n_0)}\pi_a(n_0+1/2)}\right)\pi_i({\bf n},n_0-1/2) + {\delta t\mathcal{K}_i\Big[\lbrace f_j({\bf m},n_0)\rbrace,a(n_0)\Big]\over1+{3\delta t\over2a(n_0)}\pi_a(n_0+1/2)}\,,
\label{eq:CNexpDiscreteEq1bis}\\
a(n_0+1) &=& a(n_0) + \delta t\pi_a(n_0+1/2)\,,
\label{eq:CNexpDiscreteEq2}\\
f_i({\bf n},n_0+1) &=& f_i({\bf n},n_0) + \delta t\mathcal{D}_i[\pi_i({\bf n},n_0+1/2),a(n_0+1/2)]\,.
\label{eq:CNexpDiscreteEq2bis}
\end{eqnarray}
\endgroup

\noindent where in the first line $\overline{E_K}(n_0) \equiv {1\over2}(E_K(n_0-1/2)+E_K(n_0+1/2))$, and in the last line, $a(n_0+1/2) \equiv (a(n_0)+a(n_0+1))/2$. We note that this algorithm cannot be made explicit at $\mathcal{O}(\delta t^2)$ due to the implicit relation between Eqs.~(\ref{eq:CNexpDiscreteEq1}) and (\ref{eq:CNexpDiscreteEq1bis}), so we would need a recursive solution at every time step\footnote{The algorithm remains however explicit and of $\mathcal{O}(\delta t^2)$ for a fixed background expansion, i.e.~if the scale factor is given by an external function and we do not need to evolve Eqs.~(3.109) and (3.111).}. Alternatively, if we make the substitution $\overline{E_K}(n_0) \longrightarrow E_K(n_0-1/2)$, the algorithm becomes explicit but only at the expense of reducing its accuracy to order $\mathcal{O}(\delta t)$. Because of these issues, we will not present concrete realizations of this algorithm specialized to the case of interacting fields in either Section~\ref{sec:LatScalars}, \ref{sec:LatU1} or \ref{sec:LatSUN}.

\subsection{Higher-order integrators}

~~~~ We discuss now the construction of higher-order integrators with accuracy  $\mathcal{O}(\delta t^4)$, $\mathcal{O}(\delta t^6)$, $\mathcal{O}(\delta t^8)$ and even $\mathcal{O}(\delta t^{10})$, based on the use of $\mathcal{O}(\delta t^2)$ staggered/synchronized leapfrog algorithms as building blocks (Section~\ref{subsec:Yoshida}), and based on the generalization of the previous explicit Runge-Kutta equations into an implicit problem (Section~\ref{subsec:GaussLegendre}).

\subsubsection{Yoshida method: recursive Verlet integration}
\label{subsec:Yoshida}

~~~ The $\mathcal{O}(\delta t^2)$ Verlet integration methods, introduced in Section~\ref{sec:Verletint} to solve the problem $\ddot x(t) = \mathcal{K}[x(t)]$ with initial conditions $x(t_0) = x_0$, $\dot x(t_0) = \pi_0$, can be used recursively as building blocks to conveniently construct integrators of higher (even) order $\mathcal{O}(\delta t^n)$. The idea is to decompose appropriately a single time step $\delta t$ into $s$ sub-steps $\delta t_i = w_i\delta t$ (with $\sum_{i=1}^s w_i = 1$), in such a way that the errors of the intermediate steps cancel up to order $n$. In practice, the only thing that has to be done is to iterate $s$-times the Verlet algorithm \eqref{eq:VVkickDrifKick} or \eqref{eq:PVkickDrifKick}, using each time the appropriate $\delta t_i$ sub-step. For example, using \eqref{eq:VVkickDrifKick} as the building block, one full step $\delta t$ of the algorithm must be divided in the sum of different $\delta t_i$'s as follows,
\begin{eqnarray}
\left\lbrace
\begin{array}{l}
t^{(0)} = t_n\vspace*{0.15cm}\\
\pi^{(0)} \equiv \pi_n \vspace*{0.15cm}\\
x^{(0)} \equiv x_n
\end{array}
\right.
~~~~
\Longrightarrow~~~~
\left\lbrace
\begin{array}{l}
\pi^{(i)}_{1/2} = \pi^{(i-1)} + \omega_i{\delta t\over 2}\mathcal{K}[x^{(i-1)}]\vspace*{0.15cm}\notag\\
x^{(i)} = x^{(i-1)} + \pi^{(i)}_{1/2}\omega_i\delta t \vspace*{0.175cm}\\
\pi^{(i)} = \pi^{(i)}_{1/2} + \omega_i{\delta t\over 2}\mathcal{K}[x^{(i)}]
\end{array}
\right\rbrace_{i\,=\,1,\, ...,\, s}
~~~~\Longrightarrow ~~~~
\left\lbrace
\begin{array}{l}
t_{n+1} = t_n+\delta t\vspace*{0.15cm}\\
\pi_{n+1} \equiv \pi^{(s)}  \vspace*{0.15cm}\\
x_{n+1} \equiv x^{(s)} \ .
\end{array}
\right.
\end{eqnarray}
For information about how to construct a specific algorithm, i.e.~how to find the corresponding weights $\omega_i$, we refer the interested reader to the original paper by Yoshida \cite{Yoshida:1990zz}. We collect in Table~\ref{tab:VVnCoeffs} of the Appendix, sets of $\delta t_i$'s characterizing algorithms of order $O(\delta t^4), O(\delta t^6),O(\delta t^8)$ and $O(\delta t^{10})$, see \cite{Yoshida:1990zz, Kahan:1997:CCR} for their derivation. We will refer to these algorithms as $VV4, VV6, VV8$ and $VV10$, while we will refer to the standard velocity Verlet building block as $VV2$.

Some comments are, however, in order. First, the number of steps required to reach a given accuracy grows quickly. For example, $VV4$ requires only $3$ times more operations than $VV2$, while $VV10$ requires $31$ times more operations than $VV2$. Actually, to go from one algorithm to the next, the number of steps in each iteration is slightly more than doubled every time. This gives a rule of thumb as of when it is beneficial to use the next more accurate algorithm: if in order to reach some target precision, the time step must be decreased by more than a factor two, then we should consider using the next more accurate algorithm.

This said, let us write for completeness how this algorithm reads for field theories in an expanding background, again assuming that a clever choice of $dof$ variables has been made, so that the field kernels do not depend on their conjugate momenta:
\begingroup
\allowdisplaybreaks
\begin{eqnarray}
\hspace{-0.5cm} \left.
\begin{array}{l}
\pi_i^{(0)} \equiv \pi_i({\bf n},n_0) \vspace*{0.15cm}\\
f_i^{(0)} \equiv f_i({\bf n},n_0)\vspace*{0.15cm}\\
a^{(0)} \equiv a(n_0)\vspace*{0.15cm}\\
\pi_a^{(0)} \equiv \pi_a(n_0)
\end{array}
\right\rbrace ~~~\Longrightarrow\hspace{7cm}
\nonumber\\
\left\lbrace
\begin{array}{rcl}
\pi_{a,1/2}^{(p)}&=&\pi_a^{(p-1)}+ \frac{\omega_p\delta\eta}{2}\mathcal{K}_a\left [a^{(p-1)},E_K^{(p-1)},E_K^{(p-1)},E_V^{(p-1)}\right]\\
\pi_{i,1/2}^{(p)}&=&\pi_i^{(p-1)}+ \frac{\omega_p\delta\eta}{2}\mathcal{K}_i[a^{(p-1)},\lbrace f_j^{(p-1)}\rbrace]\vspace*{0.15cm}\notag\\
a^{(p)} &=& a^{(p-1)}+  \omega_p\delta\eta\pi^{(p)}_{a,1/2}\,,\\
a_{1/2}^{(p)}&=&\frac{1}{2}\left(a^{(p)}+a^{(p-1)}\right)\\
f^{(p)}_{i} &=&f^{(p-1)}_{i}+\omega_p\delta  \eta\mathcal{D}_i\left[a_{1/2}^{(p)},\pi^{(p)}_{i,1/2}\right]\\
\pi_i^{(p)}&=&\pi_{i,1/2}^{(p)}+\frac{\omega_p\delta\eta}{2} \mathcal{K}_i[a^{(p)},\lbrace f_j^{(p)}\rbrace]\vspace*{0.15cm}\notag\\
\pi_a^{(p)}&=& \pi_{a,1/2}^{(p)}+\frac{\omega_p\delta\eta}{2}\mathcal{K}_a[a^{(p)},E_K^{(p)},E_K^{(p)},E_V^{(p)}]\\
\end{array}
\right\rbrace_{p\,=\,1,\, ...,\, s}
\Longrightarrow ~~~~
\left\lbrace
\begin{array}{l}
\pi_i({\bf n},n_0+1) \equiv \pi_i^{(s)} \vspace*{0.15cm}\\
f_i({\bf n},n_0+1) \equiv f_i^{(s)}\vspace*{0.15cm}\\
a(n_0 + 1) \equiv a^{(s)} \vspace*{0.15cm}\\
\pi_a(n_0 + 1) \equiv \pi_a^{(s)}
\end{array}
\right.\nonumber\\
\end{eqnarray}
\endgroup
\\
\noindent
Note that a similar algorithm could be constructed instead, using the position-Verlet method as the building block.

\subsubsection{Gauss-Legendre methods: Implicit Runge-Kutta}
\label{subsec:GaussLegendre}

The Runge-Kutta methods $RK2$ and $RK4$ (of order $\mathcal{O}(\delta t^2)$ and $\mathcal{O}(\delta t^4)$ respectively), previously introduced in Section~\ref{subsec:RungeKutta} to solve the problem $\ddot x(t) = \mathcal{K}[x(t),\dot x(t)]$ with initial condition $x(t_0) = x_0$ and $\dot x(t_0) = \pi_0$, are actually only representative examples of a whole family of Runge-Kutta methods. Runge-Kutta methods are characterized in general by a one-step $\delta t$ iteration algorithm of the form
\begin{eqnarray}\label{eq:RKgeneralSol}
x_{n+1} &=& x_n + \delta t\sum_{i = 1}^s c_i\pi^{(i)}\,,~~~~~\pi_{n+1} = \pi_n + \delta t\sum_{i = 1}^s c_ik^{(i)}\,,
\end{eqnarray}
with
\begin{eqnarray}
\label{eq:RKimplicitEQ}
x^{(i)} \equiv x_n + \delta t\sum_{j = 1}^s b_{ij}\pi^{(j)}\,,~~~~ \pi^{(i)} \equiv \pi_n + \delta t\sum_{j = 1}^s b_{ij}k^{(j)}\,,~~~~~~
k^{(i)} \equiv \mathcal{K}[x^{(i)},\pi^{(i)}] \ ,
\end{eqnarray}
where a single step is subdivided in $s$ sub-intervals, $\delta t = \sum_{i = 1}^s \delta t_i$, with $\delta t_i \equiv  c_i\delta t$, $c_{i} < 1$, and $\sum_{i = 1}^s c_i = 1$. Schematically, each RK algorithm can be represented by a {\it Butcher tableau} as follows
\begin{eqnarray}
\begin{array}{r|cccc|l}
\, & b_{11} & b_{12} & \cdots & b_{1s} & \,\\
\, & b_{21} & b_{22} & \cdots & b_{2s} & \,\\
\, & \ddots & \ddots & \cdots & \ddots & \,\\
\, & b_{s1} & b_{s2} & \cdots & b_{ss} & \,\\
\hline
\, & c_1 & c_2 & \cdots & c_s & \,
\end{array}\,.
\end{eqnarray}
The $RK2$ and $RK4$ algorithms are represented by the following tableaux,
\begin{eqnarray}
RK2: ~ \begin{array}{r|cc|l}
\, & 0 & 0 & \,\\
\, & 1 & 0 & \,\\
\hline
\, & 1/2 & 1/2 & \,
\end{array}~~~~~,~~~~~~~~~~
RK4:~\begin{array}{r|cccc|l}
\, & 0 & 0 & 0 & 0 & \,\\
\, & 1/2 & 0 & 0 & 0 & \,\\
\, & 0 & 1/2 & 0 & 0 & \,\\
\, & 0 & 0 & 1 & 0 & \,\\
\hline
\, & 1/6 & 1/3 & 1/3 & 1/6 & \,
\end{array}~~~~~.
\end{eqnarray}
These correspond to {\it explicit RK algorithms}, as they are characterized by $b_{ij} = 0$ $\forall~i \leq j$, which allows to compute the successive $k_i$'s, $i = 1, 2, ..., s$, as an explicit function of the previous ones. In any other circumstance, Eq.~(\ref{eq:RKgeneralSol}) corresponds to an {\it implicit RK algorithm}, as the $k_i$'s depend on the previous and following ones (even on themselves) through the implicit relations in Eq.~(\ref{eq:RKimplicitEQ}). In a seminal paper~\cite{Butcher}, J.~C.~Butcher demonstrated that $i)$ the coefficients $c_i$ and $b_{ij}$ in Eqs.~(\ref{eq:RKgeneralSol}) and (\ref{eq:RKimplicitEQ}) are unique (see appendix of~\cite{Butcher}), and $ii)$ the accuracy of the numerical solution of a method with $s$ sub-steps is of order $\mathcal{O}(\delta t^{2s})$. Furthermore, Butcher also determined the corresponding tableaux for the implicit $RK$ methods with $s = 2, 3, 4$ and $5$ sub-steps, which we reproduce in Table~\ref{tab:ButcherTables} of the Appendix.

Adapting the implicit $RK$ methods to the field theory of our interest, we obtain
\begin{eqnarray}
\label{eq:impRKfieldTheory}
\left.
\begin{array}{rcl}
\pi_i^{(l)} &\equiv& \pi_i({\bf n},n_0) + \deta\sum_{m = 1}^s b_{lm}k_i^{(m)}\,,\\
\pi_a^{(l)} &\equiv& \pi_a(n_0) + \deta\sum_{m = 1}^s b_{lm}k_a^{(m)}\,,\\
f_i^{(l)} &\equiv& f_i({\bf n},n_0) + \deta\sum_{m = 1}^s b_{lm}\pi_i^{(m)}\,,\\
a^{(l)} &\equiv& a(n_0) + \deta\sum_{m = 1}^s b_{lm}\pi_a^{(m)}\,,\\
{\it k}^{(l)}_{i} &\equiv& \mathcal{K}_i[a^{(l)},\pi_a^{(l)},\lbrace f_j^{(l)} \rbrace,\lbrace \pi_j^{(l)}\rbrace]\,,\\
{\it k}^{(l)}_{a} &\equiv& \mathcal{K}_a[a^{(l)},E_K^{(l)},E_K^{(l)},E_V^{(l)}]\,,
\end{array}
\right\rbrace~~ \Longrightarrow ~~
\left\lbrace
\begin{array}{rcl}
\Delta_0^+f_i(\bn,n_0) &=& \sum_{m = 1}^s c_{m}\pi_i^{(m)}\,,\vspace*{0.2cm}\\
\Delta_0^+a(n_0) &=& \sum_{m = 1}^s c_{m}\pi_a^{(m)}\,,\vspace*{0.2cm}\\
\Delta_0^+\pi_i(\bn,n_0) &=& \sum_{m = 1}^s c_{m}k_i^{(m)}\,,\vspace*{0.2cm}\\
\Delta_0^+\pi_a(n_0) &=&  \sum_{m = 1}^s c_{m}k_a^{(m)}\,,
\end{array}\right.\nonumber\\
\end{eqnarray}

\subsection{Integrator properties}
\label{subsec:IntegratorProperties}

Before we move into discussing the explicit adaptation of the previous algorithms for their use in interactive field theories in an expanding background, we review here the list of desired properties that we may want to demand to a good numerical integrator:

\begin{itemize}

    \item {\it Time reversal}. Dynamical processes are time-reversible if their EOM are invariant under a change in the sign of the time variable. Since this is an exact symmetry in the continuum EOM, it is desirable that a numerical integration method respects the same property: An evolution algorithm for discrete EOM respects time reversibility if we can integrate forward $p$ steps, and then reverse the direction of integration and integrate backwards $p$ steps, to arrive exactly at the original starting initial condition.

    \item {\it Symplecticness}. Dynamical processes driven by conservative forces (i.e.~from kernels that do not depend on conjugate momenta or on any time-dependent external function) respect the {\it Liouville's theorem}, which states that the infinitesimal phase-space area per degree of freedom is preserved as the system evolves. As this area-preserving property is an exact feature of the continuum EOM which we want to mimic, it is desirable that a numerical integrator respects such a conservation law. Numerical schemes that do so are referred to as {\it symplectic}. The relevance of having a symplectic integrator is that they possess a great stability: since the phase-space area is preserved during the evolution, there cannot be situations where the field amplitudes or their conjugate momenta (and hence their energy) increase without bound, because this would expand the phase-space area. Symplectic integrators offer therefore a good numerical conservation of energy\footnote{In the case of scenarios with an expanding background, by conservation of energy we actually mean the preservation of the Hubble constraint $3m_p^2H^2 = \rho$.}, with accuracy given by the accuracy $\mathcal{O}(\delta t^p)$ of the integrator itself.

    \item {\it Integration accuracy}. Depending on the nature of a given numerical integrator method, we may obtain integrated field amplitudes and conjugate momenta which differ from their continuum values by some error of order $\mathcal{O}(\delta t^p)$, typically with an even value $p = 2, 4, 6, 8$, or even $10$. Our base symplectic algorithms (Leapfrog/Verlet) have an accuracy $\mathcal{O}(\delta t^2)$. However, these can be turned into higher-order integrators using techniques due to {\it Haruo Yoshida}. Essentially, by applying the basic algorithm over a number of adjusted different timesteps chosen so that the errors cancel, far higher-order integrators can be obtained, see Section~\ref{subsec:Yoshida}. This is particularly interesting as the base methods are symplectic integrators, and hence the degree of conservation of energy (Hubble constraint for expanding backgrounds) will improve significantly as we increase the integrator accuracy. Basic Runge-Kutta methods of order $\mathcal{O}(\delta t
   ^2)$ and $\mathcal{O}(\delta t
   ^4)$ can also be generalized into implicit algorithms of higher order as demonstrated by J.~C.~Butcher, see Section~\ref{subsec:GaussLegendre}.

    \item {\it Efficiency}. We obviously want to make our numerical integration as fast as possible, so if we need to choose between two integration methods with the same accuracy $\mathcal{O}(\delta t^p)$, but different levels of energy conservation, in certain situation we might still prefer the faster integrator even if it has worse energy conservation (as long as it can be confronted and calibrated against the outcome from other integrators with better energy conservation).

\end{itemize}

\section{Lattice formulation of interacting scalar fields} \label{sec:LatScalars}

\subsection{Continuum formulation and natural variables}

Let us consider a set of interacting real scalar fields $\lbrace \phi_{a} \rbrace$ with canonically normalized kinetic terms. If they live in a FLRW background $g_\mn = {\rm diag}(-a^{2\alpha},a^2,a^2,a^2)$,  with line element $ds^2 = -a^{2\alpha}\deta^2 + a^2(\eta)d\vec x^2$ and $\alpha$-time $\eta$,
their action can be written like
\bea\label{eq:NsCanonicalScalarsAction}
S &=& - \int d^4x\, \sqrt{-g}\left(\frac{1}{2}\partial_{\mu} \phi_b \partial^{\mu} \phi_b + V(\lbrace \phi_c \rbrace) \right) = \left( \frac{f_*}{\omega_*}\right)^2\tilde S\,,
\eea
where
\bea
\tilde S = \int d^3\tilde x d \tilde\eta \left\{ \frac{1}{2} a^{3 - \alpha}\sum_b{\tilde\phi}_{b}'^{\,2} - \frac{1}{2} a^{1 + \alpha} \sum_{b,k} (\tilde\nabla_k \tilde\phi_{b})^2 - a^{3 + \alpha} \widetilde V(\lbrace \tilde\phi_{c} \rbrace) \right\} \, \label{eq:ActionScalar}
\eea
is the action expressed in the dimensionless variables
\begin{eqnarray}
\label{eq:FieldSpaceTimeNaturalVariables}
\tilde\phi_a \equiv {\frac{\phi_a}{f_*}}\,,~~~~ d\tilde\eta \equiv a^{- \alpha}  \omega_* dt\,,~~~~ d\tilde x^i \equiv \omega_* dx^i\,,
\end{eqnarray}
with $' \equiv d /d\tilde\tau$ and $\tilde\nabla_i \equiv \partial / \partial\tilde x^i$, and where a dimensionless potential has been also introduced as
\bea\label{eq:PotNat}
\widetilde V(\lbrace \tilde\phi_c \rbrace) \equiv \frac{1}{f_*^2 \omega_*^2}V(\lbrace \phi_c \rbrace)\Big|_{\phi_c = f_*\tilde\phi_c}\,.
\eea
Here, $f_*$ and $\omega_*$ are scales conveniently chosen depending on the problem at hand. See the discussion on {\it program variables} on the next page, for further clarification. Explicit examples will be provided in Sect.~\ref{sec:WorkingExample}.

The EOM in the dimensionless variables follow immediately from varying the action $\tilde S$,
\bea
\tilde\phi_a'' - a^{-2 (1  - \alpha )} \tilde\nabla^2 \tilde\phi_a + (3 - \alpha)\frac{a'}{a} \tilde\phi'_a +  a^{2 \alpha} \widetilde V_{,\tilde\phi_a} = 0\,. \label{eq:EOMscalarContinuumNat}
\eea

The expansion of the universe, on the other hand, is dictated by the Friedmann equations, sourced by the volume averaged energy and pressure densities $\langle \rho_{\phi} \rangle$, $\langle p_{\phi} \rangle$ of the fields. Writing the relevant part of Eqs.~(\ref{eq:FriedmannHubble}), (\ref{eq:FriedmannDDa}) in the dimensionless variables~(\ref{eq:FieldSpaceTimeNaturalVariables}), we have
\be a'' = a^{2 \alpha + 1} \left( \frac{f_*}{m_p} \right)^2 \frac{1}{6}[ (2 \alpha - 1) \langle \tilde\rho_{\phi} \rangle - 3 \langle \tilde p_{\phi} \rangle ]  \ , \hspace{0.5cm} a'^2 = a^{2 \alpha + 2} \left( \frac{f_*}{m_p} \right)^2 \frac{1}{3} \langle \tilde\rho_{\phi} \rangle \ , \label{eq:Friedmann-NatVar} \ee
with \textit{program energy} and {\it pressure densities} defined as
\bea
\tilde\rho_\phi \equiv \frac{\rho}{f_*^2 \omega_*^2} = \widetilde{K}_\phi + \widetilde{\rm G}_\phi + \widetilde V \,,~~~~ ; ~~~~ \tilde p_\phi \equiv \frac{p}{f_*^2 \omega_*^2} = \widetilde{K}_\phi - \frac{1}{3} \widetilde{\rm G}_\phi - \widetilde V \ , \eea
where $\tilde V$ is given by Eq.~(\ref{eq:PotNat}), and we have introduced
\bea\label{eq:KandGprogramUnits}
\widetilde{K}_\phi = \frac{1}{2 a^{2\alpha}}  \sum_i({\tilde \phi_i}')^2  \ , \hspace{0.4cm}
\tilde{\rm G}_\phi = \frac{1}{2 a^2 }  \sum_{i,k} (\widetilde\nabla_k \tilde \phi_{i})^2 \,.
\eea
As in reality we need the volume averages $\langle ... \rangle$ of the dimensionless energy density components, we define
\bea\label{eq:EK_EG_EV}
{\widetilde E}_K \equiv \frac{1}{2 a^{2\alpha}} \sum_{i}\left\langle   ({\tilde \phi_i}')^2 \right\rangle\,,~~~ {\widetilde E}_G \equiv \frac{1}{2 a^2 }  \sum_{i,k} \left\langle  (\widetilde\nabla_k \tilde \phi_{i})^2 \right\rangle\,, ~~~{\widetilde E}_V \equiv \left\langle \tilde{V}(\lbrace \tilde\phi_j\rbrace) \right\rangle\,,
\eea
so that the Friedmann equations read
\begin{eqnarray}\label{eq:NewFriedmannEQsI}
\left({a'\over a}\right)^2 &=& \frac{a^{2\alpha}}{3} \left( \frac{ f_*}{m_p} \right)^2 \Big[ {\widetilde E}_{K}  + {{\widetilde E}_{G}} + {{\widetilde E}_V} \Big] \,,\\
\label{eq:NewFriedmannEQsII}
{a''\over a} &=&  \frac{a^{2\alpha}}{3} \left( \frac{ f_*}{m_p} \right)^2 \Big[ (\alpha - 2){\widetilde E}_{K}  + \alpha {{\widetilde E}_{G}} + (\alpha + 1 ) {{\widetilde E}_V} \Big] \,.
\end{eqnarray}
If, on the contrary, the expansion of the universe is sourced by an external fluid, say with constant {\it barotropic} equation of state $w \equiv \langle p  \rangle / \langle \rho \rangle$, then we obtain the scale factor simply from the analytical expression
\be  a(\tilde\eta) = a_0\left(1 + \frac{1}{p}\widetilde{\mathcal{H}}_0 (\tilde\eta-\tilde\eta_0)\right)^p \,, \hspace{0.6cm}{\rm with}~~~ p = \frac{2}{3(1 + \omega) - 2 \alpha } \ , \ee
where we fixed the initial conditions at an initial time $\tilde\eta_0$ to $a_0 = a(\tilde\eta_0)$ and $\widetilde{\mathcal{H}}_0 \equiv \widetilde{\mathcal{H}}(\tilde\eta_0)$, and introduced a (dimensionless) \textit{program Hubble rate} $\widetilde{\mathcal{H}} = \frac{a^{\alpha} }{\omega_*} H$, where $H \equiv \dot a/a$ is the physical Hubble parameter.\\

\begin{mdframed}
{\it Program variables --.} We will refer to the dimensionless field and space-time variables in Eq.~(\ref{eq:FieldSpaceTimeNaturalVariables}) as the {\it lattice} or {\it program variables}\footnote{We will also define later on dimensionless \textit{program variables} for the charged scalars and gauge fields in Eqs.~(\ref{eq:GaugeProgramVar}) and (\ref{eq:GaugeProgramVarSU2}).}, and to the dimensionless potential in Eq.~(\ref{eq:PotNat}) as the {\it lattice} or {\it program potential}.  The values of $\alpha$, $f_*$ and $\omega_*$ can be chosen, in principle, arbitrarily. However, certain choices can be more convenient than others, depending on the form of the potential $V$. First, let us consider the choice of $\alpha$. A priori, this could be chosen at will: we could take $\alpha = 0$ if we wanted to solve our dynamics in cosmic time, whereas we could choose $\alpha = 1$ if we wanted to solve it in conformal time (up to dimensionful constant factors). However, there are many situations in which an oscillatory field dominates the energy budget of the system for a long time, with a time-dependent oscillation period $T_{\rm osc}(t)$. As the integration techniques introduced in the previous sections assume a constant time step, we would not be able to resolve later oscillations of the field with the same accuracy as early ones. If the  oscillation period decreases with time, this may cause stability problems in the simulation at late times.  Therefore, if we were in such a situation, it would be extremely convenient to choose a value of $\alpha$ that makes the oscillation period constant in the new $\alpha$-time variable. In Section \ref{sec:WorkingExample} we show an example of this in the context of a scalar field oscillating around the minimum of a monomial potential, so that we choose conveniently different values for $\alpha$, depending on the exponent of the power-law of the monomial potential.

Let us now consider the choice of $f_*$. For this, let us imagine a scenario in which one scalar field (say $\phi$) has initially a homogeneous configuration with a certain initial amplitude $\Phi_*$. A natural choice would be $f_* = \Phi_*$, so that as long as that field dominates the energy budget of the matter content in the simulation, its normalized amplitude $\tilde\phi = \phi/f_*$ will be of order unity (modulo red-shifting dilution factors due to the expansion of the universe). This is often the case in models with parametric resonance, such as the preheating scenario presented in Section \ref{sec:WorkingExample}. If on the contrary, relevant field(s) in the dynamics start with vanishing (or small) amplitude but acquire a vacuum expectation value $\langle \phi \rangle = v \neq 0$ later on, it might be convenient to take $f_* = v$. This will be the case e.g.~in models with spontaneous symmetry breaking, like in phase transitions and cosmic defect formation.

The choice of $\omega_*$ follows also a similar logic. For instance, if the dominant scalar field of the system oscillates say with a frequency $\Omega_{\rm osc}(\eta)$ (possibly time-dependent), it can be convenient to take $\omega_* = \Omega_{\rm osc}(\eta_*)$, at the time $\eta = \eta_*$ marking the onset of field oscillations. However, if the time scale $\Delta\eta_*$ of excitation of other fields is rather the relevant time scale in the problem, it might be more convenient to choose $\omega_*$ of the order of $1/\Delta\eta_*$. Another possibility would be to simply set $f_* = \omega_*$, so we prevent ratios $f_*/\omega_*$ (naturally appearing e.g.~in the initial condition of scalar field fluctuations) to become tiny or extremely large, see Section~\ref{sec:InitCond}.

In summary, if the choice of $\alpha$, $f_*$ and $\omega_*$ is made judiciously, it will lead to slowly varying dimensionless (initially of order unity) program field amplitudes and time scales. This will achieve a twofold objective: First, a better handle of the program variables in the computer, so we do not need to deal with very large nor very small numbers, and secondly, an easier and more transparent conversion of the dimensionless computer program variables into the physical mass/time scales of a given scenario. We will present practical examples of how to make optimal choices of ($\alpha$, $f_*$, $\omega_*$) in Section~\ref{sec:WorkingExample}.
\end{mdframed}

From now on, we assume that independently of the scenario under study, a convenient choice of ($\alpha$, $f_*$, $\omega_*$) has been made. In order to solve the dynamics of the interacting scalar fields in a computer, we need now to obtain some discretized version of the continuum EOM~(\ref{eq:EOMscalarContinuumNat}) expressed in the natural variables~(\ref{eq:FieldSpaceTimeNaturalVariables}), (\ref{eq:PotNat}). We need to do two things: first, to substitute somehow the time and spatial continuum derivatives by lattice operators mimicking such continuum differential operations up to some order $\mathcal{O}(\dx^\mu)$; and second, to solve the resulting discrete lattice EOM with some algorithms. Our toolkit to address these two aspects was provided in Section~\ref{sec:LatticeApproach}, where we introduced both lattice differential operators and evolution algorithms. Armed with such toolkit, we have essentially two options:\\

{\it i) Lattice action approach}. This is based on discretizing the continuum action, so that it is substituted by a lattice version. Varying such lattice action with respect to the lattice field $dof$, leads to lattice EOM enjoying whichever symmetry the lattice action enjoyed in first place. Constraint equations (expected as a consequence of the symmetries) are then automatically satisfied at the lattice level.\\

{\it ii) EOM discretization approach}. This is based on discretizing the continuum EOM directly. Here we simply substitute the partial derivatives involved in certain terms of the continuum EOM by appropriate lattice operators mimicking those continuum derivatives. This method allows to envisage lattice EOM adapted to essentially any evolution algorithm we wish to use. \\

Either approach $i)$ or $ii)$ may have its advantages and disadvantages depending on the model and circumstances. Whereas for EOM in flat space-time the two approaches are essentially very similar, this might not be case in an expanding universe, particularly in the presence of gauge fields. As for the time being, in this section, we only deal with scalar sectors in expanding backgrounds, we present next a series of algorithms based on either approach $i)$ or $ii)$, providing in each case a set of lattice EOM suitable to be solved by the evolution algorithms we previously introduced in Section~\ref{sec:LatticeApproach}.

\subsection{Lattice formulation of interacting scalar fields: $\mathcal{O}(\deta^2)$ accuracy methods}

\subsubsection{Staggered leapfrog from a lattice action}\label{subsec:StaggeredLFaction}

A lattice version of action (\ref{eq:ActionScalar}) can be written using e.g.~forward derivatives [c.f.~Eq.~(\ref{eq:forwardbackwardd})] for the time derivatives and the spatial gradients. Promoting integrals into discrete sums $\int d\eta(...) \equiv \delta\eta\sum_{n_0}(...)$, $\int dx^3(...) \equiv \dx^3\sum_{\bn}(...)$, we write
\be
\widetilde S_{\rm L} = \delta\tilde\eta\delta \tilde x^{\,3} \sum_{n_0}\sum_{\bn} \left\{\frac{ 1}{2} a_{+0/2}^{3 - \alpha}\sum_b(\widetilde\Delta_0^+ \tilde\phi^{b})^2 - \frac{1}{2} a^{1 + \alpha}   \sum_{b,k} (\widetilde\Delta_k^+ \tilde\phi^{b})^2 - a^{3 + \alpha} \widetilde V (\lbrace \tilde\phi_c \rbrace ) \right\} \, . \label{eq:ActionScDiscr}
\ee
Note that we have not determined yet at what times the scale factor 'lives in', and we have rather referred to a scale factor at integer and half-integers times, whenever appropriate. The logic to specify at what time the scale factor lives in each term of the action, is to consider the time at which the operator multiplying the scale factor, lives in. Thus, as $(\widetilde\Delta_0^+\tilde\phi^{(a)})^2$ lives at $n_0+1/2$, we write its pre-factor as $a_{+0/2}^{3-\alpha}$, whereas $(\widetilde\Delta_k^+\tilde\phi^{(a)})^2$ lives at $n_0$, so we write its pre-factor as $a^{1+\alpha}$, etc. Varying this action with respect each field $dof$, $\delta_{\phi_a}S_{\rm L} = 0$, leads to
the discrete EOM 
\bea
\widetilde\Delta_0^- [a_{+0/2}^{3 - \alpha} \widetilde\Delta_0^+ \tilde\phi_{b} ] & = & a^{1 + \alpha} \sum_k \widetilde\Delta_k^- \widetilde\Delta_k^+ \tilde\phi_{b}  -  a^{3 + \alpha} \widetilde V_{,\tilde\phi_{b}}\,,~~~~ b = 1, 2, ..., N_s\,,
\label{eq:EOMScalar-Discr}
\eea
with $N_s$ the total number of scalar fields.

Let us now deal with the expansion of the universe. We need to express the Friedmann equations as a function of program expressions of the volume averaged field energy and pressure densities $\langle \rho_\phi \rangle$, $\langle p_\phi \rangle$. We introduce first a discretized versions of the kinetic, gradiental and potential energies, c.f.~Eq.~(\ref{eq:EK_EG_EV}),
\bea\label{eq:EK_EG_EV_Discrete}
{\widetilde E}_K \equiv \frac{1}{2 a^{2\alpha}_{+0/2} }\sum_{b}\left\langle (\widetilde\Delta_0^+\tilde \phi_b)^2 \right\rangle\,,~~~ {\widetilde E}_G \equiv \frac{1}{2 a^2 }\sum_{b,k} \left\langle (\widetilde \Delta_k^+ \tilde \phi_{b})^2 \right\rangle\,, ~~~{\widetilde E}_V \equiv \left\langle \tilde{V}(\lbrace \tilde\phi_b\rbrace) \right\rangle\,
\eea
with ${\widetilde E}_G$ and ${\widetilde E}_V$ naturally living at integer times $n_0$, and ${\widetilde E}_{K}$ at semi-integer times $n_0+1/2$. We need to decide now whether we consider a scale factor living 'naturally' at integer or semi-integer times. If we consider that $a$ lives at semi-integer times, then it is natural to define the operator $b \equiv \widetilde\Delta_0^+a_{-0/2}$ living at integer times, and hence identify the first and second derivative [via the Friedmann equations in (\ref{eq:NewFriedmannEQsI}), (\ref{eq:NewFriedmannEQsII})] of the scale factor as
\bea
a' & \rightarrow &  \widetilde\Delta_0^+ a_{-0/2} \equiv b \ , \\
a '' & \rightarrow & \widetilde\Delta_0^+b =  \frac{1}{3} \left( \frac{ f_*}{m_p} \right)^2 a_{+0/2}^{1+2\alpha}\Big[ (\alpha - 2){\widetilde E}_{K}  + \alpha \overline{{\widetilde E}_{G}} + (\alpha + 1 ) \overline{{\widetilde E}_V} \,\Big] \,.
\eea
with $\overline{{\widetilde E}_{G}} \equiv ({\widetilde E}_G+{\widetilde E}_{G,+\hat0})/2$ and $\overline{{\widetilde E}_{V}} \equiv ({\widetilde E}_V+{\widetilde E}_{V,+\hat0})/2$, so that they live at semi-integer times, like the scale factor and ${\widetilde E}_K$. Alternatively, if we think of the scale factor living at integer times, we can define the operator $b_{+0/2} \equiv (\widetilde\Delta_0^+a)$ living at semi-integer times, and identify the first and second derivative of the scale factor as
\bea
a' & \rightarrow &  \widetilde\Delta_0^+ a \equiv b_{+0/2} \,,\\
a '' & \rightarrow & \widetilde\Delta_0^+b_{-0/2} =  \frac{1}{6} \left(\frac{ f_*}{m_p} \right)^2 a^{1+2\alpha}\Big[(\alpha - 2)\overline{{\widetilde E}_K}  + \alpha\,{\widetilde E}_G + (\alpha + 1 ){\widetilde E}_V \Big] \,,
\eea
with $\overline{{\widetilde E}_K} \equiv ({\widetilde E}_{K,-\hat0/2}+{\widetilde E}_{K,+\hat0/2})/2$ living at integer times, as much as $a, {\widetilde E}_G$ and ${\widetilde E}_V$.

From a practical or computational point of view, choosing a scale factor living at integer or semi-integer times, is actually irrelevant. If we choose that it lives at e.g.~integer times, we will always be forced to obtain it also, within each iteration, at semi-integer times, from the semi-sum of its two values at the closest integer times. And vice versa. In order to provide an iterative scheme, we still need to decide on the \textit{conjugate momenta} $\tilde\pi_{+0/2}^{(a)}$, which will be implemented through forward derivative operators. The question is whether to choose that $\tilde\pi_{+0/2}^{(a)}$ represents the time-derivative of each field, i.e.~$\phi_a'$, or rather represents $a^{3-\alpha}\phi_a'$, as the EOM actually naturally suggest. It turns out that depending on this choice the integrator will be accurate to order $\mathcal{O}(\delta\eta)$ or $\mathcal{O}(\delta\eta^2)$. Altogether, we can obtain the following implementations of a staggered leapfrog algorithm (here $IC$ stands for {\it Initial Condition}, and $HC$ for {\it Hubble Constraint}):\\

\begingroup
\allowdisplaybreaks
{\it I) Iterative scheme for $\tilde\pi_{+0/2}^{(a)} \equiv \widetilde\Delta_0^+\tilde\phi_a$} and scale factor $a(n_0+1/2)$:\vspace*{-0.2cm}\\
\bea
IC & : & \lbrace \tilde\phi_a,b\rbrace {\rm ~at~} \tilde\eta_0, ~~~\lbrace \tilde\pi_{-0/2}^{(a)},a_{-0/2}\rbrace {\rm ~at~} \tilde\eta_0-0.5\delta\tilde\eta\\
a_{+0/2} &=&  a_{-0/2} + b\, \delta\tilde\eta ~~~~ \longrightarrow ~~~~ a \equiv (a_{+0/2} + a_{-0/2})/2\\
\tilde\pi_{+0/2}^{(a)} & = & \left({a_{-0/2}\over a_{+0/2}}\right)^{3 - \alpha}\hspace*{-0.2cm}\tilde\pi_{-0/2}^{(a)} ~+~ a_{+0/2}^{-(3 - \alpha)}\left( a^{1 + \alpha} \sum_k \widetilde\Delta_k^- \widetilde\Delta_k^+ \tilde\phi^{(a)}  -  a^{3 + \alpha} \widetilde V_{,\tilde\phi^{(a)}}\right)\delta\tilde\eta\\
\tilde\phi^{(a)}_{+0} &=& \tilde\phi^{(a)} + \delta\tilde\eta\,\tilde\pi_{+0/2}^{(a)} \\
b_{+0} &=& b + \frac{\delta\tilde\eta}{3} \left( \frac{ f_*}{m_p} \right)^2 a_{+0/2}^{1+2\alpha}\Big[(\alpha - 2){\widetilde E}_{K}  + \alpha \overline{{\widetilde E}_{G}} + (\alpha + 1 ) \overline{{\widetilde E}_V} \,\Big]\,,\\
\label{eq:HCschemeI}
HC &:& b^2 = \frac{1}{3} \left( \frac{ f_*}{m_p} \right)^2a^{2(\alpha+1)} \Big(\,\overline{{\widetilde E}_{K}} + {{\widetilde E}_{G}} + {{\widetilde E}_V} \Big)\,.
\eea\\
\endgroup

\begingroup
\allowdisplaybreaks
{\it II) Iterative scheme for $\tilde\pi^{(a)} \equiv \widetilde\Delta_0^+\tilde\phi^{(a)}_{-0/2}$} and scale factor $a(n_0)$\vspace*{-0.2cm}\\
\bea
IC & : & \lbrace \tilde a,\tilde\pi^{(a)}\rbrace {\rm ~at~} \tilde\eta_0, ~~~\lbrace \tilde\phi^{(a)}_{-0/2},b_{-0/2}\rbrace {\rm ~at~} \tilde\eta_0-0.5\delta\tilde\eta\\
\tilde\phi^{(a)}_{+0/2} &=& \tilde\phi^{(a)}_{-0/2} + \delta\tilde\eta\,\tilde\pi^{(a)} \\
b_{+0/2} &=& b_{-0/2} + \frac{\delta\tilde\eta}{3} \left( \frac{ f_*}{m_p} \right)^2 a^{1+2\alpha}\Big[ (\alpha - 2){\widetilde E}_{K}  + \alpha \overline{{\widetilde E}_{G}} + (\alpha + 1 ) \overline{{\widetilde E}_V}\, \Big]\,,\\
a_{+0} &=&  a + b_{+0/2}\, \delta\tilde\eta ~~~~ \longrightarrow ~~~~ a_{+0/2} \equiv (a_{0} + a)/2\,,\\
\tilde\pi_{+0}^{(a)} &=& \left({a\over a_{+0}}\right)^{3 - \alpha}\tilde\pi^{(a)} + a_{+0}^{-(3 - \alpha)}\left( a_{+0/2}^{1 + \alpha} \sum_k \widetilde\Delta_k^- \widetilde\Delta_k^+ \tilde\phi^{(a)}_{+0/2}  -  a^{3 + \alpha}_{+0/2} \widetilde V_{,\tilde\phi^{(a)}}\Big|_{+0/2}\right)\delta\tilde\eta\,,\\
\label{eq:HCschemeII}
HC &:& b_{+0/2}^2 = \frac{1}{3} \left( \frac{ f_*}{m_p} \right)^2a_{+0/2}^{2(\alpha+1)} \Big(\,\overline{{\widetilde E}_{K}} + {\widetilde E}_{G} + {\widetilde E}_{V} \Big)\,,
\eea\\
\endgroup

\begingroup
\allowdisplaybreaks
{\it III) Iterative scheme for $\tilde\pi_{+0/2}^{(a)} \equiv a_{+0/2}^{3-\alpha}\widetilde\Delta_0^+\tilde\phi_a$} and scale factor $a(n_0)$\vspace*{-0.2cm}\\
\bea
IC & : & \lbrace \tilde\phi^{(a)},a,\rbrace {\rm ~at~} \tilde\eta_0, ~~~\lbrace \tilde\pi^{(a)}_{-0/2},b_{-0/2}\rbrace {\rm ~at~} \tilde\eta_0-0.5\delta\tilde\eta\\
\tilde\pi_{+0/2}^{(a)} & = & \tilde\pi_{-0/2}^{(a)} + \left( a^{1 + \alpha} \sum_k \widetilde\Delta_k^- \widetilde\Delta_k^+ \tilde\phi^{(a)}  -  a^{3 + \alpha} \widetilde V_{,\tilde\phi^{(a)}}\right)\delta\tilde\eta\\
b_{+0/2} &=& b_{-0/2} + \frac{\delta\tilde\eta}{3} \left( \frac{ f_*}{m_p} \right)^2 a^{1+2\alpha}\Big[ (\alpha - 2)\overline{{\widetilde E}_{K}}  + \alpha {{\widetilde E}_{G}} + (\alpha + 1 ) {{\widetilde E}_V} \Big]\,,\\
a_{+0} &=&  a + b_{+0/2}\, \delta\tilde\eta ~~~~ \longrightarrow ~~~~ a_{+0/2} \equiv (a_{+0} + a_{0})/2\,,\\
\tilde\phi^{(a)}_{+0} &=& \tilde\phi^{(a)} + \delta\tilde\eta\,\tilde\pi_{+0/2}^{(a)}a_{+0/2}^{-(3-\alpha)}\,,\\
\label{eq:HCschemeIII}
HC &:& b_{+0/2}^2 = \frac{1}{3} \left( \frac{ f_*}{m_p} \right)^2a_{+0/2}^{2(\alpha+1)} \Big({{\widetilde E}_{K}} + \overline{{\widetilde E}_{G}} + \overline{{\widetilde E}_{V}} \,\Big)\,,
\eea\\
\endgroup

\begingroup
\allowdisplaybreaks
{\it IV) Iterative scheme for $\tilde\pi^{(a)} \equiv a^{3-\alpha}\widetilde\Delta_0^+\tilde\phi^{(a)}_{-0/2}$} and scale factor $a(n_0+1/2)$\vspace*{-0.2cm}\\
\bea
IC & : & \lbrace \tilde\pi^{(a)},b\rbrace {\rm ~at~} \tilde\eta_0, ~~~\lbrace \tilde\phi^{(a)}_{-0/2},a_{-0/2}\rbrace {\rm ~at~} \tilde\eta_0-0.5\delta\tilde\eta\\
a_{+0/2} &=&  a_{-0/2} + b\, \delta\tilde\eta ~~~~ \longrightarrow ~~~~ a \equiv (a_{+0/2} + a_{-0/2})/2\\
\tilde\phi^{(a)}_{+0/2} &=& \tilde\phi^{(a)}_{-0/2} + \delta\tilde\eta\,\tilde\pi^{(a)}a^{-(3-\alpha)} \\
\tilde\pi_{+0}^{(a)} & = & \tilde\pi^{(a)} ~+~ \left(a_{+0/2}^{1 + \alpha} \sum_k \widetilde\Delta_k^- \widetilde\Delta_k^+ \tilde\phi^{(a)}_{+0/2}  -  a^{3 + \alpha}_{+0/2} \widetilde V_{,\tilde\phi^{(a)}}\Big|_{+0/2}\right)\delta\tilde\eta\\
b_{+0} &=& b + \frac{\delta\tilde\eta}{3} \left( \frac{ f_*}{m_p} \right)^2 a_{+0/2}^{1+2\alpha}\Big[(\alpha - 2)\overline{{\widetilde E}_{K}}  + \alpha {{\widetilde E}_{G}} + (\alpha + 1 ) {{\widetilde E}_{V}}\, \Big]\,,\\
\label{eq:HCschemeIV}
HC &:& b^2 = \frac{1}{3} \left( \frac{ f_*}{m_p} \right)^2a^{2(\alpha+1)} \Big({{\widetilde E}_{K}} + \overline{{\widetilde E}_{G}} + \overline{{\widetilde E}_V} \,\Big)\,.
\eea\\
\endgroup

While all these iterative schemes descent from the same action~(\ref{eq:ActionScDiscr}), they are truly different algorithms, based on the choice of conjugate momenta and time domain of the scale factor. In fact, iterative schemes $I$ and $II$, which are basically very similar as they are based on (discretized versions of) the same choice $\pi_a \equiv \phi_a'$, are only accurate to order $\mathcal{O}(\delta \eta)$. Iterative schemes $III$ and $IV$, also very similar to each other as they are based on (discretized versions of) the choice $\pi_a \equiv a
^{3-\alpha}\phi_a'$, are however accurate to order $\mathcal{O}(\delta \eta^2)$. This becomes manifest in numerical simulations by monitoring the Hubble constraint $3m_p^2H^2 = \rho$, which in the case of schemes $I$ and $II$ is only verified to order $\mathcal{O}(\delta \eta)$ by Eqs.~(\ref{eq:HCschemeI}), (\ref{eq:HCschemeII}), whereas in the schemes $III$ and $IV$, Eqs.~(\ref{eq:HCschemeIII}), (\ref{eq:HCschemeIV}) are verified to order $\mathcal{O}(\delta \eta^2)$. This is a clear illustration of the importance of choosing the appropriate conjugate momentum to evolve the equations.

\subsubsection{Staggered leapfrog {\it \`a la LatticeEasy}}
\label{subsec:StaggeredLF_LE}

Alternatively to discretizing action~(\ref{eq:NsCanonicalScalarsAction}), like in (\ref{eq:ActionScDiscr}), one can start from the continuum EOM for scalar fields, Eq.~(\ref{eq:EOMscalarContinuumNat}), and discretize these equations directly. Considering the EOM of $N_s$ scalar fields canonically normalized [c.f.~Eq.~(\ref{eq:EOMscalarContinuumNat})],
\bea\label{eq:NsCanonicalScalars}
\tilde{\phi}_i'' - a^{-2(1 - \alpha)} \tilde{\nabla}^{2} \tilde{\phi}_i + (3 - \alpha)\widetilde{\mathcal{H}} {\tilde{\phi}_i'} + a^{2 \alpha} \tilde{V}_{,\tilde{\phi}_i} = 0\,,~~~~ i = 1, 2, ..., N_s \ ,
\eea
where $\widetilde{\mathcal{H}} \equiv a'/a$, we could attempt to substitute here the continuum derivatives $\tilde{\partial}_\mu$ by finite difference operators $\tilde{\Delta}_\mu^\pm$, and then obtain a discretized version of the EOM. However, we would run immediately into the problem of the friction term $\propto \tilde{\phi}'$, which prevents the iterative scheme to be in the form of a Staggered leapfrog algorithm: namely, the kernel of the conjugate momenta, $\tilde{\pi}_i' = \mathcal{K}_i[\lbrace \tilde \phi_j\rbrace,\lbrace \tilde \pi_j\rbrace,...]$ , would depend on the conjugate momenta itself $\tilde{\pi}_i = \tilde{\phi}_i'$, so $\partial_{\tilde{\pi}_i}\mathcal{K}_i \neq 0$.

It turns out that the problem can be easily avoided, by making a conformal re-definition of the scalar field amplitudes $ \tilde\phi_i \longrightarrow  \tilde \varphi_i \equiv a^\beta \tilde \phi_i$, so that the EOM become
\bea
\tilde \varphi_i'' + (3 - \alpha - 2\beta)\widetilde{\mathcal{H}} {\tilde \varphi_i'} - a^{-2(1 - \alpha)} {\tilde \nabla}^{\,2} \tilde \varphi_i + a^{2(\alpha+\beta)} \tilde V_{,\tilde \varphi_i} + \widetilde{m}^2[\alpha,\beta] \tilde \varphi_i = 0\,,
\eea
with
\bea
\widetilde{m}^2[\alpha,\beta] \equiv \beta\left((\alpha+\beta-2)\widetilde{\mathcal{H}}^2-{a''\over a}\right)\,.
\eea
Choosing $\beta = (3-\alpha)/2$ leads immediately to the elimination of the friction term, so the EOM can be finally written like
\bea\label{eq:EOMafterConfReDef}
\tilde \varphi_i'' - a^{-2(1 - \alpha)} {\tilde \nabla}^{\,2} \tilde \varphi_i + a^{3+\alpha} \tilde{V}_{,\tilde \varphi_i} + \widetilde{m}^2_\alpha \tilde \varphi_i = 0\,,~~~~ \widetilde{m}^2_\alpha \equiv {(3-\alpha)\over2}\left({(\alpha-1)\over2}\widetilde{\mathcal{H}}^2-{a''\over a}\right)\, .
\eea
This corresponds to a set of equations that can be discretized with a staggered leapfrog algorithm simply by choosing $\tilde \pi_{i} \equiv \tilde \varphi_i'$, so that the kernel $\tilde \pi_{i}' = \mathcal{K}_i[...]$ depends on the field amplitudes but not on the field momenta, $\partial_{\tilde \pi_{i}}\mathcal{K}_j = 0~\forall\, i,j$. This trick however is not enough by itself, as we still have to deal with the evolution of the scale factor through the Friedmann equations, now incorporating the conformal re-scaling of the field amplitudes. Furthermore, we notice that the new mass term $m_\beta^2$ depends actually on $\mathcal{H}$ and $a''/a$, so in order for the kernel of $\tilde \pi_i$ to depend only say on integer times (assuming $\tilde \pi_i$'s live at semi-integer times), we need both $\mathcal{H}$ and $a''/a$ to be evaluated at the same integer time. Recalling Eq.~(\ref{eq:FriedmannDDa}) for  $a''$, and denoting the scale factor conjugate momentum as $b \equiv a'$, we observe that $b' = \mathcal{K}_a[a,{\widetilde E}_K,{\widetilde E}_G,{\widetilde E}_V]$ with ${\widetilde E}_K, {\widetilde E}_G$ and ${\widetilde E}_V$ representing the volume averaged kinetic, gradient and potential terms, as contributed by all scalar fields, c.f.~(\ref{eq:EK_EG_EV}). We have therefore, on the one hand, EOM dictating the evolution of the conjugate momenta $\tilde \pi_{i}'s$ via kernels depending on $a''/a$, and on the other hand, an equation for $a''/a$ depending on the conjugate momenta $\lbrace \tilde \pi_{j}\rbrace$ through the kinetic terms. This prevent us from obtaining an explicit solution. Furthermore, if we substitute the conformal redefinition of the field amplitude in the kinetic terms sourcing $a''/a$, we obtain that the kernel $b' = \mathcal{K}_a[...]$ contains terms $\propto \tilde \pi_{j}^2$, $\propto {\widetilde{\mathcal{H}}}^2 \tilde \varphi_j^2$ and $\propto \widetilde{\mathcal{H}}\tilde \varphi_j \tilde \pi_{j}$, and hence that $\mathcal{K}_a[...]$ depends also on $\widetilde{\mathcal{H}}$. We immediately understand that implementing a staggered leapfrog algorithm is still not feasible in the current system of equations, unless we perform some extra step.

The celebrated package {\it LatticeEasy} (and for this matter its parallelized version {\it ClusterEasy}), developed in the 2000's by Gary N.~Felder and Igor I.~Tkachev, circumvented the previous issue by means of the following trick: the kinetic term ${\widetilde E}_K$ in the Friedmann equation $3m_p^2b' =   f_*^2a^{1+2\alpha}[(\alpha-2){\widetilde E}_K+\alpha {\widetilde E}_G+(1+\alpha){\widetilde E}_V]$, can be substituted by its expression obtained from the other Friedmann equation $3m_p^2\widetilde{\mathcal{H}}^2 = f_*^2a^{2\alpha}({\widetilde E}_K + {\widetilde E}_G + {\widetilde E}_V)$. That is ${\widetilde E}_K = 3(f_*/m_p
)^{-2}a^{-2\alpha}\widetilde{\mathcal{H}}^2 - {\widetilde E}_G - {\widetilde E}_V$, and hence $3m_p^2b' =  3(\alpha-2)m_p
^2a\widetilde{\mathcal{H}}^2 + f_*^2a^{1+2\alpha}(2{\widetilde E}_G + 3{\widetilde E}_V)$. Choosing that the scale factor lives at integer times, and introducing $b_{+0/2} \equiv \widetilde{\Delta}_0^+a$ and $\widetilde{\Delta}_0^-b_{+0/2} \equiv \widetilde{\Delta}_0^-\widetilde{\Delta}_0^+a$ for the first and second derivative of the scale factor, the second Friedmann equation can be written as
\bea\label{eq:FriedmannLE}
\left({b_{+0/2}-b_{-0/2}\over\delta\tilde\eta}\right) =
{(\alpha-2)\over a}\left({b_{+0/2}+b_{-0/2}\over2}\right)^2 + {1\over 3}\left({f_*\over m_p}\right)^2a^{1+2\alpha}(2{\widetilde E}_G + 3{\widetilde E}_V)\,,
\eea
where we have introduced an averaged Hubble rate as  $\overline{\widetilde{\mathcal{H}}} \equiv {(b_{+0/2}+b_{-0/2})\over 2a}$. As Eq.~(\ref{eq:FriedmannLE}) is simply a quadratic equation for $b_{+0/2}$, we can re-arrange terms to write
\bea \label{eq:B2B1B0}
B_2b_{+0/2}^2 + B_1b_{+0/2} + B_0 = 0\,,~~~~~
\left\lbrace
\begin{array}{rcl}
  B_2 & \equiv & c\delta\tilde\eta\,,~~~~~ c \equiv {(\alpha-2)\over 4a} \vspace*{0.2cm}\\
  B_1 & \equiv & 2c\delta\tilde\eta b_{-0/2} - 1\\
  B_0 & \equiv & b_{-0/2}(1 + c\delta\tilde\eta b_{-0/2}) + {\delta\tilde\eta\over 3}\left({f_*\over m_p}\right)^2a^{1+2\alpha}(2{\widetilde E}_G + 3{\widetilde E}_V)
\end{array}
\right.
\eea
so that $b_{+0/2} = (-B_1 \pm \sqrt{B_1^2 - 4B_2B_0})/(2B_2)$. Only by choosing the negative sign we arrive at the correct limit $b_{+0/2} \longrightarrow b_{-0/2}$ when $\delta\tilde\eta \longrightarrow 0$, so we finally obtain
\bea
b_{+0/2} = - b_{-0/2} + {1\over 2 c \delta\tilde\eta}\left(1-\sqrt{(1-2c\delta\tilde\eta b_{-0/2})^2-4c\delta\tilde\eta B_0}\,\right)\,.
\eea
The kernel of the second Friedmann equation $a''/a = \mathcal{K}_a[...]$ does not depend, in this way, on the field conjugate momenta $\tilde\pi_j$'s, while at the same time the mass term in the EOM of the $\tilde{\pi}_{i}$'s, Eq.~(\ref{eq:EOMafterConfReDef}), can now be built from $\overline{\widetilde{\mathcal{H}}} \equiv (b_{-0/2}+b_{+0/2})/(2a)$ and $a''/a \equiv (b_{+0/2}-b_{+0/2})/(a\delta\tilde\eta)$. Hence, a consistent staggered leapfrog algorithm can be put forward [with $c, B_0$ given in Eq.~(\ref{eq:B2B1B0})] as\\

\begingroup
\allowdisplaybreaks
{\it LatticeEasy staggered leapfrog scheme}:\vspace*{-0.2cm}\\
\bea
IC & : & \lbrace \tilde\varphi_i,a\rbrace {\rm ~at~} \tilde\eta_0, ~~~\lbrace \tilde\pi_{-0/2}^{(i)},b_{-0/2}\rbrace {\rm ~at~} \tilde\eta_0-0.5\delta\tilde\eta\\
b_{+0/2} &=& - b_{-0/2} + {1\over 2 c \delta\tilde\eta}\left(1-\sqrt{(1-2c\delta\tilde\eta b_{-0/2})^2-4c\delta\tilde\eta B_0}\right)\,\\
\tilde\pi_{+0/2}^{(i)} & = & \tilde\pi_{-0/2}^{(i)} ~+~ \delta\tilde\eta\left( a^{-2(1 - \alpha)} \tilde{\nabla}^{\,2}  \tilde \varphi_i - a^{3 + \alpha} \tilde{V}_{,\tilde \varphi_i} - \widetilde{m}^2_\alpha(a,b_{-0/2},b_{+0/2}) \tilde \varphi_i\right)\\
\tilde\varphi^{(i)}_{+0} &=& \tilde\varphi^{(i)} + \delta\tilde\eta\,\tilde\pi_{+0/2}^{(i)} \\
a_{+0} &=&  a + b_{+0/2}\, \delta\tilde\eta
\,
\eea
\endgroup

As the field amplitudes are conformally transformed as $ \tilde \phi_i = a^{-\beta} \tilde{\varphi}_i$, the canonical momenta becomes $\tilde \phi_i' = a^{-\beta}(\tilde\pi_i-\beta\widetilde{\mathcal{H}}\tilde \varphi_i)$, with $\tilde \pi_i = \tilde \varphi_i'$ and $\beta = (3-\alpha)/2$. The Hubble constraint (in the continuum) then becomes
\bea\label{eq:HC_LE}
HC:~~~~~~~ \widetilde{\mathcal{H}}^2 = {1\over 3}\left({f_*\over m_p}\right)^2\left( {a^{\alpha-3}\over2}\sum_i\left\langle(\tilde \pi_i-\beta\widetilde{\mathcal{H}} \tilde \varphi_i)^2\right\rangle + {a^{3\alpha-5}\over2}\sum_i\left\langle (\tilde{ \vv{\nabla}} \tilde{\varphi}_i)^2 \right\rangle + a^{2\alpha}\left\langle \tilde{V} \right\rangle\right)\,,
\eea
which is a quadratic equation for $\mathcal{H}$. $LatticeEasy/ClusterEasy$ presented a discretized version of Eq.~(\ref{eq:HC_LE}) that is verified numerically to accuracy $\mathcal{O}(\delta\tilde\eta^2)$, as expected for a valid staggered leapfrog scheme. Discretizing the expression in Eq.~(\ref{eq:HC_LE}) requires however to deal with the fact that there are terms in its $rhs$ that live at integer times, others that live at semi-integer times, and there are even 'crossed' terms built from the product of the latter two. The solution $LatticeEasy/ClusterEasy$ adopted for this was to synchronize the field amplitudes with the conjugate momenta, just before checking the Hubble law. In this way field amplitudes are evolved backwards in time by half time step, so that they live, during the check, at the same semi-integer step as the conjugate momenta. After the check has been done, one just evolves forward by half time step the field amplitudes, so that they are back to their appropriate value within the evolution loop.

A remark is now in order. Even though the above implementation of an $\mathcal{O}(\delta\eta^2)$ staggered leapfrog algorithm for scalar fields, works fine, it is somehow more cumbersome than our proposed schemes in Section~\ref{subsec:StaggeredLFaction}. First of all, the conformal transformation of the fields leads to mix terms between amplitudes and conjugate momenta, whenever time derivatives of the original field amplitudes are calculated. Secondly, the whole method relies on the elimination of the kinetic term from the second Friedmann equation. This leads to solving the evolution of the scale factor from a quadratic equation, and generates a Hubble constraint which becomes itself also a quadratic equation for the Hubble rate $\mathcal{H}$. Even though one does not need to solve explicitly for $\mathcal{H}$ to verify the Hubble law\footnote{It is enough to check that the $lhs$ and the $rhs$ of Eq.~(\ref{eq:HC_LE}) are numerically balanced to order $\mathcal{O}(\delta\eta^2)$.}, one still needs to synchronize and desynchronize the field amplitudes just before and after checking the Hubble law. Furthermore, whenever computing observables like the fields' energy density terms or the various relevant field spectra, one needs to un-do the conformal transformation of the field variables, in order to obtain physically meaningful quantities (or at least one needs to have a very clear idea of what is being obtained, and in what variables it is written down). While none of these aspects are particularly difficult to deal with, they altogether make this prescription more complicated than e.g.~our iterative schemes $III$ and $IV$ in  Section~\ref{subsec:StaggeredLFaction}. Our schemes are simpler and more natural, not only because they do not require to re-define the field variables, but also because they are based on a lattice principle, i.e.~in writing first a correct discretized action, from which naturally follows the dynamics and the observables. The most important caveat, however, is something that has not yet become manifest, but that we can anticipate here: the trick used by $LatticeEasy/ClusterEasy$ to eliminate the kinetic term of the scalar fields from the Friedmann equation for $a''/a$, is not generalizable to gauge theories. This is simple to understand,  recalling Eqs.~(\ref{eq:FriedmannHubble}),(\ref{eq:FriedmannDDa}), we observe that the weight of the scalar and gauge field kinetic terms is the same in the Hubble constraint, but it is different in the equation for $a''/a$. This implies that it is impossible to eliminate both kinetic terms at the same time from the second Friedmann equation, and hence it is not possible to achieve a scale factor kernel $\mathcal{K}_a[...]$ free from all fields' conjugate momenta (this is actually independent of any potential conformal transformation of the gauge fields): $\mathcal{K}_a[...]$ is always left with the conjugate momenta of the scalar fields, or with the conjugate momenta of the gauge fields (played by the electric fields of the problem). Therefore, an approach of this kind to obtain a second order integrator for a gauge theory in an expanding background (no matter whether Abelian or non-Abelian), is simply not feasible.

\subsubsection{Synchronized Leapfrog: Position- and Velocity-Verlet}
\label{subsec:VerletGeneral}

While in Section~\ref{subsec:StaggeredLFaction} our starting point was a lattice action from which we derived the lattice EOM, here we will rather discretize directly the continuum EOM, without introducing a conformal rescaling, similarly to what was done in Section~\ref{subsec:StaggeredLFaction}, using a re-definition of the field variables. Considering again the EOM of $N_s$ scalar fields canonically normalized, c.f.~Eq.~(\ref{eq:NsCanonicalScalars}), we immediately conclude that the field variables' kernel depend on the conjugate momenta through the friction term $(3 - \alpha)\widetilde{\mathcal{H}} {\tilde  \phi_b'}$. This appears seemingly as an impediment to apply staggered or synchronized leapfrog methods. The EOM as derived initially from the continuum action~(\ref{eq:ActionScalar}), can however be written as
\bea\label{eq:EOMtowardsVV}
(a^{(3 - \alpha)}\tilde \phi_i')' - a^{1 + \alpha} {\tilde \nabla}^{\,2}  \tilde \phi_i  + a^{3+ \alpha} \tilde V_{,\tilde \phi_i} = 0\,,~~~~ i = 1, 2, ..., N_s \ ,
\eea
so that only when expanding the first term (and after multiplying by $a^{-(3 - \alpha)}$), the standard second derivative and friction terms ${\tilde \phi_i''} + (3 - \alpha)\widetilde{\mathcal{H}} {\tilde \phi_i'}$ become explicit. Instead of expanding such terms, the form of Eqs.~(\ref{eq:EOMtowardsVV}) invites to rather re-write them more naturally in a Hamiltonian-like scheme as
\begin{eqnarray}\label{eq:HamiltonianEqsI}
\tilde \phi_i' &=& a^{-(3 - \alpha)}\tilde \pi_i\,,\\
\label{eq:HamiltonianEqsII}
\tilde \pi_i' &=& a^{1 + \alpha} {\tilde \nabla}^{\,2} \tilde \phi_i  - a^{3+ \alpha} \tilde  V_{,\tilde \phi_i}\,,
\end{eqnarray}
where it is manifest that the kernel does not depend on the conjugate momenta. Analogously, the second Friedmann equation~(\ref{eq:NewFriedmannEQsII}) can then be written as
\begin{eqnarray}
&& a' = b \,,\\
\label{eq:FriedmannEq_withPi_I}
&& b' = \frac{a^{1+2\alpha}}{3} \left( \frac{ f_*}{m_p} \right)^2 \Big[ (\alpha - 2){\widetilde E}_{K}  + \alpha {{\widetilde E}_{G}} + (\alpha + 1 ) {{\widetilde E}_V} \Big] \,,\\
\label{eq:EK_EG_EV_withPi}
{\rm where} &&
{\widetilde E}_K \equiv \frac{1}{2 a^{6}} \sum_{i}\left\langle   ({\tilde \pi_i})^2 \right\rangle\,,~~~ {\widetilde E}_G \equiv \frac{1}{2 a^2 }  \sum_{i,k} \left\langle  (\widetilde\nabla_k \tilde \phi_{i})^2 \right\rangle\,, ~~~{\widetilde E}_V \equiv \left\langle \tilde{V}(\lbrace \tilde\phi_j\rbrace) \right\rangle\,,
\end{eqnarray}
This immediately invites for the application of either staggered or synchronized leapfrog methods. In fact, the methods $III$ and $IV$ from Section~\ref{subsec:StaggeredLFaction} corresponded precisely with the application of staggered leapfrog schemes, so we will rather focus now on the application of synchronized leapfrog schemes, i.e.~Verlet integrators, either velocity- or position-based. Following Section~\ref{subsec:VerletGeneral}, these algorithms read:\\

\begingroup
\allowdisplaybreaks
{\it I) Velocity-Verlet scheme for interacting scalar fields in an expanding background}\vspace*{-0.4cm}\\
\bea
IC & : & \lbrace \tilde \phi^{(i)},\tilde\pi^{(i)},a,b\rbrace {\rm ~at~} \tilde\eta_0\,,\\
b_{+0/2} &=& b + \frac{1}{3} \left( \frac{ f_*}{m_p} \right)^2 a^{1+2\alpha}\Big[ (\alpha - 2){{\widetilde E}_{K}}  + \alpha {{\widetilde E}_{G}} + (\alpha + 1 ) {{\widetilde E}_V} \Big]{\delta\tilde\eta\over2}\,,\\
\tilde\pi_{+0/2}^{(i)} & = & \tilde\pi^{(i)} + \left( a^{1 + \alpha} \sum_k \widetilde\Delta_k^- \widetilde\Delta_k^+ \tilde\phi^{(i)}  -  a^{3 + \alpha} \widetilde V_{,\tilde\phi^{(i)}}\right){\delta\tilde\eta\over2}\,,\\
a_{+0} &=&  a +  b_{+0/2}{\delta\tilde\eta}\,,\\
a_{+0/2} &=& \frac{a_{+0}+a}{2}\,,\\
\tilde\phi^{(i)}_{+0} &=& \tilde\phi^{(i)} + \delta\tilde\eta\,\tilde\pi_{+0/2}^{(i)}a_{+0/2}^{-(3-\alpha)}\,,\\
\tilde\pi_{+0}^{(i)} & = & \tilde\pi^{(i)}_{+0/2} + \left( a^{1 + \alpha}_{+0} \sum_k \widetilde\Delta_k^- \widetilde\Delta_k^+ \tilde\phi^{(i)}_{+0}  -  a_{+0}^{3 + \alpha} \widetilde V_{,\tilde\phi^{(i)}}\Big|_{+0}\right){\delta\tilde\eta\over2} \, \ ,\\
b_{+0} &=& b_{+0/2} + \frac{1}{3} \left( \frac{ f_*}{m_p} \right)^2  a_{+0}^{1+2\alpha}\Big[ (\alpha - 2){{\widetilde E}_{K,{+0}}}  + \alpha {{\widetilde E}_{G,{+0}}} + (\alpha + 1 ) {{\widetilde E}_{V,{+0}}} \Big]{\delta\tilde\eta\over2}\, \ , \\
\label{eq:HC_VVscheme}
HC &:& b^2 = \frac{1}{3} \left( \frac{ f_*}{m_p} \right)^2a^{2(\alpha+1)} \Big({{\widetilde E}_{K}} + {{\widetilde E}_{G}} + {{\widetilde E}_{V}} \Big)\,,
\eea\\

{\it II) Position-Verlet scheme for interacting scalar fields in an expanding background}\vspace*{-0.4cm}\\
\bea
IC & : & \lbrace \tilde \phi^{(i)},\tilde\pi^{(i)},a,b\rbrace {\rm ~at~} \tilde\eta_0\,,\\
a_{+0/2} &=&  a + b\,{\delta\tilde\eta\over 2}\,,\\
\tilde\phi^{(i)}_{+0/2} &=& \tilde\phi^{(i)} + {\delta\tilde\eta\over 2}\tilde\pi^{(i)}a^{-(3-\alpha)}\,,\\
\tilde\pi_{+0}^{(i)} & = & \tilde\pi^{(i)} +\left( a^{1 + \alpha}_{+0/2} \sum_k \widetilde\Delta_k^- \widetilde\Delta_k^+ \tilde\phi^{(i)}_{+0/2}  -  a_{+0/2}^{3 + \alpha} \widetilde V_{,\tilde\phi^{(i)}}\Big|_{+0/2}\right){\delta\tilde\eta} \,,\\
b_{+0} &=& b + \frac{1}{3} \left( \frac{ f_*}{m_p} \right)^2 a_{+0/2}^{1+2\alpha}\Big[ (\alpha - 2)\overline{{\widetilde E}_K}  + \alpha {\widetilde E}_{G} + (\alpha + 1 ) {\widetilde E}_{V} \Big]{\delta\tilde\eta}\,,\\
a_{+0} &=&  a_{+0/2} + b_{+0}\, {\delta\tilde\eta\over2} \,,\\
\tilde\phi^{(i)}_{+0} &=& \tilde\phi^{(i)}_{+0/2} + {\delta\tilde\eta\over2}\,\tilde\pi_{+0}^{(i)}a_{+0}^{-(3-\alpha)}\,.\\
\label{eq:HC_PVscheme}
HC &:& b^2 = \frac{1}{3} \left( \frac{ f_*}{m_p} \right)^2a^{2(\alpha+1)} \Big({{\widetilde E}_{K}} + {{\widetilde E}_{G}} + {{\widetilde E}_{V}} \Big)\,,
\eea
\endgroup

Both algorithms have $\mathcal{O}(\delta\eta^2)$ accuracy, 
so one can use in principle one or the other, and obtain the same results. Verlet integrators have however three steps per iteration (as they come in a {\it kick-drift-kick} or {\it drift-kick-drift} fashion) versus two steps of the staggered leapfrog integrators $III$ and $IV$ from Section~\ref{subsec:StaggeredLFaction} (which come in a $drift-kick$ or $kick-drift$ scheme). Verlet integrators are therefore some $\sim 30\%-50\%$ slower than staggered leapfrog algorithms. They can be used however to implement higher-order in $\delta\eta$ with the method presented in Section \ref{subsec:Yoshida}, see Section \ref{subsec:YoshidaScalars}.

\subsection{Lattice formulation of interacting scalar fields: $\mathcal{O}(\deta^n)$ accuracy methods}

\subsubsection{Explicit Runge-Kutta 4$th$ order}
\label{subsec:RK4scalars}

Here we just need to specialize Eqs.~(\ref{eq:RG4algorithm}) corresponding to the (explicit) Runge-Kutta method of order $\mathcal{O}(\delta\eta^4)$ to the EOM for $N_s$ interacting scalar fields dictating the expansion of the universe, Eqs.~(\ref{eq:EOMscalarContinuumNat}) and (\ref{eq:NewFriedmannEQsII}). We first re-write the continuum EOM as
\begin{eqnarray}
a' &=& b \,,\\
\tilde\phi_i' &=& \tilde\pi_i\,,\\
\tilde\pi_i' &=& \mathcal{K}_i[a,b,\lbrace \tilde\phi_j \rbrace, \tilde\pi_i]\,,\\
b'&=& \mathcal{K}_a[a,{\widetilde E}_K,{\widetilde E}_G,{\widetilde E}_V]\,,\vspace*{0.2cm}\\
{\rm where} && \left\lbrace\begin{array}{l}
\mathcal{K}_i[a,b,\lbrace \tilde\phi_j \rbrace, \tilde\pi_i] \equiv a^{-2 (1  - \alpha )}\widetilde\nabla^2 \tilde\phi_i - (3 - \alpha)\frac{b}{a} \tilde\pi_i - a^{2 \alpha} \tilde V_{,\tilde\phi_i}\,,\\
\mathcal{K}_a[a,{\widetilde E}_K,{\widetilde E}_G,{\widetilde E}_V] \equiv \frac{1}{3} \left( \frac{ f_*}{m_p} \right)^2 a^{1+2\alpha}\Big[ (\alpha - 2){\widetilde E}_{K}  + \alpha {{\widetilde E}_{G}} + (\alpha + 1 ) {{\widetilde E}_V} \Big]\,,\\
{\widetilde E}_K \equiv \frac{1}{2 a^{2\alpha} }\sum_{i}\left\langle \tilde\pi_i^2 \right\rangle\,,~~~ {\widetilde E}_G \equiv \frac{1}{2 a^2 }\sum_{i,k} \left\langle ({\widetilde\nabla}_k \tilde \phi_{i})^2 \right\rangle\,, ~~~{\widetilde E}_V \equiv \left\langle \widetilde{V}(\lbrace \tilde\phi_j\rbrace) \right\rangle\,.
\end{array}\right.
\eea
\noindent It is then straightforward to adapt the explicit RK4 algorithm based on Eqs.~(\ref{eq:RG4algorithm}), into the above system of EOM, obtaining
\begin{eqnarray}
\label{eq:RK4scalars}
\left.
\begin{array}{rcl}
\tilde\pi_i^{(p)} &\equiv& \tilde\pi_i({\bf n},n_0) + \delta\tilde\eta b_{p,p-1}k_i^{(p-1)}\,,\\
b^{(p)} &\equiv& b(n_0) + \delta\tilde\eta b_{p,p-1}k_a^{(p-1)}\,,\\
\tilde\phi_i^{(p)} &\equiv& \tilde\phi_i({\bf n},n_0) + \delta\tilde\eta b_{p,p-1}\tilde\pi_i^{(p-1)}\,,\\
a^{(p)} &\equiv& a(n_0) + \delta\tilde\eta b_{p,p-1}b^{(p-1)}\,,\\
{\it k}^{(p)}_{i} &\equiv& \mathcal{K}_i[a^{(p)},\lbrace \tilde\phi_j^{(p)} \rbrace, \tilde\pi_i^{(p)}]\,,\\
{\it k}^{(p)}_{a} &\equiv& \mathcal{K}_a[a^{(p)},{\widetilde E}_K^{(p)},{\widetilde E}_G^{(p)},{\widetilde E}_V^{(p)}]\,,
\end{array}
\right\rbrace_{p=1,2,3,4} \hspace*{-0.7cm}\Longrightarrow \hspace*{0.5cm}
\left\lbrace
\begin{array}{rcl}
\widetilde\Delta_0^+\tilde\phi_i(\bn,n_0) &=& {1\over6}(\tilde\pi_i^{(1)}+2\tilde\pi_i^{(2)}+2\tilde\pi_i^{(3)}+\tilde\pi_i^{(4)})\,,
\vspace*{0.2cm}\\
\Delta_0^+a(n_0) &=& {1\over6}(b^{(1)}+2b^{(2)}+2b^{(3)}+b^{(4)})\,,\vspace*{0.2cm}\\
\widetilde\Delta_0^+\tilde\pi_i(\bn,n_0) &=& {1\over6}(k_i^{(1)}+2k_i^{(2)}+2k_i^{(3)}+k_i^{(4)})\,,\vspace*{0.2cm}\\
\widetilde\Delta_0^+b(n_0) &=& {1\over6}(k_a^{(1)}+2k_a^{(2)}+2k_a^{(3)}+k_a^{(4)})\,,
\end{array}\right.\nonumber\\
{\rm where}~~~~~ b_{10} \equiv 0\,,~~~~ b_{21} \equiv {1\over 2}\,,~~~~ b_{32} \equiv {1\over 2}\,,~~~~ b_{43} \equiv {1}\,,\hspace*{9cm}
\end{eqnarray}
Note that here $b_{p,p-1}$'s are coefficients, whereas $b^{(p)}$'s represent the $p$-th version of the time derivative of the scale factor $b \equiv a'$.

Since both field amplitudes and conjugate momenta live at the same integer times (after each full iteration), the Hubble constraint is simply
\begin{eqnarray}
HC &:& b^2 = \frac{1}{3} \left( \frac{ f_*}{m_p} \right)^2a^{2(\alpha+1)} \Big({{\widetilde E}_{K}} + {{\widetilde E}_{G}} + {{\widetilde E}_{V}} \Big)\,,
\end{eqnarray}
evaluated at any integer time.

\subsubsection{Verlet Integration n$th$ order}
\label{subsec:YoshidaScalars}

In order to consider any of the higher-order Verlet integrators that we introduced in Section~\ref{subsec:Yoshida}, we need to re-write first the EOM~(\ref{eq:HamiltonianEqsI})-(\ref{eq:FriedmannEq_withPi_I}) for $N_s$ interacting scalar fields dictating the expansion of the universe, as follows
\begin{eqnarray}
a' &=& b \,,\\
\tilde\phi_i' &=& a^{-(3-\alpha)}\tilde\pi_i \,, \\
\tilde\pi_i' &=& \mathcal{K}_i[a,\lbrace \tilde\phi_j \rbrace] \,,\\
b'&=& \mathcal{K}_a[a,{\widetilde E}_K,{\widetilde E}_G,{\widetilde E}_V] \,,
\end{eqnarray}
where
\begin{eqnarray}
&& \mathcal{K}_i[a,\lbrace \tilde\phi_j \rbrace] \equiv a^{1  + \alpha}\widetilde\nabla^2 \tilde\phi_i - a^{3+\alpha} \tilde V_{,\tilde\phi_i} \,,\\
&& \mathcal{K}_a[a,{\widetilde E}_K,{\widetilde E}_G,{\widetilde E}_V] \equiv \frac{1}{3} \left( \frac{ f_*}{m_p} \right)^2 a^{1+2\alpha}\Big[ (\alpha - 2){\widetilde E}_{K}  + \alpha {{\widetilde E}_{G}} + (\alpha + 1 ) {{\widetilde E}_V} \Big] \,,\\
&& {\widetilde E}_K \equiv \frac{1}{2 a^{6} }\sum_{i}\left\langle \tilde\pi_i^2 \right\rangle\,,~~~ {\widetilde E}_G \equiv \frac{1}{2 a^2 }\sum_{i,k} \left\langle ({\widetilde\nabla}_k \tilde \phi_{i})^2 \right\rangle\,, ~~~{\widetilde E}_V \equiv \left\langle \widetilde{V}(\lbrace \tilde\phi_j\rbrace) \right\rangle\,.
\eea
Decomposing one time step $\delta \eta = \sum_{p = 1}^s \delta \eta_p$ into $s$ sub-steps $\delta \eta_p = w_p\delta t$, so that  $\sum_{p=1}^s w_p = 1$, the idea is to iterate $s$-times one of the Verlet algorithms,  \eqref{eq:VVkickDrifKick} or \eqref{eq:PVkickDrifKick}, using each time the appropriate $\delta \eta_p$ sub-step. For instance, using the Velocity Verlet algorithm \eqref{eq:VVkickDrifKick} as the building block, we can write
\begingroup
\allowdisplaybreaks
\begin{eqnarray}
&& \left\lbrace
\begin{array}{l}
\tilde\pi_i^{(0)} \equiv \tilde\pi_i({\bf n},n_0) \vspace*{0.15cm}\\
\tilde\phi_i^{(0)} \equiv \tilde\phi_i({\bf n},n_0)\vspace*{0.15cm}\\
a^{(0)} \equiv a(n_0)\vspace*{0.15cm}\\
b^{(0)} \equiv b(n_0)
\end{array}
\right.
~~~~
\Longrightarrow~~~~
\left\lbrace
\begin{array}{rcl}
b^{(p)}_{1/2} &=& b^{(p-1)} + \omega_p{\delta\tilde\eta\over 2}\mathcal{K}_a[a^{(p-1)},{\widetilde E}_K^{(p-1)},{\widetilde E}_G^{(p-1)},{\widetilde E}_V^{(p-1)}]\\
\tilde\pi^{(p)}_{i,1/2} &=& \tilde\pi_i^{(p-1)} + \omega_p{\delta\tilde\eta\over 2}\mathcal{K}_i[a^{(p-1)},\lbrace\tilde\phi_j^{(p-1)}\rbrace]\vspace*{0.15cm}\notag\\
a_{1/2}^{(p)} &=&  a^{(p-1)} + b_{1/2}^{(p)}\omega_p{\delta\tilde\eta\over2}\,\\
\tilde\phi^{(p)}_{i} &=& \tilde\phi^{(p-1)}_i + \omega_p\delta\tilde\eta\,\tilde\pi_{i,1/2}^{(p)}(a_{1/2}^{(p)})^{-(3-\alpha)}\,,\\
a^{(p)} &=& a^{(p)}_{1/2} +  b^{(p)}_{1/2}\omega_p{\delta\tilde\eta\over2}\,,\\
\tilde\pi_{i}^{(p)} & = & \tilde\pi^{(p)}_{i,1/2} + \omega_p{\delta\tilde\eta\over 2}\mathcal{K}_i[a^{(p)},\lbrace\tilde\phi_{j}^{(p)}\rbrace]
\\
b^{(p)} &=& b^{(p)}_{1/2} + \omega_p{\delta\tilde\eta\over 2}\mathcal{K}_a[a^{(p)},{\widetilde E}_K^{(p)},{\widetilde E}_G^{(p)},{\widetilde E}_V^{(p)}]\,
\end{array}
\right\rbrace_{p\,=\,1,\, ...,\, s}
\hspace*{-1cm}\Longrightarrow\\
&&\Longrightarrow ~~~~
\left\lbrace
\begin{array}{l}
\tilde\pi_i({\bf n},n_0+1) \equiv \tilde\pi_i^{(s)} \vspace*{0.15cm}\\
\tilde\phi_i({\bf n},n_0+1) \equiv \tilde\phi_i^{(s)}\vspace*{0.15cm}\\
a(n_0 + 1) \equiv a^{(s)} \vspace*{0.15cm}\\
b(n_0 + 1) \equiv b^{(s)} \,.
\end{array}
\right.
\end{eqnarray}
\endgroup\vspace*{0.2cm}

By choosing the appropriate weights $w_p$'s from Table~\ref{tab:VVnCoeffs} of the Appendix, the errors of the intermediate steps cancel up to order $\mathcal{O}(\delta\eta^{n})$, with $n = 4, 6, 8$ and $10$ for $s = 3, 7, 15$ and $31$, respectively.

Finally, let us look at the Hubble constraint. Like in the $RK4$ case, here both field amplitudes and conjugate momenta live at the same integer times (after each full iteration over the $s$-subintervals), so we simply write
\begin{eqnarray}
HC &:& b^2 = \frac{1}{3} \left( \frac{ f_*}{m_p} \right)^2a^{2(\alpha+1)} \Big({{\widetilde E}_{K}} + {{\widetilde E}_{G}} + {{\widetilde E}_{V}} \Big)\,,
\end{eqnarray}
evaluated at any integer time.

We note that a similar algorithm of accuracy $O(\delta t^6)$, has been previously introduced in Ref.~\cite{Huang:2011gf}.

\subsubsection{Gauss-Legendre n$th$ order}
\label{subsec:GaussLegendreScalars}

We can adapt the higher-order Gauss-Legendre integrators (based on implicit Runge-Kutta algorithms) introduced in Section~\ref{subsec:GaussLegendre}, to solve the dynamics of $N_s$ interacting scalar fields with self-consistent expansion of the universe. The continuum EOM can be written either as we did 
in Section~\ref{subsec:RK4scalars} [here referred to as {\it Scheme I}\,], or as we did 
in Section~\ref{subsec:YoshidaScalars} [here referred to as {\it Scheme II}\,]:
\begin{eqnarray}
Scheme~I:~~~ \left\lbrace\begin{array}{l}
a' = b  \, , \\
\tilde\phi_i' = \tilde\pi_i \, ,\\
\tilde\pi_i' = \mathcal{K}_i[a,b,\lbrace \tilde\phi_j \rbrace, \tilde\pi_i]\, , \\
b' = \mathcal{K}_a[a,{\widetilde E}_K,{\widetilde E}_G,{\widetilde E}_V]\, , \\
\mathcal{K}_i[a,b,\lbrace \tilde\phi_j \rbrace, \tilde\pi_i] \equiv a^{-2 (1  - \alpha )}\widetilde\nabla^2 \tilde\phi_i - (3 - \alpha)\frac{b}{a} \tilde\pi_i - a^{2 \alpha} \widetilde V_{,\tilde\phi_i}\, , \\
\mathcal{K}_a[a,{\widetilde E}_K,{\widetilde E}_G,{\widetilde E}_V] \equiv \frac{1}{3} \left( \frac{ f_*}{m_p} \right)^2 a^{1+2\alpha}\Big[ (\alpha - 2){\widetilde E}_{K}  + \alpha {{\widetilde E}_{G}} + (\alpha + 1 ) {{\widetilde E}_V} \Big]\, , \\
{\widetilde E}_K \equiv \frac{1}{2 a^{2\alpha} }\sum_{i}\left\langle \tilde\pi_i^2 \right\rangle\,,~~~ {\widetilde E}_G \equiv \frac{1}{2 a^2 }\sum_{i,k} \left\langle ({\widetilde\nabla}_k \tilde \phi_{i})^2 \right\rangle\,, ~~~{\widetilde E}_V \equiv \left\langle \widetilde{V}(\lbrace \tilde\phi_j\rbrace) \right\rangle\,.
\end{array}\right.
\end{eqnarray}

\begin{eqnarray}
Scheme~II:~~~ \left\lbrace\begin{array}{l}
a' = b \, , \\
\tilde\phi_i' = a^{-(3-\alpha)}\tilde\pi_i\, , \\
\tilde\pi_i' = \mathcal{K}_i[a,\lbrace \tilde\phi_j \rbrace]\, , \\
b'= \mathcal{K}_a[a,{\widetilde E}_K,{\widetilde E}_G,{\widetilde E}_V]\, , \\
\mathcal{K}_i[a,\lbrace \tilde\phi_j \rbrace] \equiv a^{1  + \alpha}\widetilde\nabla^2 \tilde\phi_i - a^{3+\alpha} \widetilde V_{,\tilde\phi_i}\, , \\
\mathcal{K}_a[a,{\widetilde E}_K,{\widetilde E}_G,{\widetilde E}_V] \equiv \frac{1}{3} \left( \frac{ f_*}{m_p} \right)^2 a^{1+2\alpha}\Big[ (\alpha - 2){\widetilde E}_{K}  + \alpha {{\widetilde E}_{G}} + (\alpha + 1 ) {{\widetilde E}_V} \Big]\, , \\
{\widetilde E}_K \equiv \frac{1}{2 a^{6} }\sum_{i}\left\langle \tilde\pi_i^2 \right\rangle\,,~~~ {\widetilde E}_G \equiv \frac{1}{2 a^2 }\sum_{i,k} \left\langle ({\widetilde\nabla}_k \tilde \phi_{i})^2 \right\rangle\,, ~~~{\widetilde E}_V \equiv \left\langle \widetilde{V}(\lbrace \tilde\phi_j\rbrace) \right\rangle\,.
\end{array}\right.
\eea
The Gauss-Legendre integrator works in both schemes, since it can deal with field kernels that either contain or do not contain conjugate momenta. Adapting algorithm~(\ref{eq:impRKfieldTheory}) into program variables, we arrive at
\begin{eqnarray}
\label{eq:impRKscalars}
\left.
\begin{array}{rcl}
\tilde\pi_i^{(l)} &\equiv& \tilde{\pi}_i({\bf n},n_0) + \delta\tilde\eta\sum_{m = 1}^s b_{lm}k_i^{(m)} \, ,\\
b^{(l)} &\equiv& b(n_0) + \delta\tilde\eta\sum_{m = 1}^s b_{lm}k_a^{(m)}\, ,\\
\tilde\phi_i^{(l)} &\equiv& \tilde{\phi}_i({\bf n},n_0) + \delta\tilde\eta\sum_{m = 1}^s b_{lm}\tilde\pi_i^{(m)}\, ,\\
a^{(l)} &\equiv& a(n_0) + \delta\tilde\eta\sum_{m = 1}^s b_{lm}b^{(m)}\, ,\\
{\it k}^{(l)}_{a} &\equiv& \mathcal{K}_a[a^{(l)},\bar K^{(l)},\bar G^{(l)},\bar V^{(l)}]
\vspace*{1mm}\, ,\\
{\it k}^{(l)}_{i} &\equiv&
\left\lbrace\begin{array}{l}
\hspace*{-2mm}\mathcal{K}_i[a^{(l)},b^{(l)},\lbrace \tilde\phi_j^{(l)} \rbrace, \tilde\pi_i^{(l)}] ~(Sch.~I) \, ,\vspace*{1mm}\\
\hspace*{-2mm}\mathcal{K}_i[a^{(l)},\lbrace \tilde\phi_j^{(l)} \rbrace] ~(Sch.~II) \, ,\\
\end{array}\right.
\end{array}
\right\rbrace_{l = 1, 2, ..., s}~~ \hspace*{-1.0cm}\Longrightarrow ~~~~~
\left\lbrace
\begin{array}{rcl}
\widetilde\Delta_0^+\tilde\phi_i(\bn,n_0) &=& \sum_{m = 1}^s c_{m}\tilde\pi_i^{(m)}\, ,\vspace*{0.2cm}\\
\widetilde\Delta_0^+a(n_0) &=& \sum_{m = 1}^s c_{m}b^{(m)}\, ,\vspace*{0.2cm}\\
\widetilde\Delta_0^+\tilde\pi_i(\bn,n_0) &=& \sum_{m = 1}^s c_{m}k_i^{(m)}\, ,\vspace*{0.2cm}\\
\widetilde\Delta_0^+\pi_a(n_0) &=&  \sum_{m = 1}^s c_{m}k_a^{(m)}\,,
\end{array}\right.\nonumber\\
\end{eqnarray}
where we note again that here $b_{lm}$'s are coefficients, whereas $b^{(l)}$'s represent the $l$-th version of the time derivative of the scale factor $b \equiv a'$. The coefficients $\lbrace b_{lm} \rbrace$ and $\lbrace c_m \rbrace$ are listed in Table~\ref{tab:ButcherTables} of the Appendix for the cases of $s  = 2, 3, 4$ and $5$ number of sub-steps [recall that a method with $s$-substeps has an accuracy of $\mathcal{O}(\delta\eta^{2s})$].

As both schemes $I$ and $II$ have field amplitudes and conjugate momenta living at the same (integer) time, the Hubble constraint can be written in both schemes simply as
\begin{eqnarray}
HC &:& b^2 = \frac{1}{3} \left( \frac{ f_*}{m_p} \right)^2a^{2(\alpha+1)} \Big({{\widetilde E}_{K}} + {{\widetilde E}_{G}} + {{\widetilde E}_{V}} \Big)\,,
\end{eqnarray}
evaluated at any integer time.

\subsection{Observables}

To conclude this section, we collect the main observables of interest, such as energies and power-spectra. In the case of scalar fields, we are mostly concerned with the fields and conjugate momenta themselves,
$\tilde\phi_i$ and $\tilde\pi_i$. In particular, we typically monitor their mean value, quadratic mean value, and variance,
\begin{eqnarray}
\overline{\tilde\phi_i} \equiv \left\langle \tilde\phi_i \right\rangle\,,~~~~~ \overline{\tilde\phi_i^2} \equiv \left\langle \tilde\phi_i^2 \right\rangle\,,~~~~~ \sigma_{\tilde\phi_i}^2 \equiv \left\langle \tilde\phi_i^2 \right\rangle - \left\langle \tilde\phi_i \right\rangle^2\,.
\end{eqnarray}

\subsubsection{Energy components}

We typically monitor the kinetic energy density of each field
\begin{eqnarray}
{\widetilde E}_K^{\phi_i} = \frac{1}{2a^p}\left<\left(\tilde{\pi}_{i}\right)^2\right > \ ,~~~\left[~p = 2\alpha ~~{\rm if}~~ \tilde\pi_i \equiv \tilde\phi_i'~~~; ~~~p = 6 ~~{\rm if}~~ \tilde\pi_i \equiv a^{(3-\alpha)}\tilde\phi_i'~\right]
\end{eqnarray}
and the gradient energy densities for each field,
\begin{align}
{\widetilde E}_G^{\phi_i} = \frac{1}{2a^2}\sum_j \left <( \widetilde{\Delta}^{+}_j\tilde\phi_i)^2\right > \ .
\end{align}

\noindent The total potential energy is given by
\begin{equation}
  {\widetilde E}_V=\left <\tilde V(\{\tilde\phi_i\}) \right > \ .
\end{equation}
As in most cases, the potential can naturally be written as a sum of $p$ different terms $V(\{\tilde\phi_i\})=\sum_{a} V_a(\{\tilde\phi_i\})$, which are typically the different mass terms and interactions of the fields, we also measure each term
\begin{equation}
  {\widetilde E}_{V_a}=\left <\tilde V_a(\{\tilde\phi_i\}) \right > \ .
\end{equation}

\subsubsection{Spectra}

We also consider the power spectrum of each individual field. Following our conventions in Eq.~\eqref{eq:discretePS}, we define
\begin{align}
  {\widetilde \Delta}_{\tilde\phi_i}(\tilde k(\tilde {\bf n}))
  &= \frac{\tilde k^3(\tilde {\bf n})}{2\pi^2} \left(\frac{\delta \tilde{x}}{N}\right)^3\left<|(\tilde\phi_i)(\tilde{\bf n})|^2\right>_{R(\tilde {\bf n})} \ , \\
  {\widetilde \Delta}_{\tilde\pi_i}(\tilde k(\tilde {\bf n})) &= \frac{\tilde k^3(\tilde {\bf n})}{2\pi^2} \left(\frac{\delta \tilde{x}}{N}\right)^3\left<|\tilde\pi_i(\tilde {\bf n})|^2\right>_{R(\tilde {\bf n})} \ ,
\end{align}
where
\bea
\tilde {\bf k}(\tilde {\bf n}) \equiv {{\bf k}(\tilde {\bf n})\over\omega_*} \equiv {k_{\rm IR}\over\omega_*}\cdot(\tilde n_1,\tilde n_2,\tilde n_3)\,,~~~~ \Rightarrow ~~~~ \tilde k(\tilde {\bf n}) \equiv |\tilde {\bf k}(\tilde {\bf n})| \equiv {k_{\rm IR}\over\omega_*}\sqrt{\tilde n_1^2 + \tilde n_2^2 + \tilde n_3^2}\,\,.
\eea
We note that these dimensionless power spectra are related to their dimensionful counterparts by 
${\widetilde \Delta}_{\tilde\phi_i}\equiv \Delta_{\phi_i}/f_*^2$, and 
${\widetilde \Delta}_{\tilde\pi_i} \equiv  \Delta_{\pi_i}/f_*^2\omega_*^2$.

\section{Lattice formulation of gauge fields, Part I: $U(1)$ interactions} \label{sec:LatU1}

We move now into the lattice formulation of a $U(1)$-interacting gauge sector, which amounts to developing an appropriate discretization scheme for Eqs.~\eqref{eq:higgsU1-eom} and \eqref{eq:U1eom}, together with the Friedmann equation \eqref{eq:FriedmannDDa}. In particular, we will generalize the staggered leapfrog algorithm from Section \ref{subsec:StaggeredLFaction}, the velocity-Verlet algorithm from Section \ref{subsec:VerletGeneral}, and their higher-order generalizations introduced in Section \ref{subsec:YoshidaScalars}. For simplicity, we restrict the presentation to the case of a single charged scalar field $\varphi$, coupled to a single Abelian gauge field $A_{\mu}$. Generalization to a larger number of charged scalars and/or gauge fields is straightforward. Note also that for simplicity, we present explicitly only the velocity-Verlet version of the Verlet algorithms, as the position-Verlet algorithm is straightforwardly obtained by inverting the roles of momenta and fields, as explained in Section~\ref{subsec:VerletGeneral}.

\subsection{Continuum formulation and natural variables}

We define the following $program$ variables for the $U(1)$-charged scalars and the Abelian gauge fields as
\begin{align}
  \tilde\varphi = \frac{1}{f_*} \varphi, \ \ \ \widetilde{A}_\mu=\frac{1}{\omega_*} A_\mu \ . \label{eq:GaugeProgramVar}
\end{align}
The normalization of the charged scalar is identical to the one of singlet scalars, introduced in Eq.~(\ref{eq:FieldSpaceTimeNaturalVariables}). The gauge field is however normalized with respect to $\omega_*$, so that it cancels out the normalization factor of $ \delta \tilde{x}^{\mu}$ in the argument of the link variable, i.e.~$V_{\mu} \equiv e^{-ig_AQ^{(\tilde\varphi)}_A  \delta {x}_{\mu} A_{\mu} } = e^{-i g_AQ^{(\tilde\varphi)}_A\delta \tilde{x}_{\mu} \tilde{A}_{\mu} } \equiv \tilde{V}_{\mu}$. The continuum equations of motion in these variables, in the temporal gauge $\tilde{A}_0 = 0$, are
\begin{align}
  (a^{3-\alpha}\tilde\varphi')' - a^{1 + \alpha} {\vec{\widetilde{D}}}_{\hspace{-0.5mm}A}^{\,2}\tilde{\varphi}  &= - a^{\alpha+3} \frac{\widetilde V_{,|\tilde\varphi|}}{2} \frac{\tilde\varphi}{|\tilde\varphi |} \ , \label{eq:higgsU1-eomnofriction}\\
  \tilde{\partial}_0 (a^{1-\alpha}\widetilde{F}_{0i}) - a^{\alpha - 1}\partial_j \widetilde{F}_{ji} &=
  a^{1+ \alpha}\tilde{J}^A_i \ , \label{eq:U1eomnofriction}
\end{align}
where all field and spacetime variables are program variables, and as such, are indicated with a `$\sim$'. By inspecting these equations, we can naturally identify appropriate definitions for the conjugate momenta of the field variables as
\begin{align}
\piSingl&\equiv  a^{3-\alpha}\tilde\varphi' \ , \label{eq:momU1singlet}\\
\piApar_i &\equiv a^{1-\alpha}\widetilde F_{0i}\label{eq:momU1vec} \ .
\end{align}
We define the \textit{program} kinetic energies of the fields as
\begin{align}
  {\widetilde E}_K^\varphi &= \frac{1}{a^6}\left<\piSingl^2\right > \, , \label{eq:kinU1singletdisc}\\
  {\widetilde E}_K^A &= \frac{1}{2a^4}{\omega_*^2\over f_*^2}
  \sum_{i=1}^3 \left <\piApar_i^2\right > \ , \label{eq:kinU1vecdisc}
\end{align}
where we note 
that the physical expression of the gauge kinetic energy density relates to the dimensionless one via $E_K^A = \omega_*^2 f_*^2 {\widetilde E}_K^A$. This is equivalent to the standard relation for scalar fields, which relate as e.g.~$E_K^\varphi = \omega_* ^2f_*^2{\widetilde E}_K^\varphi$). For convenience, let us also define the following kernels for each of the amplitudes and momenta,
\begin{align}
  (\piSingl)'  &= \mathcal{K}_{\varphi}[a,\tilde\varphi,\widetilde A_j] \ , \\
  \piApar'_i  &=\mathcal{K}_{A_i}[a,\tilde\varphi,\widetilde A_j] \ , \\
\mathcal{K}_{\varphi}[a,\tilde\varphi,\widetilde A_j]&\equiv  - a^{\alpha+3} \frac{\widetilde V_{,|\tilde\varphi|}}{2} \frac{\tilde\varphi}{|\tilde\varphi |} + a^{1 + \alpha} {\vec{\widetilde D}}_{\hspace{-0.5mm}A}^{\,2}\tilde\varphi  \ , \\
\mathcal{K}_{A_i}[a,\tilde\varphi,\widetilde A_j] &\equiv a^{1+ \alpha}\widetilde J^A_i + a^{\alpha - 1}\tilde{\partial}_j \widetilde F_{ji} \ ,
\end{align}
as well as the kernel for the scale factor [recall Eq.~\eqref{eq:FriedmannDDa} and that $b \equiv a'$],
\begin{eqnarray}\label{eq:KernelU1totII}
&& \hspace*{-1.5cm}b' = \mathcal{K}_a\hspace*{-1mm}\left[a,{\widetilde E}_K^\varphi,{\widetilde E}_G^\varphi,{\widetilde E}_V^\varphi,{\widetilde E}_K^A,{\widetilde E}_G^A\right]\,, \\
\label{eq:KernelU1tot}
&& \hspace*{-1.5cm}\mathcal{K}_a\hspace*{-1mm}\left[a,{\widetilde E}_K^\varphi,{\widetilde E}_G^\varphi,{\widetilde E}_V^\varphi,{\widetilde E}_K^A,{\widetilde E}_G^A\right] \\
&& ~~~~~~~ \equiv \frac{a^{2\alpha+1}}{3}{f_*^2\over m_p^2}\left[ (\alpha-2)({\widetilde E}_K^\varphi +\dots) + \alpha ({\widetilde E}_G^\varphi+\dots)+(\alpha+1) {\widetilde E}_V+(\alpha-1)
({\widetilde E}_K^A+{\widetilde E}_G^A)\right].\nonumber
\end{eqnarray}
We reproduced only the terms directly relevant to the $U(1)$ gauge sector; the dots are here simply to remind the reader that other contributions will enter if some other sectors are present (e.g.~scalar singlets). The same applies for the Gauss law
\bea\label{eq:HubbleGauge}
b^2=\frac{1}{3}\left(\frac{f_*}{m_p}\right)^2a^{2(\alpha+1)}\left[{\widetilde E}_K^\varphi+{\widetilde E}_G^\varphi+{\widetilde E}_V+
\left({\widetilde E}_K^A+{\widetilde E}_G^A\right) + \dots \right]\,.
\eea

\subsection{Non-compact Lattice formulation(s)}

We first present here a spatial discretization of the kernels using non-compact variables, which means that the variables to evolve are the field amplitudes and momenta themselves, $\{\tilde\varphi, \piSingl, \widetilde A_i, \piApar_i\}$. Using our $U(1)$-toolkit \eqref{eq:U1toolkit}, we write
\begin{align}
  \mathcal{K}_{\varphi}[a,\tilde\varphi,\widetilde A_i] &= - a^{\alpha+3}  \frac{\widetilde V_{,|\tilde\varphi|}}{2} \frac{\tilde\varphi}{|\tilde\varphi |}+ a^{1 + \alpha} \sum_i \widetilde D_i^- \widetilde D^+_i\tilde\varphi \,, \\
  \mathcal{K}_{A_i}[a,\tilde\varphi,\widetilde A_j] &= a^{1+ \alpha}\left({2g_A Q_A^{(\tilde\varphi)}\over\delta\tilde x}{f_*^2\over \omega_*^2} \mathcal{I}m [ \tilde\varphi^{*} \widetilde{V}_i\tilde\varphi] + \dots \right ) + a^{\alpha - 1}\sum_j\left(\tilde{\Delta}_j^-\tilde{\Delta}_j^+ \widetilde A_i - \tilde{\Delta}^-_j \tilde{\Delta}^+_i\widetilde A_j\right ) \,, \label{eq:nckernelU1}
\end{align}
where the dots indicate that any other $U(1)$-charged field coupled to the gauge field $A_\mu$, would contribute to its kernel through the gauge current. For example, an $SU(N)$-doublet $\widetilde \Phi$ charged under $U(1)$, would add a contribution $+2g_A Q_A^{(\tilde\Phi)} \mathcal{I}m [ \widetilde \Phi^\dag \widetilde (V_i)^{Q_A^{(\tilde \Phi)}/Q_A^{(\tilde \varphi)}} \widetilde\Phi]$. Here $Q_A^{(\tilde\varphi)}, Q_A^{(\tilde\Phi)}$ are the Abelian charges of $\varphi$ and $\Phi$, respectively. Note here a subtlety. To recover the correct covariant derivative for the doublet, one need to compute a "new" link $(V_i)^{Q_A^{(\tilde \Phi)}/Q_A^{(\tilde \varphi)}}$ as some power of the original link variable.

We discretize the gradient and potential energies as follows,
\begin{align}
{\widetilde E}_G^{\varphi} &= \frac{1}{a^2}\sum_i \left <(\tilde{ D}^{A~+}_i\tilde\varphi)^*(\tilde{D}^{A~+}_i\tilde\varphi)\right > \,,\\
{\widetilde E}_G^A &= \frac{1}{2a^4}{\omega_*^2\over f_*^2}
\sum_{i, j<i} \left <(\tilde{\Delta}^+_i \widetilde A_j - \tilde{\Delta}^+_i \widetilde A_j)^2 \right > \,, \\
{\widetilde E}_V&=\left <\widetilde V(\tilde\varphi,\dots) \right > \ ,
\end{align}
whereas the kinetic energies ${\widetilde E}_K^\varphi$, ${\widetilde E}_K^A$ are given by Eqs.~\eqref{eq:kinU1singletdisc} and \eqref{eq:kinU1vecdisc}. Note that here again, 
as for the gauge field kinetic (electric) energy densities, the dimensionless and physical gauge field gradient (magnetic) energy densities are related to each other through $E_G^A = \omega_*^4 {\widetilde E}_G^A$. Finally, a crucial quantity to monitor is the Gauss law, which must be obeyed at all times during the simulation. It is written in the continuum in Eq.~(\ref{eq:GaussU1-eom}). In terms of program variables, we can discretize it as follows,
\begin{equation}
-\sum_i \Delta_i^-\piApar_i= 2 g_A Q_A^{(\tilde\varphi)}{f_*^2\over \omega_*^2}\,\mathcal{I}m [ \tilde\varphi^*  \piSingl ] \ . \,
\label{eq:U1vector-Gauss_Discrete}
\end{equation}

\subsubsection{(Staggered) Leapfrog}

Let us now consider the time evolution of the above equations. We first present an adaptation of the staggered leapfrog algorithm of order $O(\delta\eta^2)$. We consider momenta evaluated at semi-integer times, while fields living at integer times. If needed, by means of interpolation momenta can be evaluated at integer times, and field amplitudes at semi-integer times. This is required, in particular, in the scale-factor kernels, which now depend on $\overline{{\widetilde E}_K^\varphi} \equiv ({\widetilde E}_{K,-\hat 0/2}^\varphi + {\widetilde E}_{K,+\hat0/2}^\varphi)/2$ and $\overline{{\widetilde E}_K^A} \equiv ({\widetilde E}_{K,-\hat0/2}^A + {\widetilde E}_{K,+\hat0/2}^A)/2$.

The algorithm to evolve the fields and their momenta by one time step reads then as follows:\\

\begingroup
\allowdisplaybreaks
\hspace{1cm}\text{ \it Non-Compact Staggered Leapfrog}\vspace{-0.4cm}\\
\begin{eqnarray}
  IC &:& \left \{a,\tilde\varphi,\widetilde A_i\right\} \text{ at } \tilde{\eta}_0, ~~~ \left \{b_{-1/2}, \piSinglpar_{-1/2}, \piApar_{i,-1/2}\right\} \text{ at } \tilde{\eta}_0 - {\delta\tilde\eta\over2}.\\
\piSinglpar_{+1/2} &=& \piSinglpar_{-1/2} + \delta \tilde{\eta} \mathcal{K}_\varphi[a,\tilde\varphi,\widetilde A_i] \,,\\
  \piApar_{i,+1/2} &=& \piApar_{i,-1/2} + \delta \tilde{\eta} \mathcal{K}_{A_i}[a,\tilde\varphi,\widetilde A_j] \,,\\
  b_{+1/2} &=& b_{-1/2} + \delta \tilde{\eta}\mathcal{K}_a\hspace*{-1mm}\left[a,\overline{{\widetilde E}_K^\varphi},{\widetilde E}_G^\varphi,{\widetilde E}_V^\varphi,\overline{{\widetilde E}_K^A},{\widetilde E}_G^A\right]\,,\\
  a_{+0} &=& a + \delta \tilde{\eta} b_{+1/2} \,,\\
  a_{+1/2}&=&(a_{+0}+a)/2\,,\\
  \tilde\varphi_{+0} &=&\tilde\varphi + \delta \tilde{\eta} a_{+1/2}^{-(3-\alpha)}\piSinglpar_{+1/2} \,,\\
  \widetilde A_{i,+0}&=& \widetilde A_{i} + \delta \tilde{\eta} a_{+1/2}^{-(1-\alpha)}\piApar_{i,+1/2} \,,\\
  HC&:& b^2=\frac{1}{3}\left(\frac{f_*}{m_p}\right)^2a^{2(\alpha+1)}\left[\,\overline{{\widetilde E}_K^\varphi}+{\widetilde E}_G^\varphi + {\widetilde E}_V + 
  \overline{{\widetilde E}_K^A}+{\widetilde E}^A_G\right]
  \ ,
\end{eqnarray}
\endgroup
where the last line corresponds to the Hubble constraint. We see that the scale factor needs to be interpolated, as it enters into the relation between the conjugate momenta and the fields' time derivative. Note also that this scheme can be obtained from an action principle, similar to the analogous singlet scalar case.

\subsubsection{Velocity-Verlet}

The equations can be also solved with a velocity-Verlet scheme of order $O(\delta \tilde{\eta}^2)$, similarly to the analogous singlet scalar case. The algorithm to update the system proceeds as follows,\\

\begingroup
\allowdisplaybreaks
\hspace{1cm}\text{ \it Non-Compact Velocity-Verlet VV2}\vspace*{-0.4cm}\\
\begin{eqnarray}
  IC &:&  \left \{a, b,\tilde\varphi, \piSingl,\widetilde A_i,\piApar_i\right\} \text{ at } \eta_0. \\
  \piSinglpar_{+1/2} &=& \piSingl+ \frac{\delta \tilde{\eta}}{2} \mathcal{K}_\varphi[a,\tilde\varphi,\widetilde A_j] \,,\\
  \piApar_{i,+1/2} &=& \piApar_i + \frac{\delta \tilde{\eta}}{2} \mathcal{K}_{A_i}[a,\tilde\varphi,\widetilde A_j] \,,\\
  b_{+1/2} &=& b + \frac{\delta \tilde{\eta}}{2}
  \mathcal{K}_a\hspace*{-1mm}\left[a,{\widetilde E}_K^\varphi,{\widetilde E}_G^\varphi,{\widetilde E}_V^\varphi,{\widetilde E}_K^A,{\widetilde E}_G^A\right],\\
  a_{+0} &=& a + \delta \tilde{\eta} b_{+1/2} \,,\\
  a_{+1/2}&=&\frac{a_{+0}+a}{2} \,,\\
  \tilde\varphi_{+0} &=&\tilde\varphi + \delta \tilde{\eta} \frac{\piSinglpar_{+1/2}}{a_{+1/2}^{3-\alpha}} \,,\\
  \widetilde A_{i,+0}&=& \widetilde A_{i} + \delta \tilde{\eta} \frac{\piApar_{+1/2}}{a_{+1/2}^{1-\alpha}} \,,\\
  \piSinglpar_{+0} &=& \piSinglpar_{+1/2} + \frac{\delta \tilde{\eta}}{2} \mathcal{K}_\varphi[a_{+0} ,\tilde\varphi_{+0} ,\widetilde A_{j, +0}] \,,\\
  \piApar_{i,+0} &=& \piApar_{i,+1/2} + \frac{\delta \tilde{\eta}}{2} \mathcal{K}_{A_i}[a_{+0},\tilde\varphi_{+0},\widetilde A_{j,+0}] \,,\\
  b_{+0} &=& b_{+1/2} + \frac{\delta \tilde{\eta}}{2}\mathcal{K}_a\hspace*{-1mm}\left[a_{+0},{\widetilde E}_{K,+0}^\varphi,{\widetilde E}_{G,+0}^\varphi,{\widetilde E}_{V,+0}^\varphi,{\widetilde E}_{K,+0}^A,{\widetilde E}_{G,+0}^A\right],\\
    HC&:& b^2=\frac{1}{3}\left(\frac{f_*}{m_p}\right)^2a^{2(\alpha+1)}\left[{\widetilde E}_K^\varphi+{\widetilde E}_G^\varphi+{\widetilde E}_V+
    {\widetilde E}_K^A+{\widetilde E}_G^A\right]
\ ,
\end{eqnarray}
\endgroup

\noindent where the last line is again the Hubble constraint.  Note that a similar integrator, based on the position-Verlet method, has been also presented in Ref.~\cite{Lozanov:2019jff}.

\subsubsection{Velocity-Verlet n$th$ order}

In order to construct the higher order integrators VV4, VV6, VV8 and VV10, we simply need to apply the procedure described in Section \ref{subsec:Yoshida}. Explicitly, by  choosing $\omega_p$'s in Table \ref{tab:VVnCoeffs} of the Appendix, we can write an algorithm as follows

\begingroup
\allowdisplaybreaks
\begin{eqnarray}
&& \left.
\begin{array}{r}
\piSingl^{(0)} \equiv \piSingl({\bf n},n_0) \vspace*{0.15cm}\\
\tilde\varphi^{(0)} \equiv \tilde\varphi({\bf n},n_0) \vspace*{0.15cm} \\
\widetilde A_i^{(0)} \equiv \widetilde A_i({\bf n},n_0)\vspace*{0.15cm}\\
\piApar_i^{(0)} \equiv \piApar_i({\bf n},n_0)\vspace*{0.15cm}\\
a^{(0)} \equiv a(n_0)\vspace*{0.15cm}\\
b^{(0)} \equiv b(n_0)
\end{array}\right\rbrace ~~~\Longrightarrow
\\
&&\Longrightarrow\left\lbrace
\begin{array}{rcl}
  \piSinglpar^{(p)}_{1/2} &=& \piSingl^{(p-1)} + \frac{\omega_p\delta \tilde{\eta}}{2} \mathcal{K}_\varphi\left[a^{(p-1)},\tilde\varphi^{(p-1)},\widetilde A_j^{(p-1)}\right] \\
  \piApar_{i,1/2}^{(p)} &=& \piApar^{(p-1)}_i + \frac{\omega_p\delta \tilde{\eta}}{2} \mathcal{K}_{A_i}\left[a^{(p-1)},\tilde\varphi^{(p-1)},\widetilde A_j^{(p-1)}\right] \\
  b^{(p)}_{1/2} &=& b^{(p-1)} + \frac{\omega_p\delta \tilde{\eta}}{2} \mathcal{K}_a^{(p-1)}\hspace*{-1mm}\left[a,{\widetilde E}_{K}^{\varphi},{\widetilde E}_{G}^{\varphi},{\widetilde E}_{V}^{\varphi},{\widetilde E}_{K}^{A},{\widetilde E}_{G}^{A}\right] \\
  a^{(p)} &=& a^{(p-1)} + \omega_p\delta \tilde{\eta} b^{(p)}_{1/2} \\
  a^{(p)}_{1/2}&=&\frac{a^{(p)}+a^{(p-1)}}{2}\\
  \tilde\varphi^{(p)} &=&\tilde\varphi^{(p-1)} + \delta \tilde{\eta} \frac{\piSinglpar^{(p)}_{1/2}}{\left(a_{1/2}^{(p)}\right)^{3-\alpha}} \\
  \widetilde A_{i}^{(p)}&=& \widetilde A_{i}^{(p-1)} + \omega_p\delta \tilde{\eta} \frac{\piApar^{(p)}_{1/2}}{\left(a_{1/2}^{(p)}\right)^{1-\alpha}} \\
  \piSinglpar^{(p)} &=& \piSinglpar^{(p)}_{1/2} + \frac{\omega_p\delta \tilde{\eta}}{2} \mathcal{K}_\varphi\left[a^{(p)},\tilde\varphi^{(p)},\widetilde A_j^{(p)}\right]\\
  \piApar^{(p)}_i &=& \piApar^{(p)}_{i,1/2} + \frac{\omega_p\delta \tilde{\eta}}{2}\mathcal{K}_{A_i}[a^{(p)},\tilde\varphi^{(p)},\widetilde A_j^{(p)}] \\
  b^{(p)} &=& b^{(p)}_{1/2} + \frac{\omega_p\delta \tilde{\eta}}{2} \mathcal{K}_a^{(p)}\hspace*{-1mm}\left[a,{\widetilde E}_{K}^{\varphi},{\widetilde E}_{G}^{\varphi},{\widetilde E}_{V}^{\varphi},{\widetilde E}_{K}^{A},{\widetilde E}_{G}^{A}\right]
\end{array}
\right\rbrace_{p\,=\,1,\, ...,\, s}
\hspace*{-1cm}\\
&&\Longrightarrow ~~~~
\left\lbrace
\begin{array}{l}
   \piSingl({\bf n},n_0)\equiv \piSingl^{(s)} \vspace*{0.15cm}\\
  \tilde\varphi({\bf n},n_0)\equiv \tilde\varphi^{(s)}\vspace*{0.15cm}\\
  \widetilde A_i({\bf n},n_0)\equiv \widetilde A_i^{(s)}\vspace*{0.15cm}\\
  \piApar_i({\bf n},n_0)\equiv \piApar^{(s)}_i\vspace*{0.15cm}\\
  a(n_0)\equiv a^{(s)}\vspace*{0.15cm}\\
  b(n_0)\equiv b^{(s)} \, \ ,
\end{array}
\right.\\
\notag\\
  HC&:& b^2=\frac{1}{3}\left(\frac{f_*}{m_p}\right)^2a^{2(\alpha+1)}\left[{\widetilde E}_K^\varphi+{\widetilde E}_G^\varphi+{\widetilde E}_V+
  {\widetilde E}_K^A+{\widetilde E}_G^A\right] \ .
\end{eqnarray}
\endgroup
where, to ease notation, we have used $\mathcal{K}_a^{(p)}\hspace*{-1mm}\left[a,{\widetilde E}_{K}^{\varphi},{\widetilde E}_{G}^{\varphi},{\widetilde E}_{V}^{\varphi},{\widetilde E}_{K}^{A},{\widetilde E}_{G}^{A}\right] \equiv \mathcal{K}_a\hspace*{-1mm}\left[a^{(p)},{\widetilde E}_{K}^{\varphi^{(p)}}\hspace*{-1.6mm},{\widetilde E}_{G}^{\varphi^{(p)}}\hspace*{-1.6mm},{\widetilde E}_{V}^{\varphi^{(p)}}\hspace*{-1.6mm},{\widetilde E}_{K}^{A^{(p)}}\hspace*{-1.6mm},{\widetilde E}_{G}^{A^{(p)}}\right]$.

\subsection{Compact Lattice formulation(s)}

As presented in Section \ref{subsubsec:latgaugeinvtech}, the link variables $V_i$ can also be used as the "fundamental" variables to solve for in the EOM, instead of the gauge field amplitudes $A_i$. This leads to a `compact' discretization of the $U(1)$ gauge sector. In this approach, we keep the same definitions for momenta as before, \eqref{eq:momU1singlet} and \eqref{eq:momU1vec}. However, we take for the gauge fields' kernel,
\begin{align}
  \mathcal{K}_{A_i}[a,\tilde\varphi,\widetilde V_j] &= a^{1+ \alpha}\left({2g_A Q_A^{(\tilde\varphi)}\over\delta\tilde x} {f_*^2\over \omega_*^2}\mathcal{I}m [ \tilde{\varphi}^{*} \widetilde V_i \tilde{\varphi}] + \dots \right ) + \frac{a^{\alpha - 1}}{\delta \tilde{x}^3g_A Q_A^{(\tilde\varphi)}}\sum_j\left(\widetilde V_{ij}-\widetilde V_{ij~-j}\right ) \ , \label{eq:ckernelU1}
\end{align}
where the second term in the $rhs$ is a {\it backward discretization} of $\tilde{\partial}_j\widetilde F_{ji}$, which appears naturally from a discrete action made out of a plaquettes. The kinetic energy is given by Eq.~(\ref{eq:kinU1vecdisc}), whereas the magnetic energy can be approximated by
\begin{align}
  \widetilde E_G^{A}=\frac{1}{a^4\delta \tilde{x}^4g_A^2 Q_A^{(\tilde\varphi) \,2}}
  \sum_{i, j<i}(1-  \mathcal{R}e (\widetilde V_{ij}))  \ ,
  \label{eq:magenU1compact}
\end{align}
which again represents its physical counterpart in units of $\omega_*^4$ (and not $\omega_*^2f_*^2$, like in the case of scalar field energy densities).

 The last difference with respect to the non-compact formulation is how the link variables are evolved in time. There are two different approaches which can be followed. We will see that we can compute the drifts directly from the conjugate momenta, as in the previous cases but we can also compute them by reconstructing the plaquette from the conjugate momentum. We present below both approaches as they can both be of interest when generalizing to other non-Abelian groups.\\

\noindent {\it Drifts from momenta:}\vspace*{-0.25cm}\\

In order to write down the drifts directly from the conjugate momenta
let us first compute the continuum time derivative of a link
\begin{align}
 (\widetilde V_i)' = \partial_0 e^{-i g_A Q_A^{(\tilde\varphi)}\delta \tilde{x} \tilde{A}_i} = -ig_A Q_A^{(\tilde\varphi)} \delta \tilde{x} (\widetilde{A}_i)'\widetilde V_i \,,
\end{align}
or in terms of the conjugate momenta,
\begin{align}
 (\widetilde V_i)' = -i \frac{g_A Q_A^{(\tilde\varphi)}\delta \tilde{x}}{a^{1-\alpha}} \piApar_i \widetilde V_i \ .
\end{align}

This equation can be solved explicitly by a discrete scheme, as follows. We approximate the right hand side by a centered semi-sum, and get
\begin{align}
 \frac{\widetilde V_{i,+0} - \widetilde V_{i}}{\delta \tilde{\eta}} = -i \frac{g_A Q_A^{(\tilde\varphi)}\delta \tilde{x}}{a^{1-\alpha}_{+1/2}} \piApar_{i,+1/2} \frac{1}{2} \left(\widetilde V_{i,+0} + \widetilde V_{i}\right ) \ ,
\end{align}
and the drift is then given as
\begin{align}
 \widetilde V_{i,+0} &= \left(1+i \frac{g_A Q_A^{(\tilde\varphi)}\delta \tilde{\eta}\delta \tilde{x}}{2 a^{1-\alpha}_{+1/2}} \piApar_{i,+1/2} \right )^{-1}\left(1-i \frac{g_A Q_A^{(\tilde\varphi)}\delta \tilde{\eta}\delta \tilde{x}}{2 a^{1-\alpha}_{+1/2}} \piApar_{i,+1/2} \right )  \widetilde V_{i} \\
  &=\left(1-i \frac{g_A Q_A^{(\tilde\varphi)}\delta \tilde{\eta}\delta \tilde{x}}{2 a^{1-\alpha}_{+1/2}} \piApar_{i,+1/2} \right )^2  \widetilde V_{i} +O(\delta \tilde{\eta}^2)  \ . \label{eq:CompactAbelianDriftMom}
\end{align}
Here, the last expansion is not necessary but will prove useful in the non-Abelian case.\\

\noindent {\it Drifts from plaquettes:}\vspace*{-0.25cm}\\

In some cases, it can also be useful to think directly in terms of plaquettes instead of  conjugate momenta. We first construct the plaquette $\widetilde V _{0i}$ 
from the momenta, for instance, using the relation
\begin{align}
\mathcal{I}m\lbrace \widetilde V_{0i}[\piApar_{i}] \rbrace= -g_A Q_A^{(\tilde\varphi)}\delta \tilde x \delta \tilde \eta \frac{1}{a^{1-\alpha}}\piApar_{i} + O(\delta \tilde \eta^2, \delta \tilde x^3)\,,
\end{align}
together with the constraint $|\widetilde V_{0i}|=1$, we can construct $\widetilde V_{0i} [\piApar_{i}]$.
The drift is then given using the definition of the plaquette
\begin{align}
  \widetilde V_{0i}[\piApar_{i}]
  &=\widetilde V_{i,+0}\widetilde V^*_i
  ~~\implies~~ \widetilde V_{i,+0} = \widetilde V_{0i}[\piApar_{i}]
  \widetilde V_i\,.
\end{align}

\subsubsection{(Staggered) Leapfrog}

In this case, the only difference with respect to the non-compact formulation is how the drifts are given. The algorithm is\\

\begingroup
\allowdisplaybreaks
\hspace*{1cm}\text{ \it Compact Staggered Leapfrog}\vspace{-0.4cm}\\
\begin{eqnarray}
  IC &:& \left \{a, \tilde\varphi,\widetilde V_i\right\} \text{ at } \eta_0, ~~~ \left \{b_{-1/2}, \piSinglpar_{-1/2}, \piApar_{i,-1/2}\right\} \text{ at } \eta_0 - \delta\eta/2. \\
  \piSinglpar_{+1/2} &=& \piSinglpar_{-1/2} + \delta \tilde{\eta} \mathcal{K}_\varphi[a,\tilde\varphi,\widetilde V_j] \,,\\
  \piApar_{i,+1/2} &=& \piApar_{i,-1/2} + \delta \tilde{\eta} \mathcal{K}_{A_i}[a,\tilde\varphi,\widetilde V_j] \,,\\
  b_{+1/2} &=& b_{-1/2} + \delta \tilde{\eta} \mathcal{K}_a\hspace*{-1mm}\left[a,\overline{{\widetilde E}_K^\varphi},{\widetilde E}_G^\varphi,{\widetilde E}_V^\varphi,\overline{{\widetilde E}_K^A},{\widetilde E}_G^A\right] \,,\\
  a_{+0} &=& a + \delta \tilde{\eta} b_{+1/2} \,,\\
  \tilde\varphi_{+0} &=&\tilde\varphi + \delta \tilde{\eta} \frac{\piSinglpar_{+1/2}}{a_{+1/2}^{3-\alpha}} \,,\\
  \widetilde V_{i,+0}&=& \left(1-i \frac{g_A Q_A^{(\tilde\varphi)}\delta \tilde{\eta}\delta \tilde{x}}{2 a^{1-\alpha}_{+1/2}} \piApar_{i,+1/2} \right )^2  \widetilde V_{i}  \ \ \ \ \ \text{\bf or} \ \ \ \ \ \widetilde V_{i,+0}=\widetilde V_{0i}[\piApar_{i,+1/2}]\widetilde V_i \ \ \ \, ,\\
    HC&:& b^2=\frac{1}{3}\left(\frac{f_*}{m_p}\right)^2a^{2(\alpha+1)}\left[\,\overline{{\widetilde E}_K^\varphi}+{\widetilde E}_G^\varphi+{\widetilde E}_V+
    \overline{{\widetilde E}_K^A}+{\widetilde E}_G^A\right]
\ .
\end{eqnarray}
\endgroup

\subsubsection{Velocity-Verlet }

Again, only the drifts differ in the velocity-Verlet algorithm with respect to the non-compact case. We get\\

\begingroup
\allowdisplaybreaks
\hspace{1cm}\text{ \it Compact Velocity-Verlet VV2}\vspace{-0.4cm} \\
\begin{eqnarray}
IC &:& \left \{a, b,\tilde\varphi, \piSingl,\widetilde V_i,\piApar_i\right\} \text{ at } \eta_0. \\
\piSinglpar_{+1/2} &=& \piSingl + \frac{\delta \tilde{\eta}}{2} \mathcal{K}_\varphi[a,\tilde\varphi,\widetilde V_j] \,,\\
  \piApar_{i,+1/2} &=& \piApar_i + \frac{\delta \tilde{\eta}}{2} \mathcal{K}_{A_i}[a,\tilde\varphi,\widetilde V_j] \,,\\
  b_{+1/2} &=& b + \frac{\delta \tilde{\eta}}{2} \mathcal{K}_a\hspace*{-1mm}\left[a,{\widetilde E}_K^\varphi,{\widetilde E}_G^\varphi,{\widetilde E}_V^\varphi,{\widetilde E}_K^A,{\widetilde E}_G^A\right] \,,\\
  a_{+0} &=& a + \delta \tilde{\eta} b_{+1/2} \,,\\
  a_{+1/2} &=& \frac{a_{+0}+a}{2} \,,\\
  \tilde\varphi_{+0} &=&\tilde\varphi + \delta \tilde{\eta} a_{+1/2}^{-(3-\alpha)}\piSinglpar_{+1/2} \,,\\
  \widetilde V_{i,+0}&=& \left(1-i \frac{g_A Q_A^{(\tilde\varphi)}\delta \tilde{\eta}\delta \tilde{x}}{2 a^{1-\alpha}_{+1/2}} \piApar_{i,+1/2} \right )^2  \widetilde V_{i}  \ \ \ \ \ \text{\bf or} \ \ \ \ \ \widetilde V_{i,+0}= \widetilde V_{0i}[\piApar_{i,+1/2}]\widetilde V_i \ \ \ \, ,\\  \piSinglpar_{+0} &=& \piSinglpar_{+1/2} + \frac{\delta \tilde{\eta}}{2} \mathcal{K}_\varphi[a_{+0} ,\tilde\varphi_{+0} ,\widetilde V_{j, +0}] \,,\\
  \piApar_{i,+0} &=& \piApar_{i,+1/2} + \frac{\delta \tilde{\eta}}{2} \mathcal{K}_{A_i}[a_{+0},\tilde\varphi_{+0},\widetilde V_{j,+0}]  \,,\\
  b_{+0} &=& b_{+1/2} + \frac{\delta \tilde{\eta}}{2} \mathcal{K}_a\hspace*{-1mm}\left[a_{+0},{\widetilde E}_{K,+0}^\varphi,{\widetilde E}_{G,+0}^\varphi,{\widetilde E}_{V,+0}^\varphi,{\widetilde E}_{K,+0}^A,{\widetilde E}_{G,+0}^A\right] \,,\\
  HC&:& b^2=\frac{1}{3}\left(\frac{f_*}{m_p}\right)^2a^{2(\alpha+1)}\left[{\widetilde E}_K^\varphi+{\widetilde E}_G^\varphi+{\widetilde E}_V+
  {\widetilde E}_K^A+{\widetilde E}_G^A\right]
\ .
\end{eqnarray}
\endgroup

\subsubsection{Velocity-Verlet n$th$ order}

The higher order integrators VV4, VV6, VV8 and VV10 for the compact formulation are also obtained by a simple modification of the drifts,
\begingroup
\allowdisplaybreaks
\begin{eqnarray}
&& \left.
\begin{array}{r}
\piSingl^{(0)} \equiv \piSingl({\bf n},n_0) \vspace*{0.15cm}\\
\tilde\varphi^{(0)} \equiv \tilde\varphi({\bf n},n_0)\vspace*{0.15cm}\\
\widetilde V_i^{(0)} \equiv \widetilde V_i({\bf n},n_0)\vspace*{0.15cm}\\
\piApar_i^{(0)} \equiv \piApar_i({\bf n},n_0)\vspace*{0.15cm}\\
a^{(0)} \equiv a(n_0)\vspace*{0.15cm}\\
b^{(0)} \equiv b(n_0)
\end{array}
\right\rbrace ~~~\Longrightarrow\\
&&\Longrightarrow\left\lbrace
\begin{array}{rcl}
  \piSinglpar^{(p)}_{1/2} &=& \piSingl^{(p-1)} + \frac{\omega_p\delta \tilde{\eta}}{2} \mathcal{K}_\varphi\left[a^{(p-1)},\tilde\varphi^{(p-1)},\widetilde V_j^{(p-1)}\right] \\
  \piApar_{i,1/2}^{(p)} &=& \piApar^{(p-1)}_i + \frac{\omega_p\delta \tilde{\eta}}{2} \mathcal{K}_{A_i}\left[a^{(p-1)},\tilde\varphi^{(p-1)},\widetilde V_j^{(p-1)}\right] \\
  b^{(p)}_{1/2} &=& b^{(p-1)} + \frac{\omega_p\delta \tilde{\eta}}{2} \mathcal{K}_a^{(p-1)}\hspace*{-1mm}\left[a,{\widetilde E}_K^\varphi,{\widetilde E}_G^\varphi,{\widetilde E}_V^\varphi,{\widetilde E}_K^A,{\widetilde E}_G^A\right] \\
  a^{(p)} &=& a^{(p-1)} + \omega_p\delta \tilde{\eta} b^{(p)}_{1/2} \\
  a^{(p)}_{1/2}&=&\frac{a^{(p)}+a^{(p-1)}}{2}\\
  \tilde\varphi^{(p)} &=&\tilde\varphi^{(p-1)} + \delta \tilde{\eta} \frac{\piSinglpar^{(p)}_{1/2}}{\left(a_{1/2}^{(p)}\right)^{3-\alpha}} \\
  \widetilde V_{i}^{(p)}&=& \left(1-i \frac{g_A Q_A^{(\tilde\varphi)}\delta \tilde{\eta}\delta \tilde{x}}{2 (a^{(p)}_{1/2})^{1-\alpha}} \piApar^{(p)}_{i,1/2} \right )^2  \widetilde V_{i}^{(p-1)}  \\
   && \ \ \  \ \ \ \ \ \  \ \ \ \ \  \ \ \  \ \ \ \ \ \ \ \text{\bf or} \ \ \ \ \ \widetilde V_{i}^{(p)}= \widetilde V_{0i}[\piApar^{(p)}_{i,1/2}]\widetilde V_i^{(p-1)} \ \ \ \, ,\\  \piSinglpar_{+0} &=& \piSinglpar_{+1/2} + \frac{\delta \tilde{\eta}}{2} \mathcal{K}_\varphi[a_{+0} ,\tilde\varphi_{+0} ,\widetilde V_{j, +0}] \,,\\
\\
  \piSinglpar^{(p)} &=& \piSinglpar^{(p)}_{1/2} + \frac{\omega_p\delta \tilde{\eta}}{2} \mathcal{K}_\varphi\left[a^{(p)},\tilde\varphi^{(p)},\widetilde V_j^{(p)}\right]\\
  \piApar^{(p)}_i &=& \piApar^{(p)}_{i,1/2} + \frac{\omega_p\delta \tilde{\eta}}{2}\mathcal{K}_{A_i}[a^{(p)},\tilde\varphi^{(p)},\widetilde V_j^{(p)}] \\
  b^{(p)} &=& b^{(p)}_{1/2} + \frac{\omega_p\delta \tilde{\eta}}{2} \mathcal{K}_a^{(p)}\hspace*{-1mm}\left[a,{\widetilde E}_K^\varphi,{\widetilde E}_G^\varphi,{\widetilde E}_V^\varphi,{\widetilde E}_K^A,{\widetilde E}_G^A\right] \, \ ,
\end{array}
\right\rbrace_{p\,=\,1,\, ...,\, s}\\
&&\hspace*{-1cm}\Longrightarrow ~~~~
\left\lbrace
\begin{array}{l}
   \piSingl({\bf n},n_0)\equiv \piSingl^{(s)} \vspace*{0.15cm}\\
  \tilde\varphi({\bf n},n_0)\equiv \tilde\varphi^{(s)}\vspace*{0.15cm}\\
  \widetilde V_i({\bf n},n_0)\equiv \widetilde V_i^{(s)}\vspace*{0.15cm}\\
  \piApar_i({\bf n},n_0)\equiv \piApar^{(s)}_i\vspace*{0.15cm}\\
  a(n_0)\equiv a^{(s)}\vspace*{0.15cm}\\
  b(n_0)\equiv b^{(s)} \, \ ,
\end{array}\notag
\right.\\
\notag\\
  HC&:& b^2=\frac{1}{3}\left(\frac{f_*}{m_p}\right)^2a^{2(\alpha+1)}\left[{\widetilde E}_K^\varphi+{\widetilde E}_G^\varphi+{\widetilde E}_V+
  {\widetilde E}_K^A+{\widetilde E}_G^A\right] \ ,
\end{eqnarray}
\endgroup
where, as before, we have introduced $\mathcal{K}_a^{(p)}\hspace*{-1mm}\left[a,{\widetilde E}_K^\varphi,{\widetilde E}_G^\varphi,{\widetilde E}_V^\varphi,{\widetilde E}_K^A,{\widetilde E}_G^A\right] \equiv \mathcal{K}_a\hspace*{-1mm}\left[a^{(p)},{\widetilde E}_{K}^{\varphi^{(p)}}\hspace*{-1.6mm},{\widetilde E}_{G}^{\varphi^{(p)}}\hspace*{-1.6mm},{\widetilde E}_{V}^{\varphi^{(p)}}\hspace*{-1.6mm},{\widetilde E}_{K}^{A^{(p)}}\hspace*{-1.6mm},{\widetilde E}_{G}^{A^{(p)}}\right]$, to ease the notation. As we will see in the next section, an advantage of the compact formulation is that it directly generalizes to non-Abelian groups, contrary to the non-compact one. However, before moving on, let us introduce some relevant observables for the $U(1)$ gauge sector.

\subsection{Observables}

We list here some observables with mean value of interest (as usual $\langle ... \rangle$ denotes volume averaging),
\begingroup
\allowdisplaybreaks
\begin{eqnarray}
U(1)\text{-charged~matter:}&&  \left\langle\mathcal{R}e [\tilde\varphi] \right\rangle\,,~ \left\langle \mathcal{I}m [\tilde\varphi] \right\rangle\,,~  \left\langle \mathcal{R}e [\piSingl]\right\rangle\,, ~ \left\langle \mathcal{I}m [\piSingl]\right\rangle\,, ~\left\langle |\tilde\varphi|^2\right\rangle\,, ~\left\langle |\piSingl|^2 \right\rangle\,,\\
\label{eq:U1noncompactmeans}
\text{electric fields:}&&  \left\langle \big|\widetilde{\mathcal{E}}\big|^2 \right\rangle\,, ~~{\rm with}~~\widetilde{\mathcal{E}}_i = \frac{1}{a^{1-\alpha}}\piApar_i \,,\\
\label{eq:U1compactmeans}
\text{magnetic fields:}&&  \left\langle \big|\widetilde{\mathcal{B}}\big|^2 \right\rangle\,, ~~{\rm with}~\left\lbrace
\begin{array}{ll}
\widetilde{\mathcal{B}}_i=\sum_{jk}\epsilon_{ijk}\tilde{\Delta}^+_j\widetilde A_k\,,     & {\rm non-compact} \\
\widetilde{\mathcal{B}}_i=\frac{i}{\delta \tilde{x}^2g_A Q_A^{(\tilde\varphi)}}(\mathcal{I}m (\tilde{V}_{jk}))\,, &  {\rm compact}
\end{array}\right.
\end{eqnarray}
\endgroup
Note that, as presented in the $U(1)$-toolkit in Eq.~(\ref{eq:U1toolkit}) in Section~\ref{subsubsec:latgaugeinvtech}, other expressions for the magnetic field are also possible.

\subsubsection{Energy density components}

We collect here the different expressions for the energy components of the system,
\begin{align}
U(1)\text{ matter: }& {\widetilde E}_K^\varphi = \frac{1}{a^6} \left<\piSingl^{2}\right >,\ {\widetilde E}_G^{\varphi} = \frac{1}{a^2}\sum_i \left <(\widetilde D^{A+}_i\tilde\varphi)^*(\widetilde{D}^{A+}_i\tilde\varphi)\right > \,.\\
U(1)\text{ gauge fields, non-compact: }& {\widetilde E}_K^A = \frac{1}{2a^4} {\omega_*^2 \over f_*^2} \sum_{i=1}^3 \left <\piApar_i^{2}\right >, \ {\widetilde E}_G^A = \frac{1}{2a^4} {\omega_*^2 \over f_*^2} \sum_{i, j<i} \left <(\tilde{\Delta}^+_i \widetilde A_j - \tilde{\Delta}^+_i \widetilde A_j)^2 \right > \,.\\
U(1)\text{ gauge fields, compact: }&  {\widetilde E}_K^A = \frac{1}{2a^4} {\omega_*^2 \over f_*^2} \sum_{i=1}^3 \left <\piApar_i^{2}\right >, \ {\widetilde E}_G^{A}=\frac{1}{a^4 \delta \tilde{x}^4g_A^2 Q_A^{(\tilde\varphi)\, 2}} {\omega_*^2 \over f_*^2} \sum_{i, j<i}(1-\mathcal{R}e (\widetilde V_{ij}))  . \\
\text{Potential: }& {\widetilde E}_V=\left <\widetilde V(\tilde\varphi,\dots) \right > \,.
\end{align}

\subsubsection{Spectra}

The last quantities of interest are the power spectra, which according to the discrete expression of Eq.~(\ref{eq:discretePS}), we define as follow,

\begingroup
\allowdisplaybreaks
\begin{align}
  {\widetilde \Delta}_{\tilde\varphi}(\tilde k(\tilde {\bf n})) &= \frac{\tilde k^3(\tilde {\bf n})}{2\pi^2} \left(\frac{\delta \tilde{x}}{N}\right)^3\left<|\mathcal{R}e (\tilde\varphi)(\tilde {\bf n})|^2+|\mathcal{I}m (\tilde\varphi)(\tilde {\bf n})|^2\right>_{R(\tilde{\bf n})} \, \ , \\
  {\widetilde \Delta}_{\piSingl}(\tilde k(\tilde {\bf n})) &= \frac{\tilde k^3(\tilde {\bf n})}{2\pi^2} \left(\frac{\delta \tilde{x}}{N}\right)^3\left<|\mathcal{R}e (\piSingl)(\tilde {\bf n})|^2+|\mathcal{I}m(\piSingl)(\tilde {\bf n})|^2\right>_{R(\tilde{\bf n})} \ ,  \\
   {\widetilde \Delta}_{\mathcal{E}}(\tilde k(\tilde {\bf n})) &= \frac{\tilde k^3(\tilde {\bf n})}{2\pi^2} \left(\frac{\delta \tilde{x}}{N}\right)^3\Big<\sum_i  |\widetilde{\mathcal{E}}_i(\tilde {\bf n})|^2\Big>_{R(\tilde{\bf n})} \ , \\
  {\widetilde \Delta}_{\mathcal{B}}(\tilde k(\tilde {\bf n})) &= \frac{\tilde k^5(\tilde {\bf n})}{2\pi^2} \left(\frac{\delta \tilde{x}}{N}\right)^3\Big<\sum_i  |\tilde A_i(\tilde {\bf n})|^2\Big>_{R(\tilde{\bf n})} \ \ \ \ \text{, [non-compact]} \ , \\
  {\widetilde \Delta}_{\mathcal{B}}(\tilde k(\tilde {\bf n})) &= \frac{\tilde k^3(\tilde {\bf n})}{2\pi^2} \left(\frac{\delta \tilde{x}}{N}\right)^3\Big<\sum_i  |\widetilde{\mathcal{B}}_i(\tilde {\bf n})|^2\Big>_{R(\tilde{\bf n})} \ \ \ \ \text{, [compact]} \ ,
\end{align}
\endgroup
where $\langle ... \rangle_{R(\tilde{\bf n})}$ represents angular averaging in $\bf k$-space, c.f.~Eq.~(\ref{eq:discretePS}), and the electric and magnetic fields are defined as in equations \eqref{eq:U1noncompactmeans} and \eqref{eq:U1compactmeans}. The extra powers of $k(\tilde {\bf n})$ in the non-compact magnetic field spectra come from the spatial derivative of $\tilde A_i$. We note that the dimensionless power spectra are related to their dimensionful counterparts by ${\Delta}_{\varphi} = {\widetilde \Delta}_{\tilde\varphi}f_*^2$, ${\Delta}_{\pi_\varphi} = {\widetilde \Delta}_{\tilde\pi_\varphi}f_*^2\omega_*^2$, ${\Delta}_{\mathcal{E}} \equiv {\widetilde \Delta}_{\widetilde{\mathcal{E}}}\,\omega_*^4$, and ${\Delta}_{\mathcal{B}} \equiv {\widetilde \Delta}_{\widetilde{\mathcal{B}}}\,\omega_*^4$. 

\section{Lattice formulation of gauge fields, Part II: $SU(N)$ interactions}\label{sec:LatSUN}

We introduce now a set of Gauss-preserving evolution algorithms for a $SU(N)$ gauge sector with self-consistent expansion of the universe. We will follow closely what has been done for the compact $U(1)$ formulation.  As in the previous section, when it comes to Verlet integrators, we use only the velocity-based one; the corresponding position-Verlet algorithms can be straightforwardly obtained.

\subsection{Continuum formulation and natural variables}

We define the following \textit{program variables} for a scalar doublet and a non-Abelian gauge field,
\be \widetilde{\Phi} = \frac{\Phi}{f_*} \ , \hspace{0.4cm} \widetilde B_{\mu}^a = \frac{B_{\mu}^a}{\omega_*} \ , \label{eq:GaugeProgramVarSU2} \ee
following closely the normalization criteria we used for the U(1) gauge sector, see Eq.~(\ref{eq:GaugeProgramVar}). Our lattice formulation will be based in these variables.

We start again by identifying an appropriate set of conjugate momenta. This is achieved as in the $U(1)$ case, by writing the continuum equations \eqref{eq:higgsSU2-eom} and \eqref{eq:SU2eom} appropriately. In the temporal gauge, they are
\begin{align}
  (a^{3-\alpha}\tilde{\Phi}')' - a^{1 + \alpha} {\vec{\widetilde{D}}}^{\,2}\widetilde\Phi  &= -a^{\alpha+3} \frac{\tilde{V}_{,|\widetilde\Phi|}}{2}\frac{\widetilde\Phi}{|\widetilde\Phi |} \ ,  \\
  \partial_0 (a^{1-\alpha} (\widetilde G_{0i})^a) -a^{\alpha - 1} ( \mathcal{\widetilde D}_j )_{a b} (\widetilde G_{ji} )^b  &=   a^{1+ \alpha}\widetilde J^a_i \ .
\end{align}
From here, we identify the conjugate momenta as
\begin{align}
\piDoubl&=  a^{3-\alpha}\widetilde\Phi' \ ,  \label{eq:momSU2doublet}\\
\piBpar^{a}_i &= a^{1-\alpha}\widetilde G^a_{0i}\label{eq:momSU2vec} \ .
\end{align}
The associated kinetic energies of the two field sectors become
\begin{align}
  {\widetilde E}_K^\Phi &= \frac{1}{a^6} \left<\piDoubl^\dagger \piDoubl \right > \ , \\
  {\widetilde E}_K^B &= \frac{1}{2a^4} {\omega_*^2 \over f_*^2} \sum_{a,i} \left <\left(\piBpar_i^a\right)^2\right > \, ,
\end{align}
which, as in the $U(1)$ case, are related to their physical counterparts as ${E}_K^\Phi = f_*^2\omega_*^2{\widetilde E}_K^\Phi$ and ${E}_K^B = f_*^2\omega_*^2{\widetilde E}_K^B$, respectively. Finally, we define the following kernels as
\begin{align}
(\piDoubl)'  &= \mathcal{K}_{\Phi}[a,\widetilde\Phi,\widetilde U_j] \ , \label{eq:kernelsSU21} \\
\left(\piBpar_i^a\right)' &=\kersutwoComp[a,\widetilde\Phi,\widetilde U_j] \ ,  \\
\mathcal{K}_{\Phi}[a,\widetilde\Phi,\widetilde U_j]&\equiv  - a^{\alpha+3} \frac{\widetilde V_{,|\widetilde\Phi|}}{2} \frac{\widetilde\Phi}{|\widetilde\Phi |} + a^{1 + \alpha} {\vec{\widetilde{D}}}^{\,2} \widetilde\Phi  \ ,  \\
\kersutwoComp[a,\widetilde\Phi,\widetilde U_j] &\equiv a^{1+ \alpha}\widetilde J^a_i + a^{\alpha - 1}( \mathcal{\widetilde D}_j )_{a b} (\widetilde G_{ji} )^b \ ,
\end{align}
which allows us to proceed with the discretization and time evolution of the EOM. Notice that the gauge field kernel can also be written in matrix form, omitting the internal non-Abelian indices,
\bea
\piBpar_i' =\kersutwo[a,\widetilde\Phi,\widetilde U_j] \,,
~~~~\kersutwo[a,\widetilde\Phi,\widetilde U_j] &\equiv a^{1+ \alpha}\widetilde J_i + a^{\alpha - 1}\mathcal{\widetilde D}_j\widetilde G_{ji} \ .\label{eq:kernelsSU25}
\eea

\subsection{(Compact) Lattice formulation(s)}

For non-Abelian gauge fields, we do not have the choice between compact and non-compact variables: the compact formulation is required to maintain gauge invariance. We then discretize the kernels as follows,
\begin{align}
  \mathcal{K}_\Phi[a,\widetilde\Phi,\widetilde U_j] &= - a^{\alpha+3} \frac{\tilde{V}_{,|\widetilde\Phi|}}{2} \frac{\widetilde\Phi}{|\widetilde\Phi |}+ a^{1 + \alpha} \sum_i \widetilde D_i^- \widetilde D^+_i\widetilde\Phi \ , \\
  \mathcal{K}_{B_i}[a,\widetilde\Phi,\widetilde U_j] &= a^{1+ \alpha}\left({\frac{g_B Q_B^{(\tilde\Phi)}}{2\delta\tilde x}}{f_*^2\over\omega_*^2} \mathcal{I}m [ \widetilde\Phi^{\dagger}\widetilde\sigma^a U_i\widetilde\Phi]\sigma^a + \dots \right ) + \frac{a^{\alpha - 1}}{\delta \tilde{x}^3g_B Q_B^{(\tilde\Phi)}}\sum_j\left(\widetilde U_{ij}-\widetilde U_{j,-j}^\dagger\widetilde U_{ij~-j} \widetilde U_{j,-j}\right ) \, , \label{eq:kernelSU2}
\end{align}
where the second term in the $SU(N)$ gauge field kernel is a backward finite difference approximation of the gauge covariant derivative $\mathcal{\widetilde D}_i \widetilde G_{ij}$. We use matrix notation for conciseness. Note also that we expressed the group generators $T^a=\frac{\sigma^a}{2}$ in terms of the Pauli matrices. Using our $SU(N)$-toolkit \eqref{eq:SU2toolkit}, we see that the magnetic energy can be written as
\begin{equation}
{\widetilde E}_G^B = \frac{2}{a^4 \delta \tilde{x}^4g_B^2 Q_B^{(\tilde\Phi)\, 2}}   {\omega_*^2 \over f_*^2} \left(2-\sum_{i,j<i}\left <  \mathrm{Tr}(\widetilde U_{ij}) \right >\right) \ ,
\end{equation}
again related to its physical counterpart via $E_G^B = \omega_*^2 f_*^2 {\widetilde E}_G^B$.

\noindent As in the compact $U(1)$ case, the drifts can be written down following two approaches:\\

\noindent {\it Drifts from momenta:}\vspace*{-0.25cm}\\

 We start again from the continuum relation
\begin{align}
   (\widetilde U_i)' = \partial_0 e^{-i g_B Q_B^{(\tilde\Phi)}\delta \tilde{x} \tilde B_i} = -i g_B Q_B^{(\tilde\Phi)}\delta \tilde{x}  (\tilde B_i)' \widetilde U_i + O(\delta \tilde x) \, ,
\end{align}
which in terms of the conjugate momenta is
\begin{align}
   (\widetilde U_i)' = -i \frac{g_B Q_B^{(\tilde\Phi)}\delta \tilde{x}}{a^{1-\alpha}} \piBpar_i \widetilde U_i + O(\delta \tilde x)\, , \label{eq:nonAbDrif}
\end{align}
with no sum intended. Note that due to the non-commutativity, this relation is only valid at leading order in $\delta \tilde x$. Neglecting higher order corrections, we again find an implicit equation for the drifts. Similarly as in Eq.~\eqref{eq:CompactAbelianDriftMom} of the Abelian case, Eq.~(\ref{eq:nonAbDrif}) can be solved by a discretized scheme as
\begin{align}
  \widetilde U_{i,+0}
   &=\left(1-i \frac{g_B Q_B^{(\tilde\Phi)}\delta \tilde{\eta}\delta \tilde{x}}{2 a^{1-\alpha}_{+1/2}} \piBpar_{i,+1/2} \right )^2  \widetilde U_{i} +O(\delta \tilde{\eta}^2) \ .
\end{align}

\noindent {\it Drifts from plaquettes:}\vspace*{-0.25cm}\\

In the non-Abelian case, one can also in principle evolve the links as
\begin{align}
  \widetilde U_{i,+0} = \widetilde U_{0i}\widetilde U_i \ ,
\end{align}
if one is able to construct straightforwardly the map $U_{0i}$ as a function of $\piBpar$, which needs to be done in a case by case basis. In the $SU(2)$ case can be done using the parametrization detailed in Eqs.~\eqref{eq:plaq-descomp}-\eqref{eq:plaq-descomp-mat}, together with the relation
\begin{align}
  \widetilde U_{0i}- \widetilde U_{0i}^\dagger = -2 i \frac{g_B Q_B^{(\tilde\Phi)}\delta \tilde x \delta \tilde \eta}{a^{1-\alpha}}\piBpar + O(d \tilde x_\mu^3)\ .
\end{align}
This fixes the coefficients $c_1,c_2$ and $c_3$ of $\widetilde U_{0i}$ and hence also $c_0$ as a function of the latter.

\bigskip

Finally, irrespective of the chosen drifts, a crucial quantity to monitor is the Gauss law, which must be obeyed at all times during the simulation. Based on the continuum expression in Eq.~(\ref{eq:GaussSU2-eom}), we discretize it in matrix notation as
\begin{equation}
-\sum_i \frac{\piBpar_{i,-i}-U_{i,-i}^\dagger\piBpar_{i,-i}U_{i,-i}}{\delta\tilde x} =  \frac{g_B Q_B^{(\widetilde\Phi)}}{2}{f_*^2\over\omega_*^2} \mathcal{I}m [ \widetilde\Phi^\dag\sigma^a \piDoubl ]\sigma^a\,.
\label{eq:SUNvector-Gauss_Discrete}
\end{equation}

\subsubsection{(Staggered) Leapfrog}

Let us now consider different evolution algorithms to solve the field dynamics in the compact formulation, following closely the same procedure as we did for a $U(1)$ gauge sector. We start with a straightforward generalization of the staggered leapfrog algorithm, which can be written in this context as follows\\

\begingroup
\allowdisplaybreaks
 \hspace{1cm}\text{ \it SU(N) Staggered Leapfrog}\vspace{-0.4cm}\\
\begin{eqnarray}
  IC &:& \left \{a, \widetilde\Phi, \widetilde U_i\right\} \text{ at } \eta_0, ~~~~ \left \{b_{-1/2},\piDoublpar_{-1/2}, \piBpar_{i,-1/2}\right\} \text{ at } \eta_0 - {\delta\tilde\eta\over2}.\\
  \piDoublpar_{i,+1/2} &=& \piDoublpar_{-1/2} + \delta \tilde{\eta} \mathcal{K}_\Phi[a,\widetilde\Phi,\widetilde U_j] \ , \\
  \piBpar_{i,+1/2} &=& \piBpar_{i,-1/2} + \delta \tilde{\eta} \kersutwo[a,\widetilde\Phi,\widetilde U_j] \ , \\
  b_{+1/2} &=& b_{-1/2} + \delta \tilde{\eta} \mathcal{K}_a\hspace*{-1mm}\left[a,\overline{{\widetilde E}_K^\Phi},{\widetilde E}_G^\Phi,{\widetilde E}_V^\Phi,\overline{{\widetilde E}_K^B},{\widetilde E}_G^B\right] \ , \\
  a_{+0} &=& a + \delta \tilde{\eta} b_{+1/2} \ , \\
  a_{+1/2}&=&\frac{a_{+0}+a}{2} \ , \\
  \widetilde \Phi_{+0} &=&\widetilde \Phi + \delta \tilde{\eta} \frac{\piDoublpar_{+1/2}}{a_{+1/2}^{3-\alpha}} \ , \\
  \widetilde U_{i,+0}&=& \left(1-i \frac{g_B Q_B^{(\tilde\Phi)}\delta \tilde{\eta}\delta \tilde{x}}{2 a^{\mathbf{1}-\alpha}_{+1/2}} \piBpar_{i,+1/2} \right )^2  \widetilde U_{i}  \ \ \ \ \ \text{\bf or} \ \ \ \ \ \widetilde U_{i,+0}= \widetilde U_{0i}[\piBpar_{i,+1/2}]\widetilde U_i \ \ \ \, ,\\  \piSinglpar_{+0} &=& \piSinglpar_{+1/2} + \frac{\delta \tilde{\eta}}{2} \mathcal{K}_\varphi[a_{+0} ,\tilde\varphi_{+0} ,\widetilde V_{j, +0}] \, ,\\      HC&:& b^2=\frac{1}{3}\left(\frac{f_*}{m_p}\right)^2a^{2(\alpha+1)}\left[\,\overline{{\widetilde E}_K^\Phi}+{\widetilde E}_G^\Phi+{\widetilde E}_V + \overline{{\widetilde E}_K^B}+{\widetilde E}_G^B\right] \ .
\end{eqnarray}
\endgroup
In particular, note that the scale factor kernel is evaluated using semi-sums of the different kinetic energies, as usual in the staggered leapfrog.

\subsubsection{Velocity Verlet}
Mimicking the algorithm developed for the analogous U(1) gauge sector, we obtain\\

\begingroup
\allowdisplaybreaks
\hspace*{1cm}\text{ \it SU(N) Velocity Verlet VV2}\vspace*{-4mm}\\
\begin{eqnarray}
  IC &:& \left \{a, b,\widetilde\Phi, \piDoubl,\widetilde U_i,\piBpar_i\right\} \text{ at } \eta_0. \\
  \piDoublpar_{+1/2} &=& \piDoubl + \frac{\delta \tilde{\eta}}{2} \mathcal{K}_\Phi[a,\widetilde\Phi,\widetilde U_i] \ , \\
  \piBpar_{i,+1/2} &=& \piBpar_i + \frac{\delta \tilde{\eta}}{2} \kersutwo[a,\widetilde\Phi,\widetilde U_i] \ , \\
  b_{+1/2} &=& b + \frac{\delta \tilde{\eta}}{2} \mathcal{K}_a\hspace*{-1mm}\left[a,{\widetilde E}_K^\Phi,{\widetilde E}_G^\Phi,{\widetilde E}_V^\Phi,{\widetilde E}_K^B,{\widetilde E}_G^B\right] \ , \\
  a_{+0} &=& a + \delta \tilde{\eta} b_{+1/2} \ , \\
  a_{+1/2}&=&\frac{a_{+0}+a}{2} \ , \\
  \widetilde \Phi_{+0} &=& \widetilde \Phi + \delta \tilde{\eta} \frac{\piDoublpar_{+1/2}}{a_{+1/2}^{3-\alpha}} \ , \\
  \widetilde U_{i,+0}&=& \left(1-i \frac{g_B Q_B^{(\tilde\Phi)}\delta \tilde{\eta}\delta \tilde{x}}{2 a^{1-\alpha}_{+1/2}} \piBpar_{i,+1/2} \right )^2  \widetilde U_{i}  \ \ \ \ \ \text{\bf or} \ \ \ \ \ \widetilde U_{i,+0}= \widetilde U_{0i}[\piBpar_{i,+1/2}]\widetilde U_i \ \ \ \, \\ \piDoublpar_{+0} &=& \piDoublpar_{+1/2} + \frac{\delta \tilde{\eta}}{2} \mathcal{K}_\Phi[a_{+0} ,\widetilde\Phi_{+0} ,\widetilde U_{i, +0}]\ , \\
  \piBpar_{i,+0} &=& \piBpar_{i,+1/2} + \frac{\delta \tilde{\eta}}{2} \kersutwo[a_{+0},\widetilde\Phi_{+0},\widetilde U_{i,+0}] \ , \\
  b_{+0} &=& b_{+1/2} + \frac{\delta \tilde{\eta}}{2} \mathcal{K}_a\hspace*{-1mm}\left[a_{+0},{\widetilde E}_{K,+0}^\Phi,{\widetilde E}_{G,+0}^\Phi,{\widetilde E}_{V,+0}^\Phi,{\widetilde E}_{K,+0}^B,{\widetilde E}_{G,+0}^B\right] \ , \\
    HC&:& b^2=\frac{1}{3}\left(\frac{f_*}{m_p}\right)^2a^{2(\alpha+1)}\left({\widetilde E}_K^\Phi+{\widetilde E}_G^\Phi+{\widetilde E}_K^B+{\widetilde E}_G^B+{\widetilde E}_V\right) \ .
\end{eqnarray}
\endgroup

\subsubsection{Velocity Verlet n$th$ order}

The higher-order integrators VV4, VV6, VV8 and VV10 are also obtained by a simple modification of the drift,

\begingroup
\allowdisplaybreaks
\begin{eqnarray}
&& \left.
\begin{array}{r}
\piDoubl^{(0)} \equiv \piDoublpar({\bf n},n_0) \vspace*{0.15cm}\\
\widetilde\Phi^{(0)} \equiv \widetilde\Phi({\bf n},n_0)\vspace*{0.15cm}\\
\widetilde U_i^{(0)} \equiv \widetilde U_i({\bf n},n_0)\vspace*{0.15cm}\\
\piBpar^{(0)}_i \equiv \piBpar_i({\bf n},n_0)\vspace*{0.15cm}\\
a^{(0)} \equiv a(n_0)\vspace*{0.15cm}\\
b^{(0)} \equiv b(n_0) \, \ ,
\end{array}
\right\rbrace  ~~~~\Longrightarrow
\\
&&\Longrightarrow\left\lbrace
\begin{array}{rcl}
  \piDoublpar^{(p)}_{1/2} &=& \piDoublpar^{(p-1)} + \frac{\omega_p\delta \tilde{\eta}}{2} \mathcal{K}_\Phi\left[a^{(p-1)},\widetilde\Phi^{(p-1)},\widetilde U_j^{(p-1)}\right] \\
  \piBpar_{i,1/2}^{(p)} &=& \piBpar_i^{(p-1)} + \frac{\omega_p\delta \tilde{\eta}}{2} \kersutwo[a^{(p-1)},\widetilde\Phi^{(p-1)},\widetilde U_i^{(p-1)}] \\
  b^{(p)}_{1/2} &=& b^{(p-1)} + \frac{\omega_p\delta \tilde{\eta}}{2} \mathcal{K}_a^{((p-1)}\hspace*{-1mm}\left[a,{\widetilde E}_K^\Phi,{\widetilde E}_G^\Phi,{\widetilde E}_V^\Phi,{\widetilde E}_K^B,{\widetilde E}_G^B\right] \\
  a^{(p)} &=& a^{(p-1)} + \omega_p\delta \tilde{\eta} b^{(p)}_{1/2} \\
  \widetilde\Phi^{(p)} &=&\widetilde\Phi^{(p-1)} + \delta \tilde{\eta} \frac{\piDoublpar^{(p)}_{1/2}}{a_{1/2}^{(p)\ 3-\alpha}} \\
\widetilde U_{i}^{(p)}&=&\hspace{-0.2cm} \left(1-i \frac{g_B Q_B^{(\tilde\Phi)}\delta \tilde{\eta}\delta \tilde{x}}{2 (a^{(p)}_{1/2})^{1-\alpha}} \piBpar^{(p)}_{i,1/2} \right )^2  \widetilde U_{i}^{(p-1)} \\
&& \ \ \ \ \ \ \ \ \ \ \ \ \ \ \ \ \ \ \ \ \ \ \  \ \ \ \  \text{\bf or } \ \ \ \ \  \widetilde U_{i}^{(p)}= \widetilde U_{0i}[\piBpar^{(p)}_{i,1/2}]\widetilde U_i^{(p-1)} \hspace{-0.5cm}\\
\piDoublpar^{(p)} &=& \piDoublpar^{(p)}_{1/2} + \frac{\omega_p\delta \tilde{\eta}}{2} \mathcal{K}_\Phi\left[a^{(p)},\widetilde\Phi^{(p)},\widetilde U_i^{(p)}\right]\\
  \piBpar^{(p)}_i &=& \piBpar^{(p)}_{i,1/2} + \frac{\omega_p\delta \tilde{\eta}}{2}\kersutwo[a^{(p)},\widetilde\Phi^{(p)},\widetilde U_i^{(p)}] \\
  b^{(p)} &=& b^{(p)}_{1/2} + \frac{\omega_p\delta \tilde{\eta}}{2} \mathcal{K}_a^{(p)}\hspace*{-1mm}\left[a,{\widetilde E}_K^\Phi,{\widetilde E}_G^\Phi,{\widetilde E}_V^\Phi,{\widetilde E}_K^B,{\widetilde E}_G^B\right] \ ,
\end{array}
\right\rbrace_{p\,=\,1,\, ...,\, s}
\hspace*{-1cm}\\
&&\Longrightarrow ~~~~
\left\lbrace
\begin{array}{l}
   \piDoublpar({\bf n},n_0)\equiv \piDoublpar^{(s)}_i \vspace*{0.15cm}\\
  \widetilde\Phi({\bf n},n_0)\equiv \widetilde\Phi^{(s)}\vspace*{0.15cm}\\
  \widetilde U_i({\bf n},n_0)\equiv \widetilde U_i^{(s)}\vspace*{0.15cm}\\
  \piBpar_i({\bf n},n_0)\equiv \piBpar^{(s)}_i\vspace*{0.15cm}\\
  a(n_0)\equiv a^{(s)}\vspace*{0.15cm}\\
  b(n_0)\equiv b^{(s)} \, ,
\end{array}
\right.\\
\notag\\
HC&:& b^2=\frac{1}{3}\left(\frac{f_*}{m_p}\right)^2a^{2(\alpha+1)}\left({\widetilde E}_K^\Phi+{\widetilde E}_G^\Phi+{\widetilde E}_K^B+{\widetilde E}_G^B+{\widetilde E}_V\right) \ ,
\end{eqnarray}
\endgroup
where, once again, we used $\mathcal{K}_a^{(p)}\hspace*{-1mm}\left[a,{\widetilde E}_K^\Phi,{\widetilde E}_G^\Phi,{\widetilde E}_V^\Phi,{\widetilde E}_K^B,{\widetilde E}_G^B\right] \equiv \mathcal{K}_a\hspace*{-1mm}\left[a^{(p)},{\widetilde E}_{K}^{\Phi^{(p)}}\hspace*{-1.6mm},{\widetilde E}_{G}^{\Phi^{(p)}}\hspace*{-1.6mm},{\widetilde E}_{V}^{\Phi^{(p)}}\hspace*{-1.6mm},{\widetilde E}_{K}^{B^{(p)}}\hspace*{-1.6mm},{\widetilde E}_{G}^{B^{(p)}}\right]$, to ease the notation.

\subsection{Observables}

Finally, we write here several observables that are of interest. Let us start with the following mean values (as usual $\langle ... \rangle$ denotes volume averaging)),
\begin{align}
SU(2)\text{ matter: }& \  \left\langle\widetilde\Phi^a\right\rangle, \,\,\,\,  \left\langle\piDoublpar^a\right\rangle, \,\,\,\,  \left\langle|\widetilde\Phi|^2\right\rangle, \,\,\,\, \left\langle|\piDoubl|^2\right\rangle \ .\\
SU(2)\text{ gauge fields: }& \
 \left\langle\big|\widetilde{\mathcal{E}}^B\big|^2\right\rangle=\sum_{a,i}\left\langle\left(\widetilde{\mathcal{E}}_i^{a}\right)^2\right\rangle, \,\,\,\,  \left\langle\big|\widetilde{\mathcal{B}}^B\big|^2 \right\rangle = \sum_{a,i}\left\langle\left(\widetilde{\mathcal{B}}_i^{a}\right)^2\right\rangle \ , \\
&~~\widetilde{\mathcal{E}}_i^a = \frac{1}{a^{1-\alpha}}\piBpar_i^a, \,\,\,\,  \widetilde{\mathcal{B}}^a_i=\frac{\epsilon_{ijk}}{2 \delta \tilde{x}^2g_B Q_B^{(\tilde\Phi)}} {\rm Tr} [(i T_a) (\widetilde U_{jk} - \widetilde U_{kj} ) ]. \label{eq:SU2means}
\end{align}

\subsubsection{Energy components}

The different energies associated to the $SU(N)$ gauge sector are
\begin{align}
SU(2)\text{ matter: }& {\widetilde E}_K^\Phi = \frac{1}{a^6}\left<\piDoubl^\dagger \piDoubl\right >, \,\,\,\, {\widetilde E}_G^{\Phi} = \frac{1}{a^2}\sum_i \left <(\widetilde D^{+}_i\widetilde\Phi)^*(\widetilde D^{+}_i\widetilde\Phi)\right > \ .
\\
SU(2)\text{ gauge fields: }&  {\widetilde E}_K^B = \frac{1}{2a^4} {\omega_*^2 \over f_*^2} \sum_{a,i} \left <\left(\piBpar_i^a\right)^2\right >, \,\,\,\, {\widetilde E}_G^B=\frac{2}{a^4 \delta \tilde{x}^4g_B^2 Q_B^{(\tilde\Phi)\, 2}} {\omega_*^2 \over f_*^2} \left(2-\sum_{i,j<i}\left <  \mathrm{Tr}(\widetilde U_{ij}) \right >\right) .\\
\text{Potential: }& {\widetilde E}_V=\left <\widetilde V(\widetilde\Phi,\dots) \right > \ .
\end{align}

\subsubsection{Spectra}

We also define the associated power-spectra of each field sector as follows,
\begin{align}
  {\widetilde \Delta}_{\widetilde\Phi}(\tilde k(\tilde {\bf n}))  &= \frac{\tilde k^3(\tilde {\bf n})}{2\pi^2} \left(\frac{\delta \tilde{x} }{N}\right)^3\left<\sum_{a}|\widetilde\Phi^a(\tilde {\bf n})|^2\right>_{R(\tilde{\bf n})} \ , \\
  {\widetilde \Delta}_{\piDoubl}(\tilde k(\tilde {\bf n})) &= \frac{\tilde k^3(\tilde {\bf n})}{2\pi^2} \left(\frac{\delta \tilde{x}}{N}\right)^3\left<\sum_{a}|\piDoublpar^{a}(\tilde {\bf n})|^2\right>_{R(\tilde{\bf n})} \ , \\
  {\widetilde \Delta}_{\widetilde{\mathcal{E}}}(\tilde k(\tilde {\bf n})) &= \frac{\tilde k^3(\tilde {\bf n})}{2\pi^2} \left(\frac{\delta \tilde{x}}{N}\right)^3\left<\sum_{i,a}  |\widetilde{\mathcal{E}}_i^a(\tilde {\bf n})|^2\right>_{R(\tilde{\bf n})} \ , \\
  {\widetilde \Delta}_{\widetilde{\mathcal{B}}}(\tilde k(\tilde {\bf n}))
  &= \frac{\tilde k^3(\tilde {\bf n})}{2\pi^2} \left(\frac{\delta \tilde{x}}{N}\right)^3\left<\sum_{i,a}  |\widetilde{\mathcal{B}}^a_i(\tilde {\bf n})|^2\right>_{R(\tilde{\bf n})} \ ,
\end{align}
where $\langle ... \rangle_{R(\tilde{\bf n})}$ represents angular averaging in $\bf k$-space, c.f.~Eq.~(\ref{eq:discretePS}), and the electric and magnetic fields are defined as in equations \eqref{eq:SU2means}. We note that the spectra are summed over all field-space components. As usual, these dimensionless expressions are related to the dimensionful power spectra by $\Delta_{\Phi} \equiv {\widetilde \Delta}_{\tilde\Phi}f_*^2$, $\Delta_{\pi_\Phi} \equiv {\widetilde \Delta}_{\tilde\pi_\Phi}f_*^2\omega_*^2$, $\Delta_{\mathcal{E}} \equiv {\widetilde \Delta}_{\widetilde{\mathcal{E}}}\,\omega_*^4$, and $\Delta_{\mathcal{B}} \equiv {\widetilde \Delta}_{\widetilde{\mathcal{B}}}\,\omega_*^4$. 

\section{Initial conditions}
\label{sec:InitCond}

In this section we describe how to set the  initial conditions of the different fields, both in the continuum and in the lattice. The initial condition of any field consists of a homogeneous mode, over which a given spectrum of fluctuations is added on top. In particular, let us denote the initial time of our simulations as $\eta_*$, so that we will add the subindex ``*" to any quantity evaluated at such time.  The initial conditions of e.g.~a scalar singlet can be written as
\bea
\phi ({\bf x}, \eta_* ) & \equiv & \bar{\phi}_* + \delta \phi_* ({\bf x}) \ , \label{eq:Scalar-init1} \\
\dot{\phi} ({\bf x}, \eta_* ) &\equiv & \bar{\dot{\phi}}_* + \delta \dot{\phi}_* ({\bf x}) \label{eq:Scalar-init2}\ ,
\eea
where the bar denotes here the homogeneous component of a given variable. The numerical values of $\bar{\phi}_*$ and $\bar{\dot{\phi}}_*$  depend on the details of the specific field model we want to simulate. For example, in Section \ref{sec:WorkingExample}, we take the scalar field $\phi$ as the inflaton field responsible for an accelerated expansion in the early universe, so in this context, $\bar{\phi}_*$ and $\bar{\dot{\phi}}_*$ can be conveniently chosen as the inflaton amplitude and derivative towards the end of inflation, i.e.~just before the inflaton starts oscillating around the minimum of its potential.

In this section we focus on how to set the initial fluctuations of the different fields. We first explain in Section \ref{sec:StochasticSpec} how to set a spectrum of scalar fluctuations in the lattice, so that they recover the expected distribution of fluctuations in the continuum limit. After that, we explain in Section \ref{sec:InitGauge} how to set the initial fluctuations of charged scalar fields and the corresponding (Abelian and non-Abelian) gauge fields they interact with, putting a special emphasis in preserving the Gauss constraint(s) up to machine precision.

\subsection{Stochastic spectrum of scalar fluctuations} \label{sec:StochasticSpec}

Let us consider the scalar field given in Eqs.~(\ref{eq:Scalar-init1})-(\ref{eq:Scalar-init2}). Assuming the homogeneous modes $\bar{\phi}_*$ and $\bar{\dot{\phi}}_*$ are already fixed (based on the details of the model to be simulated), we want to create an appropriate set of classical fluctuations $\delta \phi_* ({\bf x})$ and $\delta \dot{\phi}_* ({\bf x})$ at time $\eta=\eta_*$, in order to mimic some known spectrum of fluctuations. In the continuum, we write
\begin{eqnarray}\label{eq:SpectrumContinuum}
\left\langle \delta \phi^2 \right\rangle = \int d\log k~\Delta_{\delta \phi}(k)\,,~~~~~\Delta_{\delta \phi}(k) \equiv {k^3\over 2\pi^2}\mathcal{P}_{\delta \phi}(k)\,,~~~~~\left\langle {\delta \phi}_{\bf k}{\delta \phi}_{{\bf k}'} \right\rangle \equiv (2\pi)^3 \mathcal{P}_{\delta \phi} (k)\delta(\bf{k}-\bf{k}') \ ,
\end{eqnarray}
where $\langle \cdots \rangle$ represents here an ensemble average, and $\Delta_{\delta \phi}(k)$ is the power spectrum. Although these quantities must obviously be evaluated at the time $\eta=\eta_*$, here we have dropped the $``*"$ to simplify notation, as we will do in the remainder of this section. For initial conditions given by quantum vacuum fluctuations (in the continuum), we have
\begin{eqnarray}
    \mathcal{P}_{\delta \phi} (k) \equiv 
    \frac{1}{2 a^2\omega_{k,\phi}}\,,~~~~ \omega_{k,\phi} \equiv \sqrt{k^2 + a^2m_{\phi}^2} \,,~~~~ m_{\phi}^2 \equiv \frac{\partial^2 V}{\partial \phi^2}\Big|_{\phi = \bar{\phi}} \ , \label{eq:QuantumFlucts}
\end{eqnarray}
where $\omega_{k,\phi}$ is the comoving frequency of the mode, and $m_{\phi}$ is the effective mass of the field, evaluated in terms of the homogeneous field components. If we set $a_* = a(\eta_*) = 1$, as it is customary, we can simply omit the scale factors in the above expression.

In the lattice, we want to set the fluctuations of the scalar field so that expression (\ref{eq:SpectrumContinuum}) is recovered in the continuum limit. In the discrete we substitute the stochastic expectation value by a volume average as
\begin{eqnarray}
\left\langle {\delta \phi}^2 \right\rangle_V = {\delta x^3\over V}\sum_{\bf n} {\delta \phi}^2({\bf n}) = {1\over N^6}\sum_{\tilde{\bf n}} \left|{\delta \phi} (\tilde{\bf n})\right|^2 \ ,
\end{eqnarray}
where we have used Eq.~(\ref{eq:FTdiscreteDelta}). Decomposing the summation into radial and angular parts, we obtain
\begin{eqnarray}
\left\langle {\delta \phi}^2 \right\rangle_V = {1\over N^6}\sum_{|\tilde{\bf n}|} \sum_{\tilde{\bf n}' \in R(\tilde{\bf n})} \left|\delta \phi (\tilde{\bf n})\right|^2 = {4\pi\over N^6}\sum_{|\tilde{\bf n}|} |\tilde{\bf n}|^2\left\langle\left|\delta \phi (\tilde{\bf n})\right|^2\right\rangle_{R(\tilde{\bf n})} \ ,
\end{eqnarray}
where $\left\langle (\cdots) \right\rangle_{R(\tilde{\bf n})} \equiv {1\over 4\pi|\tilde{\bf n}|^2}\sum_{\tilde{\bf n}' \in R(\tilde{\bf n})} (\cdots)$ is an angular average over the spherical shell of radius $\tilde{\bf n}' \in \big[|\tilde{\bf n}|, |\tilde{\bf n}| + \Delta \tilde{\bf n}\big)$, with $\Delta \tilde{\bf n}$ a given radial binning. This leads to
\begin{eqnarray}
\left\langle \delta \phi^2 \right\rangle_V = {4\pi\over k_{\rm IR}^3N^6}\sum_{|\tilde{\bf n}|}\Delta\log k(\tilde{\bf n})~ k^3(\tilde{\bf n}) \left\langle\left| \delta \phi (\tilde{\bf n})\right|^2\right\rangle_{R(\tilde{\bf n})} = {1\over 2\pi^2}\sum_{|\tilde{\bf n}|}\Delta\log k(\tilde{\bf n})~ k^3(\tilde{\bf n}) {L^3\over N^6}\left\langle\left| \delta \phi (\tilde{\bf n})\right|^2\right\rangle_{R(\tilde{\bf n})}\,,
\end{eqnarray}
where $\Delta\log k({\tilde{\bf n}}) \equiv {k_{\rm IR}\over k(\tilde{\bf n})}$, ${\bf k}(\tilde{\bf n}) \equiv k_{\rm IR}\tilde{\bf n}$ and $k_{\rm IR} \equiv {2\pi\over L}$. In order to mimic in the lattice the continuum stochastic initial condition, we impose
\begin{eqnarray}
\left\langle \delta \phi^2 \right\rangle_V = 
\sum_{|\tilde{\bf n}|}\Delta\log k(\tilde{\bf n})~ {k^3(\tilde{\bf n})\over2\pi^2}\mathcal{P}_{\delta \phi} (k(\tilde{\bf n}))\,,
\end{eqnarray}
from where we identify
\begin{eqnarray}\label{eq:ICscalar}
\left\langle\left| \delta \phi (\tilde{\bf n} )\right|^2\right\rangle_{R(\tilde{\bf n})} = \left({N\over \delta x}\right)^3\mathcal{P}_{\delta \phi} (k(\tilde{\bf n}))\,.
\end{eqnarray}
With this choice, we reproduce the continuum correctly,
\begin{eqnarray}\label{eq:ICscalarVEV}
\left\langle \delta \phi^2 \right\rangle_V = \sum_{|\tilde{\bf n}|}\Delta\log k(\tilde{\bf n})~\Delta_{\delta \phi} (k(\tilde{\bf n})) \hspace*{0.6cm}\longrightarrow\hspace*{0.6cm} \int d\log k~\Delta_{\delta \phi} (k)\,.
\end{eqnarray}
The key point of the identification made in Eq.~(\ref{eq:ICscalar}), is that $\left\langle \delta \phi ^2 \right\rangle_V$ in Eq.~(\ref{eq:ICscalarVEV}) should not depend explicitly on the volume $V = (N\cdot \delta x)^3$, if we were to reproduce correctly the continuum limit.

The initial variance of the Fourier modes in the lattice, expressed in the program variables of Eq.~(\ref{eq:FieldSpaceTimeNaturalVariables}), must be taken therefore as
\begin{eqnarray}\label{eq:ICscalarII}
\left|\delta \tilde \phi (\tilde{\bf n})\right|^2 \equiv 
\left({\omega_*\over f_*}\right)^2\left({N\over \delta  \tilde  x}\right)^3\widetilde{\mathcal{P}}_{\delta \tilde\phi} (\tilde{k} (\tilde{\bf n}))\,,
\end{eqnarray}
where $\widetilde{\mathcal{P}}_{\delta \phi} \equiv \omega_*\mathcal{P}_{\delta \phi}$. For quantum fluctuations with variance (\ref{eq:QuantumFlucts}), we shall write
\begin{eqnarray} \label{eq:QuantumFlucts2}
    \left| \delta \tilde \phi (\tilde{\bf n})\right|^2 \equiv \left({\omega_*\over f_*}\right)^2\left({N\over \delta \tilde{x}}\right)^3{1\over 2a^2\sqrt{\tilde{k}^2(\tilde{\bf n}) + a^2\tilde{m}_{\phi}^2}} \ , \hspace{0.6cm} \tilde{m}_{\phi}^2 \equiv \frac{\partial^2 \tilde{V}}{\partial \tilde{\phi}^2 } (\tilde{\phi} = \tilde{\bar{\phi}} )  \ ,
\end{eqnarray}
where $\tilde{k} \equiv k / \omega_*$ and $\tilde{m}_{\phi} \equiv m_{\phi} / \omega_*$ are the momentum and effective mass in program variables. Let us note that the ratio $\omega_*/f_*$ enters in Eq.~(\ref{eq:QuantumFlucts2}).

The above expression Eq.~(\ref{eq:QuantumFlucts2}) gives an account of the appropriate amplitude for the modulus of the fluctuations in the lattice. We still need to consider the fluctuations of the phase of the mode, as well as of the time derivative of the mode itself. We will deal with these two problems together, as follows. We first note that field modes deep inside the horizon (i.e.~with $k/a \gg H$, $H = \dot a / a$ the Hubble rate), have a time-dependence as $\delta \phi_{{\bf k}} \sim (1/a) e^{\pm i (\omega_k/a) t}$, where $t$ is the cosmic time, and we are implicitly assuming that the initial conditions are set in an adiabatic regime $\dot{\omega}_k \ll \omega_k^2$. Taking the time-derivative of the field mode, we get $\delta \dot{\phi}_{\bf k} \approx {1\over a}(\pm i \omega_k - aH) \delta \phi_{\bf k}$. Choosing one sign in this expression is equivalent to choosing a preferred direction in position space (say a right- or left-moving wave), so even though this effect should be irrelevant once the non-linearities of the dynamics kick in, we follow the prescription of {\tt LatticeEasy} to define
'isotropic' initial conditions \cite{Felder:2000hq}, by the superposition of left- and right-moving waves. In particular, at each lattice point in momentum space, we add to the field amplitude a sum of left-moving and right-moving waves, as follows,
\bea
  \delta \tilde \phi ({  \bf \tilde{n}}) &=& \frac{1}{\sqrt{2}} (|\delta \tilde \phi^{(l)} ({  \bf \tilde{n}})|  e^{i \theta^{(l)} ({   \bf \tilde{n}}) } + |\delta \tilde \phi^{(r)} ({   \bf \tilde{n}})| e^{i \theta^{(r)} ({   \bf \tilde{n}}) }   ) \label{eq:fpr_influct} \ , \\
 \delta \tilde {\phi}' ({   \bf \tilde{n}}) &=& {1\over a^{1-\alpha}}\left[\frac{i \tilde{\omega}_k}{\sqrt{2}}  \left(|\delta \tilde \phi^{(l)} ({   \bf \tilde{n}})| e^{i \theta^{(l)} ({   \bf \tilde{n}}) } - |\delta \tilde \phi^{(r)}  ({   \bf \tilde{n}})| e^{i \theta^{(r)} ({   \bf \tilde{n}}) }   \right)\right]  - \tilde{\mathcal{H}}  \delta \tilde \phi ({   \bf \tilde{n}})\ . \label{eq:fpr_influct2} \eea
Here we have used $d\tilde\eta = a^{-\alpha}\omega_*dt$, and introduced the frequency of the mode and the Hubble rate in program units,  $\tilde{\omega}_k \equiv \sqrt{\tilde{k}^2(\tilde{\bf n}) + a^2\tilde{m}_{\phi}^2}$, and $\tilde{\mathcal{H}} \equiv a^\alpha H / \omega_*$. In these expressions, $\theta^{(l)} ({\bf \tilde{n}})$ and $\theta^{(r)} ({\bf \tilde{n}})$ are two random independent phases which vary uniformly in the range $[0, 2\pi)$ from point to point in the reciprocal Fourier lattice. It is precisely in this way that the phase of each mode is taken care of. On the other hand, $|\delta \tilde{\phi}^{(l)} ({\bf \tilde{n}})|$ and $|\delta \tilde{\phi}^{(r)} ({\bf \tilde{n}})|$ are two amplitudes that also vary from point to point, according to a {\it Rayleigh} distribution with expected square amplitude given by (\ref{eq:QuantumFlucts2}). Altogether, the uniform randomly generated phases and the moduli created according to a Rayleigh distribution\footnote{We remind the reader that if a complex variable $f_k = R_k +i I_k$ is said to be Gaussian (i.e.~both their real and imaginary parts, $R_k$ and $I_k$, follow each a Gaussian random distribution), mathematically this is equivalent to its modulus $|f_k| \equiv \sqrt{R_k^2+I_k^2}$ following a {\it Rayleigh} distribution, and its phase $\varphi = arctan(I_k/R_k)$ following a uniform distribution in the range $[0,2\pi)$.}, make $\delta \tilde \phi ({  \bf \tilde{n}})$ and $\delta \tilde \phi'({  \bf \tilde{n}})$ Gaussian random variables. Note that we draw $|\delta \tilde{\phi}^{(l)} ({\bf \tilde{n}})|$ and $|\delta \tilde{\phi}^{(r)} ({\bf \tilde{n}})|$ independently from each other, avoiding the additional constraint $|\delta \tilde{\phi}^{(l)} ({\bf \tilde{n}})| = |\delta \tilde{\phi}^{(r)} ({\bf \tilde{n}})|$ imposed by {\tt LatticeEasy} at each reciprocal lattice site: otherwise $\delta \tilde \phi ({  \bf \tilde{n}})$ will not correspond to a Gaussian random variable, as pointed out in~\cite{Frolov:2008hy}, since the left- and right-moving waves would be correlated through their modulus, and hence the resulting fluctuation $\delta \tilde \phi ({  \bf \tilde{n}})$ will not be Gaussian, as it would not be the sum of two independent Gaussian realizations. We will typically consider $a_* = a(\eta_*) = 1$ initially, so we can simply omit the scale factors in Eq.~(\ref{eq:fpr_influct2}).

\subsection{Charged scalars and gauge fields} \label{sec:InitGauge}

Let us now consider the initial conditions for the gauge fields, as well as of the charged fields coupled to them. In this work we are considering scalar fields charged under $U(1)$ and $SU(N)\times U(1)$ gauge groups, which we denote as $\varphi$ and $\Phi$ respectively. We recall that these fields are composed of multiple real components: $2$ in the case of $\varphi$, and $2N$ in the case of $\Phi$. As the potential only depends on the absolute value of these fields, we can set the same initial amplitude to the homogeneous modes of all their components. In particular, we set for each real component $\varphi_n$ of a doublet $\Phi$ ($n=0,1,\dots 2N-1$),
\bea
 \varphi_n ({\bf x}, t_* ) & \equiv& \frac{|\Phi_*|}{\sqrt{N}}  +  \delta \varphi_{n*} ({ \bf x}) \ , \label{eq:CPhi_init1}\\
  \dot{\varphi}_n ({\bf x}, t_* ) &\equiv & \frac{|\dot{\Phi}_*|}{\sqrt{N}}  +  \delta \dot{\varphi}_{n*} ({ \bf x}) \label{eq:CPhi_init2} \ ,
\eea
where $|\Phi_*|$ and $|\dot{\Phi}_*|$ are the initial homogeneous values of the multi-field norm and its time derivative (which depend on each specific model), and $\delta \varphi_{n*} ({ \bf x})$ and $\delta \dot{\varphi}_{n*} ({ \bf x}) $ are the initial mode fluctuations of each field component and its derivative. 
For simplicity, we will drop again the ``$*$" notation from now on. Following the same procedure as with the scalar singlet fluctuations (\ref{eq:fpr_influct})-(\ref{eq:fpr_influct2}), we consider the following fluctuations for charged scalar fields in a lattice,
\begin{eqnarray}
\delta \tilde{\varphi}_n({  \bf \tilde n}) &=& \frac{1}{\sqrt{2}} \left(|\delta \tilde{\varphi}_{n}^{(l)} ({  \bf \tilde n})|  e^{i \theta_{n}^{(l)} ({  \bf \tilde n}) } + |\delta \tilde{\varphi}_{n}^{(r)} ({  \bf \tilde n})| e^{i \theta_{n}^{(r)} ({  \bf \tilde n}) }   \right) \label{eq:fpr_influct3} \ , \\
\delta \tilde{\varphi}'_n ({  \bf \tilde n}) &=& a^{1-\alpha}\left[\frac{1}{\sqrt{2}} i \tilde{\omega}_{k,n} \left(|\delta \tilde{\varphi}_{n}^{(l)} ({  \bf \tilde n})| e^{i \theta_{n}^{(l)} ({  \bf \tilde n}) } - |\delta \tilde{\varphi}_{n}^{(r)}  ({  \bf \tilde n})| e^{i \theta_{n}^{(r)} ({  \bf \tilde n}) }  \right)\right]  - \tilde{\mathcal{H}} \delta \tilde{\varphi}_{n} ({  \bf \tilde n})\ ,  \label{eq:fpr_influct4}
\end{eqnarray}
with $\tilde{\omega}_{k,n}  \equiv \omega_{k,n} /\omega_* =  \sqrt{\tilde{k}^2 + a^2 (\partial^2 \tilde{V} / \partial \tilde{\varphi}_n^2)}$ the initial effective frequency in program units of each field component mode. Thus, for a charged field with $2N$ real components, there are $8N$  functions to be fixed ($|\delta \tilde{\varphi}_{n}^{(l)} ({  \bf \tilde n})|$, $|\delta \tilde{\varphi}_{n}^{(r)} ({  \bf \tilde n})|$, $\theta_{n}^{(l)}$, and $\theta_{n}^{(r)}$, with $n=0,1,\dots 2N-1$). Again, if we fixed the initial value of the scale factor to unity, we could omit it from the expression of the derivative fluctuations. In principle, all the functions $\lbrace|\delta \tilde{\varphi}_{n}^{(l)} ({  \bf \tilde n})|$, $|\delta \tilde{\varphi}_{n}^{(r)} ({  \bf \tilde n})|$, $\theta_{n}^{(l)}$, $\theta_{n}^{(r)}\rbrace$ should be randomly generated, independently at each lattice point, again by Rayleigh sampling each modulus $|\delta \tilde{\varphi}_{n}^{(i)} ({  \bf \tilde n})|$ with expected square amplitude given by Eq.~(\ref{eq:QuantumFlucts2}), while uniformly sampling each phase $\theta_{n}^{(r)}$ within the range $[0,2\pi)$. However, as we shall see, we will need to impose certain relations among these functions if we want to preserve the Gauss constraint(s) initially (we speak about the Gauss constraints, in plural, as we consider here the general case of a charged scalar under various gauge groups, e.g.~$SU(2)\times U(1)$).

Let us now consider the fluctuations of the Abelian and non-Abelian gauge fields. We will consider first the fluctuations in the continuum, and generalize to the discretized case afterwards. For the gauge fields we shall impose
\bea
A_i ({\bf x}, t_* ) & \equiv & 0 \ , \label{eq:Inflc1}\\
B_i^a ({\bf x}, t_* ) & \equiv & 0 \ ,  \label{eq:Inflc2} \\
\dot{A}_i ({\bf x}, t_* ) & \equiv & \delta \dot{A}_{i*} ({\bf x}) \ ,  \label{eq:Inflc3} \\
\dot{B}_i^a ({\bf x}, t_* ) & \equiv & \delta \dot{B}_{i*}^a ({\bf x}) \ ,  \label{eq:Inflc4} \eea
i.e.~we impose the amplitude of the gauge fields to be \textit{exactly} zero at all lattice points, while we set an initial spectrum of fluctuations to their time-derivatives (over vanishing homogeneous values).
Because of this, initially the magnetic energy is exactly zero, while there will be some amount of electric energy, due to the fluctuations of the time-derivatives. The fluctuations of the charged scalars and gauge fields must be imposed in such a way that the Gauss constraint is initially preserved. If this is achieved, then the dynamical evolution of the field EOM will guarantee the preservation of the Gauss constraint at later times. The Gauss constraints for a $SU(2)\times U(1)$ gauge-invariant theory are given by [see.~Eqs.~(\ref{eq:GaussU1-eom})-(\ref{eq:GaussSU2-eom}) and (\ref{eq:AbelianCurrent})-(\ref{eq:NonAbelianCurrent})],
\bea \partial_i F_{0i} ({\bf x}) &=&  J^A_0  ({\bf x}) \ , \hspace{0.5cm} J^A_0 ({\bf x})  \equiv  2 g_AQ_A^{(\varphi)} \mathcal{I}m [ \varphi^{*} \varphi ' ] + 2 g_AQ_A^{(\Phi)} \mathcal{I}m [ \Phi^\dag  \Phi' ]\,, \label{eq:Gauss111} \\
(\mathcal{D}_i )_{a b} (G_{0i})^b ({\bf x}) &=&  \, J^a_0 ({\bf x})  \ , \hspace{0.5cm} J_0^a ({\bf x})  \equiv  2g_BQ_B\mathcal{I}m [ \Phi^{\dag} T_a \Phi' ]\,. \label{eq:Gauss222}
\eea
where we have set the initial scale factor to $a=1$ for simplicity. By Fourier transforming both sides of these equations, we get
\be {k}^i {A}'_i ({\bf k}) = J_0^A ({\bf k}) \ , \hspace{0.4cm} {k}^i {B}_i^{a'} ({\bf k}) = J_0^a ({\bf k}) \ , \label{eq:kAi1} \ee
where $J_0^A ({\bf k})$ and $J_0^a ({\bf k})$ are the Fourier transforms of each current. 
One particular solution to each of these equations, for $\bf k \neq \bf 0$, can be written like
\be {A}'_i ({\bf k}) = i \frac{{k}_i}{{k}^2} J_0^A ({\bf k}) \ , \hspace{0.4cm} {B}^{a'}_i ({\bf k}) = i \frac{{k}_i}{{k}^2} J_0^a ({\bf k}) \label{eq:GaugeCurrentFluc} \ .\ee
The  complex scalar field fluctuations $\delta \varphi_* ({ \bf x})$ and $\delta \dot{\varphi}_* ({ \bf x})$ are given by Eqs.~(\ref{eq:fpr_influct3})-(\ref{eq:fpr_influct4}), and generate fluctuations on the currents $J_0^A ({\bf x})$, and $J_0^a ({\bf x})$. Therefore, we can impose fluctuations to the gauge fields in momentum space via Eqs.~(\ref{eq:GaugeCurrentFluc}), and then transform back to position space to obtain $\delta \dot{A}_{i*} ({\bf x})$, $\delta \dot{B}_{i*}^a ({\bf x}) $.

The above procedure should, in principle, initially preserve the Gauss constraint(s). However, we must guarantee that the imposition of Eq.~(\ref{eq:GaugeCurrentFluc}) does not add a global charge in the system: i.e.~we must impose the vanishing of the zero mode of the zero component (charge) of the gauge currents, $J_0^A ({\bf k} ={\bf 0} ) = J_0^a ({\bf k} ={\bf 0} ) = 0$ (note that this is explicitly implied by Eq.~(\ref{eq:kAi1}) for $\bf k  = 0$). For concreteness, let us consider the case of a doublet $\Phi$ charged under a $U(1)\times SU(2)$ gauge group (a $U(1)$-charged field $\varphi$ would be just a particular case, as we explain below). The homogeneous modes of the Abelian and non-Abelian currents (\ref{eq:Gauss111}) and (\ref{eq:Gauss222}) can be written in terms of the complex field fluctuations as
\bea
J_0^A ({\bf k} ={\bf 0} ) &=& \int d^3 {\bf x} J_0^A ({\bf x}) \propto \int d^3 {\bf k} \mathcal{R}e [\varphi_0^* ({\bf k}) \varphi'_{1} ({\bf k}) - \varphi'_0 ({\bf k}) \varphi_1^* ({\bf k}) + \varphi_2^* ({\bf k}) \varphi'_3 ({\bf k}) - \varphi'_2 ({\bf k}) \varphi_3^* ({\bf k}) ] = 0 \ , \nonumber \\
J_0^1 ({\bf k} ={\bf 0}) &=& \int d^3 {\bf x} J_0^1 ({\bf x}) \propto \int d^3 {\bf k}  \mathcal{R}e [\varphi_3^* ({\bf k}) \varphi'_{0} ({\bf k}) - \varphi'_3 ({\bf k}) \varphi_0^* ({\bf k}) + \varphi_1^* ({\bf k}) \varphi'_2 ({\bf k}) -  \varphi'_1 ({\bf k}) \varphi_2^* ({\bf k}) ]  = 0 \ , \nonumber \\
J_0^2 ({\bf k}={\bf 0}) &=& \int d^3 {\bf x} J_0^2 ({\bf x}) \propto \int d^3 {\bf k}  \mathcal{R}e [\varphi_0^* ({\bf k}) \varphi'_{2} ({\bf k}) - \varphi'_0 ({\bf k}) \varphi_2^* ({\bf k}) + \varphi_1^* ({\bf k}) \varphi'_3 ({\bf k}) -  \varphi'_1 ({\bf k}) \varphi_3^* ({\bf k}) ]  = 0 \ , \nonumber \\
J_0^3 ({\bf k} ={\bf 0} ) &=& \int d^3 {\bf x} J_0^3 ({\bf x}) \propto \int d^3 {\bf k}  \mathcal{R}e [\varphi_1^* ({\bf k}) \varphi'_{0} ({\bf k}) - \varphi'_1 ({\bf k}) \varphi_0^* ({\bf k}) + \varphi_2^* ({\bf k}) \varphi'_3 ({\bf k}) - \varphi'_2 ({\bf k}) \varphi_3^* ({\bf k}) ]  = 0 \ . \nonumber
\eea
A simple way of guaranteeing that these conditions hold is to set all the integrands to zero. By solving the corresponding system of linear equations, we get the following three conditions,
\be  \mathcal{R}e [ \varphi'_m ({\bf k}) \varphi_0^* ({\bf k}) - \varphi'_0 ({\bf k}) \varphi_m^* ({\bf k})] = 0 \ , \hspace{0.4cm} m=1,2,3  \ , \label{eq:zero-mode-cond} \ee
which mix the different real components of the doublet. These conditions are in general \textit{not} verified when all the functions $|\delta {\varphi}_{n}^{(l)} ({  \bf \tilde n})|$, $|\delta {\varphi}_{n}^{(r)} ({  \bf \tilde n})|$, $\theta_{n}^{(l)}$, and $\theta_{n}^{(r)}$ in the scalar fluctuations (\ref{eq:fpr_influct3}) and (\ref{eq:fpr_influct4}) are generated randomly, independently from each other. However, by substituting into Eq.~(\ref{eq:zero-mode-cond}) the expressions of each $\varphi_n$ in terms of moduli and phases of left- and right-moving waves, we can show that the conditions in Eq.~(\ref{eq:zero-mode-cond}) are  verified when the following relations hold,
\bea  |\delta \varphi_{n}^{(l)} ({\bf k})| &=& |\delta \varphi_{n}^{(r)}  ({\bf k})| \ , \hspace{3.5cm} n=0,1,2,3 \ , \label{InConstr:1} \\
\theta_{m}^{(r)} ({\bf k}) &=& \theta_{0}^{(r)} ({\bf k}) + \theta_{m}^{(l)} ({\bf k}) - \theta_{0}^{(l)} ({\bf k}) \ , \hspace{0.85cm} m=1,2,3 \ . \label{InConstr:2} \eea
The first relation imposes the same amplitude to the left- and right-moving waves of the fluctuations of each real scalar component (four in total). The second line consists in a set of three different constraints that must be imposed to the eight phases appearing in the four real components of the doublet. Therefore, in the case of a $SU(2)$-charged doublet, one can simply generate randomly $|\delta \varphi_{0}^{(l)}|$, $|\delta \varphi_{1}^{(l)}|$, $|\delta \varphi_{2}^{(l)}|$, $|\delta \varphi_{3}^{(l)}|$, and e.g.~$\theta_{0}^{(l)}$, $\theta_{0}^{(r)}$, $\theta_{1}^{(l)}$, $\theta_{2}^{(l)}$, and $\theta_{3}^{(l)}$ according to their appropriate probability distributions, and then obtain $|\delta \varphi_{0}^{(r)}|$, $|\delta \varphi_{1}^{(r)}|$, $|\delta \varphi_{2}^{(r)}|$, $|\delta \varphi_{3}^{(r)}|$, $\theta_{1}^{(r)}$, $\theta_{2}^{(r)}$ and $\theta_{3}^{(r)}$ via Eqs.~(\ref{InConstr:1})-(\ref{InConstr:2}). This procedure guarantees that the homogeneous mode of the charge vanishes initially, and hence that the Gauss laws are preserved\footnote{Another possibility to verify the constraints (\ref{eq:zero-mode-cond}) would be to just impose the relations $\varphi'_m ({\bf k}) = \varphi_0' ({\bf k}) \varphi_m^* ({\bf k})  / \varphi_0^*({\bf k})$, $m=1,2,3$, where the functions in the $rhs$ of this expression are to be generated according to the procedure explained below (\ref{eq:fpr_influct3}) and (\ref{eq:fpr_influct4}), without imposing the constraints (\ref{InConstr:1}) and (\ref{InConstr:2}). However, using this procedure, the fluctuations generated for the $0th$-component have very different amplitudes than the other components, typically one or even more orders of magnitude of difference. Furthermore, the spectra of the $0th$-component depends very much on the particular random realization of the fields. This makes us prefer the procedure described in the bulk text.}.

Let us remark that a similar procedure can be applied to the simpler case of a $U(1)$-charged field $\varphi$. In this case, to set up the initial fluctuations we proceed similarly as with the $SU(2)$ case, using (\ref{eq:fpr_influct3}) and (\ref{eq:fpr_influct4}), and following the procedure explained below those, except that there are now only two real scalar components, $\delta \varphi_0 ({\bf k})$ and $\delta \varphi_1 ({\bf k})$. This implies that now, instead of the seven constraints in Eqs.~(\ref{InConstr:1}), (\ref{InConstr:2}), there are only three constraints to be imposed: $\delta \varphi_{0}^{(l)} ({\bf k})= \delta \varphi_{0}^{(r)} ({\bf k})$,  $\delta \varphi_{1}^{(l)} ({\bf k})= \delta \varphi_{1}^{(r)} ({\bf k})$, and $\theta_{1}^{(r)} ({\bf k})= \theta_{0}^{(r)} ({\bf k})+ \theta_{1}^{(l)}({\bf k})  - \theta_{0}^{(l)} ({\bf k})$.

Finally, let us consider the translation  of this procedure developed in the continuum to the lattice. For charged scalar fields, the only difference is that the different functions $|\delta \varphi_{ab}|$ and $\theta_{ab}$ are only defined in each lattice point, instead of being continuum functions. Therefore, the randomly generated amplitudes must be imposed at each lattice site ${\bf \tilde{n}}$, according to the corresponding probability distribution, while the other amplitudes must be imposed at each lattice site via the constraint equations, e.g.~(\ref{InConstr:1})-(\ref{InConstr:2}) for $SU(2)$. On the other hand, for gauge fields we must start from the discrete Gauss law equations. As we are not imposing fluctuations to the amplitude of the gauge fields, the discrete Gauss constraints (\ref{eq:U1vector-Gauss_Discrete}) and (\ref{eq:SUNvector-Gauss_Discrete}) simply become, in position space and in physical variables,
\be \sum_i \Delta_i^- \Delta_0^+ A_i ({\bf n}) =  J^A_0  ({\bf n}) \ , \hspace{0.5cm} \sum_i \Delta_i^- \Delta_0^+ B_i^a ({\bf n})  =  J^a_0 ({\bf n}) \ , \hspace{0.4cm} (a=1,2,3) \ .
\label{eq:GaussDiscrete1} \ee
By taking a discrete Fourier transform in both sides of the equation, we can write lattice equations, analogous to (\ref{eq:GaugeCurrentFluc}) in the continuum, as
\be \Delta_0^+ A_i ({\bf \tilde{n}}) = i \frac{k_{{\rm Lat},i}^-}{(k_{{\rm Lat},i}^-)^2} J^A_0  ({\bf \tilde{n}}) \ , \hspace{0.5cm} \Delta_0^+ B_i^a ({\bf \tilde{n}})  = i \frac{k_{{\rm Lat},i}^-}{(k_{{\rm Lat},i}^-)^2} J^a_0 ({\bf \tilde{n}}) \ , \hspace{0.4cm} (a=1,2,3) \ . \label{eq:GaussDiscrete2}
\ee
Note that, as we are taking the backward spatial derivative $\Delta_i^-$ in Eq.~(\ref{eq:GaussDiscrete1}), it is the corresponding complex lattice momentum $k_{{\rm Lat},i}^-$ that must appear in Eq.~(\ref{eq:GaussDiscrete2}) after Fourier transforming, c.f.~Eq.~(\ref{eq:ForwardBackwardMomentum}). Therefore, in order to set the fluctuations of the gauge field derivatives in the lattice, we first add the fluctuations to the real components of the charged scalar fields, and compute their corresponding currents. After that, we transform these to momentum space, and impose expressions (\ref{eq:GaussDiscrete2}) to the gauge field initial derivatives. Finally, we transform the gauge field derivatives back to position space.

\section{A working example: the $SU(2)\times U(1)$ gauge invariant inflaton} \label{sec:WorkingExample}

~~~ In order to illustrate some of the techniques and concepts explained above, we dedicate this section to study the dynamics of a scalar-gauge field theory model using lattice simulations. In particular, we are going to consider an observationally viable single-field inflationary model, with monomial potential $V(\phi) \propto |\phi|^p$ around the minimum, and flattening `wings' at large field amplitudes. We will study the post-inflationary stage of preheating, which is triggered by the inflaton oscillations around the minimum of its potential. In order to apply the gauge-invariant lattice techniques presented before, we will couple the inflaton to both scalar and gauge fields, all of which will be denoted collectively as the \textit{daughter fields}.  We will study the transfer of energy from the inflaton into such daughter fields.

The structure of this section is as follows. First, in Section \ref{eq:ModelDetails} we present the details of how inflation and preheating proceed in the model under consideration. In particular, we review the basis of the two excitation phenomena that govern the post-inflationary dynamics: parametric resonance of the daughter field(s), and self-resonance of the inflaton. After that, we present the results of our lattice simulations, fully capturing the non-linear regime of the dynamics. In Section \ref{sec:U1-sims} we consider the case of a $U(1)$ gauge invariant inflaton, coupled to an Abelian gauge field through a covariant derivative. In Section  \ref{sec:SU2U1-sims} we consider the case of a $SU(2)\times U(1)$ gauge invariant inflaton, coupled simultaneously to a $SU(2)\times U(1)$ gauge sector (formed by Abelian and non-Abelian gauge fields) and a light scalar singlet.

\subsection{Model details} \label{eq:ModelDetails}

Let us begin by considering a singlet scalar field $\phi$ with the following potential energy,
\be  V(\phi) = \frac{\Lambda^4}{p} \tanh^p \left( \frac{|\phi|}{M} \right) \ , \label{eq:InflPot} \ee
where $\Lambda$ and $M$ have dimensions of energy, and $p$ is a positive number obeying $p>1$.  The particular form of this potential is based on $\alpha$-attractor models of inflation, see Ref.~\cite{Kallosh:2013hoa}. The potential is a function of the absolute value of the field (which in the case of a gauge charged field, naturally with more than one real component, it will be its modulus).
The potential has always a minimum at $\phi=0$, independently of the choice of $p$.  Similarly, the potential develops a plateau $V(\phi) \rightarrow \Lambda^4 / p$ at large field amplitudes $|\phi| \gg M$. We consider a scenario where $\phi$ is the \textit{inflaton} field responsible for the inflationary period. Without loss of generality, we will study scenarios with $\phi > 0$ during inflation. We will then study in detail the stage of preheating following inflation, during which the sign of $\phi$ will alternate within each semi-oscillation. The values of the model parameters ($\Lambda$, $M$, $p$) are related by the observed amplitude of the scalar perturbations in the CMB, i.e.~$\Lambda = \Lambda (M, N_{\rm _{CMB}},p)$, with $N_{\rm _{CMB}}=50-60$ the number of e-folds between the end of inflation and horizon crossing of the relevant perturbations.

The potential (\ref{eq:InflPot}) can be expanded around the minimum as the following monomial function,
\be V(\phi) = \frac{1}{p} \lambda \mu^{4-p} |\phi|^p \ , \hspace{0.5cm} \lambda\mu^{4-p} \equiv \Lambda^4 M^{-p} \ , \label{eq:PowerLaw-pot} \ee
where $\lambda$ is dimensionless and $\mu$ has dimensions of energy. The product of parameters  $\lambda \mu^{4-p}$ in Eq.~(\ref{eq:PowerLaw-pot}) is fixed in terms of  ($\Lambda$, $M$, $p$) to match the exact potential (\ref{eq:InflPot}) in the limit $|\phi| \ll M$. The field value that separates the monomial and plateau regimes in the exact potential can be estimated by computing its inflection point, i.e.~the amplitude at which $V_{_,{\phi \phi}} (\phi_{\rm i})= 0$. It is given by
\be \phi_{\rm i} = M {\rm arcsinh} \left( \sqrt{\frac{p-1}{2}} \right) \ .\ee
The monomial potential (\ref{eq:PowerLaw-pot}) is a very good approximation to the exact potential (\ref{eq:InflPot}) for field amplitudes $|\phi| \ll M$. In particular, in the limit $M \rightarrow \infty$, the inflaton potential (\ref{eq:InflPot}) recovers the monomial function (\ref{eq:PowerLaw-pot}) exactly, recovering this way the well-known chaotic inflation scenario.

Inflation takes place during the slow-rolling of the inflaton at large field amplitudes, heading towards the minimum of the potential. The inflaton acquires a sizable acceleration when the slow-roll conditions break down, and as a consequence, it starts oscillating around the minimum of its potential. The field amplitude $\phi_*$ when the slow-roll parameter becomes unity $\epsilon_{V}\equiv m_{\rm pl}^2 V_{,\phi}^2/(2V^2) = 1$, is
\be \phi_{*} \equiv \frac{M}{2} {\rm arcsinh} \left( \frac{\sqrt{2} p m_{\rm pl}}{M} \right) \,\, \xrightarrow[M\to\infty]{} \frac{p m_{\rm pl}}{\sqrt{2}} \ , \label{eq:sl-roll}\ee
where we have also written the corresponding value in the limit $M \rightarrow \infty$. In this model, inflation happens for field amplitudes $|\phi| \gtrsim \phi_*$, while the  oscillatory regime which follows takes place for $|\phi| \lesssim \phi_*$. Therefore, the field amplitude $|\phi| = \phi_*$ constitutes a natural initial condition for our lattice simulations. If $M \gtrsim m_{p}$, we have that $\phi_* \ll \phi_{\rm i}$, so the inflaton is already in the positive-curvature region of the potential when the slow-roll regime breaks down, and does not enter into the tachyonic region (with $V'' < 0$) during the subsequent inflaton oscillations. In that case, we can safely take the monomial potential (\ref{eq:PowerLaw-pot}) as a very good approximation to the exact potential (\ref{eq:InflPot}) during preheating. On the other hand, if $M \lesssim m_{p}$ we have that $\phi_* \gg \phi_{\rm i}$, so the inflaton enters into the tachyonic region at least during some of the first oscillations. Here we consider only the first scenario, so our results do not depend very much on the details of the transition between the monomial function and the plateau. In particular, we will fix the value $M = 10 m_p$ in the lattice simulations.

The equation of motion of the homogeneous part of the inflaton modulus in the limit $M \rightarrow \infty$ is
\be \ddot{\phi} + 3 H \dot{\phi} + \Omega^2 (|\phi|) \phi = 0 \ , \hspace{0.5cm} \Omega (|\phi|) \equiv \sqrt{\lambda} \mu^{\frac{4-p}{2}} |\phi |^{\frac{p-2}{2}} \ , \label{eq:scalar-fld}\ee
which corresponds to an oscillator with time-dependent effective frequency $\Omega (|\phi|)$ and friction term $\propto 3 H \dot{\phi}$, induced by the expansion of the universe. Note that for multi-component fields there will be an equation of this type for each component\footnote{In that case, the initial velocity of the field is typically considered only within a radial trajectory in field space, i.e.~if we decompose $\phi$ into modulus and phases, the velocity of the phases are taken to be vanishing initially, so that $|\phi|$ oscillates radially, maintaining constant phases in its own field space.}. The oscillation frequency of the field is constant for $p=2$, but depends explicitly on the field amplitude (and is time dependent) for any other value $p \neq 2$. This equation can be solved together with the Friedmann equation (\ref{eq:FriedmannHubble}) in the homogeneous approximation, with initial conditions deep inside slow-roll. During inflation we have $|\phi| \gg |\phi_*|$, or equivalently, $H (|\phi|) \gg \Omega (|\phi|)$. Eventually, when the field amplitude becomes approximately $\phi \simeq \phi_*$, the condition $H (|\phi|) = \Omega (|\phi|)$ holds, and the inflaton starts oscillating. The solution for the homogeneous inflaton can be approximated during the oscillatory regime as \cite{Turner:1983he}
\bea
\phi(t)  & \simeq &  \Phi (t) F(t) \ , \hspace{0.5cm} \Phi (t) = \Phi_{*} \left( \frac{t}{t_*} \right)^{-2/p}   \ , \label{eq:PowLaw-sol0}\\
a(t) & \propto  & a_* \left( 1 + \frac{3p}{2+p} H_* t \right)^{\frac{2+p}{3p}}  \,\, \sim \,\, t^{\frac{2+p}{3p}} \ , \label{eq:PowLaw-sol}
\eea
where we have also provided the oscillation averaged scale factor. In Eq.~(\ref{eq:PowLaw-sol0}), $F (t)$ is an oscillatory function that is periodic for $p =2$ and non-periodic for $p \neq 2$, whereas $\Phi (t)$ is a decaying amplitude that starts at the time $t_*$ with initial amplitude $\Phi_*$, so that $|\phi| = \Phi_*$ (in the case of a multi-component field with various real components, we will split evenly the initial value among the different components). The quantities $a_*$ and $H_*$ are the scale factor and Hubble parameter at time $t = t_*$. Note that for times $H_* t \gg 1 $, this field configuration gives rise, in particular, to a matter/radiation-dominated equation of state for $p=2, 4$, respectively.

In order to do lattice simulations of this system, we have to fix the program variables ($f_*$, $\alpha$, $\omega_*$), defined in Eq.~(\ref{eq:FieldSpaceTimeNaturalVariables}). First, we want to use variables that guarantee that typical numbers of certain physical quantities (such as field amplitudes or range of excited momenta) are of order unity. And second, as the evolution algorithms discussed before assume a constant time step, we want to use a program time variable that guarantees an approximately constant oscillation frequency. In this way, each oscillation period will be well resolved independently of how long the simulation takes. In this regard, we get from  Eqs.~(\ref{eq:PowLaw-sol0}) and (\ref{eq:PowLaw-sol}) that the inflaton oscillation frequency (defined in Eq.~\ref{eq:scalar-fld}) scales with the scale factor as
\be \Omega (\phi) \sim \omega_* \left(\frac{t}{t_*} \right)^{4/p - 2} \sim \left(\frac{a}{a_*} \right)^\frac{-3(p-2)}{(p+2)}  \ , \hspace{0.5cm} \omega_* \equiv \sqrt{ \lambda } \mu^{\frac{4-p}{2}} \phi_*^{\frac{p-2}{2}} \ ,  \label{eq:init-osc-fr}\ee
where $\omega_*$ is the oscillation frequency at the onset of oscillations. A convenient choice of program variables is therefore
\be \alpha = 3 \left(\frac{p-2}{p+2} \right)
\ ,  \hspace{0.5cm} f_* \equiv \phi_* \ , \hspace{0.5cm} \omega_* \equiv \Lambda^2 M^{-\frac{p}{2}} \phi_*^{\frac{p-2}{2}} \ . \label{eq:prunits-powl}  \ee
Note that for this choice of variables, program time corresponds to cosmic/conformal time for $p=2, 4$ respectively, up to a dimensionful constant factor. The corresponding program potential $\tilde{V} (\tilde{\phi})$ of our model, defined in Eq.~\eqref{eq:PotNat}, can be then written as
\bea \tilde{V} (\tilde{\phi}) \equiv \frac{1}{f_*^2 \omega_*^2} V(\tilde{\phi}) = \frac{1}{p} \left( \frac{M}{\phi_*} \right)^p \tanh^p \left( \frac{\phi_* |\tilde{\phi}|}{M}\right)  \ ,\eea
and its first and second derivatives are
\bea \frac{\partial \tilde{V}}{\partial \tilde{\phi} } &=& 2 \left( \frac{M}{\phi_*} \right)^{p-1} \frac{\tanh^p (\phi_*|\tilde{\phi}| / M )}{\sinh (2 \phi_* |\tilde{\phi}| / M )} {\rm sgn} (\tilde{\phi}) \ ,\\
\frac{\partial^2 \tilde{V}}{\partial \tilde{\phi}^2} &=& 4 \left( \frac{M}{\phi_*} \right)^{p-2} \left( p - \cosh ( 2 \phi_* |\tilde{\phi}| / M  )\right) \frac{\tanh^p (\phi_* |\tilde{\phi}| /M)}{\sinh^2 (2 \phi_*|\tilde{\phi}| /M) } \ , \eea
where ${\rm sgn} (\tilde{\phi})$ is the sign function.

\subsubsection{Preheating}

Let us now see how preheating proceeds in this model. The post-inflationary dynamics of an inflaton with potential (\ref{eq:InflPot}) has been studied with lattice simulations in the past: in the absence of inflaton interactions to other species in \cite{Lozanov:2016hid,Lozanov:2017hjm}, with interactions to a second scalar field with non-canonical kinetic terms in \cite{Krajewski:2018moi}, and more recently, with quadratic interactions to a daughter field in \cite{Antusch:2020iyq}.  In all of these studies, the fields involved were real scalars. Here, in order to illustrate the gauge-invariant lattice techniques introduced in the previous sections, we will also consider coupling the inflaton field to a gauge structure.

Let us start by considering a light scalar singlet $\chi$ as the only daughter field, coupled to the inflaton via a quadratic interaction. The potential of such a theory can be written as
\be V(\phi, \chi) = \frac{1}{p} \lambda \mu^{4-p} |\phi|^p + \frac{1}{2} g^2 \phi^2 \chi^2  \ , \label{eq:PowerLaw-pot-3} \ee
where $g$ is a dimensionless coupling constant, and we have taken the limit $M \rightarrow \infty$ in the inflaton potential. When inflation ends at the amplitude $\phi = \phi_*$, the energy budget of the universe is dominated by the homogeneous component of the inflaton. Therefore, the evolution of the inflaton amplitude and scale factor can be described approximately by Eqs.~(\ref{eq:PowLaw-sol0})-(\ref{eq:PowLaw-sol}) during the first inflaton oscillations. Using the program variables defined in (\ref{eq:prunits-powl}) in this context, the program potential (again in the limit $M \rightarrow \infty$) reads
\bea \tilde{V} (\tilde{\phi}, \tilde{\chi})  \equiv  \frac{1}{f_*^2 \omega_*^2} V(\tilde{\phi}, \tilde{\chi} ) = \frac{1}{p} |\tilde{\phi} |^p + \frac{1}{2} q_* \tilde{\phi}^2 \tilde{\chi}^2 \, , \label{eq:PowerLaw-pot-4}  \eea
where the \textit{resonance parameter} $q_{*}$ is defined as the following dimensionless ratio,
\be q_{*} \equiv \frac{g^2 \phi_{*}^2}{\omega_{*}^2} \ . \label{eq:ResPar} \ee
The first and second derivatives of the program potential with respect to the two fields are
\bea \frac{\partial \tilde{V}}{\partial \tilde{\phi}} &=& |\tilde{\phi}|^{p-2} \tilde{\phi} + q_* \tilde{\chi}^2 \tilde{\phi}  \ , \hspace{0.5cm} \frac{\partial \tilde{V}}{\partial \tilde{\chi}} = q_* \tilde{\phi}^2 \tilde{\chi}    \ , \\
 \frac{\partial^2 \tilde{V}}{\partial \tilde{\phi}^2} &=& (p-1) |\tilde{\phi}|^{p-2} + q_* \tilde{\chi}^2 \ , \hspace{0.5cm} \frac{\partial^2 \tilde{V}}{\partial \tilde{\chi}^2} = q_* \tilde{\phi}^2 \ . \eea

During the first stages of preheating, the linearized fluctuations of both fields have time-dependent effective masses, induced by the oscillations of the inflaton homogeneous mode. These masses vary non-adiabatically each time the inflaton crosses zero, which triggers an exponential growth of the amplitude of the field modes for certain bands of momenta. More specifically, the post-inflationary dynamics is governed by the interplay of two different resonant phenomena, which may or may not be present for certain choices of model parameters. These are:

\begin{itemize}

    \item \textbf{Self-resonance of the inflaton:}  The inflaton has a time-dependent effective mass $m^2_{\phi} \propto  | \phi|^{p-2}$ for $p \neq 2$, see Eq.~(\ref{eq:PowerLaw-pot-3}). In these cases, the (conformally rescaled) inflaton fluctuations can grow exponentially during this regime as $|\delta \tilde{\phi}_k |^2 \propto e^{2\nu_k z}$, where $\mathcal{R}e(\nu_k) > 0 $ for certain momenta bands, and $\nu_k  \equiv \nu_k (k;p)$ is the corresponding so-called Floquet index.  These bands are always narrow for all reasonable values of $p$, $\Delta k / \bar{k} \lesssim 0.1$ (with $\bar{k}$ the average momentum inside the band), and the maximum Floquet index within each band is $\mathcal{R}e (\nu_k) \lesssim 0.035$.

    \item  \textbf{Parametric resonance of the daughter field:} Similarly, the daughter field also has a time-dependent mass $m^2 = g^2 \chi^2$ for any of choice of $p$ as long as the quadratic interaction is present, see Eq.~(\ref{eq:PowerLaw-pot-3}). Due to this, the (conformally rescaled) daughter field fluctuations can also grow exponentially as $|\delta \tilde{ \chi}_k |^2 \propto e^{2\mu_k z}$, with $\mathcal{R}e (\mu_k) > 0$ for certain ranges of momenta, and $\mu_k \equiv \mu_k (k,q_*;p)$ the corresponding Floquet index. The key parameter signaling the strength of the resonance is the \textit{effective resonance parameter} $q_{\rm res}$, which is a time dependent function as
    \be  q_{\rm res} \equiv q_* a^{\frac{6(p-4)}{p+2}} \ , \label{eq:EffRes} \ee
    which evolves with the expansion of the universe as indicated. If $q_{\rm res} \gtrsim 1$, the parametric resonance is \textit{broad}: the width of the resonance bands for all values of $p$ is $\Delta k / \bar{k} \sim 1$, and the maximum Floquet index within those bands is typically $\mathcal{R}e(\mu_k) \sim 0.1 - 0.2$.  In this case, the maximum momenta excited by the main resonance band (before the dynamics become non-linear) scales as $k \sim q_{\rm res}^{1/4} \omega_*$, modulo some multiplicative scale factor power. On the other hand, if $q_{\rm res} \ll 1$,  the width of the bands is very small $\Delta k / \bar{k} \ll 1$,  and we say that the resonance is \textit{narrow}. This second effect cannot be typically captured in the lattice due to lack of resolution. We note that as $q_{\rm res}$ changes with time, the type of resonance may also change during preheating: decreasing with time for $p < 4$, growing for $p>4$, and remaining constant for $p=4$. Therefore, the type of parametric resonance (either broad or narrow) can change as the universe expands.
    \end{itemize}
If broad parametric resonance of the daughter field is present (i.e.~if $q_{\rm res} > 1$), it is almost always a stronger effect than the inflaton self-resonance. However, parametric resonance eventually becomes narrow for $p<4$, even if it was broad initially. This contrasts with inflaton self-resonance, which is always present independently of the value of $p$, as long as $p \neq 2$. The different behaviour of these phenomena for different model parameters is key to understand how energy distributes between the different field sectors during preheating, and the evolution of the equation of state after inflation.

Let us now consider a scenario in which the inflaton is coupled to a $SU(2)\times U(1)$ gauge sector via a gauge-invariant covariant derivative. We take the inflaton as the scalar doublet $\Phi$ with potential
\be V(|\Phi|) = \frac{2^{p/2}}{p} \lambda \mu^{4-p} |\Phi|^p \ . \ee
The prefactor $2^{p/2}$ is included to compensate the $\sqrt{2}$ factor in the definition of the doublet, see Eq.~(\ref{eq:ChargedScalars}). For example, for $p=4$ we get $V(|\Phi|) \equiv \lambda |\Phi|^4 = (\lambda / 4) (\varphi_0^2 + \varphi_1^2 + \varphi_2^2 + \varphi_3^2)^2$, which recovers the scalar singlet expression (\ref{eq:PowerLaw-pot}) for one component when the other three are set to 0, e.g.~$V(|\Phi|) = (\lambda /4 ) \varphi_0^4$ for $\varphi_1$, $\varphi_2$, $\varphi_3 \rightarrow 0$.

Fortunately, the simpler singlet scalar theory described before constitutes an excellent proxy for this more complicated gauge model, as the dominant interaction term generated by the gauge interaction is also quadratic in the inflaton and daughter (gauge) fields. In order to see this, let us consider the covariant derivative term in action (\ref{eq:Lagrangian}), which contains the interaction between the inflaton and the gauge fields. It can be expanded as
\be ( \vec{D}_{\mu} \Phi)^{\dagger} (\vec{D}^{\mu} \Phi) = (\partial_{\mu} \Phi)^{\dagger}  \partial^{\mu} \Phi +  Q_A^2 g_A^2 |\Phi |^2  |\vec{A}|^2 + \frac{1}{4} Q_B^2 g_B^2 | \Phi |^2 \sum_a |\vec{B}^a |^2 + 2 Q_A g_A Q_B g_B \sum_a \vec{A} \cdot \vec{B}^a (\Phi^{\dagger} T_a \Phi )   \dots \label{eq:GaugeCovInt} \ ,\ee
where we have ignored terms of the type $\sim (\partial_{\mu} \Phi) \Phi$, which are subdominant during the early linear regime of parametric resonance. The first term in Eq.~(\ref{eq:GaugeCovInt}) gives rise to the usual Laplacian in the field equations. The second and third terms constitute quadratic interactions between the inflaton and the Abelian and non-Abelian gauge fields, respectively. These are analogous to the quadratic interaction of Eq.~(\ref{eq:PowerLaw-pot-3}) between the inflaton and a secondary scalar field, with the identification $g \rightarrow Q_A g_A$ and $g \rightarrow Q_B g_B / 2 $ in each case. Mimicking the notation of Eq.~(\ref{eq:ResPar}), it is then natural to define the resonance parameters of the Abelian and non-Abelian gauge fields as
\be q_{A*} \equiv \frac{Q_A^2 g_A^2 \phi_*^2}{\omega_*^2} \ , \hspace{0.4cm} q_{B*} \equiv \frac{Q_B^2 g_B^2 \phi_*^2}{4 \omega_*^2} \ , \label{eq:ResParGauge}\ee
where $\phi_* \equiv \sqrt{2} |\Phi_*|$ with $|\Phi_*|$ the amplitude of the inflaton norm at the end of inflation, and $\omega_*$ is given in Eq.~(\ref{eq:init-osc-fr}). Therefore, we can use the scalar theory as a proxy to study the equivalent $U(1)$ or $SU(2)$ gauge-invariant theories, at least during the initial linear regime. In particular, in the gauge scenario the inflaton develops fluctuations via self-resonance, while the gauge fields are also excited via parametric resonance. However, once non-linearities become relevant at later times, important differences between singlet scalar and gauge theories may appear. Finally, the fourth term in Eq.~(\ref{eq:GaugeCovInt}) appears when the inflaton is coupled to a full $SU(2)\times U(1)$ gauge sector. One can prove that the effect of such term is to couple the EOM of the Abelian and non-Abelian gauge fields, so that they experience parametric resonance with a common resonance parameter $q_{{\rm eff} *} = q_{A*} + q_{B*}$. The details of the parametric resonance process in the presence of Abelian and non-Abelian gauge fields, as well as the relevance of that term, will be discussed in more detail in an upcoming work \cite{NonAbelianInpreparation}. This goes beyond the objective of this manuscript, which is mainly to illustrate lattice gauge-invariant techniques in a specific physics model.

\subsection{Lattice simulations: U(1) gauge interactions} \label{sec:U1-sims}

\begin{table}
    \begin{center}
        \begin{tabular}{|c|c |c |c | c| c|| c| c| c| }
            \hline
            Simulation & $p$ & $M /m_p$ & $\Lambda^4$ & $q_*$ &$N$ & $\tilde{k}_{\rm IR}$ & $\delta \tilde{t}$ \\
            \hline\hline
           U(1) &  2 & 10 & $1.8 \cdot 10^{65}$ & $4 \cdot 10^4$  & 128 &  4 & $5 \cdot 10^{-3}$ \\
            \hline
           U(1) &  4 & 10 & $4.3 \cdot 10^{65}$ & $10^2$ & 128 & 0.6 & $10^{-2}$ \\
            \hline
           U(1) &  6 & 10 & $6.8 \cdot 10^{65}$ & 1  & 128 & 0.15 & $7 \cdot 10^{-4}$\\ \hline
           \hline
              SU(2) $\times$ U(1) + $\chi$ &  2 & 10 & $1.8 \cdot 10^{65}$ & $4 \cdot 10^4$  & 128 & 4 & $3 \cdot 10^{-4}$\\ \hline
              SU(2) $\times$ U(1) + $\chi$ &  4 & 10 & $4.3 \cdot 10^{65}$ & $10^2$ & 128 & 0.6 & $10^{-2}$\\ \hline
        \end{tabular}
    \end{center} \label{tab:LatticeParameters}
    \caption{Benchmark model and lattice parameters used in the U(1) and $SU(2)\times U(1)$+$\chi$ gauge simulations}
\end{table}

We now proceed to discuss the results from our lattice simulations.  We start by considering the post-inflationary dynamics of a complex inflaton field $\varphi \equiv \frac{1}{\sqrt{2}} (\varphi_0 + i \varphi_1)$ with potential energy (\ref{eq:InflPot}) [where we must substitute $|\phi | \rightarrow \sqrt{2} |\varphi|$], coupled to an Abelian gauge boson $A_{\mu}$ via a gauge-invariant covariant derivative. The model and lattice parameters considered in the simulations are provided in Table \ref{tab:LatticeParameters}. We have chosen a set of three representative power-law coefficients, $p=2,4,6$. In each case, the resonance parameter $q_{A*}$ is fixed to guarantee broad parametric resonances at the onset of the inflaton oscillatory regime. We have fixed the value $M=10 m_p$ as a benchmark, which guarantees that the inflationary slow-roll condition breaks down in the positive-curvature region of the potential. As described above, the initial exponential growth of the gauge field modes during broad parametric resonance takes place mainly within an infrared band of width $p \lesssim p_* \equiv q_{A*}^{1/4} \omega_*$ (modulo a multiplying scale factor power). However, when the energy transferred to the gauge fields is large enough, they backreact onto the inflaton homogeneous condensate, which triggers a propagation of power in the spectra of all fields towards the ultraviolet. Due to this, the minimum momenta $\tilde{k}_{\rm IR}$ and number of points per lattice side $N$ are chosen, in each case, to guarantee that both the initial infrared growth and the following ultraviolet excited scales are well resolved in the lattice.

We start by showing in Fig.~\ref{fig:Table-phisfw} the evolution of the volume-averaged inflaton norm $| \varphi |$, equation of state $w\equiv p  / \rho $, and scale factor as a function of program time $\tilde{\eta}$ [$d \tilde{\eta} \equiv a^{-\alpha} \omega_* dt$, c.f.~(\ref{eq:FieldSpaceTimeNaturalVariables})], for each of the three power-law coefficients $p=2,4,6$. As described above, the inflaton can be approximated as a homogeneous condensate during its first oscillations, and the evolution of the inflaton amplitude and scale factor are approximately given by Eqs.~(\ref{eq:PowLaw-sol0})-(\ref{eq:PowLaw-sol}). From these expressions, we deduce that the amplitude of the inflaton oscillations scales initially as $| \varphi | \sim a^{\frac{-6}{p+2}}$, so $|\varphi | \sim  a^{-3/2}, a^{-1}, a^{-3/4}$ for $p=2$, 4, 6, respectively. Therefore, in the Figure we have multiplied the inflaton norm by the inverse of these factors, so the amplitude of the rescaled inflaton oscillations is initially constant. Although the inflaton homogeneous regime holds qualitatively well during the first inflaton oscillations, the energy stored in gauge fields and inflaton gradients grows exponentially due to broad parametric resonance. Eventually, the fraction of transferred energy is so large that they backreact onto the inflaton, destroying the homogeneity of the condensate. We identify this time scale as the \textit{backreaction time} $\tilde{\eta}_{\rm br}$. From the simulation, we get $\tilde{\eta}_{\rm br} \simeq 130$, 40, 70 for $p=2$, 4, 6 respectively.
\begin{figure}
    \centering
    \includegraphics[width=5.4cm]{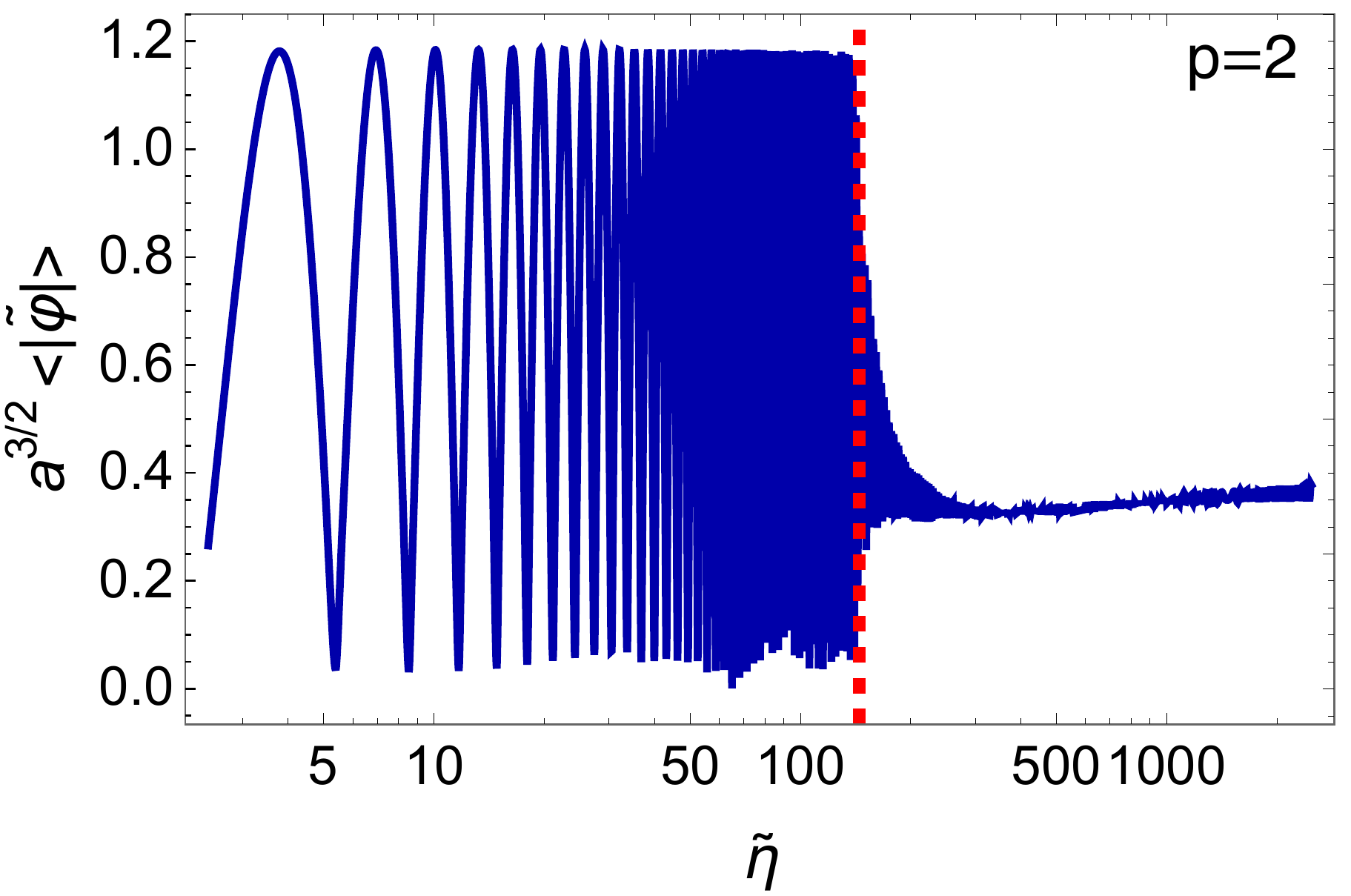} \hspace{0.05cm}
    \includegraphics[width=5.4cm]{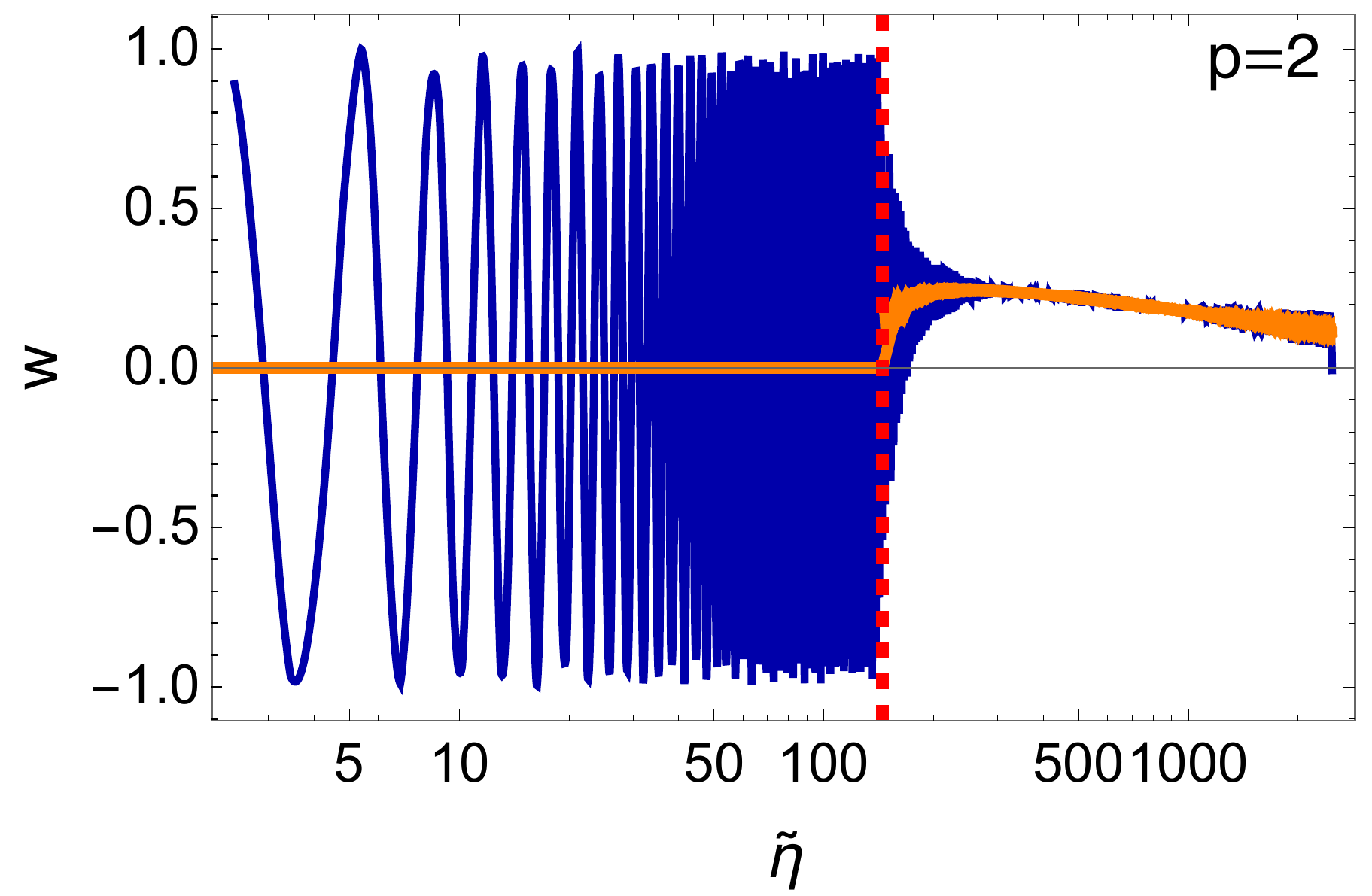} \hspace{0.05cm}
    \includegraphics[width=5.4cm]{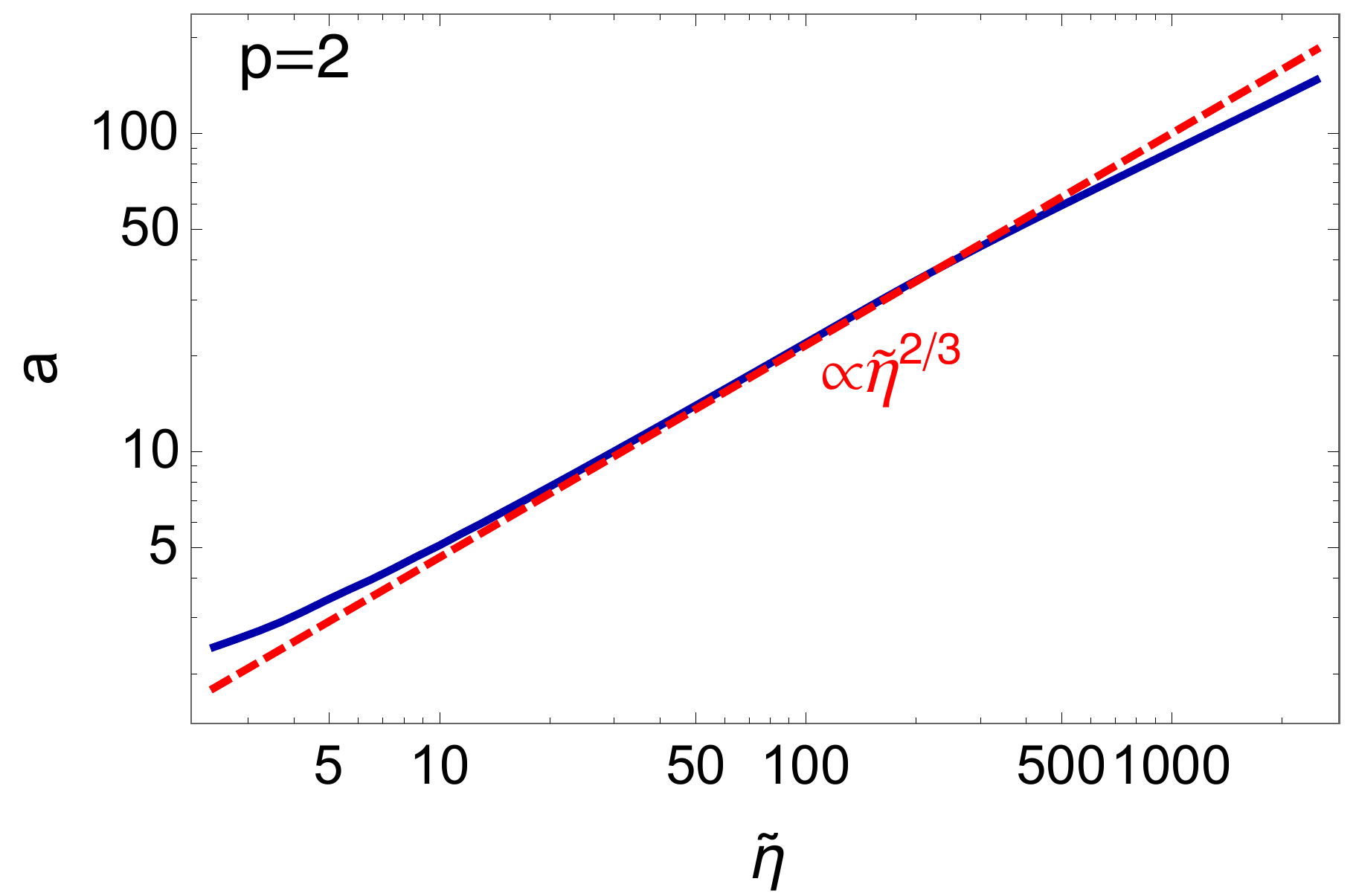} \\ \vspace*{0.1cm}
    \includegraphics[width=5.4cm]{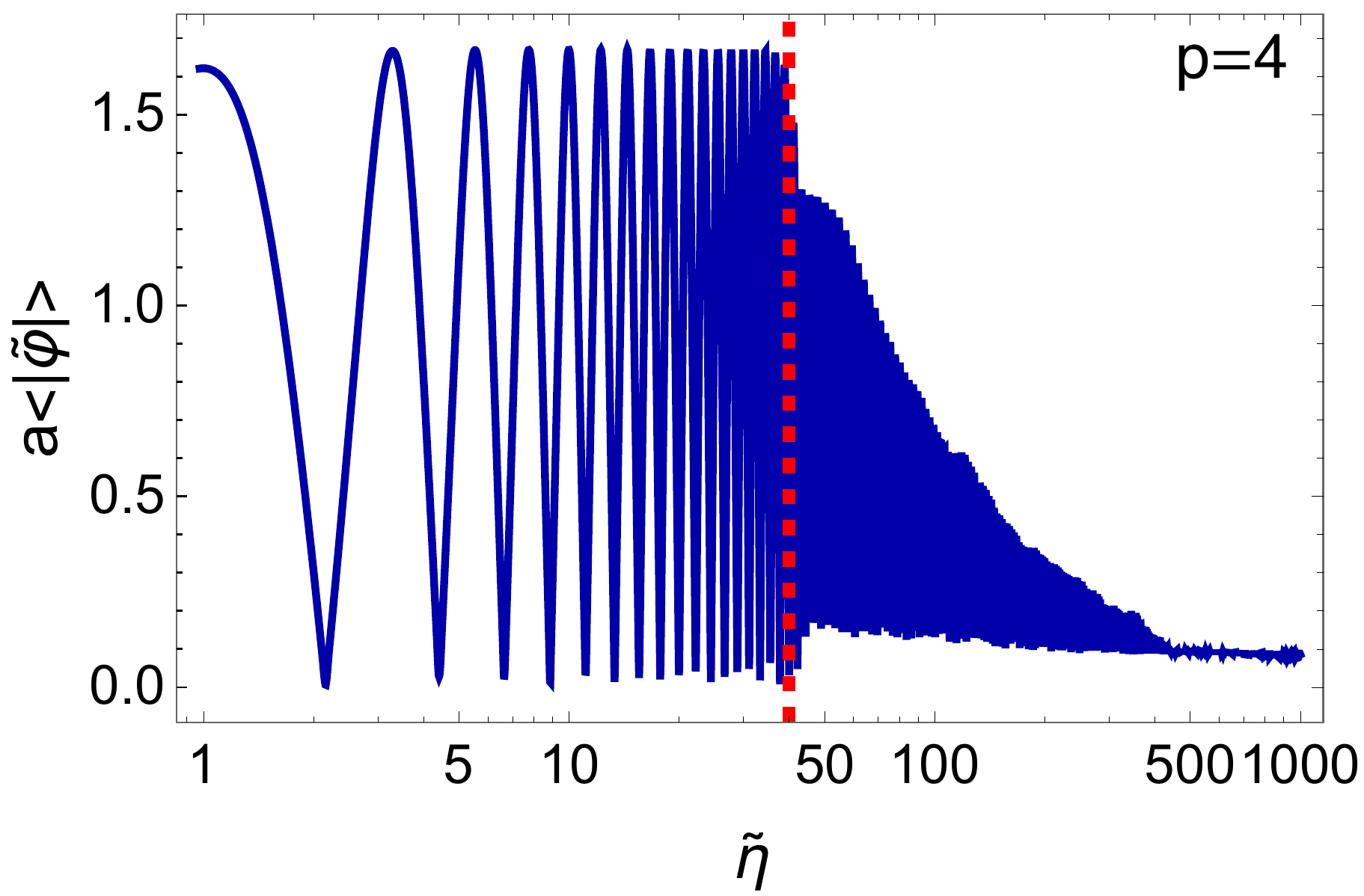} \hspace{0.05cm}
    \includegraphics[width=5.4cm]{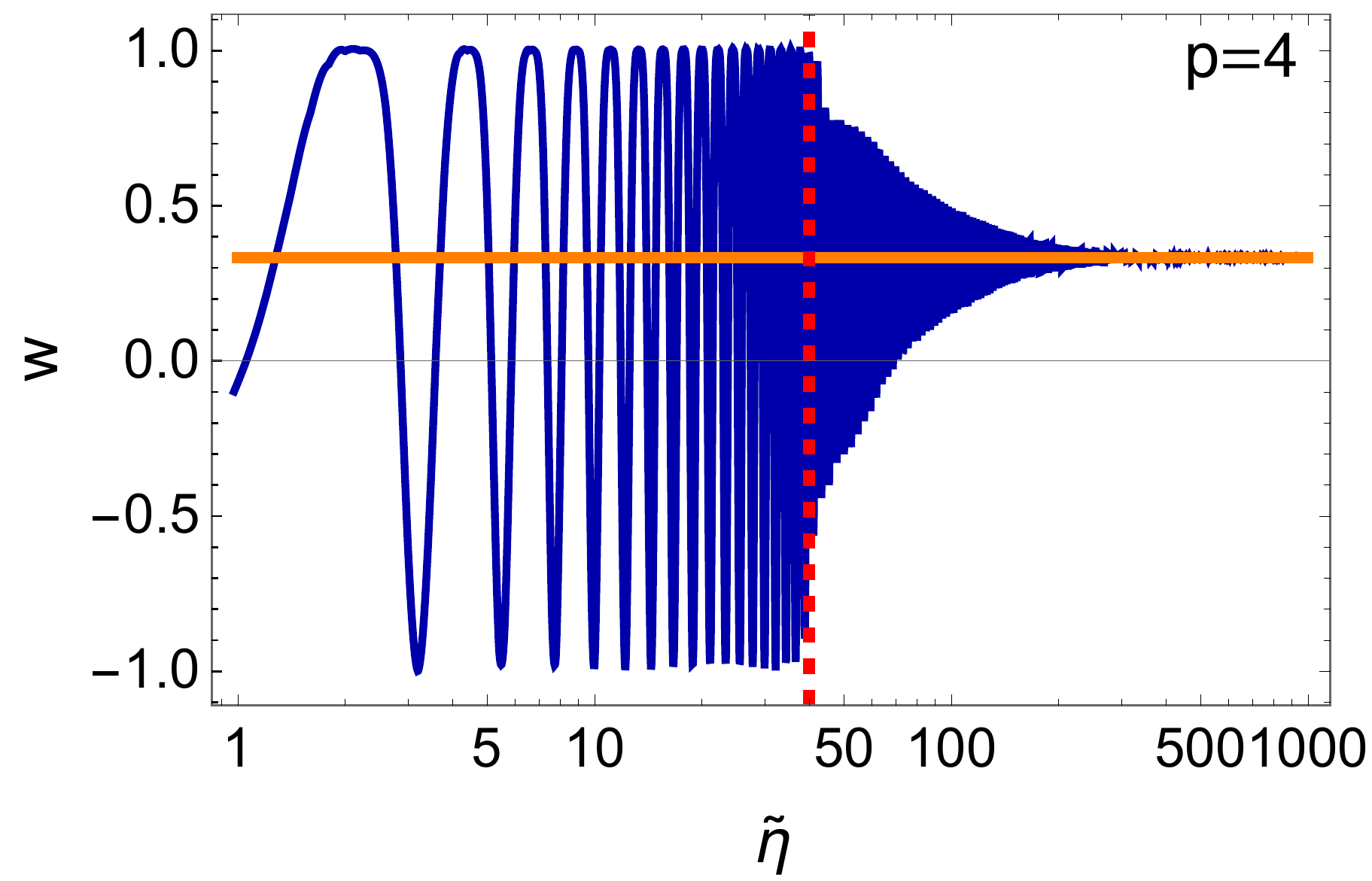} \hspace{0.05cm}
    \includegraphics[width=5.4cm]{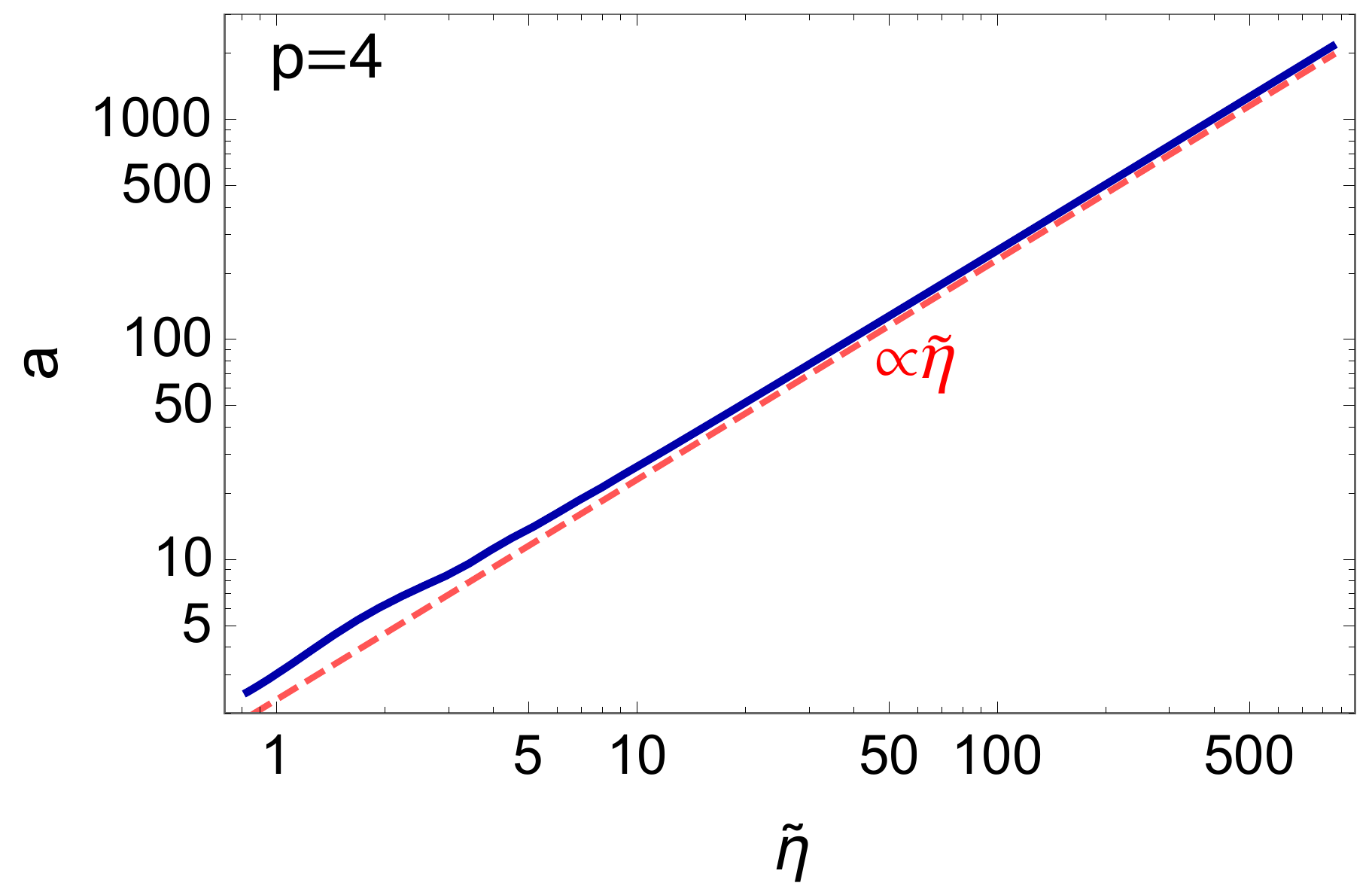}  \\ \vspace*{0.1cm}
    \includegraphics[width=5.4cm]{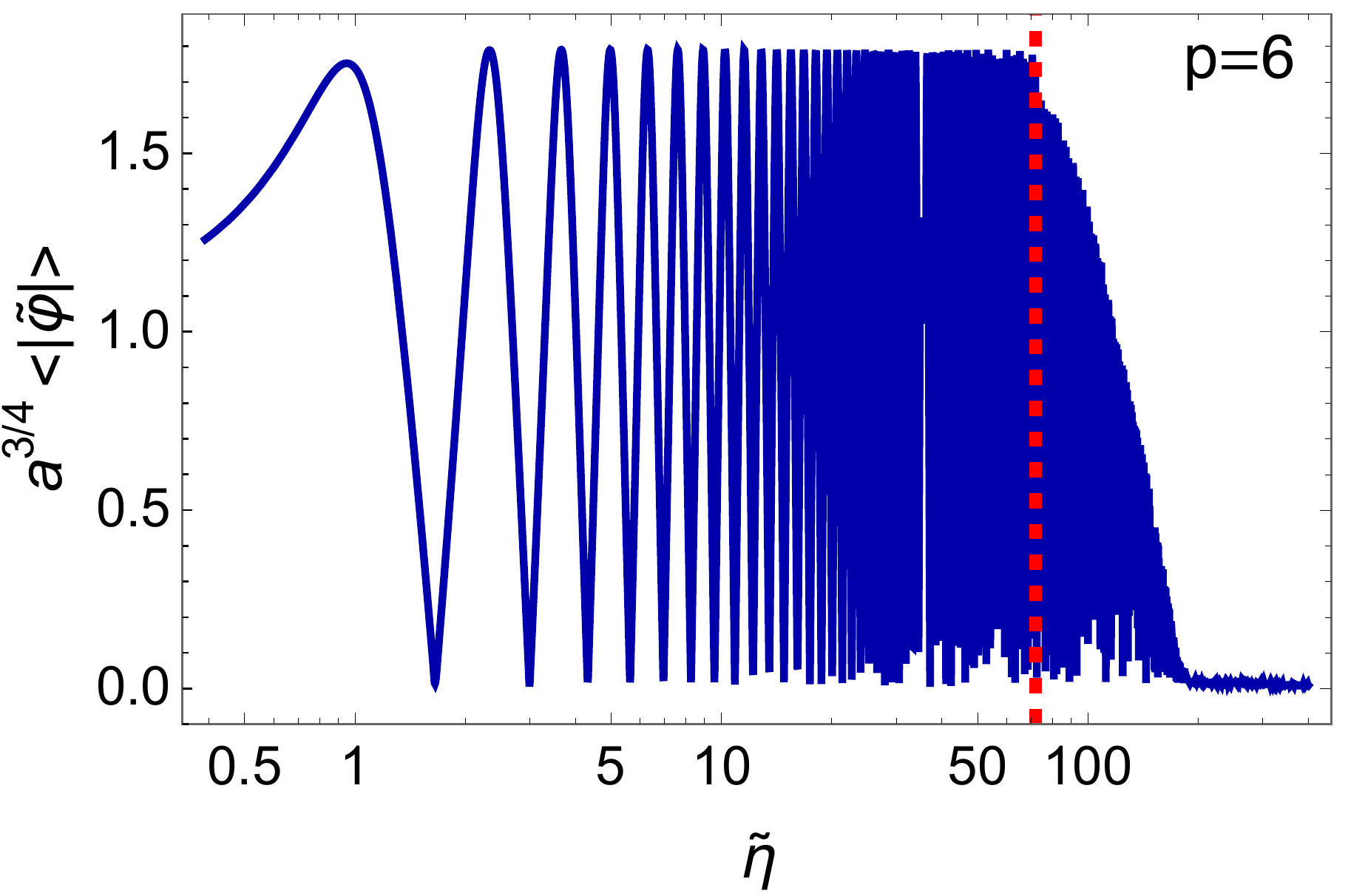} \hspace{0.05cm}
    \includegraphics[width=5.4cm]{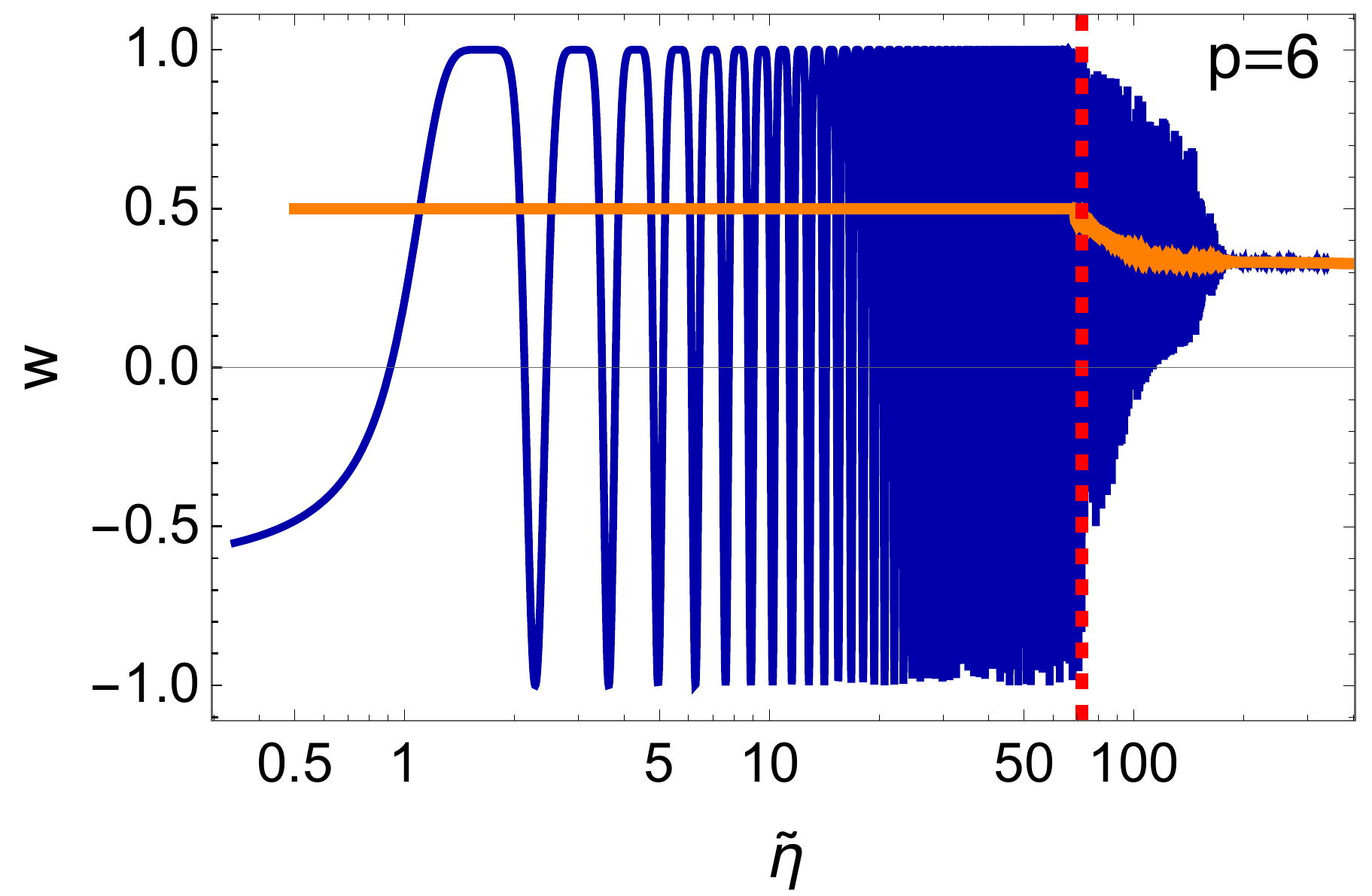} \hspace{0.05cm}
    \includegraphics[width=5.4cm]{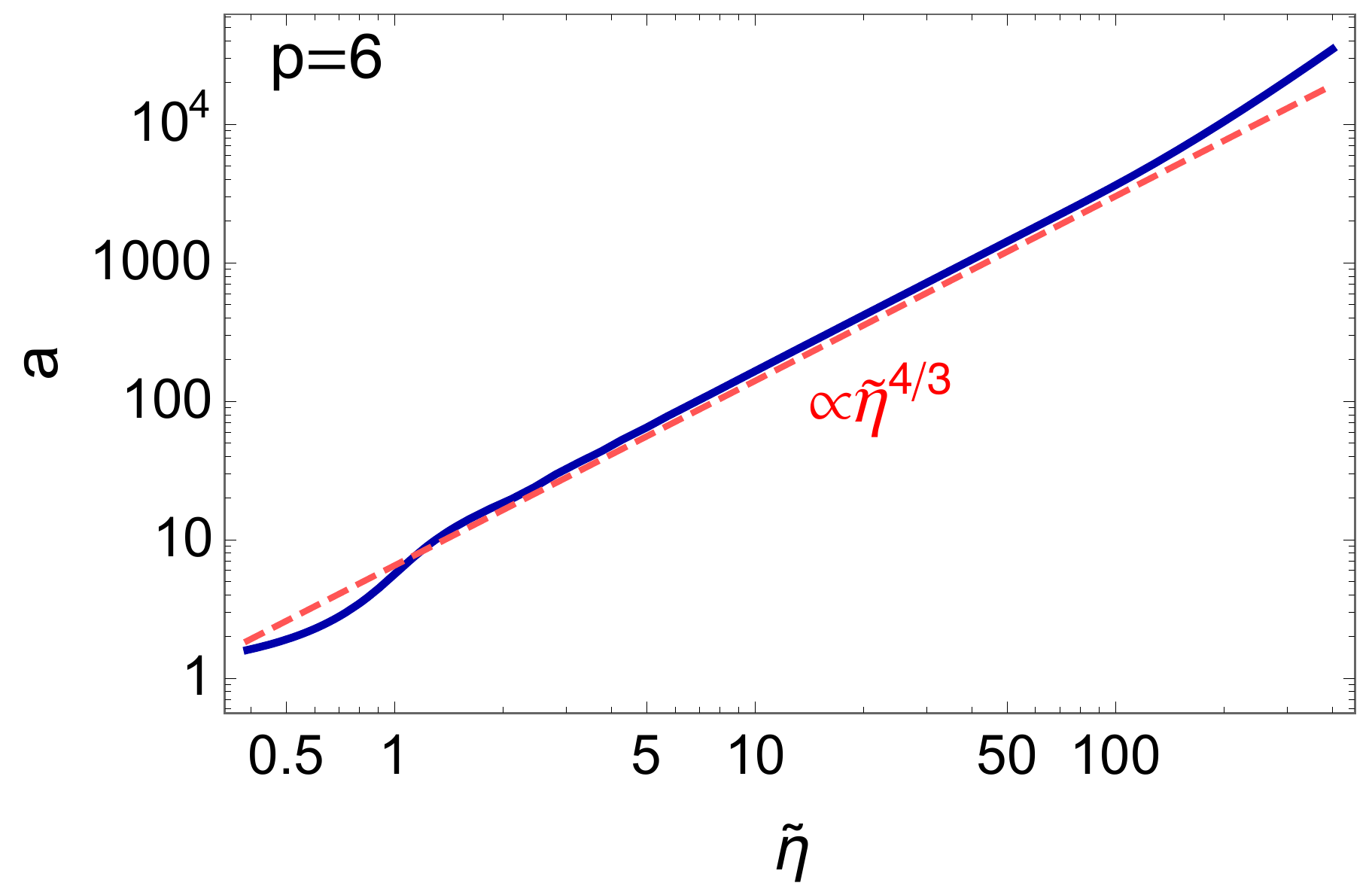}
    \caption{Average values of the scalar field norm, equation of state $w$, and scale factor as a function of time, for three $U(1)$ gauge simulations with $p=2$ (top row), $p=4$ (middle row), and $p=6$ (bottom row). The backreaction time $\tilde{\eta}_{\rm br}$ in indicated for the first two quantities with a vertical dashed line. In the scale factor panels we have added the prediction $a \sim \tilde{\eta}^{\frac{p+2}{6}}$, coming from the linear regime of homogeneous inflaton oscillations.}
    \label{fig:Table-phisfw}
\end{figure}
Let us now focus on the post-inflationary evolution of the equation of state $w \equiv p /\rho $, i.e.~the ratio between the (volume-averaged) pressure and energy densities of the system. Initially, the inflaton oscillates coherently around the minimum, which gives rise to similar oscillations in the equation of state in the range $-1 < w < 1$. From Eqs.~(\ref{eq:PowLaw-sol0})- (\ref{eq:PowLaw-sol}), we can compute that the \textit{effective} (i.e.~oscillation-averaged) equation of state in this regime is approximately $\bar{w} \equiv (p-2)/(p+2)$. This corresponds to $\bar{w} = 0$, 1/3, 1/2 for $p=2$, 4, 6 respectively, which agrees with our lattice results, see the middle column of Fig.~\ref{fig:Table-phisfw}. After backreaction, the equation of state stops oscillating, and slowly evolves towards the asymptotic values $w \rightarrow 0$ (for $p=2$) and $w \rightarrow 1/3$ (for $p=4$, 6). We will be able to understand these results better in light of the evolution of the energy distribution, which we discuss below. We also show the scale factor as a function of program time in the right panels of  Fig.~\ref{fig:Table-phisfw}. We know that during the initial linear regime, the scale factor evolves in cosmic time as $a \sim t^{\frac{2+p}{3p}}$ [c.f.~(\ref{eq:PowLaw-sol})]. By substituting this expression in the program time definition (\ref{eq:FieldSpaceTimeNaturalVariables}), we get that the scale factor evolves as $a \sim \tilde{\eta}^{\frac{p+2}{6}}$ in terms of program time, in agreement to what we see in the lattice.

We can understand better the evolution of these quantities if we focus on the evolution of the energy distribution. In the left panels of Fig.~\ref{fig:energies} we show the total energy of the system as a function of time [Eq.~(\ref{eq:rhoLocal})], for the considered power-law coefficients $p=2$, 4, 6. We also depict the evolution of each of its individual contributions: the kinetic, gradient, and potential energies of the inflaton, as well as the electric and magnetic energies of the gauge fields (see Eq.~(\ref{eq:energy-contrib}) for their exact expressions). As described above, the effective equation of state during the initial linear regime is $\bar{w} \simeq (p-2)/(p+2)$, so the total energy decays during the initial regime as $\rho \sim a^{-3 (1+\bar{w})} = a^{\frac{-6 p}{p+2}}$, which corresponds to $\rho \sim a^{-3}, a^{-4}, a^{-4.5}$ for $p=2,4,6,$ respectively. Therefore, we have multiplied the energies by the inverse of those factors, so that the rescaled total energy is constant initially. We also depict in the right panels of Fig.~\ref{fig:energies}  the evolution of the energy \textit{ratios} $\epsilon_{\rm i} \equiv \tilde{\rho_i} / \tilde{\rho}$ for the same simulations, i.e.~the relative contribution of each of the energy components to the total energy. By construction, the sum of all ratios is one.

   \begin{figure}
    \centering
    \includegraphics[width=7.5cm]{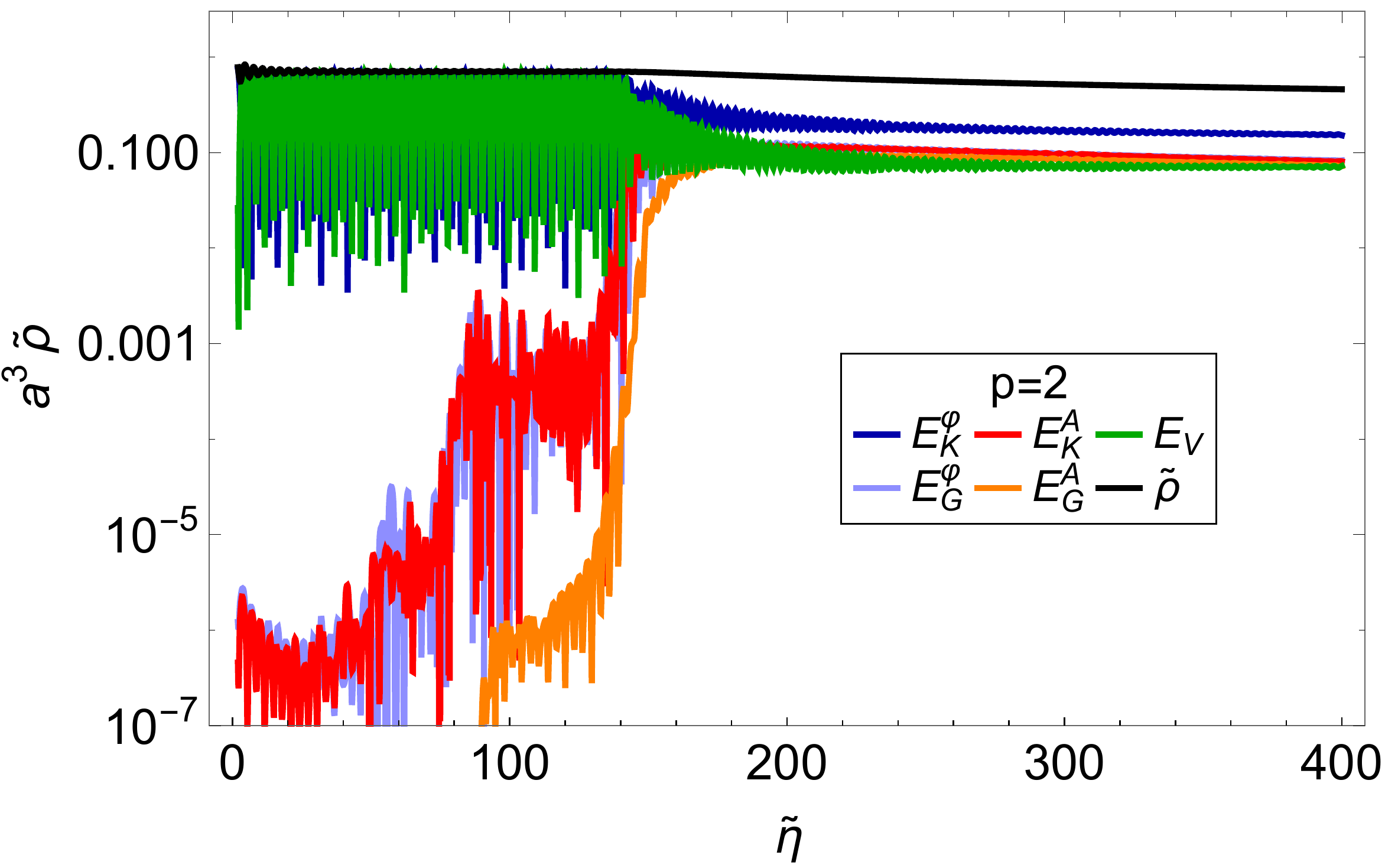} \hspace{0.3cm}
        \includegraphics[width=7.5cm]{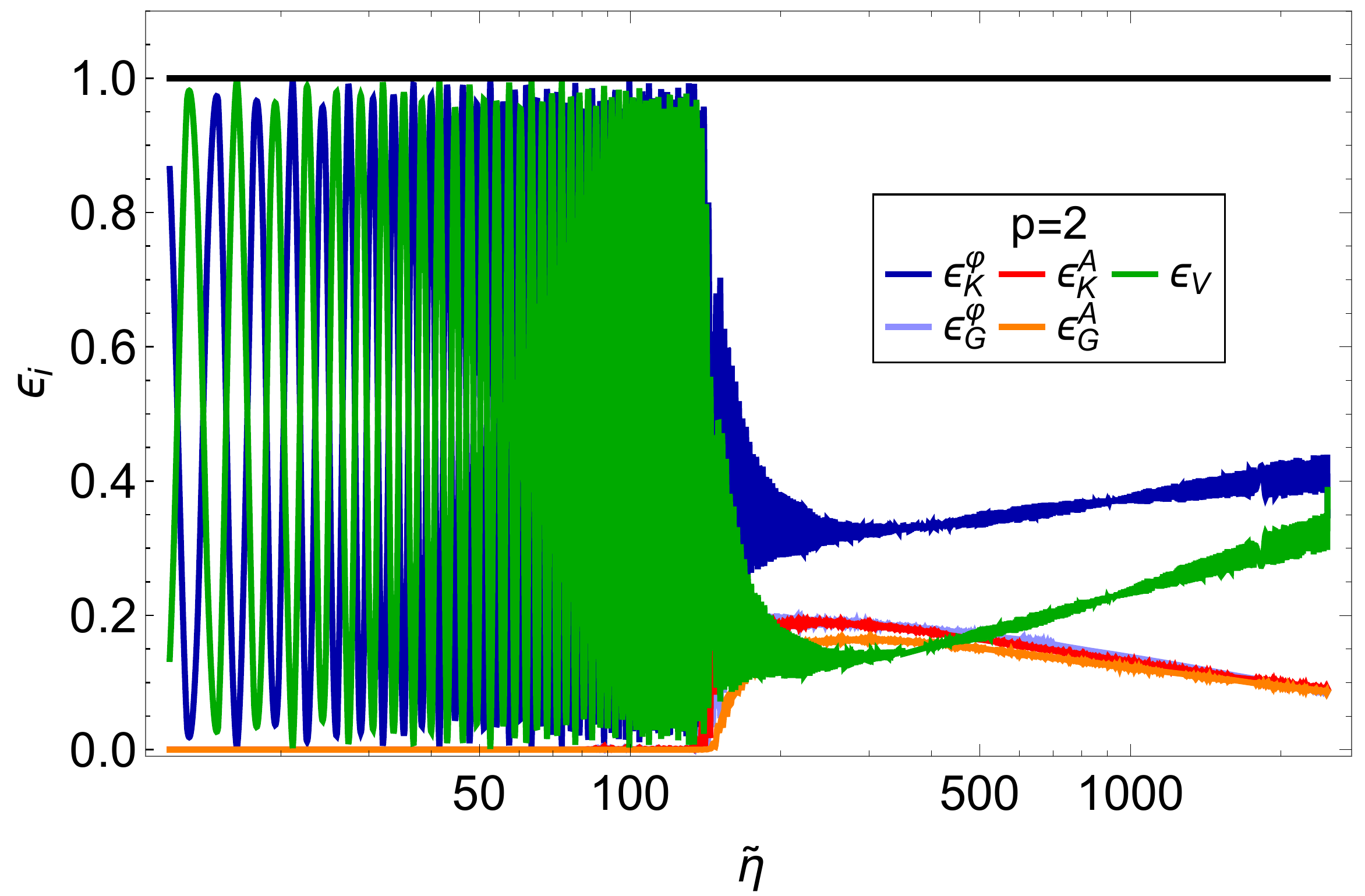} \hspace{0.3cm}
        \includegraphics[width=7.5cm]{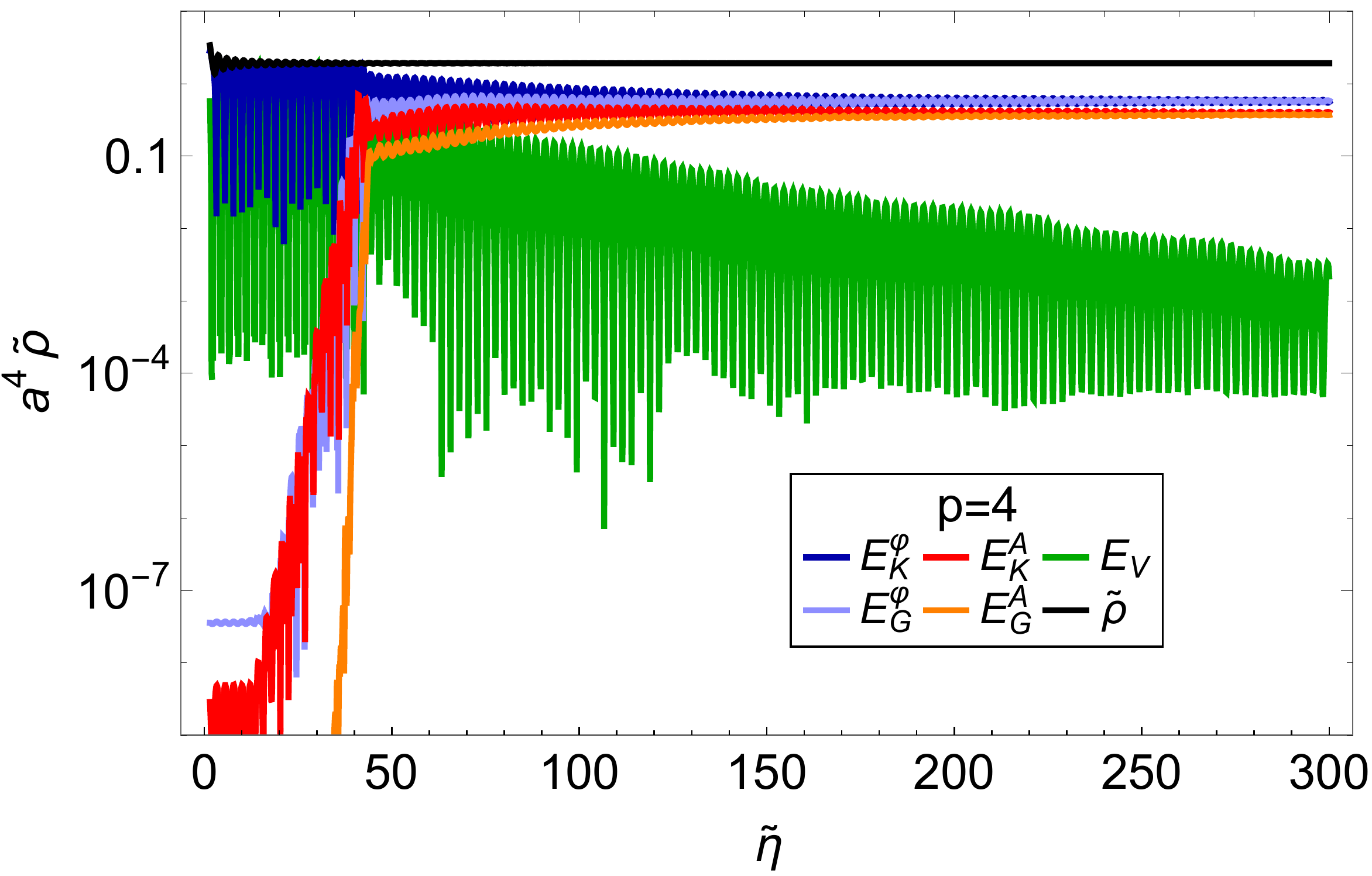} \hspace{0.3cm}
                \includegraphics[width=7.5cm]{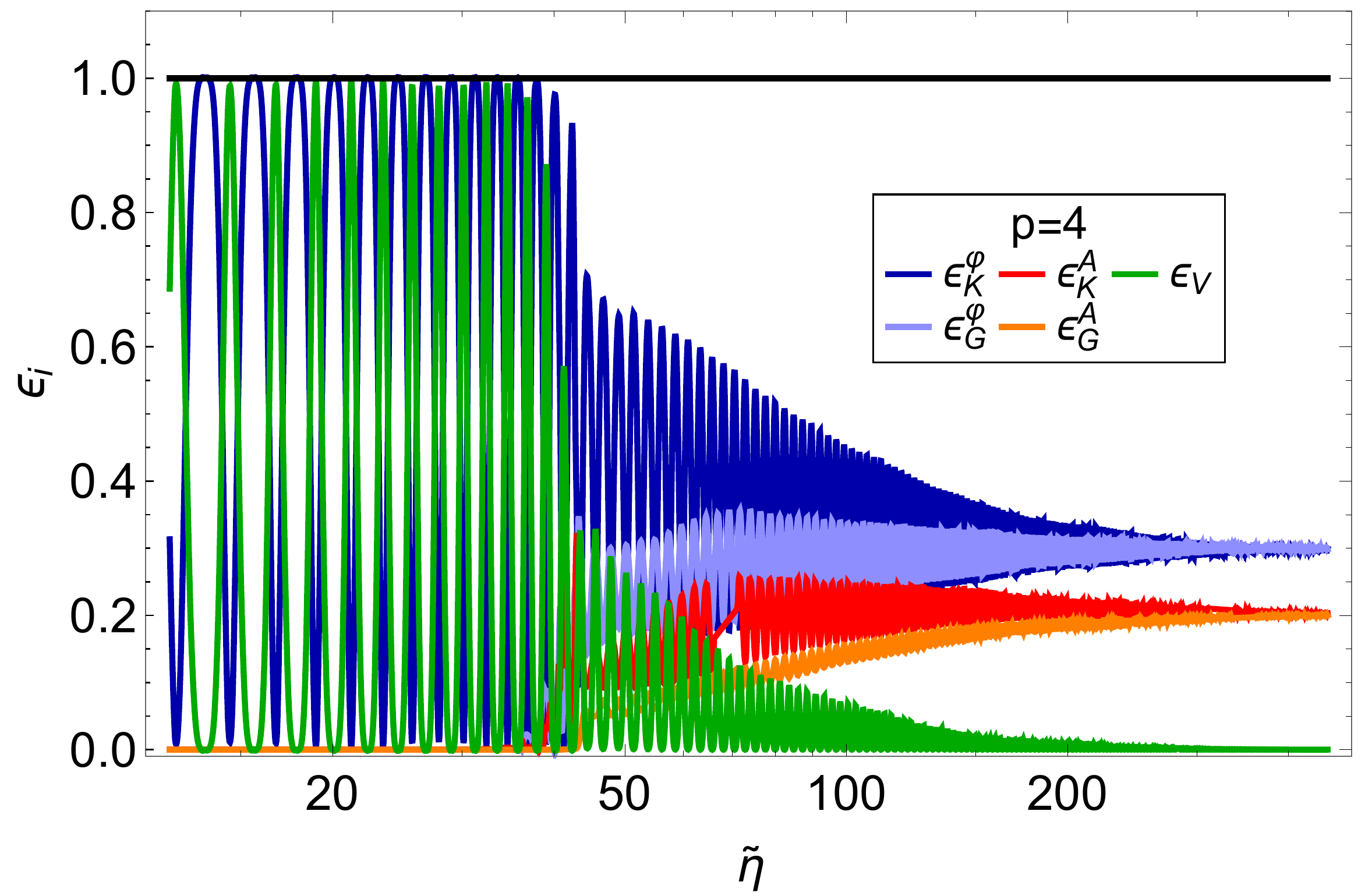} \hspace{0.3cm}
            \includegraphics[width=7.5cm]{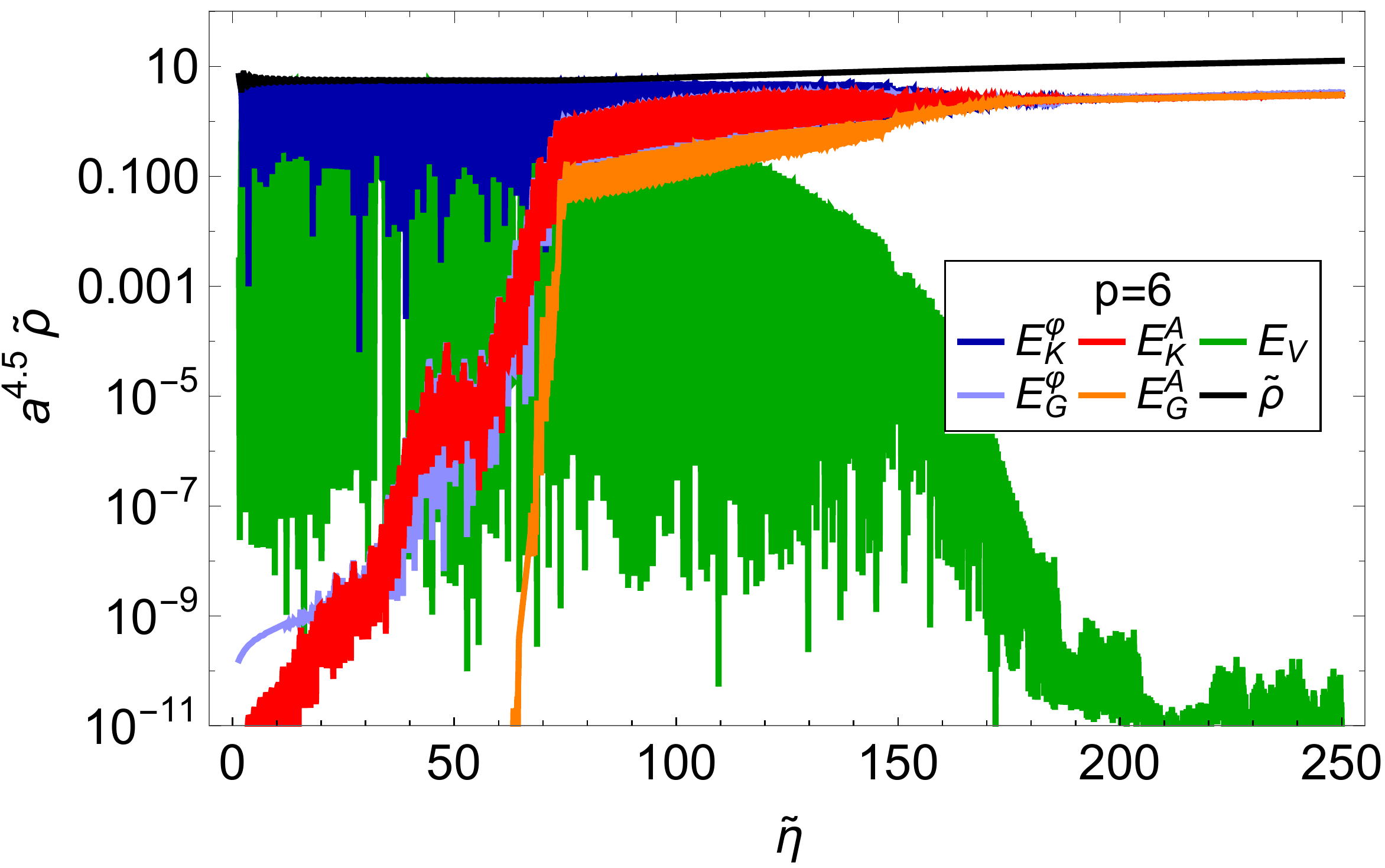} \hspace{0.3cm}
                        \includegraphics[width=7.5cm]{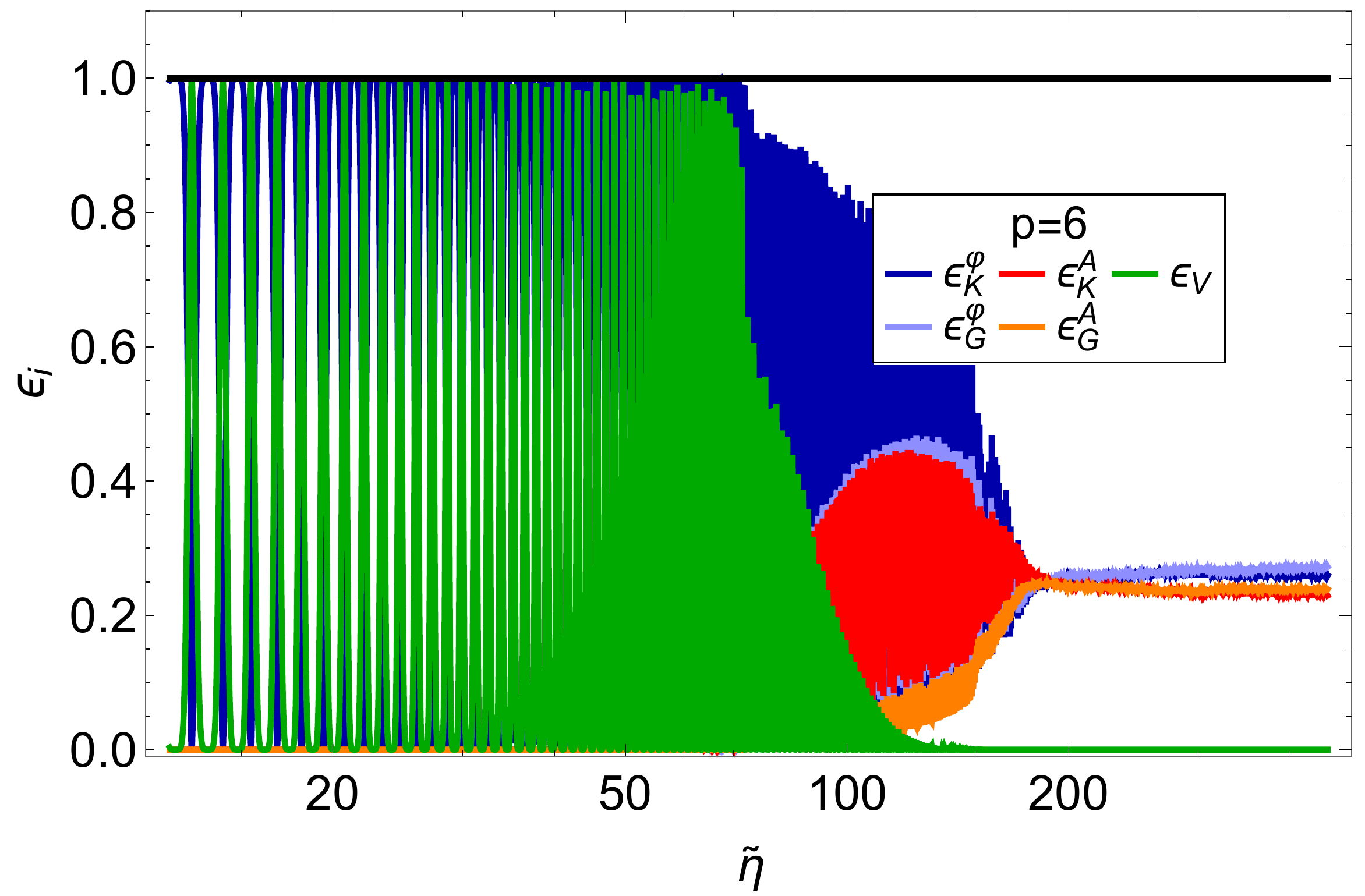} \hspace{0.3cm}
    \caption{Left panel: Evolution of the total energy $\tilde{\rho} \equiv \rho / (f_*^2 \omega_*^2)$ for the $U(1)$ gauge simulation and $p=2,4,6$, as well as of each of its individual contributions: kinetic, gradient, and potential energies of the inflaton, as well as electric and magnetic energies of the gauge field. These quantities are multiplied by the factor $\sim a^{\frac{6 p}{p+2}}$. Right panel: Evolution of the energy ratios $\epsilon_i \equiv \tilde{\rho}_i / \tilde{\rho}$ for the same simulations as in corresponding left panel. The sum of all ratios is one.}
    \label{fig:energies}
\end{figure}

As expected, the energy budget of the universe is initially dominated by the kinetic and potential energies of the inflaton, while the other energies are subdominant, i.e.~$\langle E_K^A \rangle,\langle E_G^A \rangle,\langle E_G^{\varphi} \rangle \ll \langle E_K^{\varphi} \rangle, \langle E_V \rangle$. However, a very small fraction of the initial energy is stored in the electric and inflaton gradient energies, due to the spectrum of fluctuations imposed to $\varphi$ and $\dot{A}_i$. In contrast, the initial magnetic energy is exactly zero (up to machine precision), as we do not set fluctuations to the amplitude of the gauge field $A_i$, see Eqs.~(\ref{eq:Inflc1})-(\ref{eq:Inflc4}). In any case, these energies soon start growing exponentially due to parametric resonance, as seen in the three simulations. These energies become sizable approximately at the backreaction time $\tilde{\eta} \simeq \tilde{\eta}_{\rm br}$, and the inflaton homogeneous condensate gets destroyed via backreaction effects. From then on, non-linear effects become relevant, and the system eventually achieves a stationary regime at late times. Remarkably, we observe that the inflaton gets virialized very quickly after inflation, with their oscillation-averaged energies satisfying the relation $\langle E_K
^{\varphi} \rangle \simeq  \langle E_G^{\varphi} \rangle + \frac{p}{2} \langle E_V \rangle $, for the three cases $p=2$, 4, 6. Similarly, we observe equipartition between the electric and magnetic energies at late times, $\langle E_K^A \rangle \simeq \langle E_G^A \rangle$.

\begin{figure}
    \centering
    \includegraphics[width=8cm]{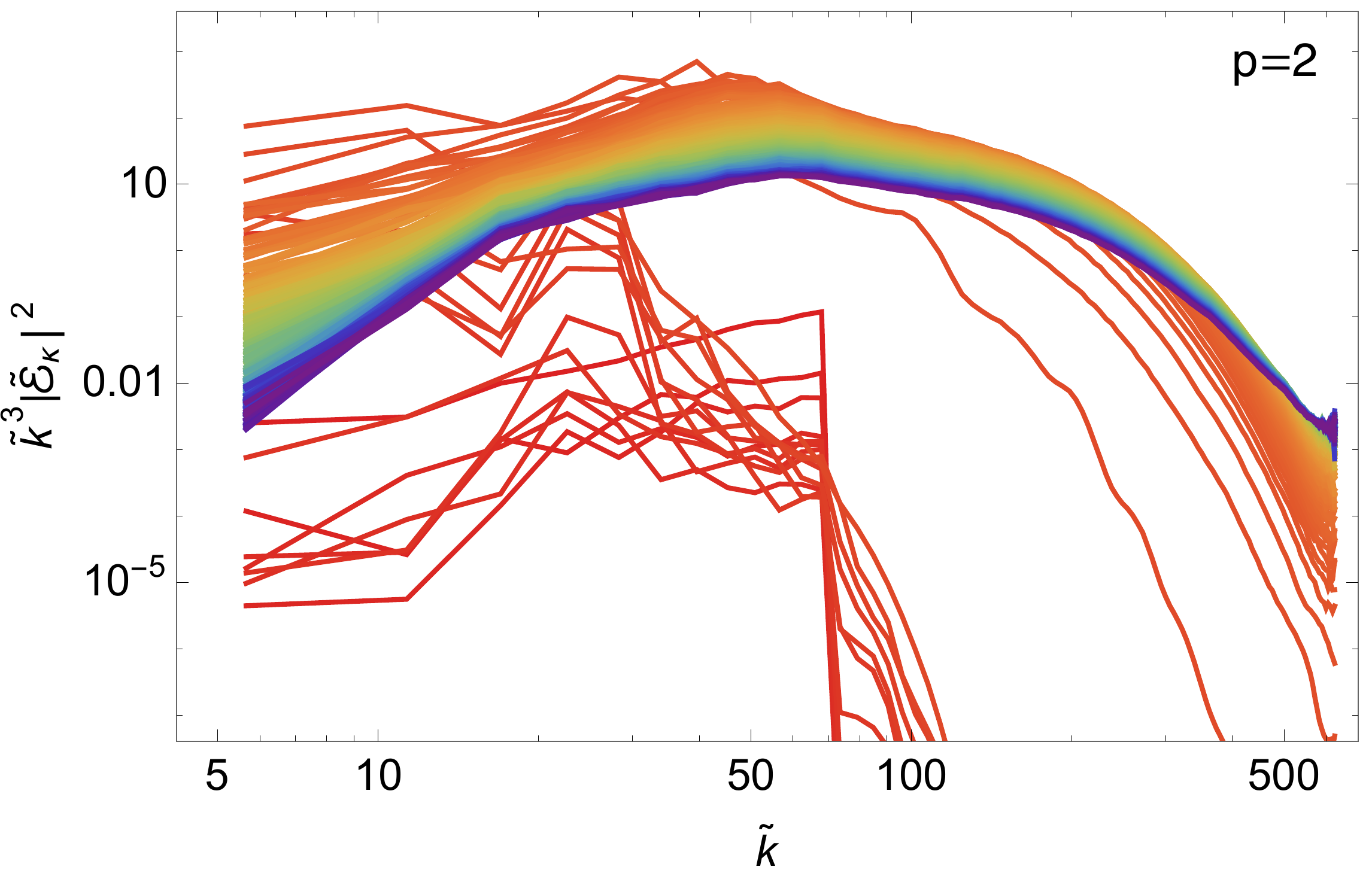} \hspace{0.3cm}
    \includegraphics[width=8cm]{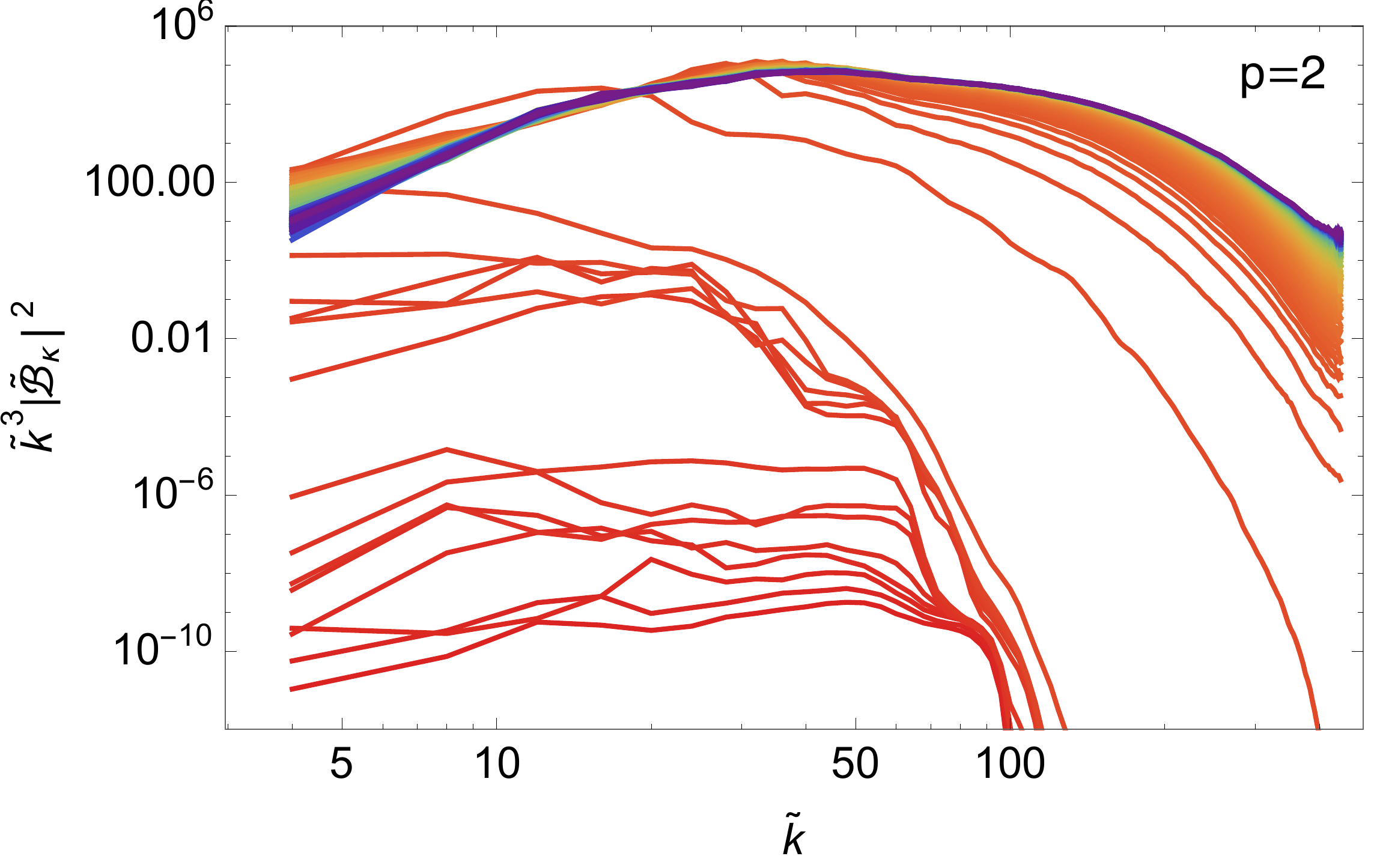} \\ \vspace*{0.2cm}
        \includegraphics[width=8cm]{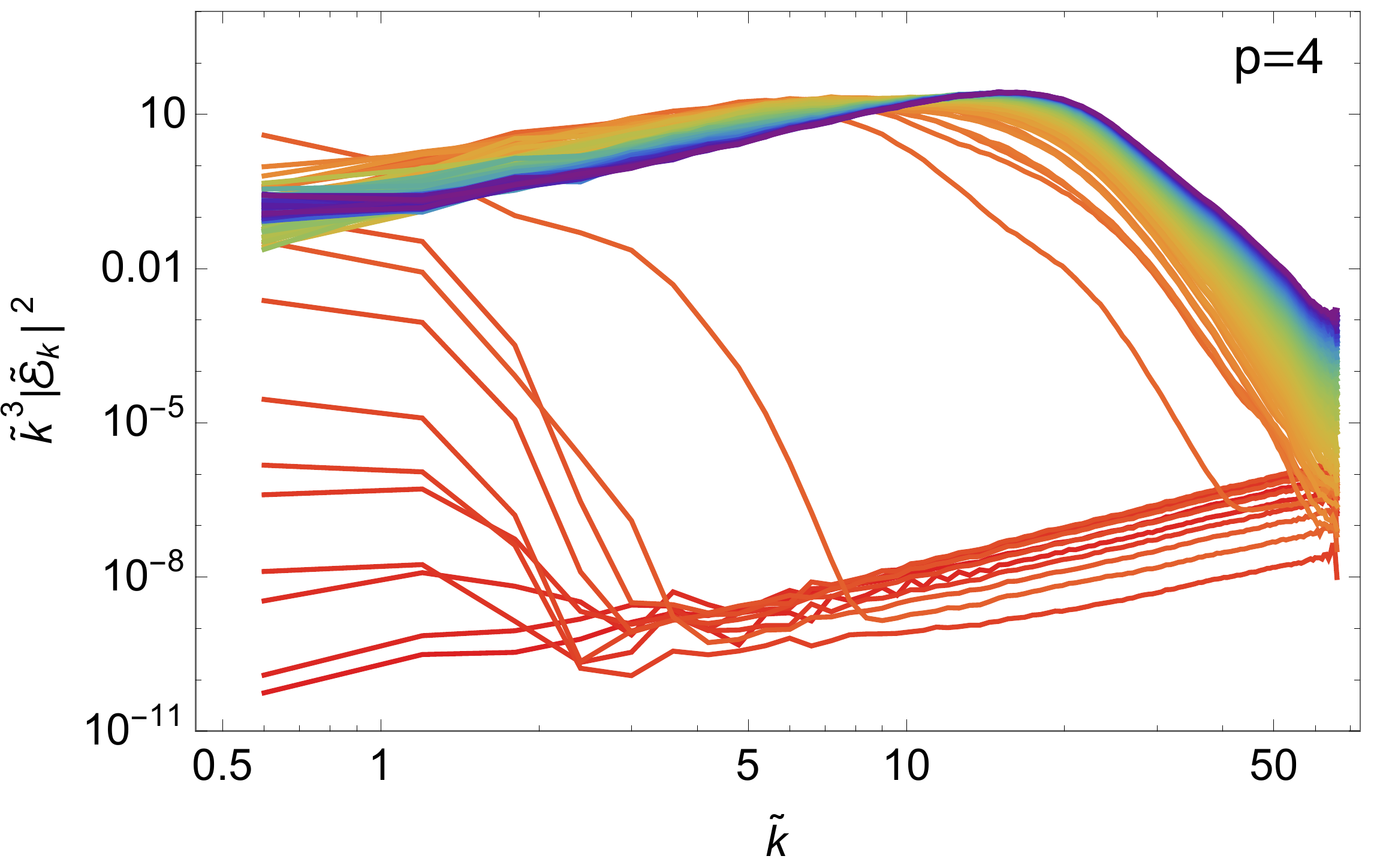} \hspace{0.3cm}
    \includegraphics[width=8cm]{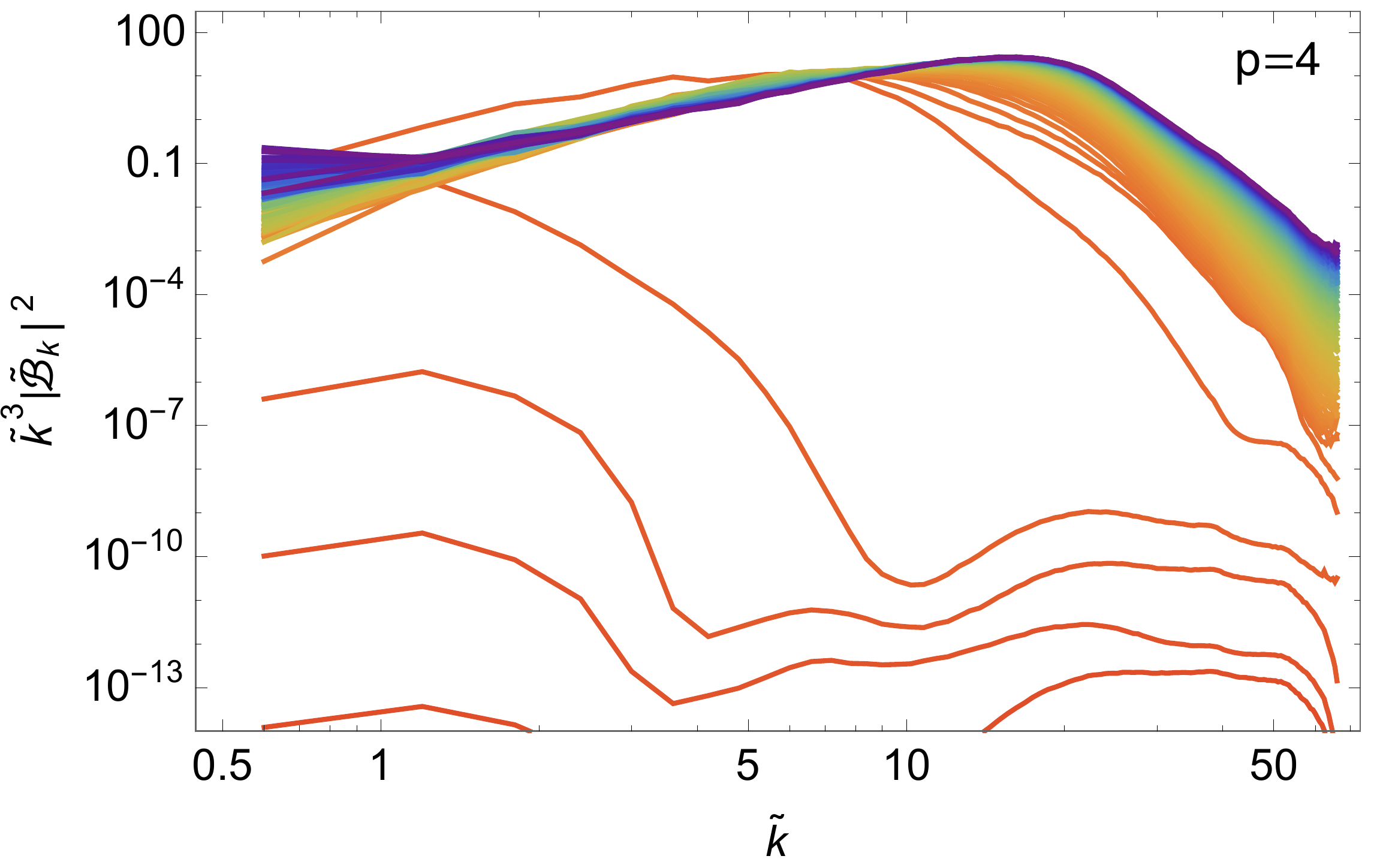} \\ \vspace*{0.2cm}
        \includegraphics[width=8cm]{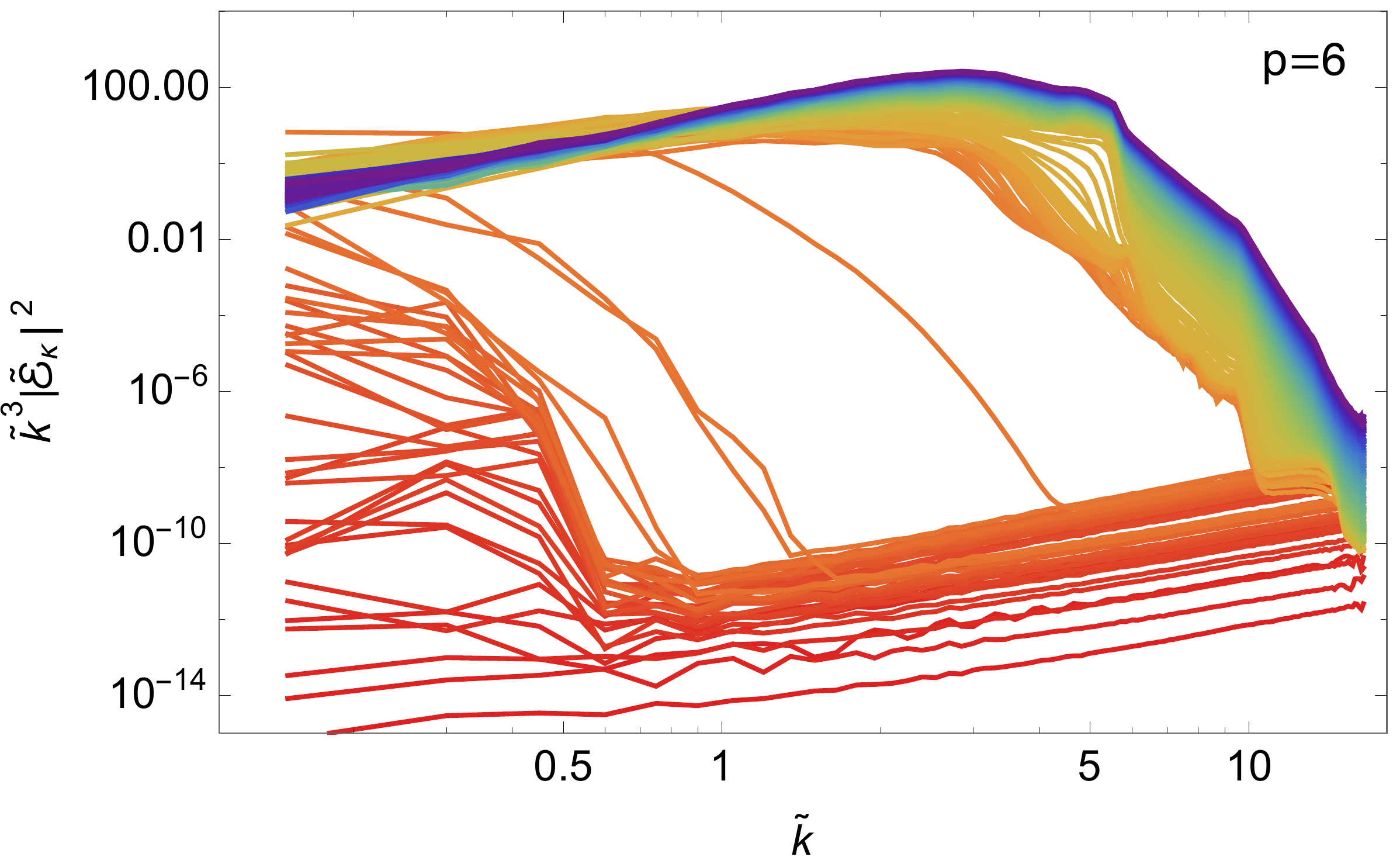} \hspace{0.3cm}
     \includegraphics[width=8cm]{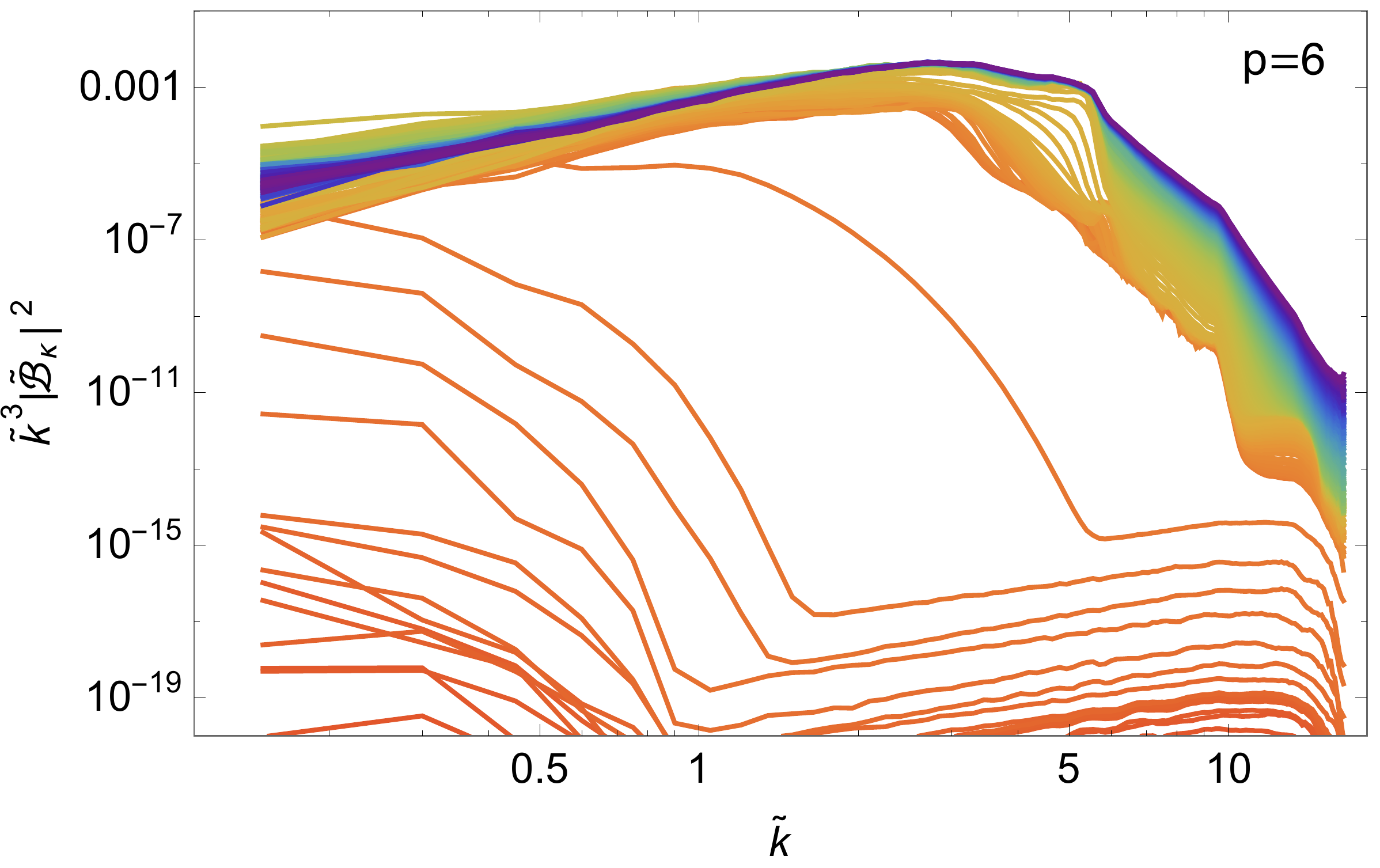} \\ \hspace{0.3cm}
    \caption{Spectral evolution of the electric field $\tilde{k}^3 |\tilde{\mathcal{E}}_k|^2$ (left panels) and magnetic field $\tilde{k}^3 |\tilde{\mathcal{B}}_k|^2 $ (right panels), as a function of $\tilde{k} \equiv k / \omega_*$, for the $U(1)$ gauge simulations with $p=2$, $4$, $6$. Each line shows the spectra at different moments of the evolution, going from red lines (early times) to purple lines (late times).}
    \label{fig:U1spectra}
\end{figure}

It is very interesting to analyze how the energy is distributed at very late times in the simulation, i.e.~well within the non-linear regime. This was studied recently in Ref.~\cite{Antusch:2020iyq} in the context of a real singlet inflaton with the same potential as here, coupled to a massless scalar singlet via a quadratic interaction. Although here we are considering  a gauge sector, the explanation developed in Ref.~\cite{Antusch:2020iyq} also applies here. In particular, we find that the energy distribution at late times is determined by the choice of $p$ in the inflaton potential. For $p=2$, the inflaton cannot get excited via self-resonance, but the daughter field does get excited via broad parametric resonance because $q_{A*}>1$. However, the effective resonance parameter (\ref{eq:EffRes}) decreases with time, so parametric resonance eventually becomes narrow. After that, the inflaton kinetic and potential energies dilute as matter, while the other ones dilute as radiation or faster. Due to this, at very late times we get the energy ratios $\epsilon_{_{K}}^{{\varphi}}, \epsilon_{_
{V}} \rightarrow 0.5$, with the other ratios becoming negligible. This explains why the effective equation of state goes to $w \rightarrow 0$ at late times in Fig.~\ref{fig:Table-phisfw}. On the other hand, for $p=4$, $6$, both the inflaton and the gauge fields are always being excited resonantly at late times: inflaton self-resonance is always present for $p > 2$, while parametric resonance is always broad at late times because $q_{\rm res}$ is either constant (for $p=4$) or grows with time (for $p > 4$). Therefore, the energy contributions of both field sectors are sizeable at late times. In the case $p=4$, the inflaton possesses 60\% of the total energy of the system  at very late times (divided by half between kinetic and gradient energy), while the gauge fields possess the other 40\% (divided also by half between electric and magnetic energy). Moreover, the inflaton potential energy becomes negligible, which explains why the effective equation of state goes to $w \rightarrow 1/3$ at late times in Fig.~\ref{fig:Table-phisfw}. We expect this final distribution to be quite independent on the choice of $q_*$, as will be seen in Ref.~\cite{NonAbelianInpreparation} in a slightly different context. This contrasts with the simulations of the analogous scalar theory simulated in Ref.~\cite{Antusch:2020iyq}, where the energy is distributed 50\%-50\% between the inflaton and the daughter field. On the other hand, in the gauge simulations for $p=6$ we do observe equipartition between inflaton and gauge energies at late times, although the simulation is not long enough in this case to determine if this distribution will hold for later times, or if it will slowly evolve towards the 60\%-40\% distribution observed for $p=4$.

Finally, we show in Fig.~\ref{fig:U1spectra} the spectra of the electric and magnetic fields for the three power-law potentials $p=2$, 4, 6 considered here. As expected from the linear analysis, mainly field modes within an infrared band $\tilde{k} \equiv k / \omega_* \lesssim q_*^{1/4}$ grow exponentially during the initial linear regime, at times $\tilde{\eta} < \tilde{\eta}_{\rm br}$. However, when backreaction happens at time $\tilde{\eta} = \tilde{\eta}_{\rm br}$, the growth of the infrared band saturates, and the different fields start populating modes of increasingly high momenta due to rescattering. The spectra eventually saturate, showing a peak at larger scales. This process is qualitatively similar for the different choices of $p$ considered here.

\subsubsection{Accuracy tests}

\begin{figure}
    \centering
    \includegraphics[width=9cm]{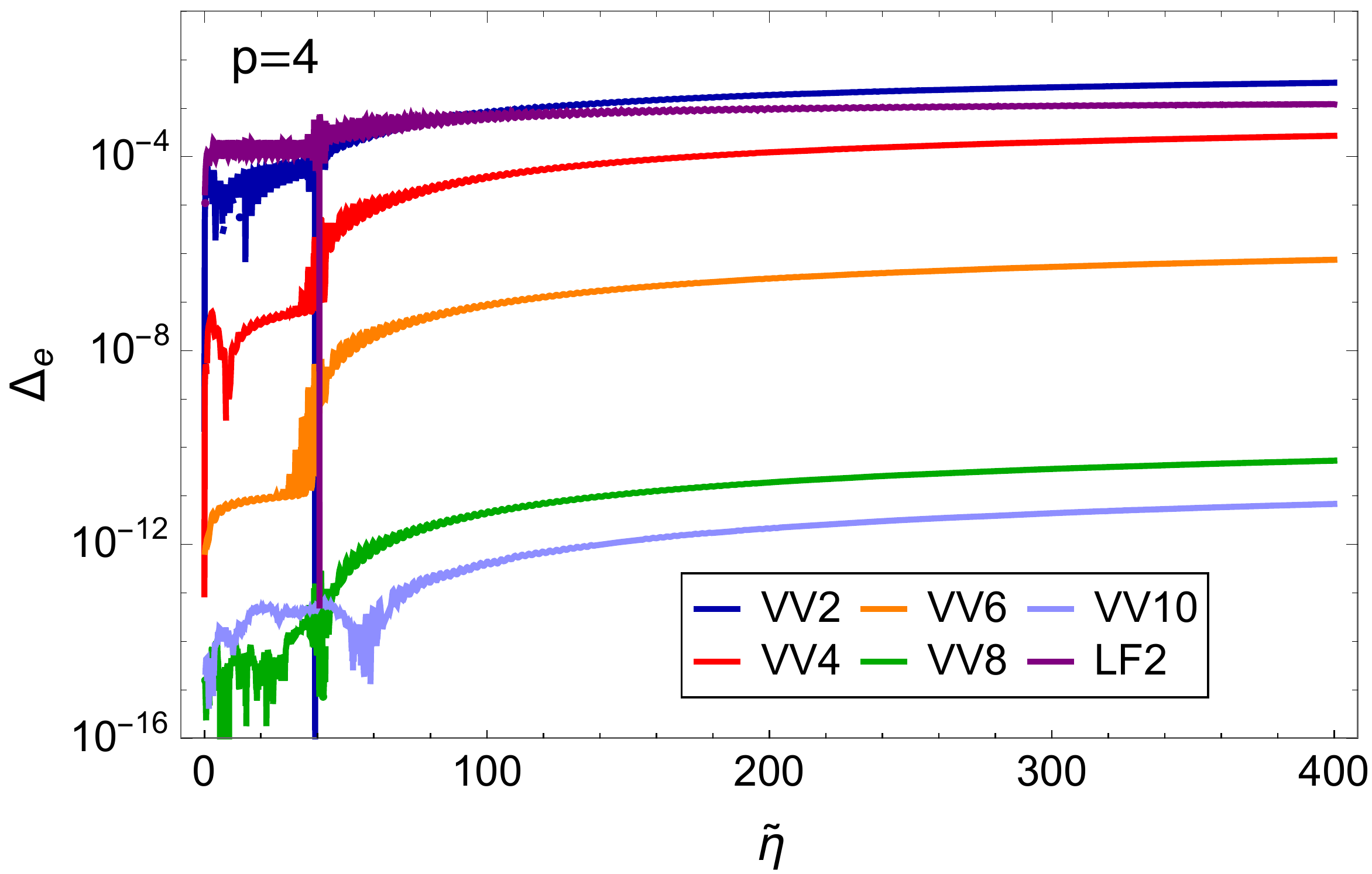} \hspace{0.3cm}
    \caption{Comparison of ``energy conservation" in the $U(1)$ gauge simulation with $p=4$, for different evolution algorithms: velocity-Verlet (orders 2, 4, 6, 8, and 10), as well as staggered leapfrog (order 2).}
    \label{fig:U1-EnergyConstraint}
\end{figure}

In flat space and in conservative systems, energy conservation can be used to monitor the precision of evolution algorithms. However, we are now considering an expanding universe, and in particular, we are using the second Friedmann equation (\ref{eq:KernelU1totII}) to evolve the scale factor. In this context we can instead check that the first Friedmann equation (\ref{eq:HubbleGauge}) holds during the evolution. However, contrary to the Gauss law (which is preserved by design when the discretized equations are gauge invariant), the first Friedmann's law will be only approximately respected. We will loosely refer to this first Friedmann's equation being respected as
``energy conservation", in analogy to the flat case. In particular, we require that the relative difference between the left and right hand sides of Eq.~(\ref{eq:HubbleGauge}), which we denote by $\Delta_{\rm e}$, obeys always $\Delta_{\rm e} \ll 1$. The better the accuracy of the evolution algorithm used to solve the lattice equations, the better the energy is ``preserved". In order to illustrate this, we show in Fig.~\ref{fig:U1-EnergyConstraint} the evolution of $\Delta_{\rm e}$ as a function of time, for the case $p=4$. The lattice equations have been solved with different accuracy orders of the velocity Verlet algorithm, introduced in Section \ref{sec:Verletint}. As expected, the higher the order, the better the
conservation of energy is preserved: the violation of energy conservation at time $\tilde{\eta} \simeq 400$ is $\Delta_{\rm e} \simeq 3\cdot 10^{-3}$, $2\cdot 10^{-4}$, $7 \cdot 10^{-7}$, $5 \cdot 10^{-11}$, and $6 \cdot 10^{-12}$, for VV2, VV4, VV6, VV8, and VV10 respectively. This means that ``energy conservation" improves by factors $\sim$12, 360, $10^4$, and 8, as we increase the order of the integrator from one to the next one, i.e.~from VV2 to VV4, VV4 to VV6, etc. Interestingly, the value of $\Delta_{\rm e}$ saturates for VV10: in that case, the error in the scale factor constraint is due exclusively to an accumulation of machine precision errors, so using velocity-Verlet algorithms of higher-orders than VV10 will not improve the energy constraint anymore. Of course, the negative side of using higher-order iterators is the increase of the required computation time. Finally, we have also solved the field dynamics with a second-order staggered leapfrog algorithm (see iterative scheme III or IV in Section \ref{subsec:StaggeredLFaction}), which we denote as LF2. Remarkably, this algorithm slightly improves ``energy conservation" at late times with respect VV2, as observed in the Figure.

Let us now focus on the conservation of the Gauss constraint, given by Eq.~(\ref{eq:GaussU1-eom}) in the continuum. As already mentioned, the Gauss constraint  must be always satisfied up to machine precision, \textit{independently of the accuracy of the integrator}, as it is a direct consequence of the lattice equations of motion: a violation of the Gauss constraints is a violation of gauge invariance. We show in Fig.~\ref{fig:GaussLawU1} the relative difference between the left and right hand sides of Eq.~(\ref{eq:GaussU1-eom}) as a function of time, which we denote as $\Delta_{\rm g}$. At the onset of the simulation we get $\Delta_{\rm g} \sim 10^{-9}$, which is explained by the $\sim 7$ orders of magnitude of difference between the amplitudes of the inflaton homogeneous mode and its fluctuations. After backreaction, the relative difference decreases down to $\Delta_{\rm g} \sim 10^{-13}$, and starts increasing slowly from then on, due to a constant accumulation of machine precision errors.

\begin{figure}
    \centering
    \includegraphics[width=8cm]{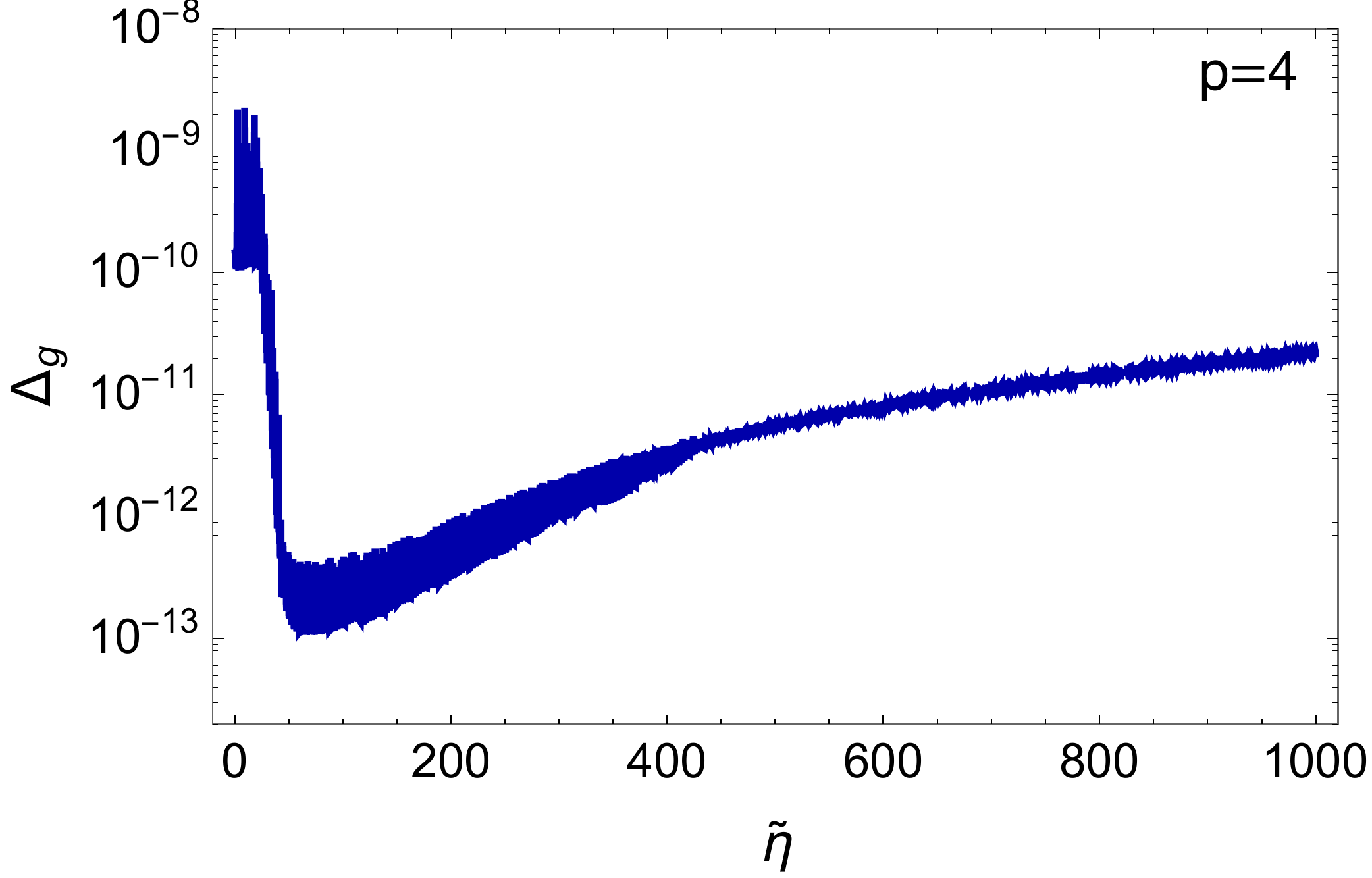} \hspace{0.3cm}
    \caption{Conservation of the Gauss law as a function of time for the $U(1)$ gauge simulation with $p=4$.}
    \label{fig:GaussLawU1}
\end{figure}

\subsection{Lattice simulations: SU(2)\texorpdfstring{$\times$}{x}U(1) gauge interactions} \label{sec:SU2U1-sims}

We now consider a scenario in which a complex doublet $\Phi$ with potential \eqref{eq:InflPot}  [where we must substitute $|\phi | \rightarrow \sqrt{2} |\Phi |$] acts as an inflaton field, simultaneously coupled to 1) a $SU(2)\times U(1)$ gauge sector via a gauge-invariant covariant derivative, and 2) a massless singlet scalar field $\chi$ via a quadratic interaction $V_{\rm int} (|\Phi|, \chi) \equiv g^2 |\Phi|^2 \chi^2$, with $g$ a dimensionless coupling constant. The strength of the parametric resonance is determined, in each case, by the corresponding resonance parameters: $q_*$ for the singlet scalar field (defined as in Eq.~(\ref{eq:ResPar}) with $\phi_* \equiv \sqrt{2} |\Phi_*|$), and $q_{A*}$ and $q_{B*}$ for the $U(1)$ and $SU(2)$ gauge fields (see Eqs.~\ref{eq:ResParGauge}). Here we fix $q_{A*} = q_{B*} = q_*$ for illustrative purposes, with $q_* > 1$ to have broad parametric resonance for all daughter field sectors. We have simulated the preheating process for the power-law coefficients $p=2$, $4$, and studied the post-inflationary dynamics of the system. The lattice and model parameters chosen for the simulations are given in Table \ref{tab:LatticeParameters}, and are similar to the analogous $U(1)$ simulations. In particular, we choose again $M=10m_p$, which ensures that the inflaton always oscillates in the positive-curvature region of its potential. Similarly, the number of points and volume of the lattice are chosen, in each case, to resolve well both the infrared resonant bands, as well as the following propagation of the spectra towards the UV after backreaction.

The evolution of the inflaton amplitude, equation of state, and scale factor are, in this case, qualitatively similar to the examples shown for the $U(1)$ gauge simulation in the previous section. Therefore, we proceed to consider directly the evolution of the energy distribution, which differs in some aspects with respect to the $U(1)$ case.  We show in Fig.~\ref{fig:energiesSU2U1} the evolution of the total energy of the system during preheating, as well as of each of its different contributions, for $p=2$, 4. These are the kinetic and gradient energies of the scalars $\Phi$ and $\chi$, the electric and magnetic energies of the $U(1)$ and $SU(2)$ gauge sectors, the inflaton potential energy $\tilde{V}_{\rm pot} \equiv \tilde{|\Phi|}^4$, and the interaction energy $\tilde{V}_{\rm int} \equiv q_* |\tilde{\Phi}|^2 \tilde{\chi}^2$ between $\Phi$ and $\chi$. As in the $U(1)$ case, we have multiplied the different energies by the appropriate scale factor term, so that the (oscillation-averaged) total energy is constant during the initial linear regime. We also show the  evolution of the energy ratios $\epsilon_{\rm i} \equiv \tilde{\rho}_i / \tilde{\rho}$, which sum one by construction.

As expected, the energy budget is dominated by the inflaton homogeneous mode initially, so the kinetic and potential energies of the inflaton dominate over all the other energy contributions. However, the kinetic and gradient energies of all daughter fields grow exponentially due to broad parametric resonance, as well as the inflaton gradient energy due to self-resonance. These contributions become sizeable enough at a certain time scale, destroying the inflaton homogeneous condensate via backreaction effects. As before, we denote this time scale as the \textit{backreaction time} $\tilde{\eta}_{\rm br}$. From the simulations, we get $\tilde{\eta}_{\rm br} \simeq 60, 40$ for $p=2$, $4$ respectively.  From then on, the non-linearities of the field EOM can no longer be ignored, as they affect the dynamics of the system, and lead eventually into achieving a stationary regime at late times. As in the $U(1)$ gauge simulation, we observe that the system gets virialized very quickly, with the inflaton energies obeying $\langle E_K
^{\Phi} \rangle \simeq  \langle E_G^{\Phi} \rangle + \frac{p}{2} \langle E_V \rangle + \langle E_{\rm int} \rangle $ when averaged over oscillations. Also, we also have  equipartition between the kinetic and gradient energies of all daughter field sectors at late times, as can be observed in the Figure.

Let us now comment about how the energy distributes at very late times. Let us consider first the case $p=2$. Here we observe a qualitatively similar behaviour than in the equivalent $U(1)$ simulation: although the inflaton kinetic and potential energy ratios decay around backreaction time, at later times they start growing again. The reason is the same as in the $U(1)$ simulations: the inflaton does not get excited via self-resonance for $p=2$, while the parametric resonance of the daughter fields eventually becomes narrow (because the effective resonance parameter (\ref{eq:EffRes}) decreases with time). Therefore, at very late times neither of the two resonant phenomena is present,  and the inflaton slowly recovers all the energy of the system due to the different dilution rates of the energy contributions (the inflaton behaves as matter, while the daughter fields as radiation). Due to this, although our simulations are not long enough to observe this effect, we expect that $\epsilon_{_{K}}^{{\varphi}}, \epsilon_{_
{V}} \rightarrow 0.5$ at asymptotically late times. Moreover, this energy configuration also gives rise to a matter-dominated equation of state at late times, $w \rightarrow 0$.

   \begin{figure}
       \centering
       \includegraphics[width=7.8cm]{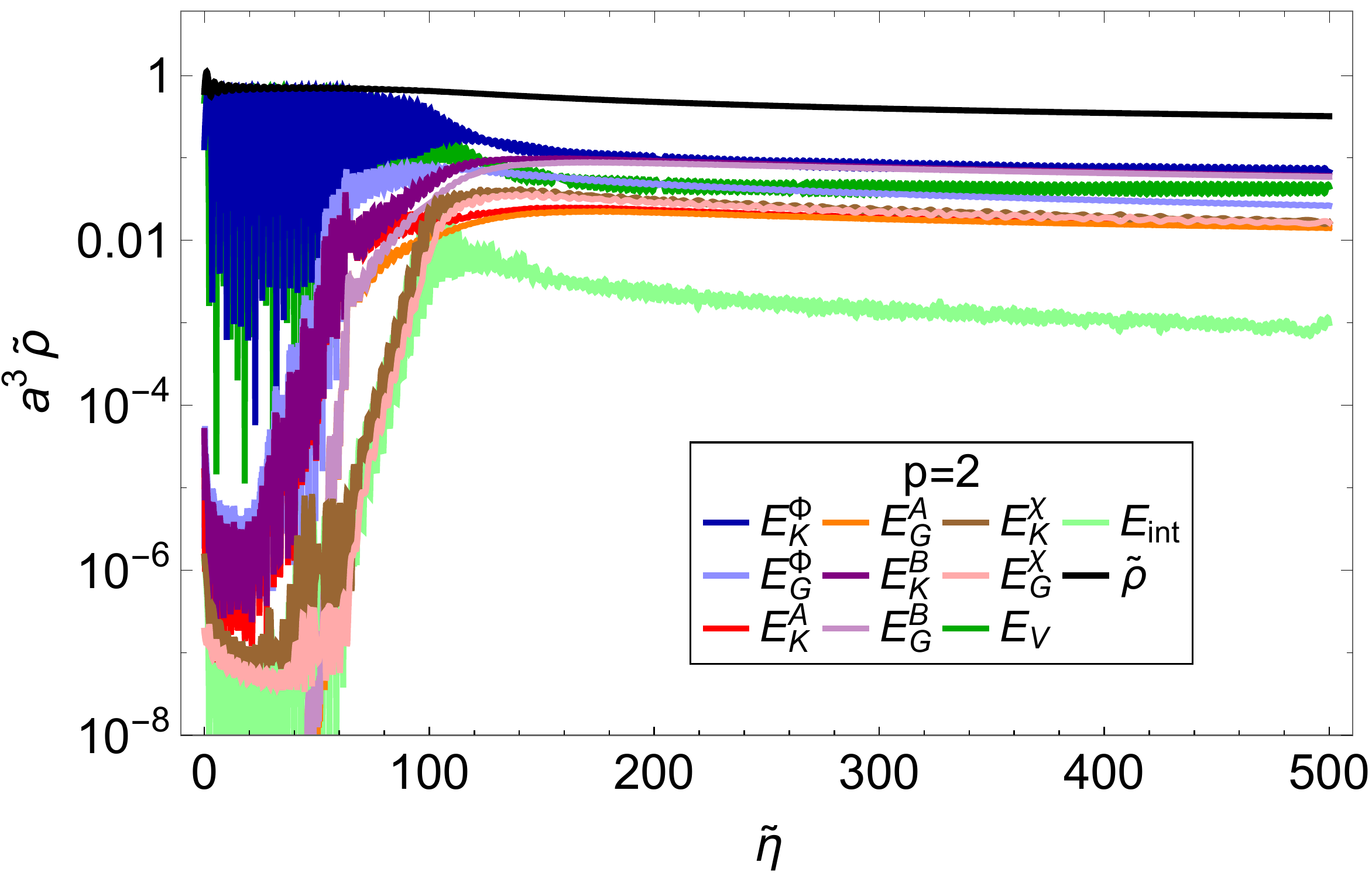} \hspace{0.3cm}
       \includegraphics[width=7.8cm]{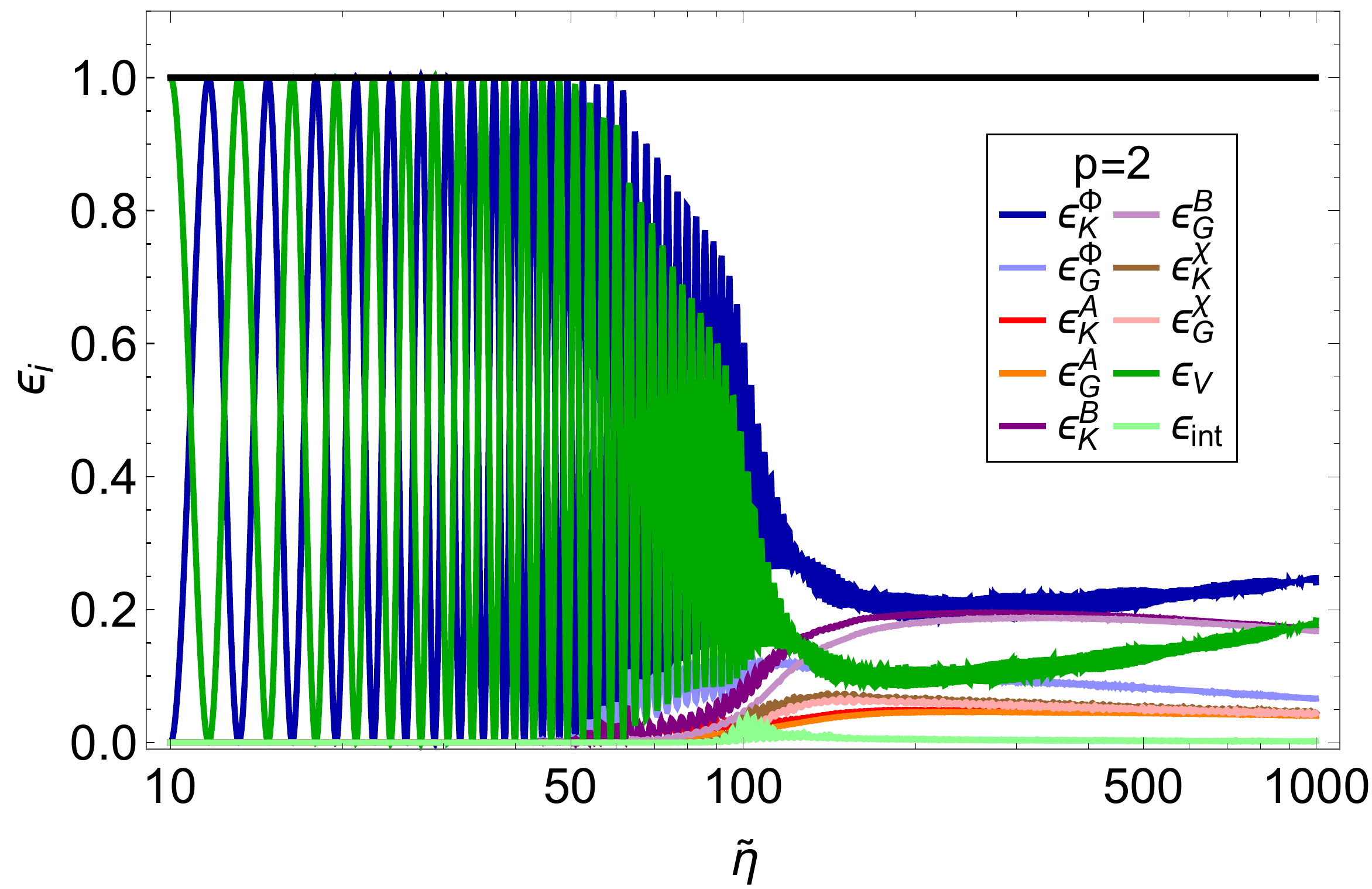} \hspace{0.3cm}
       \includegraphics[width=7.8cm]{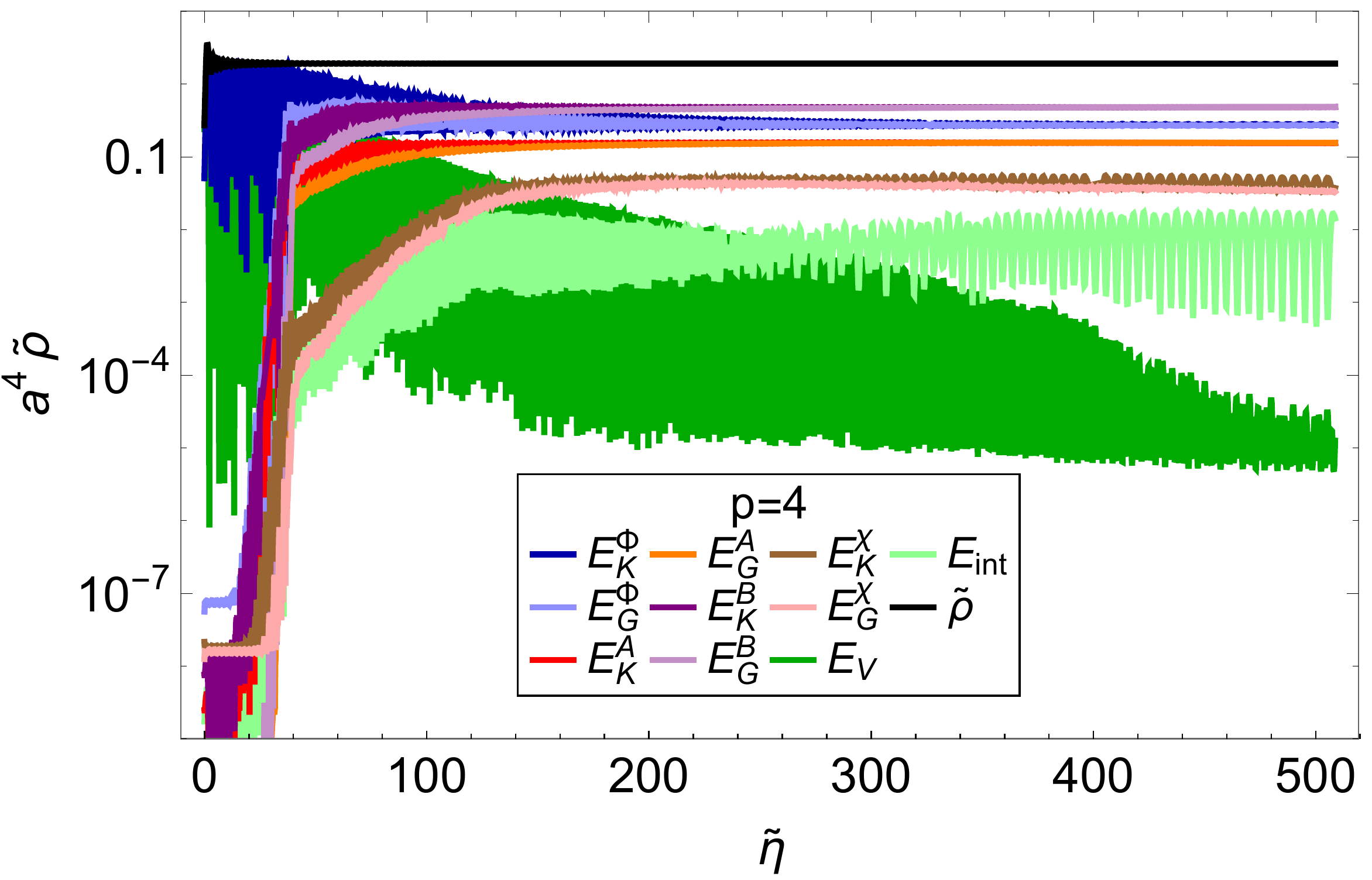} \hspace{0.3cm}
       \includegraphics[width=7.8cm]{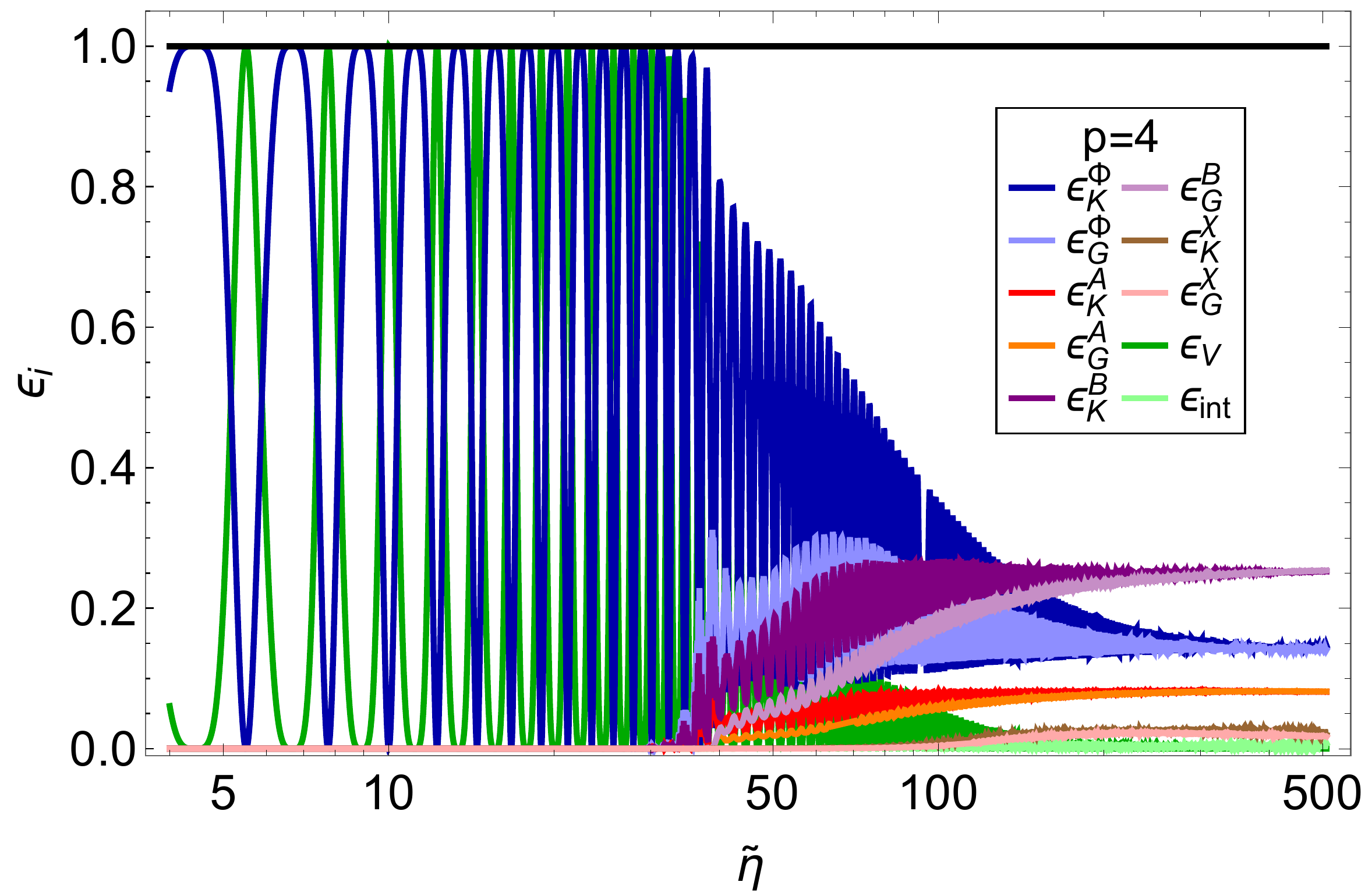} \hspace{0.3cm}
       \caption{Left panel: Evolution of the total energy in program units, $\tilde{\rho} \equiv \rho / \omega_*$, as well as of each of its energy contributions,  for the $SU(2)\times U(1)$+$\chi$ gauge simulations with $p=2,$ 4. Quantities are multiplied by the factor $\sim a^{\frac{6p}{p+2}}$. Right panel: Evolution of the energy ratios for the same simulations as in the left panels.}
       \label{fig:energiesSU2U1}
   \end{figure}

Let us focus now on the simulation with $p=4$. In this case, the effective resonance parameter (\ref{eq:EffRes}) remains constant. Therefore, as we have fixed $q_* = q_{\rm eff} > 1$ for all daughter field species (scalar $\chi$, Abelian and non-Abelian gauge bosons), they experience a broad parametric excitation during the whole time evolution of the system, including at late times. Similarly, the inflaton is also being excited due to the oscillations of its own homogeneous mode, and develops fluctuations via self-resonance. Neither of the two effects dies out, which could explain why neither of the two field sectors (inflaton or daughter fields) possesses 100\% of the total energy at asymptotically late times. In our particular scenario, we observe that at the end of the simulation, the inflaton possesses $\sim$30\% of the total energy, the scalar singlet $\sim$4\%, the $U(1)$ gauge sector $\sim$16\%, and the $SU(2)$ gauge sector $\sim$50\%. In each of the four cases, the energy is divided half and half between kinetic and gradient contributions. These results are in contrast with the analogous $U(1)$ simulation, which show that $\sim$60\% of the total energy remains in the inflaton at late times. From this result, we can conclude the (somewhat expected) result that the larger the number of daughter fields, the larger the amount of energy that gets transferred to them from the inflaton. Finally, let us also observe that the inflaton potential and inflaton-$\chi$ interaction energies go to zero at late times, $\varepsilon_V,\varepsilon_{\rm int} \rightarrow 0$, as in the analogous $U(1)$ gauge simulation. Due to this, the effective equation of state goes to $w \rightarrow 1/3$ at late times.

We also show in Fig.~\ref{fig:Plots-spectra:SU2U1} the evolution of the spectra of all fields involved: the norm of the inflaton $|\Phi|$, the scalar singlet $\chi$, and the electric and magnetic energies of the $U(1)$ and $SU(2)$ sector. We observe in all cases the same qualitative behaviour: first an exponential growth of the field modes within an infrared band, which saturates at backreaction time, followed by a propagation of the spectra towards the UV, populating modes of higher and higher momenta. The initial infrared growth of the gauge fields is in agreement with the linear analysis presented above, except in the case of the inflaton, which does not experience broad parametric resonance. The inflaton growth is, instead, triggered by backreaction effects from the daughter fields.

   \begin{figure}
       \centering
       \includegraphics[width=7.5cm]{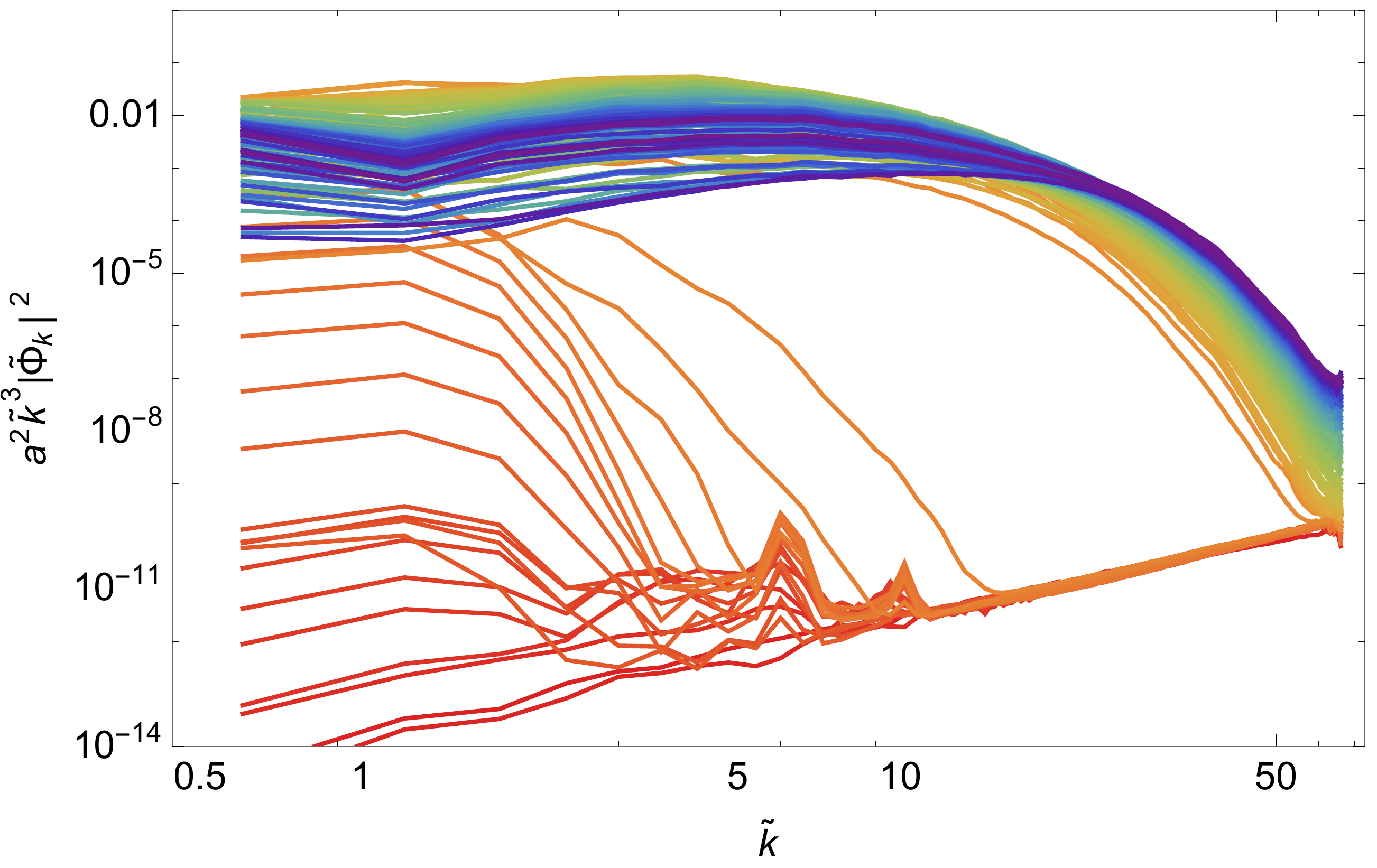} \hspace{0.3cm}
       \includegraphics[width=7.5cm]{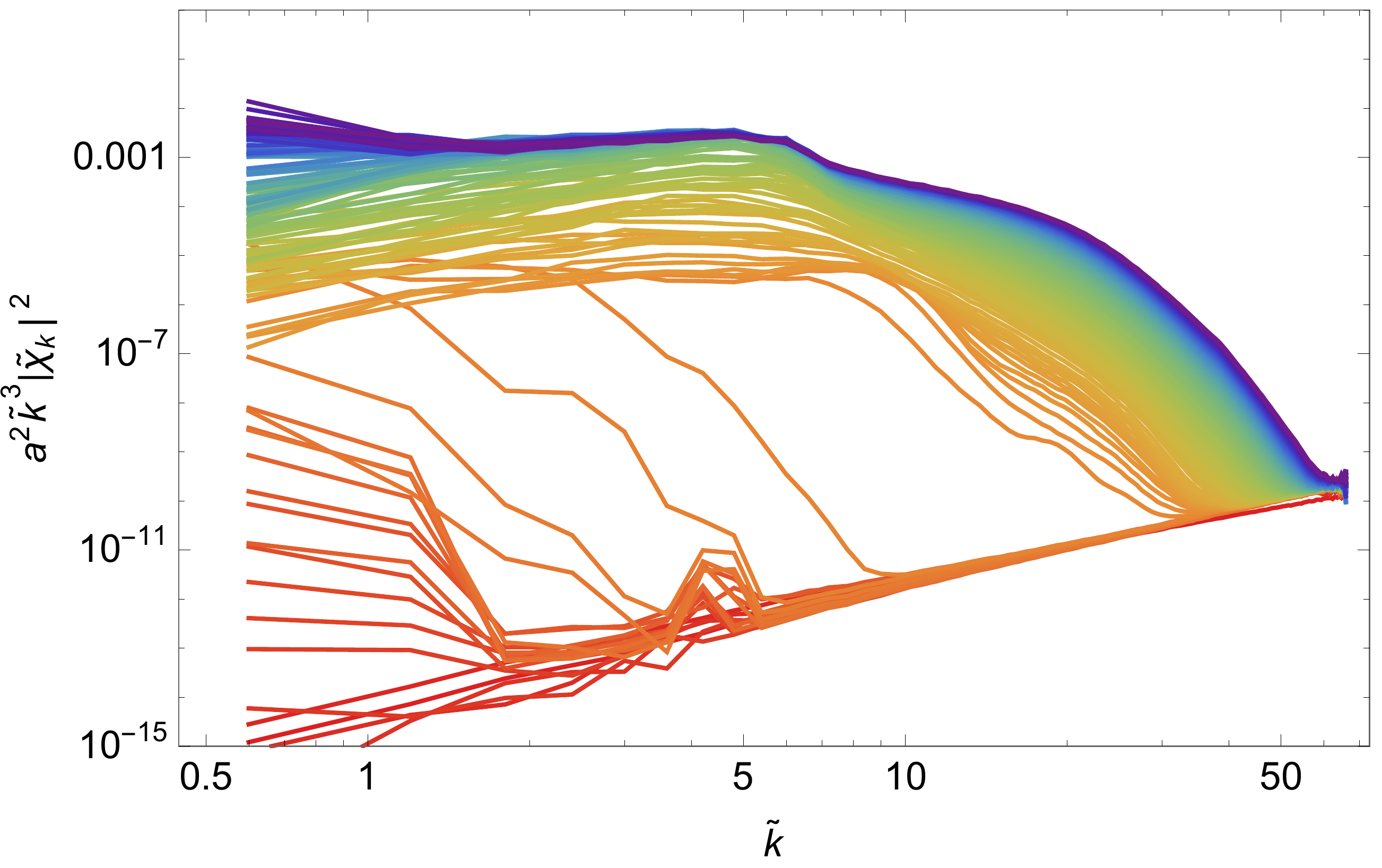} \vspace*{0.3cm} \\
       \includegraphics[width=7.5cm]{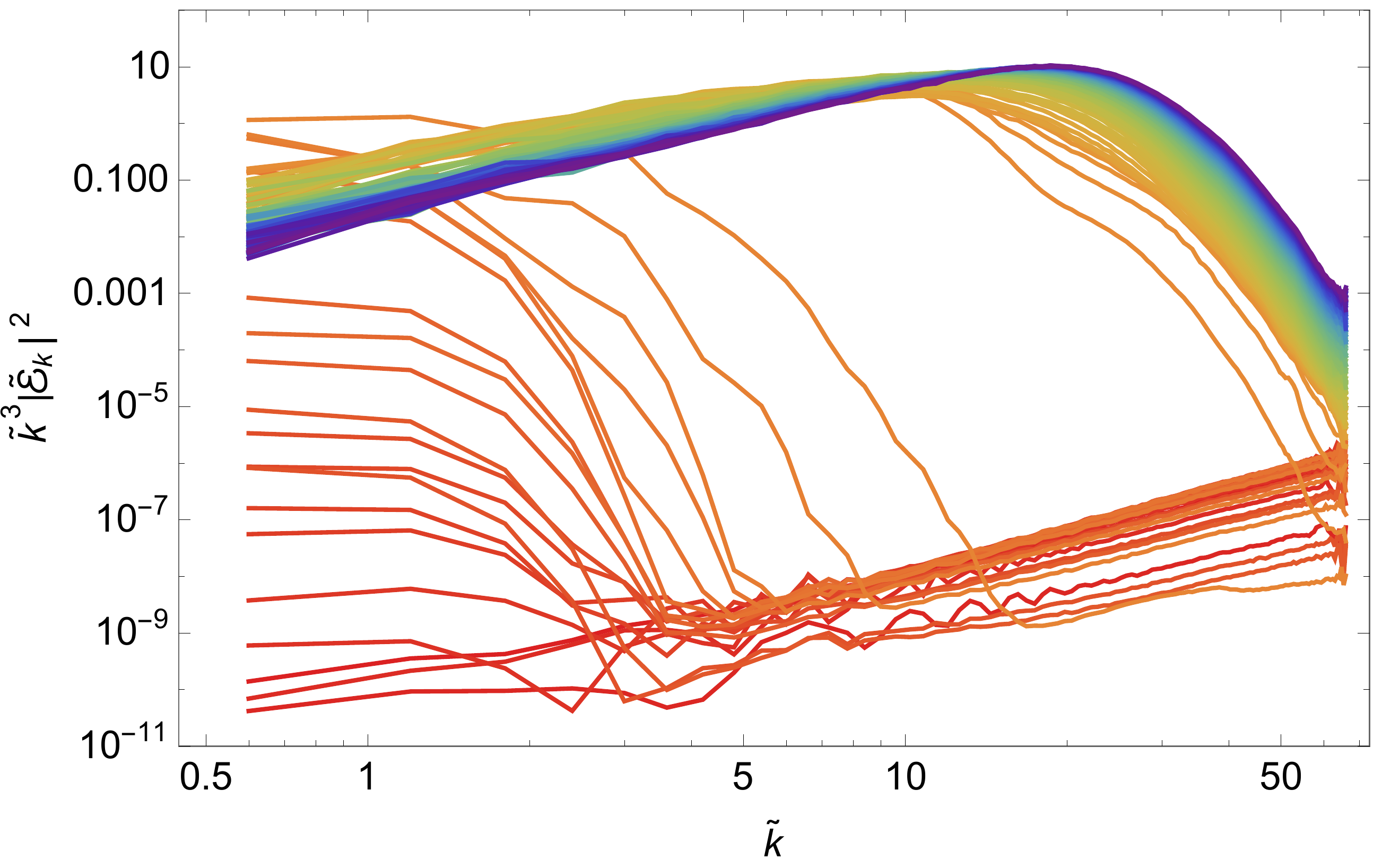} \hspace{0.3cm}
       \includegraphics[width=7.5cm]{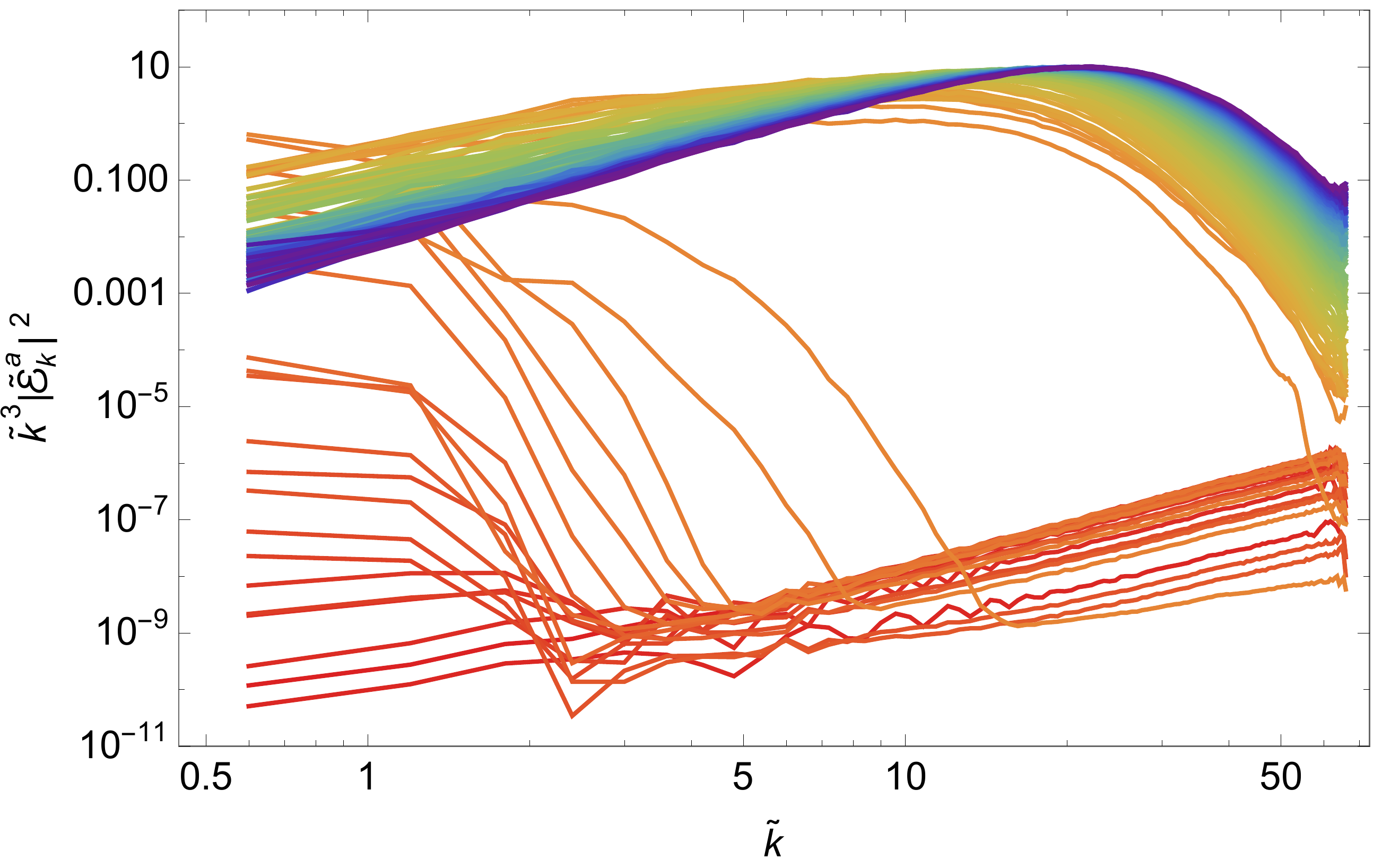}  \vspace*{0.3cm} \\
         \includegraphics[width=7.5cm]{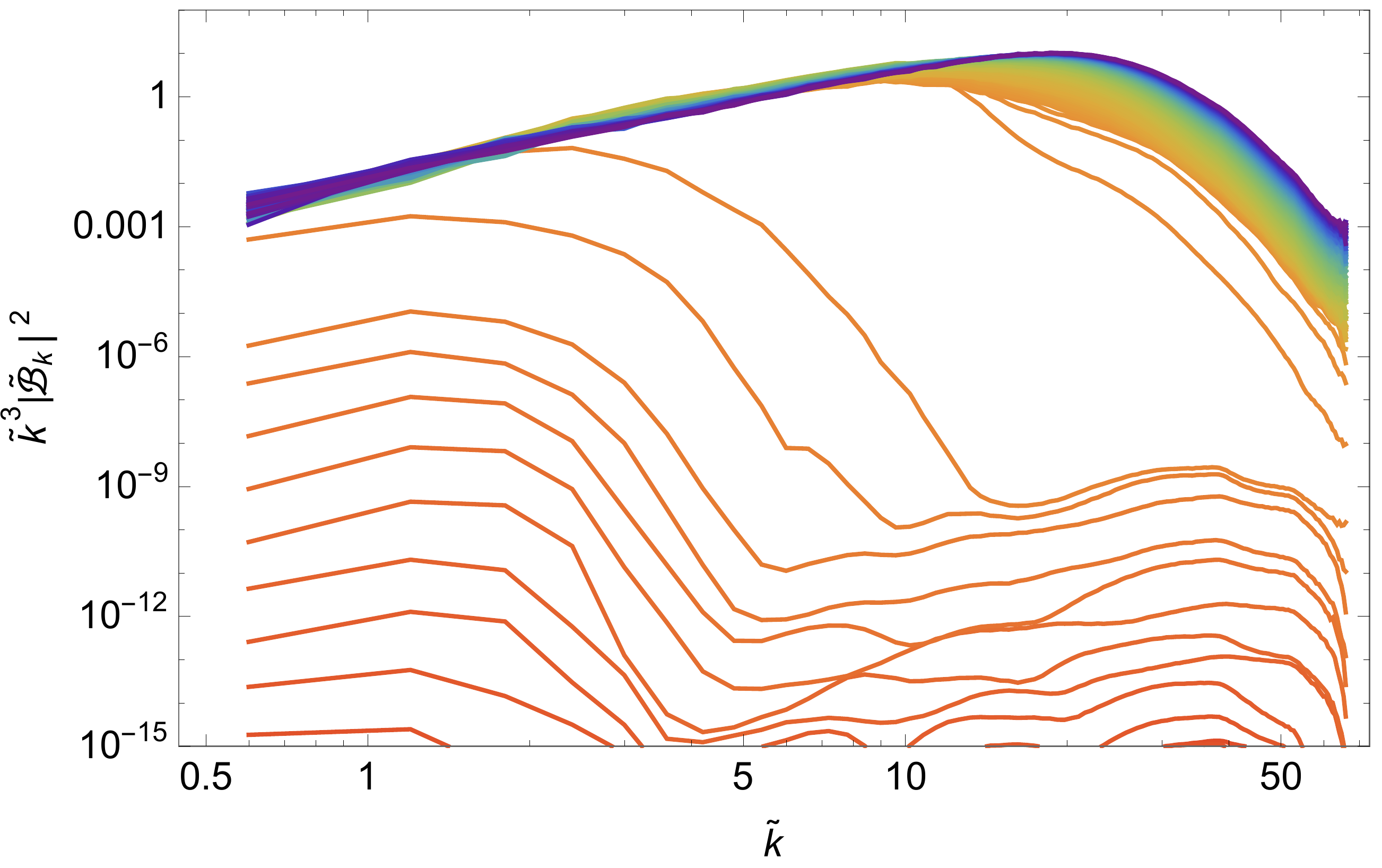} \hspace{0.3cm}
           \includegraphics[width=7.5cm]{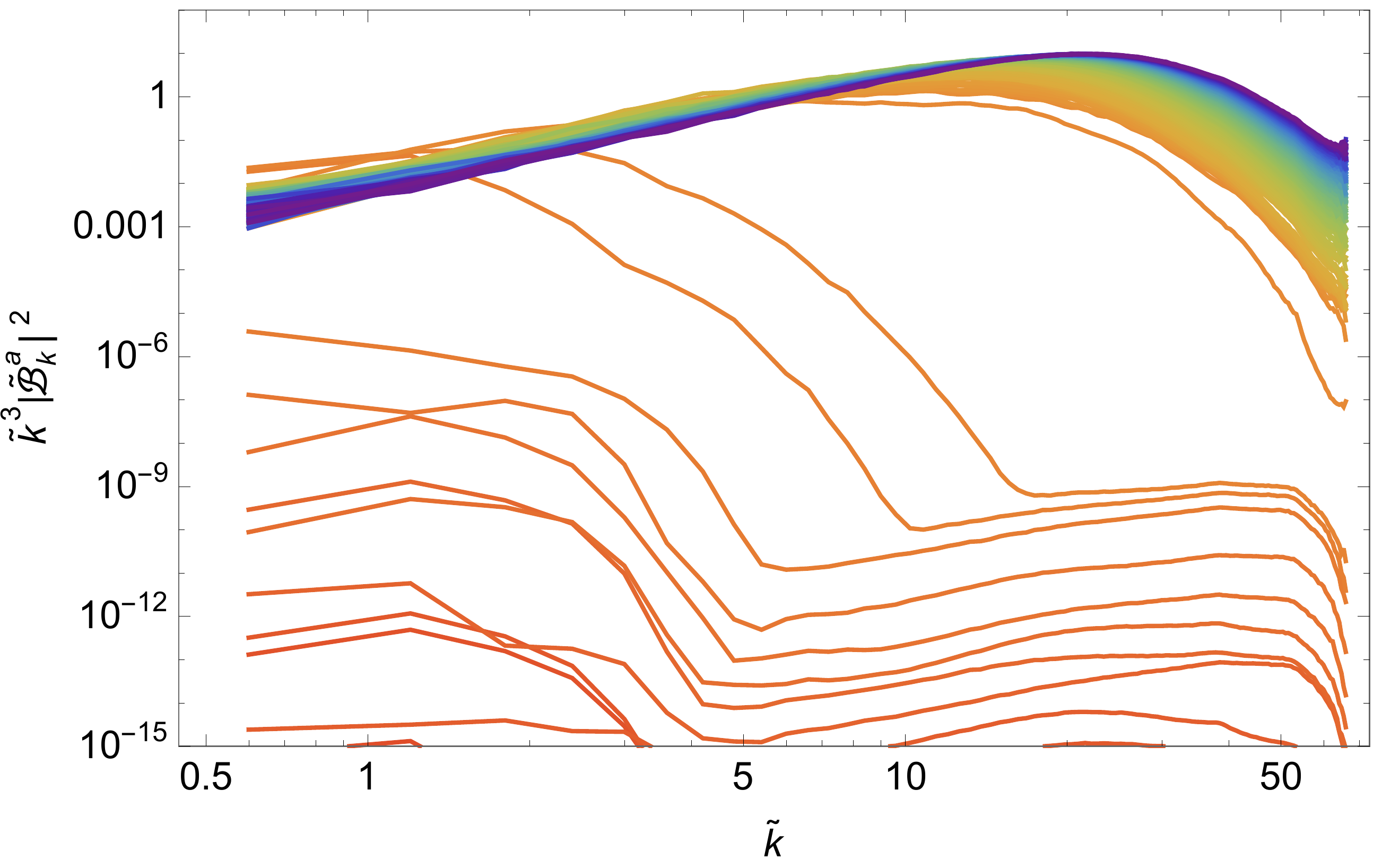}
       \caption{Spectral evolution of all the fields involved in the $SU(2)\times U(1)$+$\chi$ lattice simulations as a function of $\tilde{k} \equiv k /\omega_*$, for $p=4$. From left to right and from top to bottom, we show the inflaton, the massless scalar $\chi$, the electric and magnetic fields of the $U(1)$ gauge sector, and electric and magnetic fields of the $SU(2)$ gauge sector. Each line shows the spectra at different times during the field evolution, from red (early times) to purple (late times).}
       \label{fig:Plots-spectra:SU2U1}
   \end{figure}

\subsubsection{Accuracy tests}

Finally, it is always important to check that both ``energy conservation" and the Gauss constraints are preserved at all times during the simulation. Let us consider first the left of panel of  Fig.~\ref{fig:Gauss-energy:SU2U1}, where we show the relative difference between the left and right hand sides of the 1st Friedmann equation as a function of time (denoted as $\Delta_{\rm e}$), for $p=4$. Naturally, we require $\Delta_{\rm e} \ll 1$ in order to trust the results of our simulations. For illustrative purposes, we have solved the field dynamics with velocity-Verlet evolution algorithms of orders 2 and 4, for the same lattice and model parameters. As expected, the higher the accuracy of the integrator, the better
the energy is preserved: at time $\tilde{\eta} \simeq 340$ we get $\Delta_{\rm e} \simeq 2.2 \cdot 10
^{-3}$ for VV2, and $\Delta_{\rm e} \simeq 1.1 \cdot10^{-4}$ for VV4, i.e.~VV4 preserves energy a factor $\sim 20$ better than VV2. However, the negative side is that the required simulation time for VV4 increases with respect VV2, as expected. In principle, one should be able to improve the accuracy of the integrator arbitrarily up to machine precision, as in the analogous $U(1)$ gauge simulation shown in Fig.~\ref{fig:U1-EnergyConstraint}. This can be useful if one wants to apply this algorithm to any particular scenario requiring extremely good energy conservation. This is always at the expense, of course, of longer simulation times.

{Let us also mention a subtlety arising from our discretization and evolution schemes of a non-Abelian sector. As observed before in the absence of non-Abelian fields, the discretized Hubble law is satisfied up to corrections of $O(\delta\tilde\eta^p)$, but it is exact in the lattice spacing. Of course, the expressions differ from the continuum ones by some power of the lattice spacing, but the conservation is exact. This comes about from the fact that this discretized conservation law may be derived as a conserved quantity associated to a discrete action as written down in Sec.~\ref{subsec:StaggeredLFaction}. A discretized action can also be written for the non-Abelian case, but its obvious discretization in terms of plaquettes and lattice covariant gauge derivatives does not lead to the naive continuum discretization of the equations of motion Eqs. \eqref{eq:kernelsSU21}-\eqref{eq:kernelsSU25}. Instead it leads to a more involved discretized matter current. As we showed in this section, using the naive discretization does lead however to perfectly functional algorithms. We do expect nonetheless to have a contamination of $O(\delta \tilde x)$ in the Hubble's conservation law. This is manifested as a violation of the conservation law, which cannot reach machine precision, no matter the smallness of $\delta\tilde\eta$ or the accuracy of the evolver.}

In the right panel of Fig.~\ref{fig:Gauss-energy:SU2U1} we show how the Gauss laws are preserved during the simulation. In this case there are two Gauss laws that must be satisfied: one for the $U(1)$ sector, c.f.~Eq.~(\ref{eq:U1vector-Gauss_I}) in the continuum, and another one for the $SU(2)$ sector, c.f.~Eq.~(\ref{eq:SUNvector-Gauss_I}) in the continuum. We measure this by the parameter $\Delta_g$ which, as defined before, is the relative difference between the left and hand sides of the corresponding Gauss constraints.  As explained before, these constraints must be preserved up to machine precision independently of the chosen evolution algorithm, as they are a direct consequence of the gauge invariance that our careful discretization techniques maintain in the lattice equations. We observe a similar behaviour as in the analogous $U(1)$ gauge simulations: before backreaction we have $\Delta_g \sim 10^{-9}$ for both gauge sectors, due to the large relative difference between the amplitudes of the inflaton homogeneous mode and its fluctuations. After backreaction we get $\Delta_g \simeq 10^{-13}$, and from then on, the error slowly grows due to a constant accumulation of machine precision errors in each time step. At time $\tilde{\eta} \simeq 340$ we get $\Delta_g \sim 10^{-12}$ for both $U(1)$ and $SU(2)$ gauge sectors, which shows that both Gauss constraints are exceptionally well preserved during the simulation.

\begin{figure}
    \centering
     \includegraphics[width=7.5cm]{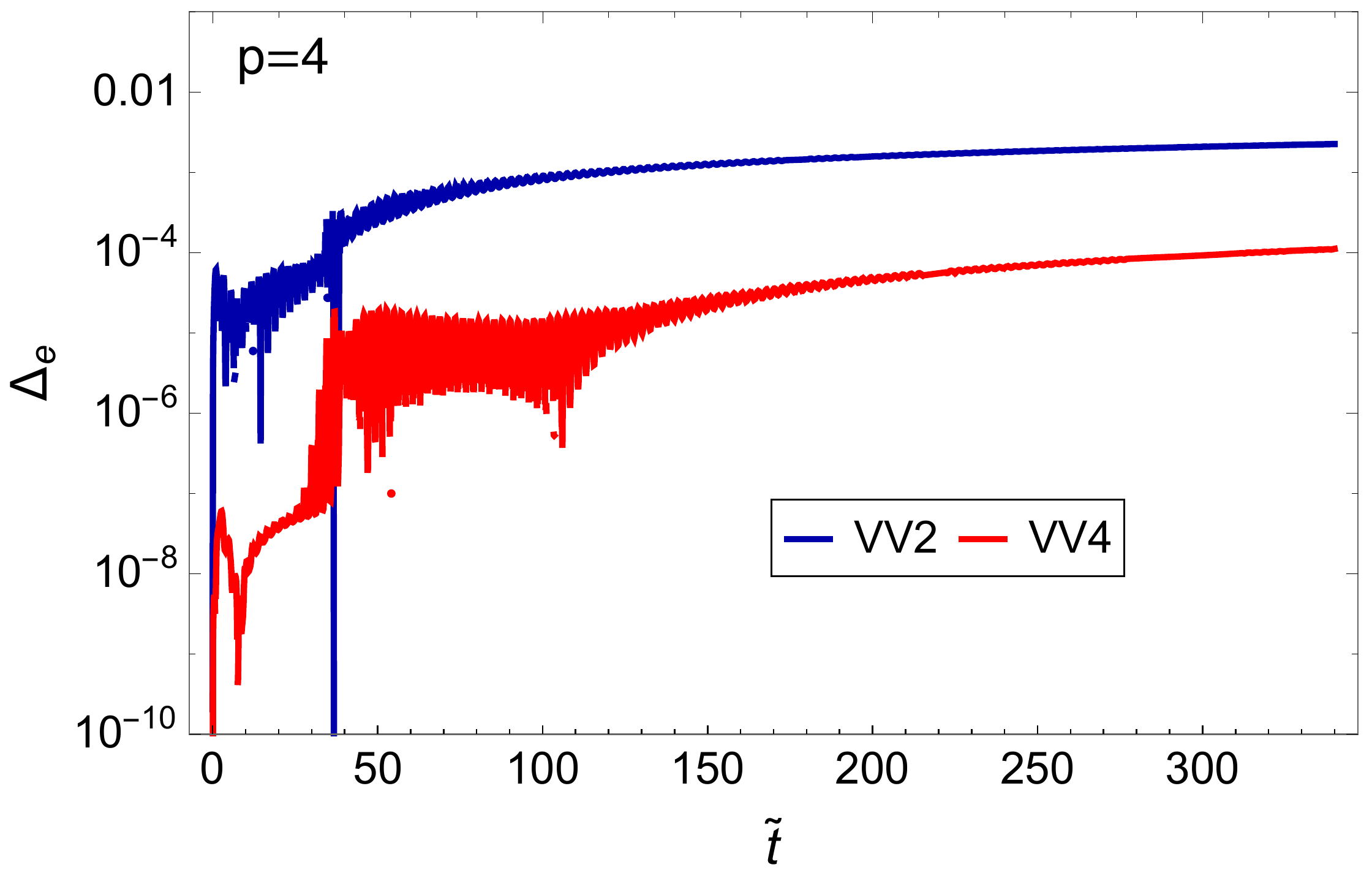}\,\,\,\,\,\,
    \includegraphics[width=7.5cm]{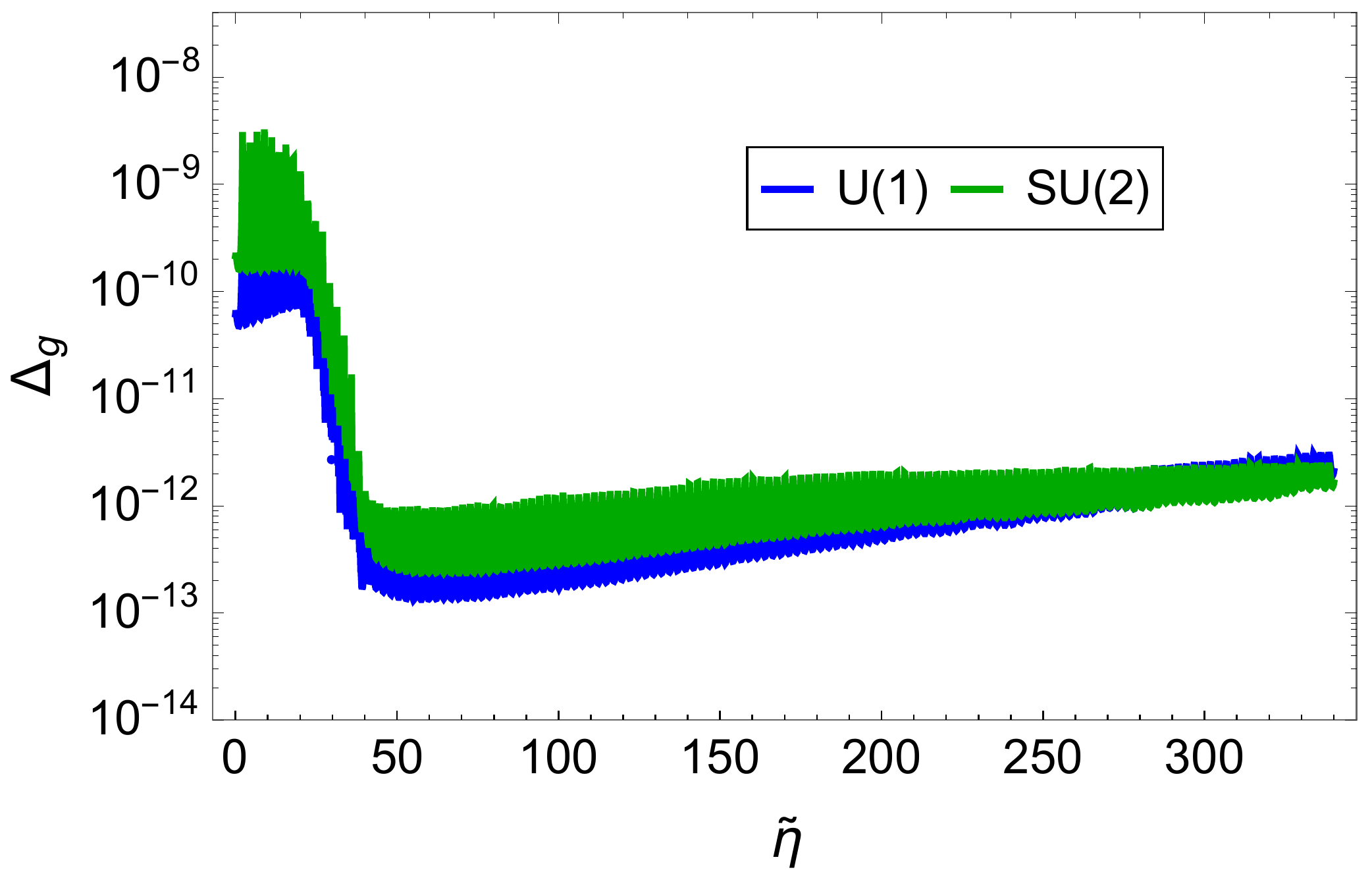}
    \caption{Left panel: Evolution of the relative error in ``energy conservation" for the $SU(2)\times U(1)$+$\chi$ simulation with $p=4$, obtained for the VV2 and VV4 algorithms. Right panel: Evolution of the relative error in the Gauss constraints of the $U(1)$ and $SU(2)$ gauge sectors, for the same simulation.}
    \label{fig:Gauss-energy:SU2U1}
\end{figure}

\section{Summary and outlook}

The present document represents {\it Part I} of our dissertation on lattice techniques for the simulation of non-linear dynamics in the early universe. Here we have focused on the lattice treatment of canonical scalar-gauge field theories in an expanding universe, considering an arbitrary number of interacting (real and complex) scalars and (Abelian and non-Abelian) gauge fields. This suffices to describe the majority of physically relevant scenarios in the early universe. In addition, we plan to discuss methods for non-canonical interactions in an upcoming {\it Part II}~\cite{PartII}, like those in theories with non-minimal gravitational couplings, or in general with non-minimal kinetic terms, as well as non-canonical interactions defined by the product between field variables and their conjugate momenta.

We summarize now the content of the present work. In Section~\ref{sec:ContinuumFldDynamics} we reviewed first the field dynamics of scalar-gauge theories in a continuum space-time, both with and without expansion of the universe. We considered a theory containing different kinds of scalars (singlets, $U(1)$-charged, and $SU(N)\times U(1)$-charged scalars) and (Abelian and non-Abelian) gauge fields. We wrote explicitly the EOM of such theory, as well as introduced the notation later used throughout the document. We then introduced in Section \ref{sec:LatticeApproach} basic concepts of lattice techniques, with a special emphasis on how to discretize appropriately gauge theories to preserve gauge invariance on the lattice. We then introduced basic evolution algorithms for the integration of the field EOM: staggered leapfrog and Verlet methods, with accuracy $\mathcal{O} (\delta t
^2)$, and (explicit) Runge-Kutta methods with accuracy up to $\mathcal{O} (\delta t^4)$. We also showed how some of these basic algorithms can be used as building blocks for the higher-order Yoshida and Gauss-Legendre integrators, with accuracy up to $\mathcal{O} (\delta t^{10})$.

In the following three sections we focused on developing lattice formulations for the different field sectors of the canonical theories considered here. In Section~\ref{sec:LatScalars} we considered the case of multiple interacting (singlet) scalar fields. We introduced a set of dimensionless field and spacetime variables, which we have called the \textit{lattice} or \textit{program} variables. When thoughtfully defined, these variables can be very useful when working on a lattice. Our lattice algorithms are therefore written in terms of these variables. We explained how to apply different evolution algorithms to solve the scalar EOM, as well as define different useful observables. In Section~\ref{sec:LatU1} we developed the same ideas for gauge theories with $U(1)$ interactions, and in Section~\ref{sec:LatSUN} for gauge theories with $SU(N)$ interactions. In Section~\ref{sec:InitCond} we described how to set the initial conditions for the different type of fields, both in the continuum and in the lattice. For scalar fields, we imposed a spectrum of vacuum fluctuations, which mimics the expected spectrum of quantum fluctuations in a FLRW universe. For gauge fields, we discussed how to set their initial conditions so that the Gauss constraint is preserved from the beginning.

Finally, in Section \ref{sec:WorkingExample} we simulated the dynamics of a specific scalar-gauge field model with \CLns, to illustrate some of the techniques presented in the previous sections. In particular, we considered the preheating dynamics of a charged inflaton, with monomial shape around the minimum of its potential. We considered two different scenarios: 1) a $U(1)$-charged scalar coupled to an Abelian gauge field, and 2) a $SU(2)\times U(1)$ charged scalar coupled to Abelian and non-Abelian gauge fields simultaneously, as well as to a scalar singlet. We considered different model parameters, and in particular, we studied different power-law coefficients in the monomial function. We studied the evolution during preheating of several relevant observables: the inflaton mean amplitude value, the evolution of the scale factor and of the equation of state, the energy distributions among field components, and the relevant field spectra. We showed explicitly how  each Gauss constraint is preserved to machine precision during the evolution of the system. We also demonstrated the power of the higher-order Verlet evolution algorithms implemented in \CLns, which can be used to obtain energy conservation up to machine precision in simulations of scalar-gauge theories in an expanding universe.

Let us emphasize here that, to the best of our knowledge, we are presenting for the first time an algorithm for non-Abelian $SU(N)$ gauge theories, which is symplectic, explicit in time, and preserving exactly the Gauss constraint, while solving for the expansion of the universe self-consistently. Furthermore, it can be made of arbitrary order. Besides, we also present higher-order integration algorithms for Abelian $U(1)$ gauge theories, also demonstrating explicitly their numerical implementation for the highest orders, including $\mathcal{O}(\delta t^{4})$, $\mathcal{O}(\delta t^{6})$, $\mathcal{O}(\delta t^{8})$ and $\mathcal{O}(\delta t^{10})$. Similarly, we also present higher-order integration algorithms for interacting scalar theories, going also to higher-orders, again including explicit implementations for $\mathcal{O}(\delta t^{4})$, $\mathcal{O}(\delta t^{6})$, $\mathcal{O}(\delta t^{8})$ and $\mathcal{O}(\delta t^{10})$.

The concepts and techniques discussed in this dissertation, in particular the explicit-in-time algorithms presented in Sections~\ref{sec:LatScalars}\,-\,\ref{sec:LatSUN}, are already implemented in \CLns, a \textit{user-friendly} and highly modular C++ MPI-based code, for lattice simulations of non-linear classical dynamics in an expanding universe. \CL is publicly available at \href{http://www.cosmolattice.net}{\color{blue} http://www.cosmolattice.net}. Most of the algorithms presented in this work are bundled in a high-level interface which allows the user to add almost effortlessly models with different interaction potentials, and easily add new integration algorithms. Moreover, the library has been designed in such a way to allow the user to use complex, vectorial and matricial representation of fields, to keep the lattice equations resembling as much as possible to the continuum ones. This level of abstraction is achieved through the use of compile-time code generation, using C++ expression templates, so that performance is never sacrificed.

The aim of this manuscript has been to illustrate different concepts of lattice gauge-invariant techniques and of general integration methods, which we have then specialized and adapted for their use in the context of canonical scalar-gauge field theories. We expect that the work we have developed so far here (soon to be complemented by {\it Part II}~\cite{PartII}) shall be useful for a large fraction of the research community interested in the early universe, let it be completely inexperienced researchers in lattice field theory simulations, or very experienced ones.
\vspace*{0.2cm}

To conclude, we comment on several aspects that we plan to explore in forthcoming works (either in \textit{Part II} or elsewhere), both in the near and mid-term future:
\begin{itemize}
\item  The development of lattice techniques for the discretization of theories with non-minimal kinetic terms. As the {\it drift} and {\it kick} functionals in these theories typically contain a linear combination of  conjugate momenta of other fields, explicit-in-time symplectic integrators (such as staggered leapfrog or Verlet integrators) are not appropriate. However, one can resort to explicit (non-symplectic) Runge-Kutta methods like those presented in Section~\ref{subsec:RungeKutta} (this has been done in~e.g.~\cite{Child:2013ria}), or even to higher-order implicit (yet symplectic) integrators like the Gauss-Legendre methods, presented in Section~\ref{subsec:GaussLegendre}. As mentioned before, we postpone a specialized discussion of these problems to {\it Part II} of our dissertation on lattice techniques~\cite{PartII}. The implementation of the corresponding algorithms in \CL will also be made publicly available in that moment.

\item  In a similar spirit, the axial couplings of a pseudo-scalar field with a gauge sector is also of great interest. An implicit method for the interaction of an axion-like field $\phi$ with a $U(1)$-gauge sector through a shift invariant coupling $\phi F^{\mu\nu}\tilde F_{\mu\nu}$, has been in fact explored in~\cite{Adshead:2015pva,Adshead:2016iae,Figueroa:2017qmv,Adshead:2018doq,Cuissa:2018oiw,Adshead:2019lbr,Adshead:2019igv}. In particular, an exactly lattice-shift-symmetric formulation was developed in~\cite{Figueroa:2017qmv}, and was later on generalized to an expanding background in~\cite{Cuissa:2018oiw}. We would like to revisit and generalize this kind of approaches in light of the algorithms presented here in Section~\ref{sec:LatticeApproach}, coming possibly with many potential outlooks. We plan to present a specialized discussion on these interactions in {\it Part II}~\cite{PartII}.

\item  The creation of tensor perturbation representing gravitational waves~\cite{Khlebnikov:1997di,Easther:2006gt,Easther:2006vd,GarciaBellido:2007af,Dufaux:2007pt,Dufaux:2008dn,Dufaux:2010cf,Zhou:2013tsa,Bethke:2013aba,Bethke:2013vca,Antusch:2016con,Antusch:2017flz,Antusch:2017vga,Liu:2018rrt,Figueroa:2017vfa,Fu:2017ero,Caprini:2018mtu,Lozanov:2019ylm,Adshead:2019lbr,Adshead:2019igv,Armendariz-Picon:2019csc}, as well as the dynamics of scalar metric perturbations~\cite{Bassett:1998wg,Bassett:1999mt,Bassett:1999ta,Finelli:2000ya,Chambers:2007se,Bond:2009xx,Imrith:2019njf,Musoke:2019ima,Giblin:2019nuv,Martin:2020fgl} (possibly leading to the formation of primordial black holes~\cite{Cotner:2019ykd,Martin:2019nuw,GarciaBellido:1996qt,Green:2000he,Hidalgo:2011fj,Torres-Lomas:2014bua,Suyama:2004mz,Suyama:2006sr,Cotner:2018vug}) are all topics of great interest in recent times. In the case of tensor perturbations, we plan to follow~\cite{Figueroa:2011ye} (based on the technique proposed in~\cite{GarciaBellido:2007af}), as this methodology allows for considering generic gravitational wave sources independently of the field content of the theory studied. Regarding a general solver for scalar metric perturbations, one possibility would be to follow~\cite{Huang:2011gf}.


\item  The inclusion of fermions in the simulations. Of course, the notion of  `classical fermions' does not exist due to Pauli-blocking, and hence a straightforward discretization and evolution of the Dirac equation would not be useful. However, as first realised in \cite{Aarts:1998td}, one can still study real-time fermions' dynamics in a semi-classical formulation of the out-of-equilibrium Schwinger-Keldysh formulation, see also references \cite{Kasper:2014uaa, Buividovich:2015jfa, Mace:2019cqo}. Combining the lattice implementation proposed in~\cite{Aarts:1998td}, with the  `low cost' fermions introduced in~\cite{Borsanyi:2008eu}, Refs.~\cite{Saffin:2011kc,Saffin:2011kn,Mou:2013kca,Mou:2015aia} have succeeded in simulating out-of-equilibrium dynamics of classical scalar fields coupled to quantum fermions. These simulations are however very costly in terms of computer memory, and only very small lattices have been considered until now. 

\item  The addition of other initialization procedures. Depending on the problem, initializing fields in real space might be more convenient than imposing a certain mode spectrum in Fourier space, as we did in Section \ref{sec:InitCond}. As mentioned in the introduction, to simulate e.g.~the dynamics of field string networks or any other type of cosmic defects, one needs to create in first place the defect network in configuration space, see e.g.~\cite{Vincent:1997cx,Bevis:2006mj,Figueroa:2012kw,Hindmarsh:2014rka,Daverio:2015nva,Lizarraga:2016onn,Lopez-Eiguren:2017dmc,Hindmarsh:2018wkp,Eggemeier:2019khm,Hindmarsh:2019csc,Gorghetto:2018myk}, and then evolve the field configuration from then onward (typically after a diffusion phase to force the system to reach a scaling regime as fast as possible). Although different problems may require completely different initializers, it might be useful to consider making a library for specialized ones, such as cosmic strings, other topological defects, and other circumstances.

\item  The addition of `cooling' algorithms for the initialization of gauge fields.  Instead of imposing the Gauss constraints from the beginning as we have done in Section~\ref{sec:InitGauge}, one can impose completely unconstrained fluctuations to the gauge fields, and then remove the unwanted transverse degrees of freedom by a minimization procedure~\cite{Ambjorn:1990pu,Moore:1996qs}. For thermal initial conditions, one can also thermalize the system while exactly preserving the Gauss law through some Langevin dynamics \cite{Krasnitz:1995xi}. Studying such algorithms will allow us to consider different initial conditions and thus study another variety of scenarios. 
\end{itemize}

\section*{Acknowledgments}

The authors would like to thank Andrei Frolov for useful discussions, and in particular for pointing us to Gauss-Legendre methods. We also thank Joanes Lizarraga for his careful check of some of the algorithms. D.G.F.~and F.T.~acknowledge hospitality from KITP-UCSB, where part of this work was carried out under support by the National Science Foundation under Grant No. NSF PHY-1748958. F.T.~thanks M.~Shaposhnikov for his kind invitation to EPFL during the development of this work. A.F. wants to thank S.~Antusch for his kind invitation to Basel. DGF (ORCID 0000-0002-4005-8915) is supported by a Ram\'on y Cajal contract by Spanish Ministry MINECO, with Ref. RYC-2017-23493, and by the grant "SOM: Sabor y Origen de la Materia”, from Spanish Ministry of Science and Innovation, under no. FPA2017-85985-P. A.F.~acknowledges support from the Swiss National Science Foundation. F.T.~also acknowledges support from the Swiss National Science Foundation (project number 200020/175502).

\appendix
\section*{Appendix: Coefficient tables for higher order integrators}

We provide in Table \ref{tab:VVnCoeffs} the weights of the time steps required in the construction of higher orders of the velocity Verlet algorithms, see Section \ref{subsec:YoshidaScalars}. We also show in Table \ref{tab:ButcherTables} the \textit{Butcher tableaux} used in the implicit Runge-Kutta methods, see Section \ref{subsec:GaussLegendreScalars}.

\begin{table}[H]
    \centering
\begin{tabular}{|c|c|c|c|}
\hline
    Name & Order & $w_i = \frac{\delta t_i}{\delta t}$ & $q$  \\
     \hline
    $VV4$ & $O(\delta t^4)$ & \begin{tabular}{c}
          $w_1 = w_3 = 1.351207191959657771818$  \\
          $w_2 = -1.702414403875838200264$
     \end{tabular} & 3 \\
     \hline
     $VV6$ & $O(\delta t^6)$ &  \begin{tabular}{c}
          $w_1 = w_7 = 0.78451361047755726382$  \\
          $w_2 = w_6 =  0.23557321335935813368$ \\
          $w_3 = w_5 = -1.1776799841788710069$ \\
          $w_4 = 1.3151863206839112189$
     \end{tabular} & 7 \\
     \hline
     $VV8$ & $O(\delta t^8)$ & \begin{tabular}{c}
          $w_1 = w_{15} = 0.74167036435061295345$  \\
          $w_2 = w_{14} = -0.40910082580003159400$  \\
          $w_3 = w_{13} = 0.19075471029623837995$  \\
          $w_4 = w_{12} = -0.57386247111608226666$  \\
          $w_5 = w_{11} = 0.29906418130365592384$  \\
          $w_6 = w_{10} = 0.33462491824529818378$  \\
          $w_7 = w_9 = 0.31529309239676659663$  \\
          $w_8 = -0.79688793935291635402$
     \end{tabular} & 15 \\
     \hline
      $VV10$ &  $O(\delta t^{10})$ & \begin{tabular}{c}
          $w_1 = w_{31} = -0.48159895600253002870$  \\
          $w_2 = w_{30} = 0.0036303931544595926879$  \\
          $w_3 = w_{29} = 0.50180317558723140279$  \\
          $w_4 = w_{28} = 0.28298402624506254868$  \\
          $w_5 = w_{27} = 0.80702967895372223806$  \\
          $w_6 = w_{26} = -0.026090580538592205447$  \\
          $w_7 = w_{25} = -0.87286590146318071547$  \\
          $w_8 = w_{24} = -0.52373568062510581643$  \\
          $w_9 = w_{23} = 0.44521844299952789252$  \\
          $w_{10} = w_{22} = 0.18612289547097907887$  \\
          $w_{11} = w_{21} = 0.23137327866438360633$  \\
          $w_{12} = w_{20} = -0.52191036590418628905$  \\
          $w_{13} = w_{19} = 0.74866113714499296793$  \\
          $w_{14} = w_{18} = 0.066736511890604057532$  \\
          $w_{15} = w_{17} = -0.80360324375670830316$  \\
          $w_{16} = 0.91249037635867994571$  \\
     \end{tabular} & 31 \\
     \hline
\end{tabular}
    \caption{Weights of the time steps required to construct higher-order velocity Verlet algorithms. A given algorithm requires $q$ iterations. The coefficients are symmetric, in each case, with respect to the intermediate $\omega_i$ parameter. Note that we reported here only the algorithms of a given order with the minimal number of steps. For others, see Ref.~\cite{Kahan:1997:CCR}. }
    \label{tab:VVnCoeffs}
\end{table}

\begin{table}[H]
    \centering
    \begin{tabular}{rcc}
    s = 2:  & \, & $\tableauII$ \vspace*{0.5cm}\\
    s = 3:  & \, & $\tableauIII$ \vspace*{0.5cm}\\
    s = 4:  & \, & $\tableauIV$ \vspace*{0.25cm}\\
    \, & \, & $\coeffTableauIV$
    \vspace*{0.5cm}\\
    s = 5:  & \, & $\tableauV$
    \vspace*{0.25cm}\\
    \, & \, & $\coeffTableauV$
    \end{tabular}
    \caption{{\it Butcher tableaux} for the implicit RK methods with $s$ sub-intervals and accuracy $\mathcal{O}(\delta t^{2s})$.}
    \label{tab:ButcherTables}
\end{table}

 \newpage

\begin{multicols}{2}
\footnotesize
  \bibliography{biblio,extra}
\bibliographystyle{h-physrev4}
\end{multicols}

\end{document}